\documentclass[a4paper,11pt,openbib]{book}

\usepackage{oldgerm}

\usepackage{graphicx}
\usepackage{amssymb}
\usepackage{ifthen}

\usepackage{natbib}
\bibpunct[,]{(}{)}{;}{a}{}{,}

\newsavebox{\tempbox}
\newlength{\captionwidth}
\makeatletter
\renewcommand{\@makecaption}[2]{%
  \vspace{10pt}\sbox{\tempbox}{\small{\sc #1:\ } \sffamily #2}%
  \ifthenelse{\lengthtest{\wd\tempbox > \linewidth}}%
   {\sbox{\tempbox}{\small\sc #1:\ }%
    \addtolength{\captionwidth}{-\wd\tempbox}%
    \mbox{\small\sc #1:\ }\parbox[t]{\captionwidth}{\small \sffamily #2}}%
   {\centering \small{\sc #1:\ } \sffamily #2}%
}
\makeatother

\newcommand{\boldvec}[1]{ {\bf #1} } 

\def\G{{\sc Gadget}}
\def\g{\G\ }
\def\YK3{YK$^3$}

\def\Madd{M_{\rm add}}
\def\LCDM{$\Lambda$CDM}
\def\lcdm{\LCDM\ }
\def\Om{\Omega_{\rm m}}
\def\OL{\Omega_\Lambda}
\def\Ok{\Omega_{\rm k}}
\def\Ob{\Omega_{\rm b}}

\def\Msun{M$_\odot$}
\def\msun{\Msun\ }
\def\oldMsun{M_{\odot}}
\def\Lx{L_{\rm X}}

\def\Tx{T_{\rm X}}
\def\TX{$\Tx$}
\def\tx{\TX\ }
\def\Am{\alpha_M}
\def\AM{$\Am$}
\def\am{\AM\ }
\def\BM{$\beta$-model}
\def\bm {\BM\ }

\def\Rv{r_{200}}
\def\RV{$\Rv$}
\def\rv{\RV\ }
\def\Mv{M_{200}}
\def\MV{$\Mv$}
\def\mv{\MV\ }
\def\Tv{T_{200}}
\def\TV{$\Tv$}

\def\Rs{r_{\rm s}}

\def\rrs{\frac{r}{\Rs}}
\def\be{\begin{equation}\displaystyle}
\def\ee{\end{equation}}
\def\bea{\begin{eqnarray}\displaystyle}
\def\eea{\end{eqnarray}}
\def\dd{{\rm d}}
\begin{document}

%-----------------------------------------------------------------------------
   \pagenumbering{arabic}
   \pagestyle{empty}
   \includegraphics[width=\textwidth]{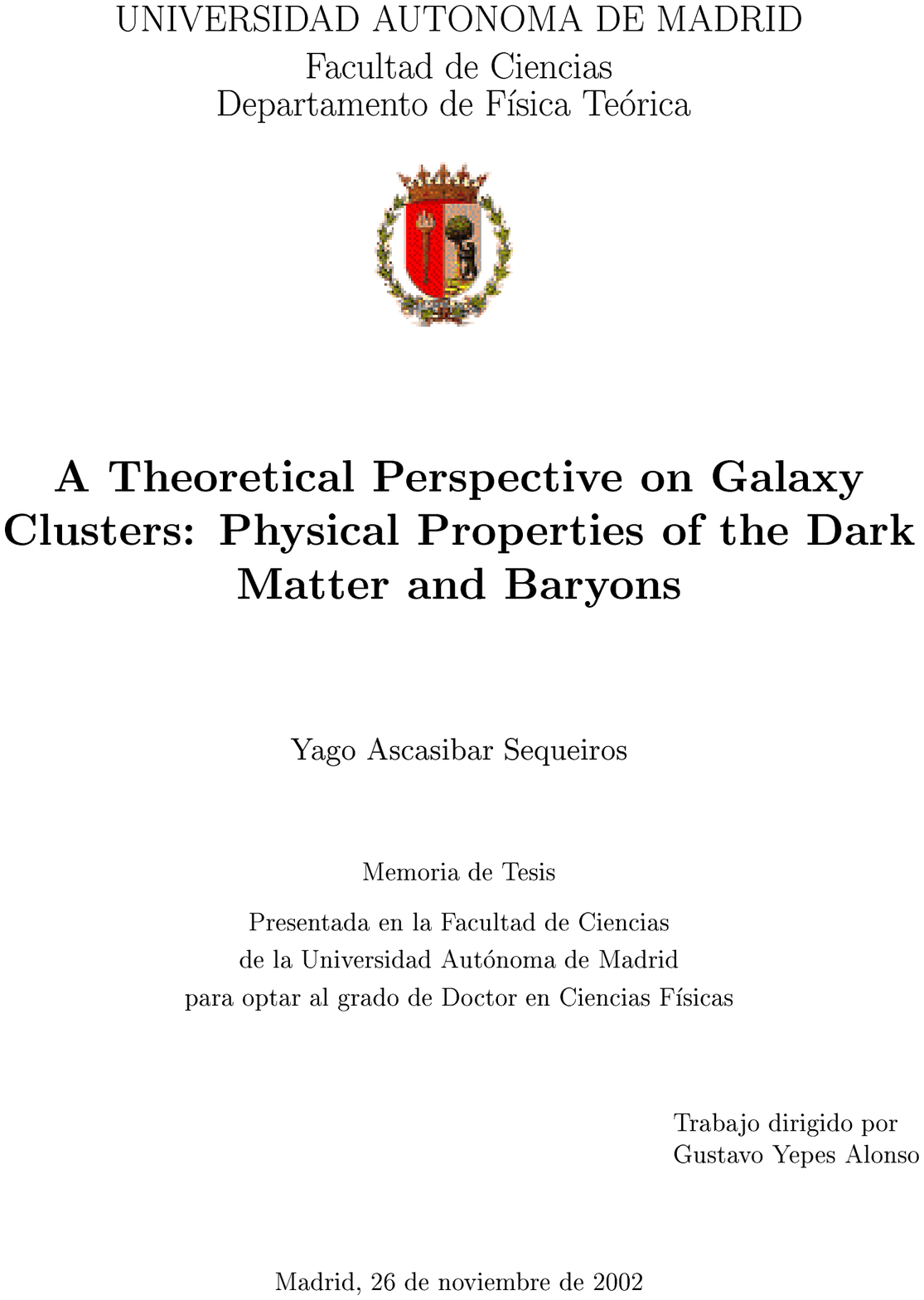}
%-----------------------------------------------------------------------------
\newpage\ \newpage\ \vspace*{4cm}

\hfill A mis padres

\vfill
... bueno, <y a mi hermano tambi\'en!

\newpage\ \newpage
%-----------------------------------------------------------------------------
   \pagenumbering{roman}
   \pagestyle{plain}

   \tableofcontents

   \listoffigures
   \addcontentsline{toc}{chapter}{\ ~~ List of Figures}

   \listoftables
   \addcontentsline{toc}{chapter}{\ ~~ List of Tables}

%-----------------------------------------------------------------------------
   \chapter*{Agradecimientos}
   \addcontentsline{toc}{chapter}{\ ~~ Agradecimientos}
%-----------------------------------------------------------------------------

\begin{quote}{\em
What would you think if I sang out of tune?\\
Would you stand up and walk out on me?\\
Lend me your ears and I'll sing you a song\\
and I'll try not to sing out of key}

-- The Beatles : {\em With a little help of my friends} (1967) --\\
\end{quote}
%-----------------------------------------------------------------------------

{\gothfamily {\Huge Y}{\large a}} veremos a quien pongo... pero de momento, no pueden faltar

\begin{itemize}
  \item Gustavo
  \item El Equipo
  \item Rafa y Pilar
  \item Y ar Natxo
\end{itemize}

%-----------------------------------------------------------------------------
   \chapter*{Resumen}
   \addcontentsline{toc}{chapter}{\ ~~ Resumen}
%-----------------------------------------------------------------------------
\begin{quote}{\em
>Qu\'e son los agujeros negros?\\
>Se expande el universo? >Es c\'oncavo o convexo?\\
>Qui\'enes somos? >De d\'onde venimos? >A d\'onde vamos?\\
>Estamos solos en la galaxia o acompa\~nados?}

-- Siniestro Total : {\em >Qui\'enes somos? >De d\'onde venimos? >A d\'onde vamos?} (1984) --\\
\end{quote}
%-----------------------------------------------------------------------------

{\gothfamily {\Huge L}{\large os}} c\'umulos de galaxias ocupan una posici\'on relevante en la jerarqu\'\i a
de la estructura cosmol\'ogica por muchas razones diferentes. \'Estos,
adem\'as de ser una de las estructuras limitadas m\'as grandes en el universo, 
contienen cientos de galaxias y gas caliente que emite rayos X, por
lo que pueden ser detectando por su alto corrimiento hacia el rojo.
Por lo tanto, estos objetos nos permiten estudiar la estructura
a gran escala del universo, comprobar las teor\'\i as de la 
formaci\'on de estructura as\'\i\  como extraer informaci\'on 
cosmol\'ogica dif\'\i cil de evaluar por otros m\'etodos (ver v.g.
Borgani y Guzzo, 2001; Rosati et al. 2002).

En el marco del modelo de jerarqu\'\i as para la formaci\'on de estructuras
a escala c\'osmica, se piensa que los c\'umulos de galaxias se formar por
acumulaci\'on de unidades m\'as pequen~as (galaxias, grupos, etc.).
Despu\'es de una \'epoca de agregaci\'on (que depende del modelo
cosmol\'ogico), le sigue un proceso de relajaci\'on violenta 
que tiende a virializar los c\'umulos produciendo sistemas
regulares y quasi-esf\'ericos, con suaves perfiles de densidad
tipo King. Las recientes observaciones a bajo corrimiento al rojo
basadas en las orientaciones relativas de las subestructuras dentro
de los c\'umulos, y en la relaci\'on entre el estado din\'amico
y el entorno a gran escala (Plionis y Basailakos, 2002) apoyan
este escenario jer\'arquico.

En la pasada d\'ecada, debido al incremento de la resoluci\'on espacial
en las im\'agenes de rayos X (ROSAT/PSPC \& HRI) y a la disponibilidad
de las c\'amaras de gran campo, muchos de los c\'umulos que anteriormente
se pensaban que eran regulares presentan en realidad alguna 
subestructura. Esto ha sido evidente gracias a la aparici\'on de
los sat\'elites Chandra (Weissjopf et al. 2000) y XMM (Jansen et al. 2001),
un hecho que podr\'\i a tener consecuencias importantes para las teor\'\i as
de formaci\'on de estructuras. Las propiedades f\'\i sicas de los c\'umulos
de galaxias, tales como la fracci\'on de c\'umulos j\'ovenes,
las funciones de luminosidad y temperatura, la estructura radial
de la materia oscura y bari\'onica, as\'\i\  como las relaciones complejas
entre todas estas magnitudes, constituyen un desaf\'\i o para
nuestro conocimiento actual sobre c\'omo esos objetos crecieron
desde las fluctuaciones en la densidad primordial

Durante los \'ultimos 20 a\~nos, el paradigma de la materia oscura fr\'\i a (CDM) ideada por
 Peebles (1982), Blumenthal et al. (1984) and Davis et al. (1985) se ha convertido en el
 modelo est\'andar que explica la formaci\'on y evoluci\'on de la estructura c\'osmica. \'Este 
supone que el contenido de materia de universo est\'a dominado por una part\'\i cula masiva
d\'ebilmente interactuante, que todavi\'a no ha sido identificada, y cuyos efectos gravitacionales 
son los responsables de la din\'amica de las galaxias y los c\'umulos.

Un concepto clave en esta cosmogon\'\i a es la formaci\'on de los halos de la materia oscura. Estos 
sistemas casi en equilibrio de part\'\i culas de materia oscura se forman gracias a un colapso
gravitacional no lineal que implica periodos de agregaci\'on estacionaria de masa, as\'\i\  como
episodios de fusi\'on violenta, en los cuales dos halos de masas comparables colisionan para 
dar lugar a un halo m\'as grande.

Se piensa que las galaxias y otros objetos luminosos nacen por el enfriamiento y condensaci\'on
de bariones dentro de pozos de potencial, creados por lo halos de materia oscura (White and
Rees 1978).En los cúmulos de galaxias, la mayor parte de los bariones se encuentran en un plasma 
caliente difuso e ionizado que constituye el medio intracumular (ICM).

La evoluci\'on din\'amica de las galaxias y del gas ICM plantea un problema dif\'\i cil para
los modelos de formaci\'on y evoluci\'on de c\'umulos (v.g. Kaiser 1986). Aunque la f\'\i sica
asociada a los procesos hidrodin\'amicos (c.f. Sarazin 1988) es m\'as complicada que un
colapso gravitacional, los modelos simples pueden ser aplicados para relacionar
el estado de los bariones con las condiciones particulares de la distribuci\'on subyacente
de la materia oscura.. Incluso, se han encontrado algunas correlaciones en las propiedades de los
rayos X en el medio intracumular, que sugiere que la estructura de densidad y temperatura
de los c\'umulos puede ser predicha por modelos te\'oricos autoconsistentes.

El presente trabajo intenta desvelar algunos de los mecanismos que dan lugar a las propiedades
observadas en los c\'umulos de galaxias. Nos hemos acercado al problema desde diferentes puntos de
vista haciendo uso de aproximaciones anal\'\i ticas y simulaciones num\'ericas, con el fin de
comprender la compleja interacci\'on entre las fuerzas gravitacionales a gran escala, que gu\'\i a 
la formaci\'on de estructura, y la f\'\i sica bari\'onica, que domina la hidrodin\'amica sobre
escalas locales.

En el cap\'\i tulo uno describimos brevemente el modelo cosmol\'ogico asumido
en nuestro estudio. La secci\'on 1.1.1. ofrece una somera descripci\'on de los
conceptos matem\'aticos b\'asicos necesarios para definir el escenario cosmol\'ogico.
La secci\'on 1.1.2 se centra en los valores concretos de los par\'ametros que determinan
la geometr\'\i a del universo, su composici\'on y su evoluci\'on temporal. La principal
evidencia observacional que apoya el modelo \lcdm tambi\'en ser\'a descrita. 

Hemos hecho una descripci\'on detallada de los experimentos num\'ericos en el
cap\'\i tulo 2. Nuestras muestras de los c\'umulos simulados son el resultado de tres
c\'odigos independientes basados en muy diferentes escenarios. La implementaci\'on
de los procesos f\'\i sicos en estos algoritmos es brevemente resumida, as\'\i  como
el procedimiento de an\'alisis que ha sido seguido para obtener los resultados
que se muestran. Finalmente damos una descripci\'on de las muestras y explicamos
los problemas que aparecen en cada uno de ellos.

El cap\'\i tulo 3 est\'a dedicado a la distribuci\'on de masas que esperamos encontrar
en los c\'umulos de galaxias. Especial incapi\'e hacemos en la estructura radial
de los halos de materia oscura, as\'\i\  como en su origen f\'\i sico. Para ser m\'as
preciso: nos centraremos en los perfiles de densidad de nuestras simulaciones de los
halos y en su dependencia con el estado din\'amico. La densidad en el espacio de fases
ser\'a tambi\'en investigada, comparando los resultados num\'ericos con trabajos
anteriores. El origen f\'\i sico de la estructura de la materia oscura en los c\'umulos
ser\'a tratado en el marco del formalismo de colapso esf\'erico (Gunn and Gott, 1972;
Gunn 1977). Veremos los perfiles predichos por el modelo del colapso esf\'erico
que proporciona una descripci\'on fiel de los experimentos num\'ericos realizados.
Intentaremos entender las razones por las cuales se producen tal acuerdo, a pesar
del hecho de que el crecimiento de los halos de materia oscura siga una serie de
fusiones no lineales. 

Las propiedades f\'\i sicas del ICM ser\'an investigadas en el cap\'\i tulo 4. En primer
lugar, comprobaremos la exactitud de las distintas implementaciones de las hidrodin\'amica del gas. Luego
testearemos la validez de que los c\'umulos de galaxias est\'an en un equilibrio
hidrodin\'amicos soportado t\'ermicamente, as\'\i\  como en una ecuaci\'on de estado apropiada
(politr\'opica o isot\'ermica) que permite describir mejor la componente de
gas difuso. Despu\'es, intentaremos predecir la densidad radial y los perfiles de
temperatura en nuestros simulaciones de c\'umulos, relacionando \'estas con las
formas fenomenol\'ogicas simples (v.g. Navarro et al. 1997; Moore et al. 1999)
propuestas para describir los halos de materia oscura. La relaci\'on de escala
entre las propiedades globales de los c\'umulos, observados en diferentes
densidades, son detalladas en una secci\'on diferente. Intentaremos entender
el origen de la relaci\'on masa-temperatura, as\'\i\  como analizaremos la discrepancia
entre la relaci\'on de autosimilaridad en la luminosidad-temperatura y la que se
observa (v.g. David et al. 1993; Ponman et al. 1996).

La formaci\'on estelar y los efectos de realimentaci\'on van a ser estudiados en el
cap\'\i tulo 5. Simulamos la historia de la formaci\'on estelar cosmol\'ogica 
para una serie de escenarios cosmol\'ogicos y los comparamos con las observaciones
a diferentes frecuencias. Evaluamos los efectos de esta realimentaci\'on en las explosiones
de supernovas y la foto-ionizaci\'on de las estrellas j\'ovenes, as\'\i\  como
la variancia c\'osmica inducida cuando un pequen\~no volumen es considerado.
Finalmente, tratamos la cuesti\'on de c\'omo el entorno altera la 
actividad de la formaci\'on estelar en las galaxias individuales.
La historia de la formaci\'on estelar de los c\'umulos de galaxias es comparada
con las de campos de galaxias, e incluimos una breve discusi\'on de
los mecanismos responsables para de la baja formaci\'on de estrellas 
halladas en esos ambientes tan densos. 

Las principales conclusiones que pueden ser extra\'\i das de nuestro resultados
son expuestas en la secci\'on 6. El ap\'endice A contiene un atlas de c\'umulos.
\'Este ense\~na las caracter\'\i sticas individuales de los 15 c\'umulos que 
reunen la principal caracter\'\i stica analizada en este trabajo. Tambi\'en
dibujamos algunas im\'agenes similares a las observadas, en las que la
diferencia entre la cantidad de estructura en gas en materia oscura
se muestra con claridad, as\'\i\  como el conjunto de figuras en el que las
predicciones discutidas en el texto principal son aplicadas en una base
de c\'umulo a c\'umulo.

%-----------------------------------------------------------------------------

%-----------------------------------------------------------------------------
   \chapter{Introduction}
   \pagenumbering{arabic}
   \pagestyle{headings}
%-----------------------------------------------------------------------------

\begin{quote}{\em
Exit light\\
Enter night\\
Take my hand\\
We're off to never-never land!}

-- Metallica : {\em Enter Sandman} (1991) --\\
\end{quote}
%-----------------------------------------------------------------------------

{\gothfamily {\Huge G}{\large alaxy}} clusters occupy a special position in the hierarchy of cosmic structure in many respects. Being the largest bound structures in the universe, they contain hundreds of galaxies and hot X-ray emitting gas and thus can be detected at large redshifts. Therefore, they appear to be ideal tools for studying large-scale structure, testing theories of structure formation and extracting invaluable cosmological information \citep[see e.g.][]{BorganiGuzzo01,Rosati02}.

In the framework of the hierarchical model for the formation of cosmic structures, galaxy clusters are supposed to form by accretion of smaller units (galaxies, groups etc). After the epoch of aggregation (which depends on the cosmological model), violent relaxation processes will tend to virialise the clusters producing `regular', quasi-spherical systems, with smooth King-like density profiles. Recent $low-z$ observations based on the relative orientation of substructures within clusters \citep{West95} and on the relation between their dynamical state and the large-scale environment \citep{PlionisBasilakos02} do support the hierarchical scenario.

In the last decade, due to the increased spatial resolution in X-ray imaging (ROSAT/PSPC \& HRI) and to the availability of wide-field cameras, many of the previously thought 'regular' clusters have shown to be clumpy to some level. This is even more so in the {\em Chandra} \citep{Weisskopf00} and {\em XMM} \citep{Jansen01} era, a fact that could have important consequences for structure formation theories. The physical properties of galaxy clusters, such as the fraction of dynamically young clusters, the luminosity and temperature functions, the radial structure of both dark and baryonic components, as well as the complex relationships between all these quantities, constitute a challenging test for our current understanding of how these objects grow from primordial density fluctuations.

Over the last 20 years, the Cold Dark Matter (CDM) paradigm set out by \citet{Peebles82}, \citet{Blumenthal84} and \citet{Davis85} has become the standard model to explain the formation and evolution of cosmic structures. It assumes that the matter content of the universe is dominated by an as yet unidentified, weakly interacting massive particle, whose gravitational effects are responsible for the dynamics of both galaxies and clusters.

A key concept in this cosmogony is the build-up of dark matter haloes. These quasi-equilibrium systems of dark matter particles are formed through non-linear gravitational collapse, involving both periods of steady accretion of mass, as well as violent merging episodes, in which two haloes of comparable mass collide to give birth to a larger halo.

Galaxies and other luminous objects are assumed to form by cooling and condensation of baryons within the gravitational potential wells created by the dark matter haloes \citep{WhiteRees78}. In galaxy clusters, most baryons can be found in a hot, diffuse, ionised plasma that constitutes the intracluster medium (ICM).

The dynamical evolution of member galaxies and the ICM gas poses a difficult problem for models of cluster formation and evolution \citep[e.g.][]{Kaiser86}. Nevertheless, although the physics associated to gasdynamical processes \citep[c.f.][]{Sarazin88} is more complicated than pure gravitational collapse, simple models can be applied in order to relate the state of the baryons to the particular conditions of the underlying dark matter distribution. Indeed, several correlations have been observed in the X-ray properties of the intracluster medium, hinting that the density and temperature structure of clusters may be predicted by self-consistent theoretical models.

The present work attempts to unveil some of the mechanisms that give rise to the observed properties of galaxy clusters. We have approached the problem from different points of view, resorting to analytical approximations and direct numerical simulations in order to gain some understanding of the complex interplay between the large scale gravitational forces driving structure formation and the baryonic physics that dominates gas dynamics on local scales.

In this chapter, we briefly describe the cosmological model that will be assumed in our study. Section~\ref{secIntroGR} offers an overview of the basic mathematical concepts required to define a cosmological scenario. Section~\ref{secIntroObs} focuses on the actual values of the parameters that determine the geometry of the universe, its composition, and its temporal evolution. The main observational evidence supporting the \lcdm model will be succinctly described.

A thorough discussion of the numerical experiments that have been performed is given in Chapter~\ref{chapExper}. Our samples of simulated clusters are the outcome of three independent codes, based on very different integration schemes. The implementation of physical processes within these algorithms is briefly summarised, as well as the analysis procedure that has been followed in order to obtain the results reported here. Finally, a description of the samples themselves is given, explaining the problems that are to be addressed within each one of them.

Chapter~\ref{chapDM} is devoted to the mass distribution that we expect to find in clusters of galaxies. Special emphasis is put in the radial structure of the dark matter haloes and its physical origin. To be more precise, we will study the density profiles of our simulated haloes and the dependence on their dynamical state. The phase-space density will also be investigated, comparing the numerical results with previous work. The physical origin of the structure of dark matter in clusters will be addressed within the framework of spherical collapse formalism \citep{GG72,Gunn77}. We will show that the profiles predicted by the spherical infall model provide an accurate description of those found in the numerical experiments. We will try to understand the reasons for such an agreement, albeit the fact that the growth of numerical dark matter haloes proceeds through a non-linear series of mergers.

The physical properties of the ICM will be investigated in Chapter~\ref{chapGas}. First of all, the reliability of current numerical implementations of gasdynamics will be assessed. Then, we will test the validity of the assumption that clusters of galaxies are in thermally-supported hydrostatic equilibrium, as well as the appropriate equation of state (polytropic or isothermal) that better describes the diffuse gas component. Then, we try to predict the radial density and temperature profiles of our simulated clusters, relating them to the simple phenomenological forms \citep[e.g.][]{NFW97,Moore99} proposed to describe the dark matter haloes. The scaling relation between the global properties of the clusters, observed at different overdensities, are dealt with in a separate section. We try to understand the origin of the mass-temperature relation, as well as of the reported discrepancy between the self-similar luminosity-temperature relation and the observed one \citep[e.g.][]{David93,Ponman96}.

Star formation and feedback are studied in Chapter~\ref{chapStars}. The cosmic star formation history is computed for a variety of cosmological scenarios and compared with observational estimates based on different wavelengths. The effects of feedback from supernova explosions and photoionisation by young stars are evaluated, as well as the cosmic variance induced when a small volume is considered. Finally, we address the question of how does the environment alter the star formation activity of individual galaxies. The star formation history of clusters of galaxies is compared to that of field galaxies, and a brief discussion on the mechanisms responsible for the lower star formation found in dense environments is also included.

The main conclusions that can be drawn from our results are highlighted in Section~\ref{chapConclus}. We also quote some of the aspects that in our opinion deserve a much deeper investigation than it is possible within the scope of the present thesis.

Appendix~\ref{apIndiv} contains a cluster atlas. It shows the individual characteristics of the 15 clusters that comprise the main sample analysed in this work. Mock observational images are plotted, in which the difference between the amount of structure in gas and dark matter is clearly seen, as well as a set of figures in which the prescriptions discussed in the main text are applied in a cluster-to-cluster basis.

%-----------------------------------------------------------------------------

\section{The cosmological model}
\label{secIntroCosmo}

This work (as any other), relies on a series of axioms that it does not intend to verify. In our case, maybe the most important of them is the cosmological model that is assumed to describe the universe beyond the region of influence of our clusters of galaxies.

The consequences of the chosen cosmological scenario on the formation and evolution of galaxy clusters are twofold: On one hand, it sets the power spectrum of primordial fluctuations that give rise to cosmological structure. On the other, the expansion rate of the universe is largely determined by the contribution of its different components to the total energy density. The expansion factor, in turn, controls the density of cosmological objects, their accretion rate, or the probability that they undergo a merging event.

%___________________________________________________

\subsection{Theory}
\label{secIntroGR}

The basic tenet that governs our current view of the universe is the cosmological principle, which is a generalisation of the Copernican idea that the Earth does not necessarily have to play a central role in the solar system. In general terms, the cosmological principle states that the universe is homogeneous and isotropic on large scales, which seems to be corroborated by observations of the Large Scale Structure (LSS) or the Cosmic Microwave Background (CMB). Furthermore, we will assume that the evolution of the universe on such large scales is dominated by gravity, and therefore it can be described by the General Theory of Relativity.

Although it is evident that the universe is anything but homogeneous on small scales, the usual approach consists in separating the problem into two components: Cosmic structures are considered to grow as a perturbation field superimposed to an otherwise homogeneous and isotropic space-time fabric. The evolution of the universe as a whole is reduced, in this simple approximation, to the dynamics of cosmological expansion, which constitutes one of the most fundamental applications of General Relativity.

%----------------------------------

\subsubsection*{Friedmann equations}

It can be shown \citep[e.g.][]{Weinberg72} that the assumptions of homogeneity and isotropy are equivalent to the requirement that the metric tensor of the universe takes the form of the Robertson-Walker metric
\be
 ds^2=dt^2-a(t)^2\left[\frac{dr^2}{1-kr^2}+r^2(d\theta^2+\sin\theta d\phi)\right]
\ee
where the dimensionless function $a(t)$ is the cosmic expansion factor.

According to General Relativity, the geometry of the space-time background is related to its matter-energy content by virtue of the Einstein field equations,
\be
  G_{\mu\nu}\equiv R_{\mu\nu}-\frac{1}{2}Rg_{\mu\nu}=8\pi G T_{\mu\nu}+\Lambda g_{\mu\nu}
  \label{ecEin}
\ee

The stress-energy tensor of the universe is taken to be, also from the cosmological principle, that of a perfect fluid (such that a comoving observer would observe an isotropic universe),
\be
  T_{\mu\nu}=pg_{\mu\nu}+(p+\rho)u_{\mu}u_{\nu}
  \label{temunu} 
\ee

Inserting the perfect fluid tensor and the Robertson-Walker metric into the Einstein field equations above, we can apply the stress-energy conservation law ($T^{\mu\nu}_{;\nu}=0$) to get the Friedmann equations, which govern the dynamics of the cosmic scale factor $a(t)$:
\be
  \left(\frac{\dot a}{a}\right)^2=\frac{8\pi G}{3}\rho +\frac{\Lambda}{3}- \frac{k}{a^2} ~~;~~
  \frac{\ddot a}{a}=-\frac{4\pi G}{3}(\rho+3p)+\frac{\Lambda}{3}
  \label{ecFri1}
\ee

The solutions of these equations are known as Lema\^\i tre-Friedmann-Robertson-Walker (LFRW) universes, and their characteristic properties vary greatly as a function of the several densities that can appear in (\ref{ecFri1}) and their corresponding equations of state.

%----------------------------------

\subsubsection*{Density parameters}

Before the 1990's, little could be said about the matter-energy composition of the universe. The only points in which a general agreement existed were the validity of the cosmological principle, the finiteness of the age of the universe, and its expansion from a gas cloud of extremely high density and temperature.

This 'hot Big-Bang' scenario had been induced and confirmed by several independent observational evidences, namely the measurement of redshift in nearby galaxies \citep{Hubble29}, the relative abundances of light elements \citep[explained through a primordial nucleosynthesis phase,][]{Gamow46}, and the detection of the CMB radiation by \citet{PenziasWilson65}. With very few ingredients, this simple model was able to explain satisfactorily enough the main features of the observable universe from its very first seconds of existence.

Nevertheless, LFRW universes have a set of free parameters that only in the last decade has it been possible to bound.  On one hand, the pressure and density terms that appear in (\ref{ecFri1}) corresponding to the different types of energy density that may exist in the universe, and, on the other hand, the actual value of the expansion rate at the present time, usually known as Hubble constant and expressed as
\be
  H(t)\equiv\frac{\dot a(t)}{a(t)} ~~;~~
  H_0\equiv H(t_0)\equiv 100\ h ~~{\rm (km\ s^{-1} Mpc^{-1})}
\ee

It is useful, in order to classify the different terms contributing to the stress-energy tensor, to work with the density parameters, defined as fractions of the critical density required for the universe to have a spatially-flat geometry ($k=0$):
\be
  \Omega_i(t)\equiv\frac{\rho_i(t)}{\rho_c(t)}\equiv\frac{8\pi G}{3H^2(t)}\rho_i(t)
\ee

In analogy with ordinary matter, we define an effective density due to the curvature of space-time and to the cosmological constant term (which can be interpreted as a uniform vacuum energy density $\rho_\Lambda=\OL\rho_c$).
\be
  \Ok(t)\equiv\frac{-k}{a^2H^2(t)} ~~;~~
  \OL(t)=\frac{\Lambda}{3H^2(t)}
\ee

With this definitions, the first Friedmann equation transforms into the following closure relation:
\be
  \sum_i\Omega_i(t)=1
  \label{cierre}
\label{ecCierre}
\ee

\begin{table}
\begin{center}
\begin{tabular}{llcc}
 \multicolumn{2}{c}{\sc Component} & {\sc Density par.} & {\sc State eq.} \\ \hline \\[-2mm]
  ~~Matter         & Relativistic     & $\Omega_\gamma+\Omega_{\rm HDM}$ & $p=1/3\rho$ \\
                   & Non-relativistic & $\Ob+\Omega_{\rm CDM}$           & $p=0$       \\
  ~~Space-time     & Vacuum           & $\OL$                            & $p=-\rho$   \\
                   & Curvature        & $\Ok$                            & $p=-1/3\rho$\\
\end{tabular}
\end{center}
\caption{Possible energy densities in the universe}
\label{tablaCosmo}
\end{table}

Table \ref{tablaCosmo} summarises a classification of the possible components that may contribute to the total energy density of the universe, according to the equation of state relating their density an pressure. These equations also define the relationship between the different densities and the cosmic scale factor $a(t)$. To obtain the time-dependence $\Omega_i(a)$ we just have to insert the corresponding equation of state in (\ref{ecFri1}):
\be
  p_i(t)=\omega_i\rho_i(t) ~~\Rightarrow~~ \rho_i(t)=\rho_i(t_0)\ a^{-3(1+\omega_i)}
\ee

As a result, the current values of the density parameters completely determine the solutions of the Friedmann equations. From now on, if there is no explicit time dependence of any density, we will take $\Omega_i$ to denote the present-time value (unless otherwise noted). The first Friedmann equation may then be written as
\be
  \left(\frac{H}{H_0}\right)^2=\Omega_{\rm r}a^{-4}+\Omega_{\rm nr}a^{-3}+\Omega_\Lambda+\Omega_ka^{-2}
\label{ecFried}
\ee
where it becomes evident that the properties and evolution of a homogeneous and isotropic universe depend only on the four density parameters (subject to the closure relation~\ref{ecCierre}) plus the Hubble constant.

%__________________________________
\begin{figure}
  \centering \includegraphics[width=8cm]{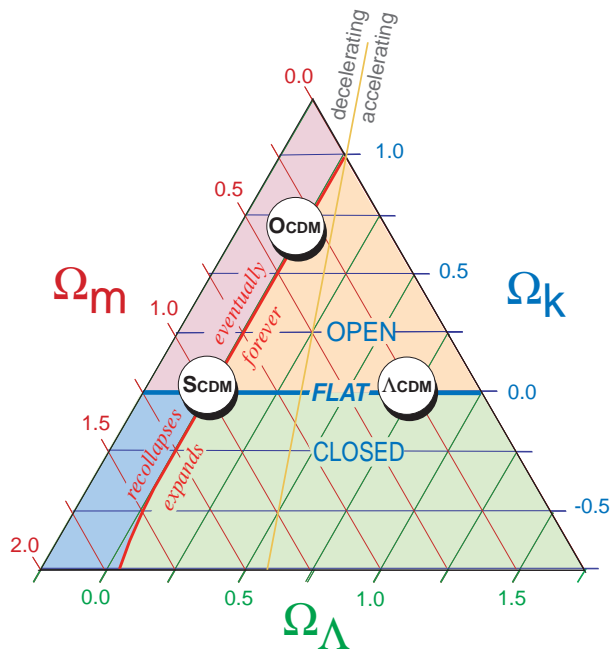}
\caption{Cosmic triangle}
\label{figCosmic3}
\end{figure}
%__________________________________

This is very aesthetically summarised in Figure~\ref{figCosmic3} \citep{Bahcall99}, where the parameter space is represented by the position of the model universe in the 'cosmic triangle'. Each point within the triangle satisfies the closure relation (\ref{ecCierre}). The horizontal line (marked 'flat') corresponds to $k=0$, separating an open universe from a closed one. The red line, nearly along the $\lambda=0$ line, separates a universe that will expand forever (approximately $\Lambda>0$) from other that will eventually re-collapse (approximately $\Lambda<0$). And the yellow, nearly vertical line separates a universe with an expansion rate that is currently decelerating from one that is accelerating.

The locations of three key models are highlighted in the figure: The Standard Cold Dark Matter (SCDM) model, dominated by non-relativistic non-baryonic matter, $(\Om,\OL)=(1,0)$; an open universe with no cosmological constant, $(\Om,\OL)=(0.3,0)$; and a flat universe with a $\Lambda$-term, defined by $(\Om,\OL)=(0.3,0.7)$.

%___________________________________________________

\subsection{Observations}
\label{secIntroObs}

From a theoretical point of view, there are no preferred values of the density parameters. Therefore, constraints on their actual values in the real universe must come from observational estimates. The last ten to fifteen years have witnessed a tremendous improvement in observational instrumentation. These novel technologies have opened a plethora of new possibilities to measure the contribution of the different types of energy densities to the total cosmological budget.

While in the 1960s Sandage's goal was to 'search for two numbers', it is now appreciated that even the basic cosmological model requires about a dozen or so parameters, which are linked in non-trivial ways. Combinations of several independent experiments must be used to break these degeneracies. In this section, we review some recent results aimed to find the values of the cosmic density parameters that best fit the whole set of available observations. As we shall see below, all observational evidence\footnote{At the moment of writing this thesis.} seem to favour the so-called \lcdm cosmological scenario.

%----------------------------------

\subsubsection*{Baryon density}

The standard theory of Big-Bang Nucleosynthesis (BBN) predicts the abundances of the light element nuclei H, D, $^3$He, $^4$He, and $^7$Li as a function of the cosmological baryon-to-photon ratio, $\eta=n_b/n_\gamma$ \citep[see e.g.][]{KolbTurner90}. A measurement of the ratio of any two primordial abundances gives $\eta$, and hence the baryon density, while a second ratio tests the theory.

However, it is extremely difficult to measure primordial abundances because in most places gas ejected from stars has enriched the medium. \citet{Adams76} suggested that it might be possible to measure the primordial D/H ratio in absorption-line systems toward QSOs. Several estimations \citep[e.g.][]{Tytler96,BurlesTytler98a,BurlesTytler98b,OMeara01} have been made thereafter. The baryon density reported by \citet{OMeara01} amounts to
\be
\Ob h^2=0.0205\pm0.0018
\ee

%----------------------------------

\subsubsection*{Hubble constant}

In standard Big Bang cosmology, the universe expands uniformly; and locally, according to the Hubble law, $v=H_0d$, where $v$ is the recession velocity of a galaxy at a distance $d$, and $H_0$ is the Hubble constant, the expansion rate at the current epoch. More than seven decades have now passed since \citet{Hubble29} initially published the correlation between the distances to galaxies and their recession velocities, thereby providing evidence for the expansion of the universe. But pinning down an accurate value for the Hubble constant has proved extremely challenging. There are many reasons for this difficulty, but primary among them is the basic difficulty of establishing accurate distances over cosmologically significant scales.

Measuring an accurate value of $H_0$ was one of the motivating reasons for building the NASA/ESA {\em Hubble Space Telescope} (HST). The overall goal of the $H_0$ Key Project was to measure the Hubble constant based on a Cepheid calibration of a number of independent secondary distance determination methods. To extend the distance scale beyond the range of the Cepheids, a number of methods that provide relative distances were chosen, including the Type Ia supernovae, the Tully-Fisher relation, the fundamental plane for elliptical galaxies, surface-brightness fluctuations, and Type II supernovae.

A recent paper by \citet{Mould00,Mould00err} combines the results of the HST with these secondary methods based on different standard candles. The final results of the $H_0$ Key Project \citep{Freedman01} yield the following value:
\be
H_0=72\pm8\ {\rm km\ s^{-1}\ Mpc^{-1}}
\ee

%----------------------------------

\subsubsection*{Decelration parameter}

Just as large distance measurements on Earth show us the curvature on Earth's surface, so do large distance measurements in cosmology show us the geometry of the universe. Since the geometry of space-time is given by the values of the cosmic density parameters, measurements of the luminosity distance at high redshifts can be very help to constrain those values.

In these sense, observations of Type Ia supernovae at $z\ge3$ by the {\em Supernova Cosmology Project} \citep{Perlmutter99} and the {\em High-Z Supernova Search Team} \citep{Riess98,Schmidt98} have provided an extremely useful benchmark to test the different cosmological scenarios. These measurements allowed for the first time an accurate determination of the deceleration parameter $q_0\equiv-a\ddot a/\dot a^2=\Om/2-\OL$. The most striking conclusion drawn from these experiments was that the expansion of the universe is nowadays accelerating ($q_0<0$), which hints the presence of a non-zero positive cosmological constant,
\be
\Om=0.28_{-0.08}^{+0.09}~~;~~\OL=0.72_{-0.08}^{+0.09}
\ee

%----------------------------------

\subsubsection*{Curvature}

One of the most promising techniques to determine the geometry of the universe is the study of the amplitude fluctuations in the cosmic microwave background. Early detection of a peak in the region of the so-called first acoustic peak \citep{Netterfield97}, as well as the availability of fast codes to compute theoretical amplitudes \citep{SeljakZaldarriaga96}, have provided a first constraint on the value of $\Ok$ \citep{Lineweaver97,Hancock98}.

The spectacular results of Boomerang and Maxima balloon experiments have firmly established that geometry of the universe is very close to flat \citep{deBernardis00,Hanany00,Balbi00,Lange01}. Recently, \citet{benoit} combined the anisotropies measured by the Archeops experiment with data from COBE \citep{Tegmark96}, Boomerang \citep{Netterfield02}, Dasi \citep{Halverson02}, Maxima \citep{Lee01}, VSA \citep{scott} and CBI \citep{pearson}. The combination of all these CMB experiments provides an estimation of the total energy density of the universe
\be
\Ok=1.04_{-0.17}^{+0.12}
\ee
as well as a measure of the baryon density completely independent (but consistent) with the nucleosynthesis estimate:
\be
\Ob h^2=0.022_{-0.004}^{+0.003}
\ee

Adding other constraints, such as those imposed by the observed large scale structure \citep[see e.g. the analysis based on the 2dF Galaxy Redshit Survey by][]{Efstathiou02} or the observed baryon fraction in galaxy clusters \citep[e.g.][]{SadatBlanchard01}, the values of the density parameters are still compatible with those listed above. The so-called 'concordance' model emerges as an apparently solid observational measurement, confirmed by several independent methods.

Given the amount of observational evidence that currently supports the \lcdm cosmology, we will base most of our study on the assumption that the space-time background is well described by a homogeneous and isotropic LFRW universe with $\Om=0.3$, $\OL=0.7$, $\Ob=0.04$ and $h=0.7$. In those cases where some other cosmology is used, it will be explicitly noted.

%-----------------------------------------------------------------------------

%-----------------------------------------------------------------------------
   \chapter{Numerical experiments}
   \label{chapExper}
%-----------------------------------------------------------------------------
   
% \begin{quote}{\em
% Lights that flash, wheels that go round\\
% Digital display\\
% Fresh silicon chips to enjoy\\
% And I need them}
%   
% -- Jethro Tull : {\em Batteries Not Included} (1980) --\\
% \end{quote}

\begin{quote}{\em
One more, one more, one more, one more...\\
Addict with an Apple Mac, megabyte maniac\\
Moron with a mouse mat, a junky\\
Distressed, you've guessed, obviously obsessesed\\
I, NEED MORE MEMORY!}

-- Toy Dolls : {\em One More Megabyte} (1997) --\\
\end{quote}
%-----------------------------------------------------------------------------

{\gothfamily {\Huge N}{\large umerical }}  simulations of three-dimensional self-gravitating fluids have become an indispensable tool in cosmology. They are now routinely used to study the non-linear gravitational clustering of dark matter, the formation of galaxy clusters, the interactions of isolated galaxies, and the evolution of the intergalactic gas. Without numerical techniques the immense progress made in these fields would have been nearly impossible, since analytic calculations are often restricted to idealised problems of high symmetry, or to approximate treatments of inherently non-linear problems.

The advances in numerical simulations have become possible both by the rapid growth of computer performance and by the implementation of ever more sophisticated numerical algorithms. The development of powerful simulation codes, with high level of parallelism, still remains a primary task in order to take full advantage of new computer technologies.

Early simulations \citep[e.g.][]{White76,Fall78,Aarseth79} largely employed the direct summation method for the gravitational N-body problem, which remains useful in collisional stellar dynamical systems, but it is inefficient for cosmological simulations with large $N$ due to the rapid increase of its computational cost with $N$ \citep[see e.g.][ for a review on numerical methods for cosmological simulations]{Yepes01}.

In order to overcome the problems inherent to the direct summation N-body methods, a large number of groups have developed new techniques for collisionless dynamics that compute the large-scale gravitational field  in a (regular or irregular) grid. In the simplest implementation of grid-based N-body methods, the \emph{Particle-Mesh}(PM) N-body technique \citep[see e.g.][]{HockneyEastwood81}, the Poisson equation is solved using Fast Fourier Transform (FFT) in a regular grid of the density field, that is constructed by interpolation from the particle positions to the grid nodes. Newtonian forces are computed by differentiating the potential on the grid and interpolating back to the particle positions. The time for force computation scales as ${\cal O}(N\log N)$ in these methods, which makes them able to treat a large amount of particles. In fact, this is the N-body  method that can handle the largest number of particles, and it has been used widely to investigate the non-linear clustering of different cosmological models since the early 1980's \citep[e.g.][]{KlypinShandarin83,White83}.

The lack of resolution below the grid-size is one of the major problems of the PM model. \citet{Hockney74} proposed an hybrid scheme to increase the numerical resolution of the PM method, consisting in the decomposition of the force acting on each particle into a long-range force, computed by the particle-mesh interaction, and a short-range force due to nearby particles.
%-% :
%-% \def\bF{{\bf F}}
%-% \def\bxi{{\bf x_i}}
%-% \begin{equation}
%-%   \label{gus:eqn:p3m1}
%-%    {\bf F (x_i)}= ({\bf F}_{PM}(\bxi)-{\bf F}_{PM}^{PP}(\bxi)) +
%-%      \bF^{\small neigh}_{PP}(\bxi)
%-% \end{equation}
%-%  
%-% \citet{Efstathiou85} showed that the contribution of the nearby particles to the PM force ${\bf F}_{PM}^{PP}$ is negligible, so
%-% \begin{equation}
%-%   \label{gus:eqn:p3m2}
%-%    {\bf F (x_i)}\sim  {\bf F}_{PM}(\bxi) +
%-%      \bF^{\small neigh}_{PP}(\bxi)
%-% \end{equation}
%-%  
%-% The short-range Particle-Particle force acting on each particle is computed by a direct summation method. To look for neighbours, particles are classified in a coarser grid, usually $\sim3$ times bigger than the one used for the PM algorithm. Those particles lying in the same cell are taken into account to calculate the direct PP force. When clustering develops, many particles will be in the same cells and the PP force computation will dominate the computing time. Therefore, the P$^3$M algorithm will behave as the direct summation technique in highly clustered situations. For this reason, this method cannot treat as many particles as the simpler PM.
Modern versions of these codes use finer grids on highly clustered regions to recursively apply the P$^3$M algorithm. Therefore, the number of neighbours per particle is considerably reduced, which makes the code much faster with respect to the original implementation. The publicly available {\em Adaptive} P$^3$M (AP$^3$M) code developed by \citet{hydra91} is a good example of this technique and has helped the development of other codes based on this integration scheme.

Another approach to increase resolution of PM methods is to use non-uniform grids and algorithms to adapt the computational mesh to the structures formed by gravitational clustering. The Poisson equation can be solved on a hierarchically refined mesh by means of finite-difference relaxation methods, an approach taken in the Adaptive Refinement Tree (ART) code by \citet{ART97}, the MLAMP code \citep{mlapm01}, and more recently by \citet{ramses02} in the RAMSES AMR code.

An alternative to these schemes are the so-called tree algorithms, pioneered by \citet{App85}. Tree algorithms arrange particles in a hierarchy of groups, and compute the gravitational field at a given point by summing over multipole expansions of these groups. In this way the computational cost of a complete force evaluation can be reduced to a ${\cal O}(N\log N)$ scaling. The grouping itself can be achieved in various ways, for example with Eulerian subdivisions of space \citep{BarnesHut86}, or with nearest-neighbour pairings \citep{Press86,JerniganPorter89}. A technique related to ordinary tree algorithms is the fast multipole-method \citep[e.g.][]{Greengard88}, where multipole expansions are carried out for the gravitational field in a region of space.

While mesh-based codes are generally much faster for nearly homogeneous particle distributions, tree codes can adapt flexibly to any clustering state without significant losses in speed. This Lagrangian nature is a great advantage if a large dynamic range in density needs to be covered. Here tree codes can outperform mesh based algorithms. In addition, tree codes are basically free from any geometrical restrictions, and they can be easily combined with integration schemes that advance particles on individual timesteps.

%-% Recently, PM and tree solvers have been combined into hybrid Tree-PM codes \citep{Xu95,Bag99,BOX00}. In this approach, the speed and accuracy of the PM method for the long-range part of the gravitational force are combined with a tree-computation of the short-range force. This may be seen as a replacement of the direct summation PP part in P$^3$M codes with a tree algorithm. 

%-% Yet another approach to the N-body problem is provided by special-purpose hardware like the GRAPE board \citep{MIE90,Ito91,Fuk91,MF93,Ebi93,Ok93,Fuk96,Ma97,K00}. It consists of custom chips that compute gravitational forces by the direct summation technique. By means of their enormous computational speed they can considerably extend the range where direct summation remains competitive with pure software solutions. A recent overview of the family of GRAPE-boards is given by \citet{HM99}. The newest generation of GRAPE technology, the GRAPE-6, will achieve a peak performance of up to 100 TFlops \citep{Mak2000}, allowing direct simulations of dense stellar systems with particle numbers approaching $10^6$. Using sophisticated algorithms, GRAPE\ may also be combined with P$^3$M \citep{Brieu95} or tree algorithms \citep{Fuk91,Mak91,At97} to maintain its high computational speed even for much larger particle numbers.

An independent way to speed up force computations for large number of particles, is to use efficient parallel algorithms to distribute the work load  among many processors. A big effort has been done recently in developing parallel N-body codes, using either direct summation \citep{Makino02}, PM \citep{cosmos00}, P$^3$M \citep{p3mpar98} or tree methods \citep{Dubinski96,LiaCarraro01,gadget01}. The advantage of parallel N-body codes is the portability. They can run from cheap Beowulf PC clusters to very expensive supercomputers.

In recent years, collisionless dynamics has also been coupled to gas dynamics, allowing a more direct link to observable quantities. Traditionally, hydrodynamical simulations have usually employed some kind of (Eulerian or Lagrangian) mesh to represent the dynamical quantities of the fluid. Modern grid-based computational fluid dynamical codes are based on Godunov methods: i.e. gasdynamical equations are integrated in each volume element of the computational mesh. The evolution of volume-averaged fluid quantities can be computed from fluxes through the cell boundaries, which are estimated from approximate Riemann solvers. 

A particular strength of these codes is their ability to accurately resolve shocks without artificial viscosity terms that might introduce numerical dissipation. A regular mesh also imposes restrictions on the geometry of the problem, as well as onto the dynamic range of spatial scales that can be simulated. New adaptive mesh refinement codes \citep{NormanBryan99,ARThydro02,ramses02} have been developed to provide a solution to this problem.

In cosmological applications, it is often sufficient to describe the gas by smoothed particle hydrodynamics (SPH), as invented by \citet{Lucy77} and \citet{GingoldMonaghan77}. The particle-based SPH is extremely flexible in its ability to adapt to any given geometry. Moreover, its Lagrangian nature allows a locally changing resolution that `automatically' follows the local gas  density. This convenient feature helps to save computing time by focusing the computational effort on those regions that have the largest gas concentrations. Furthermore, SPH ties naturally into the N-body approach for self-gravity, and can be easily implemented in three dimensions.

These advantages have led a number of authors to develop SPH codes for applications in cosmology in combination with different flavours of N-body codes, to the point that nowadays these hydrodynamic codes are by far the most numerous in cosmological simulations.
%-% Among them are {\small TREESPH} \citep[e.g.][]{He89,Ka96,Carraro98,gadget01}, {\small GRAPESPH} \citep{St96}, {\small HYDRA} \citep{Cou95,Pe97}, and codes by \citet{Ev88,Na93,Hu97,Da97,Ca98}.
%-% See \citep{Ka94} and
A comparison study can be found in \citet{SB99}, where the major differences between SPH and grid-based gasdynamical codes were found in the central part of the radial entropy and temperature profiles of the galaxy cluster used for the comparison. The AMR gasdynamical codes tend to produce an isentropic gas profile in the inner regions, while SPH codes predict an isothermal profile with a decreasing entropy towards the centre of cluster-size haloes. We will come back to this problem in Section~\ref{secEulerLagrange}, showing that results from SPH and grid-based codes can be reconciled when an appropriate treatment of entropy conservation in SPH codes is used \citep{gadgetEntro02,deva}.

The present chapter is devoted to the description of the numerical experiments reported in this thesis. First, the different codes that have been used are described in some detail, focusing on the specific merits of each one depending on the sort of problem that we wanted to tackle. The general procedure followed to analyse the results is discussed on a separate section, and finally a description of the actual numerical experiments is given.

%___________________________________________________

\section{Description of the codes}

In order to study the internal structure of dark matter haloes, a high-performance N-body algorithm is required to accurately resolve their very central parts. For this purpose we used two independent state-of-the-art N-body codes; the parallel Tree-SPH code \g and the pure N-body implementation of the AMR code ART. Results based on these two integration schemes have been compared in order to assess the reliability of our conclusions (see Section~\ref{secDMsims}).

Inclusion of gas dynamics in N-body simulations is even more demanding from the computational point of view, enforcing the use of efficient parallel codes for this purpose. The dynamical evolution of the gas component within dark matter haloes has been investigated by means of a series of simulations accomplished with \G. Because of the problems with entropy conservation in SPH codes, a new implementation of \g has been used in which the entropy (rather than energy) conservation equation is integrated \citep{gadgetEntro02}. As a test of the consistency of our results, we have compared in Section~\ref{secEulerLagrange} one of the galaxy clusters simulated by means of the SPH technique with an independent numerical experiment by \citet{cluster6} using the new hydrodynamical version of the ART code \citep{ARThydro02}.

Finally, to address the problem of star formation in different environments, it does not suffice to take into account only gravity and gasdynamics, but also all the complex physics related to star-gas interactions. In this regard, we have used a Particle-Mesh code for N-body coupled with an Eulerian hydrodynamical code based on the higher order Godunov {\em Piecewise Parabolic Method} (PPM). The code is supplemented with a model for cooling, star formation and feedback from supernovae. From now on, we will refer to this code as YK$^3$, which stands for the initials of the authors \citep{YK3}.

All the codes employed in this work have been parallelised to run efficiently either in distributed (\G) or Shared Memory Processor systems (ART and  YK$^3$). Numerical experiments have been performed in a variety of supercomputers and centres:

\begin{itemize}
\item SGI Origin 2000 at the CEPBA (Barcelona, Spain)
\item SGI Origin 2000 at the NCSA (Chicago, USA)
\item SGI Origin 3800 at the CIEMAT (Madrid, Spain)
\item Hitachi SVR at the AIP (Potsdam, Germany)
\item Hitachi SVR at the LRZ (Munich, Germany)
\end{itemize}

%___________________________________________________

\subsection{\YK3}
\label{secYK3}

This code implements the most complete treatment of the physical processes, at the expense of a much lower spatial resolution than that achieved by ART or \G.

The matter in the simulated Universe consists of four phases. 1) The
dark matter (labelled by a subscript ``dm'') in th form of weakly
interacting collisionless particles is the main contribution to the
mean density of the universe ($\Omega_{\rm dm}=1-\Omega_{\rm b})$. The
baryonic component is described as a medium consisting of three
interacting phases: 2) hot gas (labelled by subscript h, $T_{\rm h}>2
\times 10^4$ K), 3) gas in the form of cold dense clouds (subscript c,
internal temperature $T_{\rm c}~=~10^4$ K) resulting from cooling of
the hot gas, and 4) ``stars'' (subscript $*$), formed inside cold
clouds and treated as collisionless particles.  Thus, the total
density $~(\boldvec r)$ is the sum of four components:
\begin{equation}
  \rho = \rho_{\rm dm} +\rho_{\rm h} +\rho_{\rm c} +\rho_*.
\end{equation}

This picture is consistent with work on models
of galaxy formation and evolution (see for example
Nulsen 1986; Thomas, 1988a,b; Hensler and Burkert, 1990; Daines et al
1994; Nulsen ~ Fabian, 1995)  and in this context appears to be superior
to a treatment of the gas component as a one-phase medium.

%----------------------------------

\subsubsection*{Gravity and Gas Dynamics}

Gravity is described by the standard Particle-Mesh  algorithm,
i.e. density is computed  at the grid points by {\em Cloud in Cell}
interpolation from the particle positions. Poisson's equation is the
solved on the grid  by FFT and forces are obtained by differentiating
the potential  in the grid.  Forces acting on each particle are found
by   interpolation from the grid to the  particle positions 
 using the same kernel (CIC). Finally,  Newton equations are integrated
  for each particle using a Leap-frog scheme, to advance them.

Gas dynamics is described by the high order Godunov 
 PPM algorithm \citep{ColellaWoodward84}. 
As we said at the beginning of this section, 
the main advantage  of this Eulerian code is that it
 does not involve an artificial viscosity, but still treats
 shocks very accurately. This method is very fast and highly
parallelisable.  This algorithm is applied to the  gas-dynamical
equations in one dimension at a time.  Multiple dimensions are treated
by directional timestep splitting, whereas local processes (heating,
cooling, star formation, etc.), which involve various components of
collisional matter, as well as self-gravitation and gravitational
interaction with dark matter, are treated using process timestep
splitting (Oran \& Boris 1986).

%----------------------------------

\subsubsection*{Additional physics}

A proper treatment of cooling and star formation requires consideration
of processes occurring on scales well below the numerical resolution of
the code. Despite certain simplifications, the approach taken in \YK3
relies heavily on the picture of the interstellar medium by McKee \& 
Ostriker (1977), who provided a detailed description of the physics in
dense regions. This theory assumes the existence of cold clouds forming
due to thermal instability in approximate pressure equilibrium with the
surrounding hot gas.

Despite the existence of multiple phases with different temperatures,
the cooling rate can still be computed from the mean parameters of the
hot gas. Let us denote the {\it local} cooling rate of the plasma due
to radiative processes by $dE/dt=-\Lambda_r(\rho,T) $, where $\rho $ and $T$ here
represent the {\it true} local values of the gas density and
temperature. According to the theory of McKee and Ostriker, the rate of
energy loss expressed in terms of {\it average} gas density $\rho_h$ and
temperature $T_h$ will be higher than the nominal rate  $-\Lambda_r(\rho_h,T_h $)
by a {\it cooling enhancement factor} $C$, which is taken to be $C=10$
to account for all the effects resulting from unresolved density
inhomogeneities.

Cooling depends also on the chemical composition of the gas. In order
to incorporate the effects of metallicity into the code, solar
abundance is assumed in regions where either thermal instability is
present or previous star formation has occurred.  Otherwise, the region
is considered not to be enriched by metals, and the cooling rate for a
gas of primordial composition (Fall \&  Rees 1985) is used.

The ionisation of the intergalactic and interstellar medium by UV photons emitted by quasars, AGNs, and by nonlinear structures in the process of formation has two important consequences: First, gas outside dense regions ($\rho_{\rm gas}< 2\langle \rho_{\rm gas} \rangle$, where $\langle\rho_{\rm gas}\rangle$ is the mean cosmological gas density) is ionised and has a temperature
of $T=10^{3.8}$K (Giroux \& Shapiro, 1996; Petitjean, M\"ucket, \& Kates, 1996; M\"ucket et al., 1996). Second, in low to moderate-density regions, heating due to the ionising flux suppresses the thermal instability (M\"ucket \& Kates, 1996) and thus the formation of cold clouds and ultimately stars. Hence, cold clouds are allowed to form only if $\rho_{\rm gas}/\rho_{\rm cr} > {\cal D}\cdot\Omega_{\rm bar}$, with ${\cal D}\sim 50-100$.

In addition to the above processes, energy loss due to Compton cooling is also included, given by $\Lambda_{\rm Comp}=7\times 10^{-36}n_HT_ea^{-4}$, where $n_H$ is the number density of hydrogen ions, $T_e$ is the electron temperature, and $a$ is the expansion parameter.

In this model, there are two mechanisms for gas to leave the hot phase and enter the cold phase.  First, if the temperature of the hot gas drops below a threshold temperature $T_{\rm lim}=2\times 10^4$ K, and $\rho_{\rm gas}/\rho_{\rm cr} > {\cal D}\cdot\Omega_{\rm bar}$ all the hot gas is transferred immediately to the cold phase, thus making it available for star formation. This process gives the following terms in the continuity equations for the gas:

\be
      \left({d\rho_{\rm h} \over dt}\right)_{{\rm hot}\to {\rm cold}}
      = -\frac{\rho_{\rm h}} {t_{\rm cool}},
   \hskip 1.5em t_{\rm cool} = {\epsilon_{\rm h} \over 
                                   \partial \epsilon_{\rm h}/ \partial t},
    \hskip 1.5em \frac{\rho_{\rm gas}}{\rho_{\rm cr}}> {\cal D}\cdot\Omega_{\rm bar}    \hskip 1.5em T < T_{\rm lim},
\ee
where $t_{\rm cool}$ is the cooling time, and $\epsilon_{\rm h}$ is the thermal energy per unit mass. The transfer may be treated as immediate if the cooling time is very short compared to a timestep, which typically happens at  $T\approx 2\times 10^4$ K.

A second way of transferring gas to the cold phase is by sufficiently rapid thermal instability. The temperature range for creation of cold clouds depends in reality on the ionising flux and the local density (M\"ucket \& Kates, 1996), but the simplified restriction $T<T_{\rm inst} = 2\times 10^5$ is used in the code. To estimate the rate of growth of mass in cold clouds, the energy emitted by the hot gas is actually lost. If $\epsilon_{\rm h}$ and $\epsilon_{\rm c}$ are the internal thermal energies per unit mass of hot and cold gas, then the change of energy of the system due to radiative cooling is 
\be
\label{eq:EQA}
\left( \frac {d(\rho_{\rm h} \epsilon_{\rm h} 
+\rho_{\rm c} \epsilon_{\rm c})}{dt}\right)_{\rm cooling}
   =        -C\Lambda_r(\rho_h, T_h),
\ee
where $C$ is the enhanced cooling factor due to unresolved clumpiness. We take $\epsilon_{\rm c}= {\rm constant}$, which means that the cold gas cannot cool below $T_{\rm lim}$.  For an ideal gas, these assumptions imply extra terms in the continuity equations:

\be
\label{eq:EQB}
\left({d \rho_{\rm h} \over dt}\right)_{\rm therm.inst}
   = -\left({d\rho_{\rm c} \over dt}\right)_{\rm therm.inst} = -{C
   \Lambda_r(\rho_h, T_h) \over \gamma \epsilon_{\rm h}
   -\epsilon_{\rm c}}, 
\ee
where $\gamma$ is the ratio of specific heats.

Because of the frequent exchange of mass between the hot and cold gas phases, it is reasonable to consider the hot gas and cold clouds as {\em one} fluid with rather complicated chemical reactions going on within it. Thus, only the motion of the hot component is computed in the code, integrating the change of the {\em total} density of the gas ($\rho_{\rm gas}=\rho_{\rm c} +\rho_{\rm h}$). 

Both phases are assumed to be in pressure equilibrium, $P_{\rm gas}=P_{\rm c}=P_{\rm h}\equiv (\gamma-1)u_{\rm h}$, where $u_{\rm h}=\rho_{\rm h} \epsilon_{\rm h}$ is the internal energy per unit volume. The velocity associated with the fluid involves an average over a cell size, but in a multiphase medium with supernovae, the hot gas is actually ``windy'', and the cold clouds have a significant velocity dispersion (McKee \& Ostriker, 1977; Cowie, et al., 1981; Hensler and Burkert, 1990.).  Therefore, the code includes an effective pressure $P_{\rm disp}\propto \rho_cT_h$ associated with the cold gas component.

Star formation takes place in the cold clouds, leading to a decrease in the density given by
 \begin{equation}
 \left(\frac{d\rho_{\rm c}}{dt}\right)_{\rm star-formation}
 = -\frac{\rho_{\rm c}}{t_*},
 \end{equation}
where $t_*\approx 10^8$ yr is a fixed characteristic star formation time. The actual time-scale for star formation can exceed $t_*$ and will depend strongly on the rate of the conversion of hot gas to the cold phase. The lifetime of massive stars is about $10^7$ yr or even shorter; thus most of stars with $M_*$ heavier than $(10-20)\oldMsun$ will explode as supernovae during one timestep. Stars in the mass interval from (5--7)$\oldMsun$ to 10$\oldMsun$ will explode on a longer time-scale, but they produce less energy, and therefore in view of the uncertainties in supernova energy, their energy input is neglected in the model. Due to the explosion of massive stars, the stellar growth rate is decreased by the fraction $\beta$ of stars that explode as supernovae:

\begin{equation} 
 \label{eq:srate} 
  \frac{d\rho_*}{dt}=\frac{(1-\beta)\rho_{\rm c}}{t_*}. 
\end{equation}

The precise value of $\beta$ is very sensitive to to the form of the initial mass function (IMF), particularly its low-mass limit. The principal source of uncertainty is the dependence of the IMF on the abundance of heavy elements, which can be of critical importance at very early epochs of galaxy formation. In view of these uncertainties, $\beta$ is taken to be constant in time. For a Salpeter IMF, the fraction of mass in stars with mass larger than $10\oldMsun$ is $\beta=0.12$. 

Evaporation of cold clouds is an important effect of supernovae on the interstellar medium (McKee \& Ostriker 1977, Lada 1985). The total mass of cold gas heated and transferred back to the hot gas phase is assumed to be a factor $A$ higher than that of the supernova itself. The {\em supernova feedback parameter} $A$ could depend on, among other things, the energy of the supernovae, the cloud spectrum, and the ambient density. McKee and Ostriker (1977) give an estimate of the evaporated mass which scales with the energy of the supernova $E$ and the gas density as $E^{6/5}\cdot n_{\rm h}^{-4/5}$. However, at present we do not include the dependence on the gas density, because here $n_{\rm h}$ refers to the local density, which is relevant for the propagation of the supernova shock, whereas our $n_{\rm h}$ is the mean density of the hot gas averaged over quite large a volume.

Assuming that the energy input due to supernovae is proportional to their total mass, we obtain for the evaporation rate
\begin{equation}
\left(\frac{d\rho_h}{dt}\right)_{\rm evap}=-\left(\frac{d\rho_c}{dt}
\right)_{\rm evap}= \frac{A\beta \rho_{\rm c}}{t_*}
\end{equation}

Correspondingly, the net energy supplied to the hot gas phase by supernovae is:
\begin{equation}\label{eq:super}
\left(\frac{d\rho_h\epsilon_h}{dt}\right)_{\rm SN} 
   =  \frac{\beta\rho_{\rm c}}{ t_*}
      [\epsilon_{\rm SN}+A \epsilon_{\rm c}] .
\end{equation}
where $\epsilon_{\rm SN}=\langle E_{\rm SN}\rangle/\langle M_{\rm SN}\rangle$, $\langle E_{\rm SN}\rangle =10^{51}{\rm erg}$, and $\langle M_{SN}\rangle = 22\oldMsun$ (Salpeter IMF). Admissible values of the supernova feedback parameter $A$ are constrained by the condition that the energy of a supernova explosion must be larger than the energy required for the evaporation of cold clouds. This restriction leads to $\epsilon_{\rm SN} > A\cdot (\epsilon_{\rm h}-\epsilon_{\rm c})$. For the above supernova energy and mass and for hot gas temperature $T_{\rm h} \approx 10^6$ K, this condition implies the restriction $1 \leq A < 250$.

Therefore, for large values of $A$,  cloud evaporation dominates over
thermal re-heating, which translates into less pressure gradients and
more mass transfer between cold and hot phases.

%___________________________________________________

\subsection{ART}
\label{secART}

The spatial resolution of particle-mesh codes like \YK3 can be substantially improved by means of adaptive grids that provide more resolution in the high density regions where galaxies form. The {\em Adaptive Mesh Refinement} (AMR) method increases the dynamical range with respect to \YK3 by about two orders of magnitude without loss of mass resolution or computational speed.

%----------------------------------

\subsubsection*{Gravity}

The {\em Adaptive Refinement Tree} (ART) N-body code developed by \citet{ART97} is based on this technique. The computational volume is covered by a cubic regular grid that defines the minimum resolution, where the Poisson equation is solved with a traditional {\em Fast Fourier Transform} (FFT) technique using periodic boundary conditions, as in the PM algorithm. The difference is that this grid is {\em refined} (i.e. its cells are divided) wherever the density exceeds a predefined threshold.

Any cell can be subject to further refinements; the local refinement process continues recursively until the density criterion is satisfied. Once constructed, the mesh is adjusted at each timestep to the evolving particle distribution. Finer meshes are built as collections of cubic, non-overlapping cells of various sizes organised in {\em octal threaded trees} (i.e. each node may be split in eight sub-cells).

This is the same construct used in standard Tree codes, but there is an important difference in that cells are connected to their 6 neighbours in a {\em Fully Threaded Tree} \citep[FTT,][]{ftt} on all levels of refinement. In addition, cells that belong to different trees are connected to each other across tree boundaries. We can consider all cells as belonging to a single threaded tree with a root being the entire computational domain and the base grid being one of the tree levels.

The refinement procedure can be used either to construct the mesh hierarchy from scratch or to modify the existing meshes. However, in the course of a simulation the structure is neither constructed nor destroyed. Instead, the existing meshes are modified every computational cycle to account for the changes in particle distribution. Therefore, the computational mesh is adapted automatically to the structures developed as clustering proceeds.

 Poisson equation must be solved on every refinement level. Therefore, the value of the density must be computed for every cell regardless of whether or not it is a leaf. On each level, starting from the finest one and up to the zeroth level, the density is assigned using the standard CIC interpolation from the particle positions. For the  zeroth level of the mesh hierarchy (regular grid of fixed resolution), Poisson equation is solved by the standard FFT method. In the refinement sub-meshes, the {\em relaxation} method \citep {HockneyEastwood81,Press86} is used. Masses and timesteps are independent for each particle.

%----------------------------------

\subsubsection*{Gas dynamics}

The ART code achieves high spatial resolution by adaptively refining regions of interest using an automated refinement algorithm. Due to the use of an Eulerian mesh hierarchy in the FTT algorithm, the inclusion of Eulerian gasdynamics is a natural extension to the pure N-body part.

As in \YK3, the equations of gasdynamics and particle motion are integrated in 'supercomoving' variables \cite{martelshapiro98,Yepes01}. These variables are remarkable as their use almost completely (completely for ideal gas with $\gamma =5/3$) eliminates explicit dependence on cosmology in the model equations. The equations can therefore be integrated using standard solvers used in computational fluid dynamics, with no need for additional coefficients and corrections.

The main features of the gasdynamics implementation follow closely the algorithm used in \YK3. A second-order Godunov-type solver \citep{ColellaWoodward84} is used to compute numerical fluxes of gas variables through each cell interface, with 'left' and 'right' states estimated using piecewise linear reconstruction \cite{...vanLeer79}. For cells that have neighbours at the same tree level, the entire integration procedure is formally second order accurate both in space and in time. For cells that have neighbouring leaves at different levels, the accuracy reduces to first order.

Gasdynamics is coupled to the dynamics of dark matter through the common potential, which is used to compute accelerations for both the DM particles and the gas. The code employs the same data structures and similar refinement strategy as the FTT used in the $N$-body part. However, in addition to the DM density criteria it also allows for additional refinement based on the local density of gas, as well as shock or sharp gradient indicators. The refinement criteria can be combined with different weights allowing for a flexible refinement strategy that can be tuned to solve the Euler equations for an inviscid flow according to the needs of a particular simulation.

%___________________________________________________

\subsection{\G}
\label{secGadget}

In recent years there has been considerable effort in developing tree
codes that could run in parallel computers. This can alleviate the two
most important drawbacks of these algorithms: Large memory requirements
and long computations per timestep. In the present work we have
extensively employed the Tree-SPH code \g (a rather strange acronym for
\emph{GA}laxies with \emph{D}ark matter and \emph{G}as
int\emph{E}rac\emph{T}), recently developed by \citet{gadget01} and
publicly available from this Internet web page

{\tt http://www.mpa-garching.mpg.de/gadget/}

\G~ computes gravitational forces with the hierarchical octal-tree algorithm proposed by \citet{BarnesHut86}. The collisional fluid (gas) is  represented  by means of Smoothed Particle Hydrodynamics (SPH). The code is fully adaptive both in force computation and in time stepping (i.e. each particle has its individual mass and time step). Five different particle species can be defined by the user, each of them with its own gravitational smoothing, which is implemented as a spline-softened potential for point masses. Also, there are different criteria for selecting the optimal timestep. In what follows we will briefly describe how this code deals with gravity and gas dynamics.

%----------------------------------

\subsubsection*{Gravity}

As we said above,  gravity forces are computed by means of the BH tree algorithm. In this scheme, the particles are arranged in a hierarchy of groups according to their position in space. When the force on a particular particle is computed, the force exerted by distant groups is approximated by their lowest multipole moments. In this way, the computational cost for a complete force evaluation can be reduced to order ${\cal O}(N\log N)$. The forces become more accurate if the multipole expansion is carried out to higher order, but eventually the increasing cost of evaluating higher moments makes it more efficient to terminate the multipole expansion and rather use a larger number of smaller tree nodes to achieve a desired force accuracy. As a compromise between accuracy and speed, the multipole expansion is terminated after the quadrupole moments have been included.

Force computation then proceeds by walking the tree, and summing up appropriate force contributions from tree nodes. In the standard BH tree walk, the multipole expansion of a node of size $l$ is used only if 
\be r>\frac{l}{\theta} \, ,
\label{eqopen}
\ee where $r$ is the distance of the point of reference to the centre-of-mass of the cell and $\theta$ is a prescribed accuracy parameter. If a node fulfills the criterion (\ref{eqopen}), the tree walk along this branch can be terminated, otherwise it is `opened', and the walk is continued with all its daughter nodes. For smaller values of the opening angle, the forces will in general become more accurate, but also more costly to compute.

The standard BH opening criterion tries to limit the relative error of every particle-node interaction by comparing a rough estimate of the size of the quadrupole term, $\sim Ml^2/r^4$, with the size of the monopole term, $\sim M/r^2$. The result is the purely geometrical criterion of equation (\ref{eqopen}). In the \G~ code, the acceleration of the previous timestep is used as a handy approximate value for the accuracy of force computation. Therefore, it is required  that the estimated error of an acceptable multipole approximation is some small fraction of this $\alpha$ total force.

The tree-construction can be considered very fast in both cases, because the time spent for it is negligible compared to a complete force walk for all particles. However, in the individual time integration scheme only a small fraction $M$ of all particles may require a force walk at each given timestep, and hence the full tree is only reconstructed after a fixed number of timesteps has elapsed.

Time integration of Newton's equations is a variant of the classical Leap-frog scheme in which an explicit prediction step is introduced to accommodate individual timesteps for particles. There are different possibilities for the choice of timestep size in \G. One is based on the second order displacement of the kinetic energy, assuming a typical velocity dispersion $\sigma^2 $ for all particles. This results is $\Delta t = \alpha \sigma /|{\bf a}|$. The other method is to constraint the second-order particle displacement ($\Delta t \propto1/\sqrt{|a|}$), or a more sophisticated one (and more computational expensive) based on an estimation of the local dynamical time ($\Delta t \propto1/\sqrt{G\rho }$). According to the convergence studies recently carried out by \citet{power}, we have chosen the timestep criterion based on second order particle displacement, with the optimal parameters as given in that study.

%----------------------------------

\subsubsection*{Gas dynamics}

\G~ uses an implementation of SPH to estimate pressure forces acting on the collisional fluid. SPH is a powerful Lagrangian technique to solve hydrodynamical problems with an ease that is unmatched by grid based fluid solvers \citep[see][for an excellent review]{Mo92}. In particular, SPH is very well suited for three-dimensional astrophysical problems that do not crucially rely on accurately resolved shock fronts because, as we mentioned earlier, numerical artificial viscosity introduces some unphysical dissipation in the fluid.

Unlike other numerical approaches for hydrodynamics, the SPH equations do not take a unique form. Instead, many formally different versions of them can be derived. Furthermore, a large variety of recipes for specific implementations of force symmetrisation, determinations of smoothing lengths, and artificial viscosity, have been  proposed.

Most of the SPH implementations integrate the standard mass, momentum and energy conservation equations for the fluid. The computation of the hydrodynamic force and the rate of change of internal energy proceeds in two phases. In the first phase, new smoothing lengths $h_i$ are determined for the {\em active} particles (these are the ones that need a force update at the current timestep, see below), and for each of them, the neighbouring particles inside their respective smoothing radii are found. The Lagrangian nature of SPH arises when this number of neighbours is kept approximately constant. This is achieved by varying the smoothing length $h_i$ of each particle accordingly, leading to a constant mass resolution independent of the density of the flow. 

From the  neighbours of each {\em active} particle, the density and the rest of quantities can be found using an spline interpolation kernel 
 \be \rho_i=\sum_{j=1}^N m_j W(\boldvec{r}_{ij};h_i), 
\ee
where $\boldvec{r}_{ij}\equiv \boldvec{r}_i - \boldvec{r}_j$. For the rest of the particles, values of for density, internal energy, and smoothing length are predicted at the current time based on the values of the last update of those particles. An equation of state for an ideal gas gives the pressure as a function of density and internal energy: $P_i=(\gamma-1)\rho_i u_i$.

SPH is a slowly converging method with the number of particles. Theoretically, SPH equations will be exact in the limit $N_{SPH}\to \infty $. In cosmological problems, in which large volumes are often simulated, the mass resolution is rather coarse. \citet{Hernquist93} has shown that while integration of energy equation with SPH formalism results in rather good energy conservation, the entropy is not conserved even for purely adiabatic flows. He also showed that the cause of the errors can be found in the use of variable smoothing and the neglect of relevant terms in the dynamical equations associated with the space dependence of $h_i$.

However, even if fixed smoothing is used in a simulation, simultaneous conservation of entropy and energy can only be achieved in the limit of large $N_{SPH}$. Most people believe that violation of entropy is a fair price to pay for energy conservation. Therefore, most of SPH codes developed so far do show this problem. As have been shown recently \citep{gadgetEntro02}, integrating the thermal energy equation with SPH with not enough number of particles can lead to significant violations of entropy conservation in situations which are important for cosmological simulations of galaxy formation. As we will discuss in detail in Section~\ref{secEulerLagrange}, this problem with entropy dissipation of SPH codes could be the reason of the discrepancies that showed up in the Santa Barbara cluster comparison \citep{SB99}.

Two possibilities have been proposed to overcome this problem: One is to include in the formulation of the dynamical equations for SPH particles the $\nabla h$ terms associated with spatial derivatives of the smoothing length \citep[e.g. the AP$^3$M-SPH code DEVA][]{deva}. The other solution, proposed by \citet{gadgetEntro02}, is an elegant formulation of SPH which employs variable smoothing  lengths and explicitly conserve both energy and entropy. To this end, the equations of motion for the fluid particles were derived from the Lagrangian defined as:
 \be 
L(\boldvec{q},\dot \boldvec{q} )= \frac{1}{2}\sum_{i=1}^N m_i
\dot\boldvec{r}_i^2 - \frac{1}{\gamma -1 }\sum_{i=1}^N m_i A_i
\rho_i^{\gamma -1} 
\ee
 in the independent variables $\boldvec{q}=(\boldvec{r}_1,\ldots,\boldvec{r}_N, h_1,\ldots,h_N)$, where the thermal energy acts as the potential generating the motion of SPH particles. The densities $\rho_i$ are functions of $\boldvec{q}$ defined by the standard SPH kernel interpolation from particle positions. The  internal energy is  computed in terms of the specific entropy of
fluid elements, $A(s)$, which  for an ideal gas is $u_i=\frac{A_i}{\gamma-1}\rho_i^{\gamma -1}$. The quantities $A_i$ are treated as constants (i.e. the flow is assumed to be strictly adiabatic).

The  smoothing lengths $h_i$ are selected  by requiring that a fixed mass is contained within a smoothing volume, which provides $N$ constraints
\be
\phi_i(\boldvec{q})\equiv
\frac{4\pi}{3} h_i^3 \rho_i - M_{\rm sph} = 0
\label{eqnconstr}
\ee
on the coordinates of the Lagrangian (note that, for SPH particles of the same mass, the requirement of constant $M_{\rm sph}$ is equivalent to a fixed number of neighbours). The equations of motion reduce thus to
 \be \frac{\dd \boldvec{v}_i}{\dd t} = -
\sum_{j=1}^N m_j \left[ f_i \frac{P_i}{\rho_i^2} \nabla_i W_{ij}(h_i)
+ f_j \frac{P_j}{\rho_j^2} \nabla_i W_{ij}(h_j) \right],
\label{eqnmot} 
\ee
where the $f_i$ are defined by $f_i\equiv \left[ 1 +\frac{h_i}{3\rho_i}\frac{\partial \rho_i}{\partial h_i} \right]^{-1}$ and the abbreviated form $W_{ij}(h)\equiv W(|\boldvec{r}_{i}-\boldvec{r}_{j}|, h)$ has been used.

Because the potential (thermal) energy of the Lagrangian depends only on coordinate differences, the pairwise force in equation (\ref{eqnmot}) is automatically anti-symmetric. Total energy, entropy, momentum, and angular momentum are therefore all manifestly conserved, provided that the smoothing lengths are adjusted locally to ensure constant mass resolution as defined by equation (\ref{eqnconstr}). Moreover, the $\nabla h$ terms are implicitly included in the equations through the $f_i$ factors.

As in the standard (energy) implementation of SPH, artificial viscosity must be included to allow for the handling of shocks. A viscous force
\be
 \left. \frac{\dd \boldvec{v}_i}{\dd t}\right|_{\rm visc.}  =
-\sum_{j=1}^N m_j \Pi_{ij} \nabla_i\overline{W}_{ij} \, ,
\label{eqnvisc}
\ee
is added  to the acceleration given by equation (\ref{eqnmot}). The resulting dissipation of kinetic energy is exactly balanced by a corresponding increase in thermal energy if the entropy is evolved according to
\be
\frac{\dd A_i}{\dd t} =
-\frac{\gamma-1}{\rho_i^\gamma} {\cal L}(\rho_i,u_i) \; +\;
\frac{1}{2}\frac{\gamma-1}{\rho_i^{\gamma-1}}\sum_{j=1}^N m_j \Pi_{ij}
\boldvec{v}_{ij}\cdot\nabla_i \overline{W}_{ij} \,,
\label{eqnentropy}
\ee
which assumes  that entropy is {\em only} generated by the artificial viscosity in shocks, and by possible external sources of heat ($({\cal L}(\rho_i,u_i)$), if they are present (e.g. feedback from stars).

The modifications needed in the original \g code to account for this new version of SPH are not very important. A slight alteration is required for the algorithm which updates smoothing lengths, since it is now necessary to ensure that a `constant mass' resides within a smoothing volume rather than a constant number of neighbours.

Tests  performed by \citet{gadgetEntro02} with this new implementation of SPH show that entropy is much better conserved, avoiding the substantial overcooling produced in poorly resolved haloes. In the standard SPH implementation based on the thermal energy equation, the lack of resolution can lead to insufficient compressional heating in the accretion flow, resulting in severe entropy losses.

For our numerical experiments of cluster formation we have used the new version of \g, based on the entropy-conserving implementation of SPH. Nevertheless, in order to check the differences with respect to the standard SPH technique, we have run several test simulations of cluster formation with both versions of the parallel code: The  standard, public available, version and the entropy implementation of SPH, kindly provided by Volker Springel. Moreover, to compare SPH results with other numerical integration schemes, some simulations of the same  initial conditions have been rerun with the Eulerian gasdynamical code ART by Andrey Kravtsov \citep{ARThydro02,cluster6}. The outcome of this
comparison is reported in Section~\ref{secEulerLagrange}.

%-----------------------------------------------------------------------------

\section{Analysis}
\label{secAnalysis}

The output of a numerical experiment is a data file with the positions
and velocities of dark matter and gas  particles, gas densities and temperatures, and so on. In order to make a physical interpretation of these data, the analysis procedure is as important as the numerical simulation itself.

In this section, we describe the conventions followed in the analysis of the results of our numerical simulations. We address the questions of how to set the resolution limit, the definition of objects from particle data, and the computation of several magnitudes of physical interest, as well as their radial profiles. Throughout the present work, all magnitudes are expressed in units of $H_0\equiv100h$ km s$^{-1}$ Mpc$^{-1}$.

%___________________________________________________

\subsection{Resolution limit}
\label{secResolution}

Maybe the most common problem that must be faced by numerical simulations is the lack of mass and force resolution. Spurious effects arising from a low number of particles constitute a major source of uncertainty in any numerical experiment. In galaxy clusters, poor resolution leads to the so-called overmerging problem \citep[see e.g.][]{White87,Klypin99}, namely the loss of substructure within the dark matter halo. Moreover, resolution is of critical importance in determining the radial profiles of both dark matter and baryons at very small radii.

It is not clear which numerical effects may determine the minimum scale above which the results of a given code can be trusted, but it is quite likely that this convergence scale is determined by a complex interplay of all possible numerical effects. Although there have been some recent attempts at unravelling the role of numerical parameters on the structure of simulated dark matter haloes \citep[e.g.][]{Moore98,Knebe00,Klypin99,Klypin01,Ghigna00,power}, the conclusions from these works are still preliminary and, in some cases, even contradictory.

To cite an example, \citet{Moore98} argue that the smallest resolved
scales correspond to about half the mean inter-particle separation
within the virial radius, and conclude that many thousands of particles
are needed to resolve the inner density profile of dark matter
haloes. \citet{Klypin01}, on the other hand, conclude that the
convergence scale is $\sim4$ times the formal force resolutions, or the radius containing 200 dark matter particles, whichever is larger. \citet{Ghigna00} suggest a similar criterion based on the gravitational softening length, and argue that convergence is only achieved on scales that contain many particles and that are larger than about $\sim 3$ times the scale where pairwise forces become Newtonian.

Understanding the precise role of numerical parameters is clearly needed before a firm theoretical prediction for the structure of CDM haloes on $\sim$kpc scales may emerge. This question is particularly difficult because of the lack of a theory with which the true structure of dark haloes may be predicted analytically, so the best that can be done is to establish the conditions under which the structure of a simulated dark halo is independent of numerical parameters.

Furthermore, we note that most convergence criteria have been derived
considering the minimum scale at which the density profiles of
different resolution simulations are consistent among themselves. Other
properties of dark matter haloes, such as their three-dimensional
shape, their detailed orbital structure, or the mass function of
substructure haloes, may require different resolution limits. For the
radial profiles related to gasdynamical quantities, the convergence
scale has been investigated in several studies
\citep[e.g.][]{Borgani02}. The number of SPH neighbours plays a crucial
role in the accuracy of results \citep[see e.g.][]{power}. In all our SPH simulations, this parameter has been set to 40 particles.

Having all these considerations into account, we adopted the following
criteria to set the resolution limit of the numerical experiments using
particle-based codes (\G) described in this chapter:
\begin{itemize}
  \item 100 gas particles contained within $r_{\rm min}$
  \item 200 dark matter particles
  \item $r_{\rm min}\geq4\epsilon$, where $\epsilon$ is the gravitational softening
    length\footnote{In \G, using a spline-based smoothing potential for
      point particles, pairwise force becomes Newtonian for  $r\sim 2.3\epsilon$}
\end{itemize}

Usually, the first condition imposes the most stringent constraints on our resolution, since gas tends to be substantially less concentrated than the dark matter due to the repulsive pressure force. Hence, the radial profiles computed for \g feature an apparent lower spatial resolution than those of the N-body code ART, but this only reflects the presence of gas in the former simulations.

This resolution limit is only applied to the minimum scale that has been considered reliable in the radial profiles. Global quantities (see Section~\ref{secGprop} below) have been computed from all the particles within the virial radius of the clusters.

%___________________________________________________

\subsection{Object finding}
\label{secObjectFind}

Obviously, the first step that must be taken to analyse a cluster of
galaxies (or any other object) is to locate it. Identification of
haloes in high density environments, such as groups and clusters of
galaxies, is a challenging problem. Currently, there are several
methods to accomplish this task, each one with its own merits and
drawbacks. To mention some of the most popular, the {\em Friends Of
  Friends} \citep[FOF, see e.g.][]{Davis85} algorithm connects those
particles that are separated less than a given linking length. A
hierarchical version of this method has been proposed \citep{Klypin99}
to overcome the merging of apparently distinct haloes when the linking
length is set too large and the missing  of many particles when it is too small. An alternative is the 'spherical overdensity' scheme \citep[e.g.][]{LaceyCole94,Klypin96}, which looks for local maxima of the density field above some threshold defined as the 'virial' overdensity. \citet{Springel01} use a more elaborate algorithm that limits the objects by the isodensity contour that traverses a saddle point.

For our analysis, we have chosen the {\em Bound Density Maxima }\citep[BDM, see e.g.][]{Colin99,Klypin99} galaxy finding algorithm to identify bound structures from particle positions and velocities. This approach is based on the 'spherical density' outlined above, but particles that are not bound to the identified halo are removed recursively until a self-consistent definition of the object is found. The BDM algorithm proves extremely helpful in finding substructure within the dark matter haloes of galaxy clusters. As we discuss in Section~\ref{secSample} below, the presence of significant substructure within the virialised region of a cluster is used as an indicator of dynamical activity.

Another important issue is the determination of the centre of
mass. This question is particularly  relevant in those cases where the cluster halo is not exactly spherically symmetric, or when the study focuses on the detailed properties of the innermost regions of the simulated clusters.

We use as a first approximation the position given by the BDM algorithm, and then an iterative technique is employed in which the centre of mass is found for particles contained within a shrinking sphere. At each step of the iteration, the centre of the sphere is reset to the last computed barycenter (as a starting point, we use the BDM value) and its radius is reduced by $5\%$. The centre of mass is thus computed recursively until a convergence criterion is met. In our case, the iteration is stopped at the resolution limit described above, checking that the variation in the final centre of mass is less than $1\ h^{-1}$ kpc.

Halo centres identified with this procedure are quite independent of the parameters chosen to start the iteration, provided that the initial sphere is large enough to encompass a large fraction of the system. In a multi-component system, such as a dark halo with substructure, this algorithm isolates the densest region within the largest subcomponent. In more regular systems, the centre so obtained is in good agreement with centres obtained by weighing the centre of mass by the local density or gravitational potential of each particle.

%___________________________________________________

\subsection{Radial profiles}
\label{secRprof}

Throughout most part of this work, our analysis of the physical properties of galaxy clusters will focus on the spherically-averaged radial profiles of several quantities of physical interest. Although the detailed structure of clusters displays significant departures from spherical symmetry, radial profiles offer a simple description of the dynamical state of the system, providing a great deal of helpful information about both the dark and baryonic components.

This section is devoted to clarify those aspects of our analysis that are not straightforward. The spherically-averaged profiles are measured by binning the particles according to their distance from the centre of mass of the at $z=0$. Logarithmic bins in steps of 0.05 dex are used in all cases, except in the computation of the radial profiles of projected quantities, for which a constant step of $10 \ h^{-1}$ kpc is used. In this case, the cluster properties are computed within cylindrical shells oriented along the coordinate axes, truncated at a distance of $3\ h^{-1}$ Mpc. The final profile is taken to be the mean value of the three projections in order to reduce noise.

%----------------------------------

\subsubsection*{Dark matter}

Density, cumulative mass and circular velocity ($V_c^2(r)=GM(r)/r$) are computed the usual way. In order to study the kinetical structure of our clusters of galaxies, we have obtained several radial profiles related to the velocity of the particle with respect to the centre of mass. For each bin, the movement have been decomposed into the radial and tangential directions, as well as in bulk (i.e. average) and random motions ($\sigma_i^2(r)=\left<v_i^2\right>-\left<v_i\right>^2$). Angular momentum has been computed from the tangential component of the velocity, assuming that the radius is constant for all the particles in the bin. Ordered rotation is computed from $\left<v_{\rm tg}\right>$, while the contribution of random motions comes from the tangential velocity dispersion.

The orbits of individual particles have also been classified according to their apocentric radii (see Section~\ref{secJsims}). This is done by computing the gravitational potential $\phi(r)-\phi_0=\int_0^r GM(x)/x^2 dx$, assuming spherical symmetry. This approximation incurs in some uncertainty due to the time variation in the mass distribution of the cluster. This effect is more important in the outer parts, where the cluster is still accreting a significant amount of matter at the present epoch.

Once the gravitational potential is known, the pericentric and
apocentric radii that a particle reaches during its orbit are easily
computed from its position and velocity. A useful quantity is the
eccentricity, defined as $e\equiv(r_{\rm max}-r_{\rm min})/(r_{\rm
  max}+r_{\rm min})$. It measures the depth that particles are able to
penetrate into the cluster potential well; a value $e=0$ corresponds to
perfectly circular orbits, whereas $e=1$ indicates that the particle
trajectory goes  through the centre.

%----------------------------------

\subsubsection*{Gas}

Gas density and mass are computed in an analogous manner as for the dark matter. We will use the notation $F_b$ to denote the \emph{global} baryon fraction, $F_b\equiv M_g/M$. The \emph{local} baryon fraction will be referred to as $f_b(r)\equiv\rho_g(r)/\rho_{dm}(r)$. Note that the total mass is used in the definition of $F_b$ while $f_b$ is the ratio of the baryonic to the dark matter density. The difference, though, is quite small.

Temperature of gas particles is inferred from its internal energy. Bolometric X-ray luminosity is approximated by pure bremsstrahlung emission, where the emissivity is given by $n_e^2\Lambda(T)$, that is, the square of the electron density $n_e=\rho_g/(\mu m_p)$ times the cooling function $\Lambda(T)$ ($m_p$ is the proton mass and $\mu=0.65$ is the mean molecular weight of a fully ionised plasma). Following \citet{NFW95}, we will assume that, at the temperatures relevant for galaxy clusters, the cooling function can be well approximated by
\be
\Lambda(T)\simeq1.2\times10^{-24}\left(\frac{T}{1\ {\rm keV}}\right)^{1/2} {\rm erg\ s}^{-1}\ {\rm cm}^3
\ee
and hence the luminosity of a set of SPH particles can be computed as
\be
\Lx=1.2\times10^{-24}\sum_{i=1}^N\frac{m_i\rho_i}{(\mu m_p)^2}T_i^{1/2}
\ee
where $m_i$, $\rho_i$ and $T_i$ are the mass, density and temperature of each gas particle.

Other quantities can be obtained from the radial density and temperature profiles, such as the polytropic index $\gamma(r)=1+\frac{\dd \log(T)}{\dd \log(\rho)}$ or the gas entropy, defined as $S(r)=Tn_e^{-2/3}$.

%----------------------------------

\subsubsection*{Average profiles}

A convenient way to represent our results is to combine the information obtained from a certain number of clusters into a single plot. Our approach is based on the following scheme:
\begin{enumerate}
  \item Compute individual profiles
  \item Define radial bins (points to plot)
  \item For each bin, interpolate the value of every profile at that point
  \item Compute mean value and standard deviation
\end{enumerate}

Averaged profiles reported in the present work have been computed according to this simple prescription. The quoted error bars correspond to the standard deviation (note that, in some cases, this quantity can be dominated by Poisson noise due to the low number of clusters considered).

%___________________________________________________

\subsection{Global properties}
\label{secGprop}

In many occasions, we will be interested not only in the radial
structure of galaxy clusters, but also in their bulk
properties. Therefore, we must define their average value up to some
radius that we define as the  cluster limit.

As it happens  in observations, there is no unambiguous prescription to define this limit. In a Standard Cold Dark Matter (SCDM) cosmology, spherical collapse theory predicts that dark matter haloes should be in virial equilibrium at densities of roughly $200\rho_c$. The radius at which the cumulative overdensity reaches 200 has often been used to mark the boundary of both observed and simulated objects. However, this estimate is a very simple approximation to the non-linear process of virial relaxation. As we will show in Section~\ref{secHydroPolyt}, the assumption that clusters of galaxies are in equilibrium at these large radii holds only marginally.

In a \lcdm universe, this simple prescription based on spherical collapse yields an even lower density for virialisation ($\sim100\rho_c$). However, for historical reasons, it is not uncommon that the value corresponding to the SCDM model is used to define the limit of clusters. Since the difference in most physical quantities computed at $r_{100}$ or \rv is not large, and the latter is most often quoted in observational studies, throughout this thesis we will use the term 'virial' to denote quantities defined at $\Delta=200$, that is,
\be
    M_{200}\equiv\frac{4\pi}{3}200\rho_c\Rv^3
\ee

In any case, the overdensity at which a magnitude is evaluated will
always be indicated by a numerical subscript. For instance,  
in Section~\ref{secScalingRelations}, the scaling relations between the total mass, luminosity and temperature are investigated at several values of the overdensity threshold $\Delta$, similar to those observed by the current generation of X-ray satellites.

In order to make a more consistent comparison with these observations, we compute the emission-weighted temperature of our galaxy clusters as
\be
\displaystyle \Tx=\frac{\sum_{i=1}^N\rho_i T_i^{3/2}}{\sum_{i=1}^N\rho_iT_i^{1/2}}
\ee

Unless stated otherwise, 'temperature' or '$T$' will always refer to the mass-weighted temperature, $T=\sum_{i=1}^NT_i/N$. All other quantities are computed in the obvious way.

%-----------------------------------------------------------------------------

\section{Description of the simulations}

Most of the numerical analyses  reported in this work  is  based on 
the study of a sample of 15 galaxy clusters simulated both with the adaptive Eulerian code ART (including only dark matter particles) and the explicit entropy-conserving implementation of the Lagrangian code \g \citep{gadgetEntro02}, which uses a variant of the SPH technique to account for the gasdynamics of the baryonic component of the intracluster medium.

In addition to this cluster sample, we have considered also the initial
conditions used in the Santa Barbara Cluster Comparison Project
\citep{SB99}, as well as a different set of simulations (accomplished
with the Eulerian \YK3) that allow us to investigate the
 star formation and feedback effects in cluster and group environments.

%___________________________________________________

\subsection{Cluster sample}
\label{secSample}

Our sample  comprises a total number of 15 clusters of galaxies, selected from a low-resolution simulation with $128^3$ dark matter particles. In total, we have performed 7 independent numerical experiments running \g on a SGI Origin 3800 parallel computer at {\sc Ciemat} (Spain), using 32 CPU simultaneously. The average computing time needed to run  each  simulation was $\sim 8$ CPU  days ($6\times 10^5$ s). The same clusters have also been simulated with the N-body version of ART on the Hitachi SVR at the LRZ (Germany).

The properties of the cluster sample are thoroughly discussed in Section~\ref{secDMsims}, devoted to the dark matter distribution within the simulated haloes. In Chapter ~\ref{chapGas}, the structure and scaling relations of the ICM gas for these clusters are investigated.

%----------------------------------

\subsubsection*{Initial conditions}

In a  cubic volume of 80 $h^{-1}$ Mpc  on a side, an unconstrained realisation of the  power spectrum of density fluctuations corresponding to the most favoured \lcdm model ($\Omega_m=0.3$, $\Omega_\Lambda=0.7$, $h=0.7$ and $\sigma_8=0.9$) was generated for a total of $1024^3$ Fourier modes. The  density field was then re-sampled to a grid of $128^3$  particles which were displaced from their Lagrangian positions according to Zeldovich approximation up to $z=49$. From this initial conditions, ART (dark matter only) and \g (dark and gas) codes were used to evolve the particles up to $z=0$.

Selected clusters have been re-simulated with higher resolution by means of the multiple mass technique \citep[see][for further details]{Klypin01}. For each one of them, we have computed the particles in a spherical region around the centre of mass of the $128^3$ low-resolution counterpart at $z=0$. Mass resolution is then increased by adding  smaller particles in the Lagrangian volume depicted  by these particles, including the additional small-scale waves from the \lcdm power spectrum in the new initial conditions.

%__________________________________
\begin{figure}
  \centering \includegraphics[width=8cm]{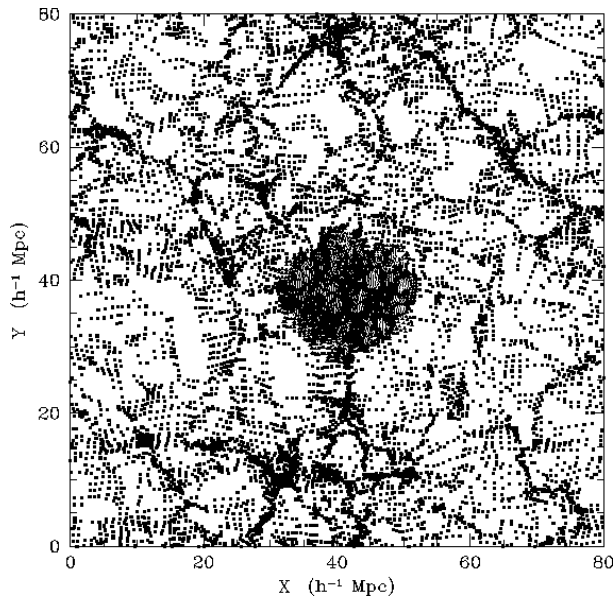}
  \caption{Mass refinement technique}
  \label{figMM}
\end{figure}
%__________________________________

This is illustrated in Figure~\ref{figMM}, where the colour code is used to indicate the refinement level. A buffer of medium-resolution particles (blue) avoids undesired two-body encounters between high-resolution (red) particles and the coarsest level (black) of refinement.

Some of the clusters in the sample described in the present section were so close to each other that their Lagrangian spaces overlapped at high $z$. Hence, these objects have been simulated together and only 7 experiments have been necessary in order to re-simulate the full sample with high resolution. Individual images of every cluster can be found in Appendix~\ref{apIndiv}.

\begin{table}
\begin{center}
\begin{tabular}{ccccc}
{\sc Clusters} & $N$ & $N_{\rm hi}$ & $M_{\rm DM}^{\rm hi}$ & $M_{\rm gas}^{\rm hi}$ \\ \hline\\[-2mm]
   A                  & 4504797 & 1190016 & 29.6 & 2.12 \\
   B                  & 4719315 & 1291008 & 29.6 & 2.12 \\
   C                  & 5454182 & 1651648 & 29.8 & 2.01 \\
   D, H               & 4411785 & 1136064 & 29.6 & 2.12 \\
   E, L, M            & 4941262 & 1389440 & 30.0 & 1.80 \\
   F, G, K$_1$, K$_2$ & 4585734 & 1217088 & 29.6 & 2.12 \\
   I, J$_1$, J$_2$    & 5032957 & 1438848 & 29.6 & 2.12 \\
\end{tabular}
\end{center}
\caption[\g simulation specifics]{Total number of particles in each simulation, number of high resolution particles (N$_{hi}$ dark matter + N$_{hi}$ gas) and their mass in $10^7 h^{-1}$ \Msun.}
\label{tabSims}
\end{table}

Parameters of these simulations are summarised in Table~\ref{tabSims}, where the first column indicates the clusters that where refined in that run. In the second column, the total number of particles in the 80 $h^{-1}$ Mpc box is given, and the third column states the number of high resolution particles for each type of matter (i.e. N$_{hi}$ dark matter particles {\em and} N$_{hi}$ gas particles have been used in each experiment). The last two columns show the mass resolution in units of $10^7 h^{-1}$ \Msun, corresponding to the three levels of mass refinement that have been employed in all cases.

The gravitational softening length has been set to $\epsilon=2-5\ h^{-1}$ kpc, depending on the number of particles within the virial radius of each individual cluster \citep{power}. The minimum smoothing length for SPH was fixed to the same value as $\epsilon$.

%----------------------------------

\subsubsection*{Physical properties}

Table~\ref{tabSample} displays the physical properties of the simulated clusters at $z=0$. As discussed in Section~\ref{secGprop}, the subscript '200' refers to the overdensity with respect to the critical value $\rho_c\simeq2.8\times 10^{11} h^2$ \msun Mpc$^{-3}$ (not to be confused with the mean density of the universe, $\rho_m\equiv\Omega_m\rho_c$).

\begin{table}
\begin{center}
\begin{tabular}{lccrrrr}
{\sc Cluster} & {\sc State} & \RV & \MV & $L_{200}^{\rm X}$ & \TV & $T_{200}^{\rm X}$ \\ \hline\\[-2mm]
~~   A    &  Minor  & 931 & 18.96 & 70.06 & 2.000 & 2.873 \\
~~   B    &  Minor  & 871 & 15.57 & 20.30 & 1.860 & 2.620 \\
~~   C    & Relaxed & 871 & 15.53 & 42.65 & 1.958 & 2.810 \\
~~   D    &  Minor  & 771 & 10.77 & 11.49 & 1.301 & 1.642 \\
~~   E    & Relaxed & 719 &  8.74 & 12.53 & 1.287 & 1.866 \\
~~   F    &  Major  & 661 &  6.79 & 12.47 & 1.059 & 1.195 \\
~~   G    &  Major  & 638 &  6.10 &  4.50 & 0.844 & 0.818 \\
~~   H    & Relaxed & 618 &  5.56 & 16.60 & 1.107 & 1.614 \\
~~   I    &  Major  & 581 &  4.60 &  2.34 & 0.666 & 0.666 \\
~~  J$_1$ & Relaxed & 584 &  4.67 &  9.99 & 0.937 & 1.498 \\
~~  K$_1$ & Relaxed & 557 &  4.06 &  4.98 & 0.690 & 1.043 \\
~~   L    &  Minor  & 547 &  3.84 &  1.21 & 0.624 & 0.779 \\
~~   M    &  Major  & 503 &  2.99 &  0.64 & 0.545 & 0.610 \\
~~ K$_2$  &  Minor  & 497 &  2.89 &  3.08 & 0.470 & 0.808 \\
~~ J$_2$  & Relaxed & 491 &  2.77 &  8.22 & 0.673 & 0.979 \\
\end{tabular}
\end{center}
\caption[Physical properties of the clusters at $z=0$]{Physical properties of the clusters at $z=0$. \rv in $h^{-1}$ kpc, enclosed mass in $10^{13} h^{-1}$ \Msun, bolometric X-ray luminosity in $10^{25} h$ erg s$^{-1}$ and average temperatures (mass and emission-weighted) in keV. One or two asterisks beside the cluster name indicate minor or major merging activity (see text).}
\label{tabSample}
\end{table}

Clusters have been sorted (and named) in Table~\ref{tabSample} according to their virial mass at the present day. Clusters J$_2$ and K$_2$ are an exception to this rule, since they are two small groups falling into J$_1$ and K$_1$ respectively. As a matter of fact, the smallest objects in the present sample could as well be considered rich groups instead of poor clusters. Not existing a clear-cut distinction between these categories, we will apply the term 'clusters' to all the objects regardless of their virial mass.

In order to study the effect of mergers and close encounters, we labelled as {\em minor merger} any cluster in which we are able to identify a companion structure within \rv whose mass is greater than 0.1 \MV; if the mass of the companion is above 0.5 \MV, then we classify the merger as {\em major}; otherwise, the cluster is assumed to be a {\em relaxed} system in virial equilibrium.

The results of this classification scheme are quoted in the second column of Table~\ref{tabSample}. Clusters named J$_i$ and K$_i$ are relatively close pairs, but they are separate enough ($\sim2-3$ Mpc) not to be considered as mergers. In any case, it is remarkable the amount of dynamical activity found in our sample, specially for low-mass systems. This is in agreement with the results of \cite{Gottloeber01}, who found that the typical merging rate in groups is much higher than in clusters, even for galaxies of the same mass.

Another interesting issue is that minor mergers tend not to show significant features in the X-ray images, although some of their physical properties can be noticeably affected (see Appendix~\ref{apIndiv}). High-resolution observation (such as those based on {\em XMM} or {\em Chandra} satellites) are required in order to detect asymmetries in the inner parts of most of our merging clusters. The consequences concerning observational results based on the assumption of hydrostatic equilibrium are difficult to evaluate.

%___________________________________________________

\subsection{Santa Barbara Cluster}
\label{secExperSB}

The Santa Barbara Cluster Comparison Project \citep{SB99} was a compilation of the results of different hydrodynamical codes, applied to the formation of an X-ray cluster of galaxies in a SCDM universe. In order to test the two different formulations of SPH implemented in \G, we have used both versions of the code to carry out simulations with these initial conditions.

Numerical experiments described in this section have been carried out in the SGI Origin 2000 at the Centro Europeo de Paralelismo de Barcelona (CEPBA), the Astrophysicalisches Institut Potsdam (AIP) and the National Center for Supercomputing Applications (NCSA). Our findings are presented in Section~\ref{secEulerLagrange}.

%----------------------------------

\subsubsection*{Initial conditions}

The initial conditions for the Santa Barbara cluster experiments have
been supplied by the authors,both in Eulerian and Lagrangian
format. They are publicly available on the Internet, at

{\tt http://star-www.dur.ac.uk/$\sim$csf/clusdata/}

The initial fluctuation spectrum was taken to have an asymptotic spectral index $n=1$, and shape parameter $\Gamma=0.25$. The cosmological parameters are those of the SCDM model, that is $\Om=1$, $\OL=0$, $h=0.5$ and $\Ob=0.1$. The cluster perturbation was chosen to correspond to a $3\sigma$ peak of the primordial density field smoothed with a Gaussian filter of radius $r_0=10$ Mpc (in $\exp[-0.5(r/r_0)^2]$). The perturbation was centred on a cubic region of side $L=64$ Mpc.

\begin{table}
\begin{center}
\begin{tabular}{ccccc}
{\sc Version} & $N_{\rm eff}$ & & MM & $\epsilon$ [$h^{-1}$ kpc] \\ \hline\\[-2mm]
 E &  $64^3$ & &  No & 20 \\
 E & $128^3$ & &  No & 20 \\
 E & $128^3$ & &  No &  5 \\
 E & $128^3$ & & Yes & 15 \\
 S & $128^3$ & &  No & 20 \\
 S & $128^3$ & &  No &  5 \\
 S & $128^3$ & & Yes & 15 \\
 S & $256^3$ & & Yes &  2 \\
\end{tabular}
\end{center}
\caption{Santa Barbara Cluster simulations}
\label{tabSBsims}
\end{table}

This cluster has been simulated several times, varying the formulation of SPH used in the integration of the gasdynamic equations, the total number of particles, and the gravitational softening length.

Table~\ref{tabSBsims} summarises the values of these parameters. The implementation of SPH is quoted by the letter in the first column, where 'E' stands for 'energy', and 'S' for 'entropy'. In each case, the corresponding conservation equation is used to evolve the gas temperature in time. The second column quotes the effective resolution of the simulation, and the third, whether multi-mass techniques have been applied or not. When multiple masses have been used, the effective resolution correspond to the highest level of refinement; otherwise, $N_{\rm eff}$ equals the total number of particles in the simulated box (actually, one half, because there are as many gaseous as dark matter particles). The gravitational softening length is quoted in the last column.

%----------------------------------

\subsubsection*{Physical properties}

Quite expectedly, given that all the simulations correspond to the same cluster, the final global properties of the system are similar in all the experiments quoted in Table~\ref{tabSBsims}: $\Rv\simeq1.35\ h^{-1}$ Mpc, $\Mv\simeq0.58\times10^{15}\ h^{-1}$ Mpc, $T_{\rm 200}=4.6$ keV, $L_{200}^{\rm X}\simeq1.27\times10^{45}\  h$ erg s$^{-1}$

These figures are consistent with the values reported in \citet{SB99} for the average of all the numerical experiments presented in the Santa Barbara study. However, the detailed structure of the radial profiles of the diffuse gas component shows systematic trends as a function of the SPH implementation. A detailed discussion of this effect, as well as that of poor resolution, will be given in Section~\ref{secEulerLagrange}.

%___________________________________________________

\subsection{Star-forming clusters}
\label{simyk3}

In order to investigate the formation of stars and the effects of the environment on the star formation rate (SFR) of the simulated galaxies, we have performed a series of numerical experiments with the code \YK3, described in Section~\ref{secYK3}. The numerical code has been parallelised using OpenMP compiler directives. Therefore, it runs very efficiently in parallel computers with either Shared or Non Uniform Memory Access (NUMA) architectures. The simulations have been performed on a variety of machines.

These simulations were aimed in principle to a broader goal than the study of galaxy clusters, namely the evolution of the SFR density in different cosmologies and environments. Therefore, they include many small-volume numerical experiments in which no clusters of galaxies are formed. These experiments, however, offer interesting insights on the star formation history of the universe, as well as the dramatical changes that the star formation activity of a galaxy undergoes when it is accreted into a massive group or cluster of galaxies. These issues are covered in detail in Chapter~\ref{chapStars}.

%----------------------------------

\subsubsection*{Initial conditions}

As has been discussed in Section~\ref{secIntroCosmo}, the \lcdm model has proven to be very successful in describing most of the observational data at both low and high redshifts. However, in order to study the influence of cosmology on the global SFR, we have investigated two other scenarios: the standard CDM model (dominated by dark matter density), and the BSI model with a {\em Broken Scale Invariant} perturbation spectrum \citep[as predicted by double inflation, see][]{Gottloeber91}. The main parameters describing these cosmologies are summarised in Table~\ref{tabSFRcosmo}.

\begin{table}
\begin{center}
\begin{tabular}{cccccc}
{\sc Model} & $\Omega_{\rm DM}$ &$\Omega_{\rm b}$ & $\Omega_\Lambda$ & $h$ & $\sigma_8$ \\ \hline \\[-2mm]
\LCDM & 0.325 & 0.025 & 0.65 & 0.7 & 0.9 \\
 SCDM & 0.95  & 0.05  & 0    & 0.5 & 1.2 \\
  BSI & 0.95  & 0.05  & 0    & 0.5 & 0.6
\end{tabular}
\end{center}
\caption{Cosmological models}
\label{tabSFRcosmo}
\end{table}

The BSI model assumes a dark matter-dominated universe similar to the SCDM cosmology, but primordial fluctuations on small scales are significantly suppressed from its power spectrum (the overall normalisation, as in the other two models, is set by COBE data on large scales). The resulting $P(k)$ is almost the same spectrum as in the $\tau$CDM model, in which the primordial perturbation spectrum has been changed due to the hypothetical decay of massive $\tau$ neutrinos \citep{Efstathiou92}.

All the experiments described in this section are based on completely independent initial conditions (i.e. they do not belong to a consistent sample, as in Section~\ref{secSample}). As will be discussed in more detail in Sections~\ref{secSFRresol} and~\ref{secSFRclus}, the effects of cosmic variance due to a small simulated (or observed) volume are not negligible on the estimation of the mean star formation rate density.

%----------------------------------

\subsubsection*{Physical properties}

Table \ref{tableSFRsims} summaries the set of experiments performed with \YK3. In order to improve statistics, several realisations have been run when the number of particles was low enough ($N=128^3$), changing the random seeds used to generate the initial conditions.  The number of realisations is given by $n_{\rm r}$ in the second column. Next 4 columns show, respectively, the number of particles, box size, cell size (which determines spatial resolution in these simulations) and mass of each dark matter particle (mass resolution). In the last two columns, we quote the values of the supernova feedback parameter $A$ and the overdensity threshold ${\mathcal D}$ required for star formation (details on the role of these parameters in \YK3 can be found in Section~\ref{secYK3}).

\begin{table}
\begin{center}
\begin{tabular}{cccccccc}
{\sc Experiment} & $n_{\rm r}$ & $N$ & $L_{\rm box}$ & $L_{\rm cell}$ & $M_{\rm DM}$ & $A$ & ${\mathcal D}$ \\ \hline \\[-2mm]
\LCDM 1 &  1 & $350^3$ & 30 & 85.7 & 28.1 & 200 & 100\\
\LCDM 2 &  1 & $270^3$ &21.4& 79.3 & 22.2 & 200 & 100 \\
\LCDM 3 &  1 & $300^3$ & 12 & 40   & 2.85 & 200 & 100 \\
\LCDM 4 &  9 & $128^3$ &  5 & 39.1 & 2.22 & 200 & 100,500,1000 \\
\LCDM 5 & 18 & $128^3$ &  5 & 39.1 & 2.65 & 0,50,200 & 100 \\
 SCDM   & 18 & $128^3$ &  5 & 39.1 & 3.96 & 0,50,200 & 100 \\
  BSI   & 18 & $128^3$ &  5 & 39.1 & 3.96 & 0,50,200 & 100 \\
\end{tabular}
\end{center}
\caption[Resolution and feedback parameters]{Resolution and feedback parameters. Number of realisations, total number of particles/cells, box length in $h^{-1}$ Mpc, cell length in $h^{-1}$ kpc, mass of dark matter particles in $10^6$ \Msun, supernova feedback parameter and overdensity threshold for star formation.}
\label{tableSFRsims}
\end{table}

The comparison of experiments $\Lambda$CDM5, SCDM and BSI allows us to determine the effect of cosmology and feedback from supernovae on the global SFR history. The photoionisation prescription can be tested basing on the results of experiment $\Lambda$CDM4, and the dependence on resolution and volume will be thoroughly studied from the whole set of $\Lambda$CDM experiments.

Finally, experiments $\Lambda$CDM 1 and 2 simulate a volume large enough to contain a cluster of galaxies. We will use their results to investigated the star formation history of clusters in Section~\ref{secSFRclus}, focusing on the alterations of the star formation activity of galaxies when they fall into the potential well of the cluster.

%-----------------------------------------------------------------------------

%%% Local Variables: 
%%% mode: latex
%%% TeX-master: t
%%% End: 

%-----------------------------------------------------------------------------
  \chapter{Dark matter}
  \label{chapDM}
%-----------------------------------------------------------------------------

\begin{quote}{\em

If the sun refuse to shine\\
I don't mind\\
I don't mind}

-- Jimi Hendrix : {\em If 6 was 9} (1968) --\\
\end{quote}
%-----------------------------------------------------------------------------

{\gothfamily {\Huge A}{\large ccording}} to the current cosmological
paradigm, most mass in the universe must be in the form of non-baryonic
cold dark matter particles, whose nature is yet an unknown that has
been puzzling both astronomers and particle physicists during the past
few decades. As far as we are concerned, cold dark matter can be
considered as a perfect fluid with negligible pressure, thus obeying
the collisionless Boltzmann equation. Not being coupled to the
electromagnetic field, CDM particles do not emit or absorb light, and
hence (neglecting some form of self-interaction) gravity is the only physical
process that must be accounted for.

In clusters of galaxies, the baryon fraction is expected to be close to the cosmic value \citep[e.g.][]{WhiteFrenk91}, which implies that the amount of cold dark matter is higher by approximately one order of magnitude than the mass in baryonic form. Structure formation will be thus driven almost entirely by the CDM component. Galaxies and clusters are assumed to form by cooling and condensation of baryons into the potential wells created by dark matter haloes, which were able to collapse during the radiation-dominated epoch.

In this chapter, we will study the gravitational and kinetic structure
of clusters of galaxies. The radial distribution of mass, density,
velocity dispersion and angular momentum will be investigated in
detail, focusing on the existence of common patterns usually known in
the literature as 'universal profiles'. We will try to understand the
physical origin of these patterns in the dark matter distribution by
means of a simple analytical treatment based on the secondary infall
model. 

Most of the work refers to the cluster sample described in
Section~\ref{secSample}. Results concerning the density profiles of
these haloes and the dependence on environment can be found in
\citet{clusters02}. The orbital structure of galaxy clusters, as well
as the numerical treatment of spherical collapse, are to going to 
 be published in \citet{sc02}.

%-----------------------------------------------------------------------------

\section{The case for a universal density profile}

One of the most interesting properties of the dark matter haloes found in numerical simulations is the apparent universality of their radial density profiles, valid over several orders of magnitude in mass.

Contrary to the early numerical work of \citet{QSZ86} and
\citet{Frenk88}, in which dark matter haloes showed an isothermal
structure (i.e. $\rho(r)\propto r^{-2}$), \citet{DubinskiCarlberg91} and
\citet{Crone94} had enough resolution to detect the first evidence of
non-power-law density profiles. Building upon these results,
\citet[hereafter NFW]{NFW96,NFW97} found that haloes in their numerical
simulations could be fitted by a simple analytical function with only
two parameters (e.g. a characteristic density and radius). 

The main result of NFW experiments was that the radial density profile was steeper than isothermal ($\rho\propto r^{-3}$) for large radii, and shallower (but still diverging as $\rho\propto r^{-1}$) near the centre. The corresponding logarithmic slope
\be
\alpha(r)\equiv\frac{\dd\log \rho(r)}{\dd\log r}
\ee
varied smoothly between these two extreme values, being equal to isothermal ($\alpha=-2$) at the characteristic radius only.

\cite{NFW97} further show that the characteristic parameters of a dark
matter halo seem to correlate to a certain extent. Should  this  be
true, the final mass distribution of objects of different mass could be
described in terms of a one-parameter family of analytical
profiles. This also implies that many relevant quantities, such as
virial mass, central density, or formation time, must  be
correlated.

Similar results have been found in several independent studies, using much higher mass and force resolution than the original NFW paper. However, there is still some disagreement about the innermost value of the logarithmic slope $\alpha$ and its dependence on resolution. \citet{Moore98,Moore99,Ghigna98,Ghigna00} and \citet{FukushigeMakino97,FukushigeMakino01} find even steeper density profiles near the centre ($\alpha\sim-1.5$), whereas other authors \citep{JingSuto00,Klypin01} claim that the actual value of $\alpha$ may depend on halo mass, merger history and substructure.

%___________________________________________________

\subsection{Models}
\label{secNFWmoo}

In order to investigate the existence and shape of a 'universal' density profile, we will compare the mass distribution of our simulated clusters with the functional forms advocated by NFW and \citet{Moore99}. Although these are not the only profiles proposed in the literature\footnote{For example, \citet{JingSuto00} use the expression $\rho(r)\propto r^{-1.5}(r+r_s)^{-1.5}$, very similar to that advocated by Moore et al.}, they are by far the most widely referenced. Below, we describe the main features of both profiles, as well as some derived quantities that we will use in the comparison. Further details concerning these functions, such as the associated gravitational potential or the radial velocity dispersion can be easily computed analytically \citep[see e.g.][]{LokasMamon01,Suto98}.

%----------------------------------

\subsubsection*{NFW}

The radial density profile proposed by NFW
\be
\rho(r)=\frac{\rho_s}{\rrs\left(1+\rrs\right)^2}
\label{ecNFW}
\ee
depends on the characteristic density $\rho_s$ and the scale radius $r_s$ of the dark matter halo. This profile is often written in terms of the {\em concentration} parameter
\be
c\equiv\frac{\Rv}{\Rs}
\ee
as
\be
\rho(r)=\frac{200\rho_c\ c^2 g(c)/3}{s(1+cs)^2}
\ee
where we have used the definitions
\be
s\equiv\frac{r}{\Rv} ~~;~~ g(c)\equiv\left[\ln(1+c)-\frac{c}{1+c}\right]^{-1}
\ee
and \rv follows the convention given in Section~\ref{secGprop} (i.e. 200 times the \emph{critical} density of the universe).

This profile steepens monotonically with radius. The logarithmic slope
is  given by the expression
\be
\alpha(r)=-1-2\frac{\rrs}{1+\rrs}=-1-2\frac{cs}{1+cs}
\label{ecANFW}
\ee

The mass enclosed within radius $r$ increases as
\be
M(r)=4\pi\rho_s r_s^3\left[\ln\left(1+\rrs\right)-\frac{\rrs}{1+\rrs}\right]
\label{ecNFWm}
\ee
or, equivalently,
\be
M(r)=\Mv\ g(c)\left[\ln(1+cs)-\frac{cs}{1+cs}\right]
\label{ecNFWmC}
\ee

Instead of the logarithmic slope (\ref{ecANFW}) of the radial density profile, in order to compare with the results of numerical simulations, we would rather use the corresponding value for the mass distribution
\be
\Am(r)\equiv\frac{\dd\log M(r)}{\dd\log r}
\label{ecAM}
\ee

We decided to do so because the density profile is prone to significant Poisson noise when a narrow radial binning is set, and hence the number of particles in each bin is not very high ($\leq100$). Since the numerical evaluation of the derivative involved in the computation of $\alpha$ is extremely sensitive to these errors, expression (\ref{ecAM}) constitutes a much more stable choice to be checked against the analytic results.

In the case of a pure power law, $\Am=\alpha+3=$ constant. However, the NFW profile (\ref{ecNFWm}) yields
\be
\Am(r)=\frac{\left(\frac{\rrs}{1+\rrs}\right)^2}{\ln\left(1+\rrs\right)-\frac{\rrs}{1+\rrs}}
\label{ecAnfwM}
\ee

The limit of this expression as $r\to0$ is $\Am(0)=2$, i.e. $M(r)\propto r^2$. We see that, although density diverges as $1/r$, the enclosed mass has always a finite value, tending to zero when we choose a sphere of null radius.

%----------------------------------

\subsubsection*{Moore et al.}

Based on a set of high resolution simulations, \citet{Moore99} claim that the density profile of dark matter haloes can be fit by the function
\be
\rho(r)=\frac{\rho_{\rm m}}{(r/r_{\rm m})^{3/2}\left[1+(r/r_{\rm m})^{3/2}\right]}
\label{ecMoo}
\ee

This expression is very similar to the form advocated by
\citet{JingSuto00} for galaxy clusters. The logarithmic slope of this
density profile is
\be
\alpha(r)=-\frac{3}{2}\left[1+\frac{(r/r_{\rm m})^{3/2}}{1+(r/r_{\rm m})^{3/2}}\right]
\label{ecAmoo}
\ee
reaching an asymptotic value $\alpha=-1.5$ near the centre.

The cumulative mass is given by
\be
M(r)=\Mv\frac{\ln\left[1+(r/r_m)^{3/2}\right]}{\ln\left[1+(\Rv/r_m)^{3/2}\right]}
\label{ecMooM}
\ee
whose logarithmic slope is
\be
\Am(r)=\frac{3}{2}\frac{(r/r_m)^{3/2}}{\left[1+(r/r_m)^{3/2}\right]\ln\left[1+(r/r_m)^{3/2}\right]}
\label{ecAmooM}
\ee
implying a growth $M(r)\propto r^{1.5}$ at small radii.

%___________________________________________________

\subsection{Central density}
\label{secCentralDens}

Obviously, every analytical fit to the mass distribution found in
numerical simulations must have approximately the same shape. NFW and
Moore et al. formulae are almost indistinguishable over most part of
the halo ($r>r_s$) as long as $r_m\simeq1.7r_s$. The only difference between these two profiles is the asymptotic behaviour as $r\to0$.

However, we would like to stress at this point that the density profile
at such small radii is \emph{not} a prediction of CDM simulations, as
it is often stated in the literature. The finite number of particles
used in numerical N-body experiments sets an upper limit to the depth of dark matter potential, and the mass distribution at $r\to 0$ is completely dominated by numerical effects. Therefore, any
 extrapolation of the phenomenological fits to the density profile
 below the resolution limit must be taken with extreme caution, since
 there is no \emph{a priori} reason why the profile should follow any
 of the proposed formulae at such small distances. 

The minimum radius down to which the mass distribution can be reliably
estimated in numerical simulations has been recently investigated by
\citet{Moore98,Knebe00,Ghigna00,Klypin01} and \citet{power} as a
function of particle number, softening length or time integration
scheme. The first studies by \citet{DubinskiCarlberg91} and NFW
contained less than $10^5$ particles within the virial radius,
resolving scales $\sim0.1\Rv$. Nowadays, high-resolution numerical
simulations of galaxy clusters attain particle numbers as high as
$\sim2\times10^7$ \citep{Springel01} inside the  dark matter halo,
which allows to probe the density profile at distances smaller than
$0.01\Rv$. 

An important fact, pointed out by \citet{power}, is that the
logarithmic slope becomes increasingly shallow inwards, with little
sign of approaching an asymptotic value at the resolved radii. In that
case, the precise value of $\alpha$ at some definite cut-off scale (either a
fixed physical distance or relative to \RV) would not be particularly
meaningful. A similar argument has been used by \citet{Klypin01} to
explain the different inner slopes found by \citet{JingSuto00} in terms
of the $c-\Mv$ relation. 

%----------------------------------

\subsubsection*{Observational constraints at galactic scales}

It is somewhat ironic to think that the cold dark matter scenario, originally proposed to explain the observed flat rotation curves of spiral galaxies, is nowadays extremely successful in describing the large-scale structure of the universe and even the whole process of structure formation, but faces important problems when confronted with the shape of observed rotation curves at sub-galactic scales.

Although rotation curves of nearby galaxies have been routinely measured since the early 1970s \citep{RubinFord70,RogstadShostak72}, the analysis of observational data has continued to evolve as larger telescopes and improved detectors became available for optical, radio and millimetre wavelengths \citep[see e.g.][]{SofueRubin01}.

The gravitational mass can be easily inferred from the position-velocity diagram, just equating
\be
\frac{GM(r)}{r}=\left[\frac{v(r)-v_{sys}}{\sin i}\right]^2
\ee
where $i$ is the inclination angle and $v_{sys}$ the systemic velocity of the galaxy.

Combined measurements of HI and H$\alpha$ or CO emission lines are the best tools to probe the dark matter content at galactic scales. Optical rotation curves provide high spatial resolution near the centre, while only the neutral Hydrogen gas extends far enough in radius to trace the outer parts.

Observed rotation curves of dwarf spiral and LSB galaxies \citep{FloresPrimack94,Moore94,Burkert95,KKBP98,BorrielloSalucci01,Blok01,BlokBosma02,Marchesini02} seem to indicate that the shape of the density profile at small scales is significantly shallower than the cusps predicted by both fitting models.

This discrepancy has been often signalled as a genuine crisis of the CDM scenario, and several alternatives, such as warm \citep{Colin00,SommerLarsenDolgov01}, repulsive \citep{Goodman00}, fluid \citep{Peebles00}, fuzzy \citep{Hu00}, decaying \citep{Cen01}, annihilating \citep{Kaplinghat00} or self-interacting \citep{SpergelSteinhardt00,Yoshida00,Dave01} dark matter, have been invoked.

Unfortunately, it has been proved remarkably hard to establish the inner slope of the dark matter distribution observationally \citep[see e.g.][]{swaters}. Some authors \citep{BoschSwaters01,blais01,jimenez,swaters} claim that a cuspy density profile with $\alpha\leq1$ is consistent with current observations, although a shallower mass distribution is able to explain them as well. Yet, a value as steep as $\alpha=1.5$ can be confidently ruled out in most cases.

%----------------------------------

\subsubsection*{Observations of galaxy clusters}

Historically, most estimates of the masses of clusters were made from optical studies of their galaxy dynamics, wherein the motions of individual galaxies were used to trace the cluster potentials by virtue of the virial theorem \citep{zwicky37,smith36}. Nonetheless, these studies were sensitive to systematic uncertainties due to velocity anisotropies, substructure and projection effects \citep{Lucey83,Frenk90,Haarlem97}. More recent work based on larger galaxy samples and employing careful selection techniques has lead to significant progress \citep{Carlberg96,HartogKatgert96,Fadda96,Mazure96,Borgani99,Geller99,KoranyiGeller00}.

Accurate measures of cluster masses are accomplished via X-ray
observations of the hot intra-cluster gas and analysis of distortion of
background galaxies due to gravitational lensing by the cluster mass.

X-ray mass measurements rely on the assumption that the emitting gas which pervades clusters is in hydrostatic equilibrium. Using spherical symmetry and considering purely thermal support,
\be
\frac{1}{\rho}\frac{\dd P}{\dd r}= -\frac{GM}{r^2}
\ee
where the total mass profile is determined once the radial profiles of
the gas density and temperature are known (see e.g. \citet{sar88}). The gas
density can be directly obtained from X-ray images, while
temperature requires detailed spatially-resolved spectroscopy. 
The advent of \emph{Chandra} and \emph{XMM-Newton} satellites allows, for the first time, sufficiently good spatial and spectral resolution for self-consistent determinations of the density, temperature and mass profiles of galaxy clusters.

In contrast to the aforementioned optical galaxy dispersion and X-ray techniques, gravitational lensing offers a method for measuring the projected surface density of matter along the line-of-sight which is essentially free from assumptions about the composition or dynamical state of the gravitating material \citep[e.g.][]{Mellier99}. While strong lensing requires compact, dense cluster cores and thus probes their central mass distribution, weak lensing arclets can be found everywhere across clusters and allow their mass to be mapped even at clustercentric distances comparable to the virial radius.

Clearly, the best approach is to combine these three methods. The first analysis combining X-ray and lensing studies \citep{MiraldaBabul95} suggested that strong lensing masses within 200 kpc of the cluster centre might be typically higher by a factor of $\sim2-3$ than the X-ray inferred mass. Later work \citep[e.g.][]{Allen98} highlighted that this discrepancy only occurs for non-cooling flow clusters, which typically appear to be more dynamically active than cooling-flow systems.

Both X-ray data \citep{EttoriFAJ02,PrattArnaud02,SAF01,AEF01,Arabadjis02,lewis}, strong \citep{Oguri01} and weak lensing observations \citep{dahle} find steep slopes consistent with the CDM paradigm (i.e. NFW or Moore et al. profiles). However, some authors \citep{Tyson98,Sand02} report evidence of soft cores based on lensing measurements, and some of the X-ray derived profiles are not able to rule out the possibility of a shallow (or even null) inner slope. The size of a possible core is constrained to be less than $\sim50$ kpc, but, for the moment, it seems that current statistics do not allow any firm conclusion to be drawn on cluster scales.

%___________________________________________________

\subsection{Phase-space structure}
\label{secPhaseTh}

Since CDM obeys the collisionless Boltzmann equation, the velocity dispersion is related to the density field by the Jeans equation. Assuming spherical symmetry and neglecting bulk velocities (i.e. infall and rotation), the Jeans equation reads \citep[see e.g.][]{BT}
\be
\frac{1}{\rho}\frac{\dd}{\dd r}(\rho\sigma_r^2)+\frac{2\beta\sigma_r^2}{r}= -\frac{GM}{r^2}
\label{ecJeans}
\ee
where $\sigma_r$ is the one-dimensional velocity dispersion and $\beta$ is the anisotropy parameter, defined as $\beta\equiv1-\sigma_\theta^2/\sigma_r^2$. If random motions are considered to be isotropic ($\beta=0$), the radial velocity dispersion is entirely determined by the mass distribution. For a NFW profile,
\be
\sigma_r^2(r)= -\frac{r}{\Rs}\left(1+\frac{r}{\Rs}\right)^2
\int_0^{\rrs}{ \frac{4\pi G\rho_s\Rs^3}{x^3(1+x)^2}
\left[\ln(1+x)-\frac{x}{1+x}\right]
{\dd x}}
\label{ecJeansNFW}
\ee
from which the corresponding 'universal' profile for $\sigma_r$ can be obtained numerically.

Following a more empirical approach, \citet{TN01} realised that the coarse-grained phase-space density of their galaxy-sized haloes followed a power law
\be
\frac{\rho}{\sigma^3}\propto r^{-\alpha}
\label{ecTN01}
\ee
with $\alpha\simeq1.875$ over more than two and a half decades in radius. This
statement is incompatible with a NFW profile and the Jeans equation,
since (\ref{ecTN01}), once substituted into the Jeans equation, fixes
both the velocity dispersion and the mass distribution of the dark
matter halo. The density profile thus obtained depends on the
normalisation of the velocity dispersion, becoming unphysical
(decreasing towards the centre) beyond a critical value. For this
critical value, the density profile is very similar to NFW, although
the asymptotic slope as $r\to0$ is $\alpha=0.7$ instead of $\alpha=1$.  

Concerning ordered motions, \citet{BarnesEfstathiou87} found in their early work an indication for a 'universal' angular momentum profile, showing a rough proportionality between $j$ (the specific angular momentum, computed in spherical shells) and the radial coordinate. More recently, \citet{Bullock01} claimed that the cumulative mass with projected\footnote{When comparing the projected and total angular momentum in cells throughout the halo volume, the profiles were found to be very similar (within a factor of less than 2).} angular momentum less or equal to $j$ can be described by the function
\be
\frac{M(j)}{\Mv}=\frac{\mu j}{j_0+j}
\label{ecBullockJ}
\ee
where $j$ reaches a maximum value $j_{200}=j_0/(\mu-1)$ and $\mu>1$ is a shape parameter. If the angular momentum distribution is assumed to be spherically symmetric and monotonically increasing with $r$, expression (\ref{ecBullockJ}) yields
\be
j(r)=\frac{j_0}{\mu\Mv/M(r) - 1}
\ee

Although \citet{Bullock01} claim that the specific angular momentum
distribution in their sample of simulated dark matter haloes is 
closer to being cylindrically symmetric than spherically symmetric,
 they address the question of the behaviour of $j(r)$ averaged over
 spherical shells, finding $j\propto M^{1.3}$ or, equivalently, $j\propto r^{1.1}$. 

%-----------------------------------------------------------------------------

\section{Simulation Results}
\label{secDMsims}

In order to study the mass distribution in cluster-size haloes, we ran a series of high-resolution numerical simulations, using two completely independent codes. Since dark matter is the main contributor to the mass budget of galaxy clusters, we used the N-body version of the adaptive mesh code ART (see Section~\ref{secART}), including dark matter only. Furthermore, we also simulated the same clusters of galaxies with the Tree-SPH code \g (Section~\ref{secGadget}), which allows us to
\begin{enumerate}
\item Test the consistency of both integration schemes.
\item Quantify the effects of a baryonic gas on the final density profiles.
\end{enumerate}

The main properties of our sample of galaxy clusters is briefly summarised in Section~\ref{secSample} for both ART and \g simulations. Individual profiles of all clusters can be found in Appendix~\ref{apIndiv}.

%----------------------------------

\subsubsection*{ART vs \g}

\begin{figure}
  \centering \includegraphics[width=10cm]{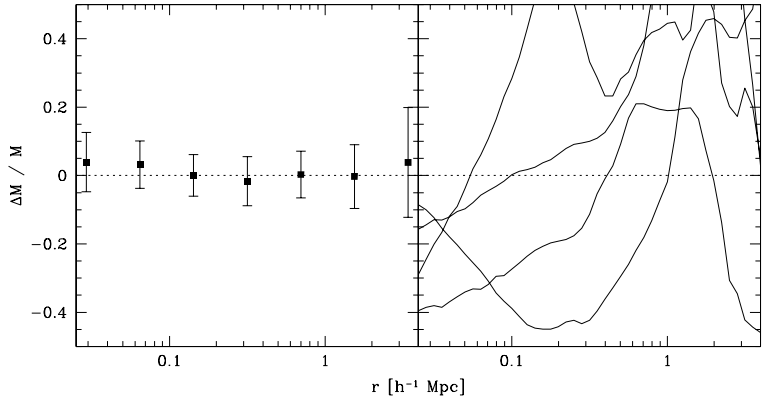}
  \caption[Comparison between ART and \g]{Comparison between ART and \g dark matter distributions. The relative difference in the cumulative mass is in most haloes (left panel) less than 15\%. Clusters F, I, J and K (right panel) display bigger discrepancies (see text).}
  \label{figARTvsG}
\end{figure}

Figure~\ref{figARTvsG} shows the ratio between the cumulative dark matter masses found in ART and \g as a function of clustercentric distance (for details on the computation of the centre of mass and radial profiles, see Section~\ref{secAnalysis}). Due to the presence of baryons in \G, we must rescale the masses by the appropriate value of $\Omega_{\rm CDM}=\Om-\Ob$ in order to compare results:
\be
\frac{\Delta M}{M}\equiv\frac{M_{\rm ART}}{M_{\rm Gadget}}\left(1-\frac{\Ob}{\Om}\right)-1
\ee

%__________________________________
%__________________________________

We compute this quantity for each cluster and average the individual profiles as discussed in Section~\ref{secRprof}. Clusters F, I, J$_1$ and K$_1$ are plotted separately on the right panel of Figure~\ref{figARTvsG}, and were not taken into account in the overall comparison. These clusters show discrepancies as high as 40$\%$ between both codes, but it cannot be attributed to inconsistencies in the simulation techniques but to slight differences in timing. Since all these clusters are undergoing major merging processes, small differences in the time of observation can have dramatic consequences on the morphology and dynamical state of the system. Indeed, clusters J$_2$ and K$_2$ are barely detectable in ART; in these simulations, we would simply have labelled clusters J and K as major mergers (obviously, clusters J$_2$ and K$_2$ have been excluded from these comparison, too).

As can be seen in the left panel of Figure~\ref{figARTvsG}, most
individual profiles of our dark haloes are consistent within $\sim10\%$
accuracy. The mean deviation is null for most of the radial range, and
it is only slightly biased towards larger masses in ART. However, this
deviation is small compared to the scatter that it is quite likely to
be a spurious effect. The fact that ART clusters appear to be a little
bit heavier both near the centre and beyond the virial radius gives
further support to this conclusion. 

%----------------------------------

\subsubsection*{Baryonic gas}

Another interesting issue is whether baryons affect the shape of the dark matter distribution. Non-adiabatic baryonic processes such as cooling, star formation and feedback can change the phase-space distribution of the gas component, making it gravitationally dominant at the inner parts of galaxy-sized haloes. In fact, results from some gasdynamical simulations of galaxy formation seem to indicate that galactic dark matter density profiles could be as steep as $\rho(r)\propto r^{-2}$ for $r\to 0$ \citep{TisseraRosa98,ThackerCouchman01}.

However, most baryons in clusters of galaxies are in the form of a hot plasma and their dynamical effects on the host dark matter halo are less important even when cooling is considered \citep[e.g.][]{Loken02}. Our results seem to support this view, but a word of caution must be said here because our simulations take into account adiabatic physics only. In the central regions of clusters, the typical cooling times are less than a Hubble time, and hence a certain amount of cooling is expected to take place. Indeed, cooling flows (see \citet{Fabian94} for a recent review) have been observed in many clusters of galaxies\footnote{Nevertheless, high-resolution \emph{Chandra} images and \emph{XMM-Newton} spectra show that the physics of the ICM within the central 100 kpc can be extremely complex \citep{fabian}, and it is likely that some form of heating reduces the mass cooling rates by a factor up to ten.}, as well as in numerical simulations including radiative cooling \citep[e.g.][]{Pearce00}.

If baryons played a significant role in the dynamics of the cluster, one would expect that they produced an adiabatic contraction of the dark matter halo, as it occurs on galactic scales, and hence the mass computed by \g should be appreciably smaller than that obtained by ART at small radii. Since we do not observe such a systematic trend in Figure~\ref{figARTvsG}, we conclude that the presence of a {\em hot} intracluster gas has no effect on the total mass distribution of the CDM component (or, at least, this effect is smaller than the $10\%$ scatter we find between the results of simulating the same cluster with both codes).

%___________________________________________________

\subsection{Universality}
\label{secDMprof}

In this section, we address the question of whether the radial dark
matter density profile of our clusters follows a universal form,
comparing the simulated mass distributions with the analytical profiles
proposed by NFW and \citet{Moore99}. If   the clusters  were described
by a universal two-parameter function, then  all the radial profiles 
should {\em exactly} overlap when they are properly rescaled.
 In the case of NFW formula (\ref{ecNFW}), 
the scale factors are the characteristic density
$\rho_s$ in one axis and the characteristic radius $r_s$ in the other. 

%----------------------------------

\subsubsection*{Fitting procedure}

The values of $\rho_s$ and $r_s$ are obtained by a $\chi^2$ fit to the
cumulative dark matter mass. As we discussed in
Section~\ref{secNFWmoo}, the density profile is much noisier than the
cumulative mass, and hence much more sensitive to the chosen radial
binning. We took equally-spaced logarithmic bins in order to give more
statistical weight to the inner regions, considering only radii between
0.05 \rv and \rv to compute the r.m.s deviations from the fit.
 In the less massive haloes, our resolution limit of 200 dark matter particles
(plus 100 gas particles in the case of \g simulations) placed a
stronger constraint on the lower limit. In total, 26 points have been
used for the fit in all clusters. 

\begin{table}
\begin{center}
\begin{tabular}{lcrrrr}
{\sc Cluster} & {\sc State} & $\rho_s$~ & $r_s$~ & $c$\ ~~ &
$\sqrt{\left<\left(\frac{\Delta M}{M}\right)^2\right>}$  \\ \hline\\[-2mm]
~~   A   &  Minor  &  5.565 & 123.3 &  7.164 &  0.061 \\
~~   B   &  Minor  &  2.878 & 159.8 &  5.401 &  0.038 \\
~~   C   & Relaxed &  5.501 & 119.8 &  7.129 &  0.007 \\
~~   D   &  Minor  &  1.993 & 159.7 &  4.595 &  0.052 \\
~~   E   & Relaxed &  5.370 &  99.7 &  7.057 &  0.012 \\
~~   F   &  Major  &  1.718 & 155.5 &  4.301 &  0.042 \\
~~   G   &  Major  &  0.971 & 176.5 &  3.315 &  0.120 \\
~~   H   & Relaxed &  8.082 &  73.9 &  8.375 &  0.019 \\
~~   I   &  Major  &  0.661 & 183.3 &  2.766 &  0.178 \\
~~ J$_1$ & Relaxed &  9.600 &  63.1 &  8.994 &  0.024 \\
~~ K$_1$ & Relaxed &  7.366 &  64.9 &  8.058 &  0.037 \\
~~   L   &  Minor  &  2.622 &  97.1 &  5.186 &  0.072 \\
~~   M   &  Major  &  2.401 &  86.1 &  4.989 &  0.138 \\
~~ K$_2$ &  Minor  &  9.871 &  49.9 &  9.098 &  0.061 \\
~~ J$_2$ & Relaxed & 12.510 &  46.3 & 10.026 &  0.042 \\
\end{tabular}
\end{center}
\caption[Best-fit values of NFW parameters]{Best-fit values of NFW parameters $\rho_s$ ($10^{15} h^2$ \msun Mpc$^{-3}$), $r_s$ ($h^{-1}$ kpc) and $c\equiv \Rv/r_s$. Merging haloes are typically more extended than relaxed ones, and they are less accurately fit by a NFW profile.}
\label{tabDM}
\end{table}

Actually, we fitted the mass distribution to expression (\ref{ecNFWmC}) in terms of \rv and $c$, and then $\rho_s$ and $r_s$ were computed from the best values of these parameters. We imposed the physical constraints that $c$ has been let to vary from $c=1$ to $c=15$, while a first guess for \rv was obtained from the mass profile itself (see Table~\ref{tabSample}). The best-fit value of this quantity was allowed to fluctuate between 0.8 and 1.2 times the initial guess.

Table~\ref{tabDM} summaries the values obtained by this method. The
accuracy of each profile is expressed as the root mean square deviation
of the numerical profiles from the analytical fit.
 For relaxed clusters, this quantity is always within a few percent, the quality of
the fit being slightly better for more massive clusters. Minor mergers
deviate about $6\%$ and the discrepancy rises up to more than $10\%$ in
the major mergers. 

At first sight, the well known $c-\Mv$ relation \citep[e.g.][]{NFW97} is not apparent in this table. The scatter in the concentration parameter seems to be quite large in the mass range covered by our sample \citep[see also][]{Jing00}. However, there is a trend in the sense that $c$ seems to be correlated with the dynamical state of the cluster. A more detailed discussion on this issue is given below, in connection with the effects of merging and environment.

%----------------------------------

\subsubsection*{Mass profiles}

\begin{figure}[!t]
  \centering \includegraphics[width=12cm]{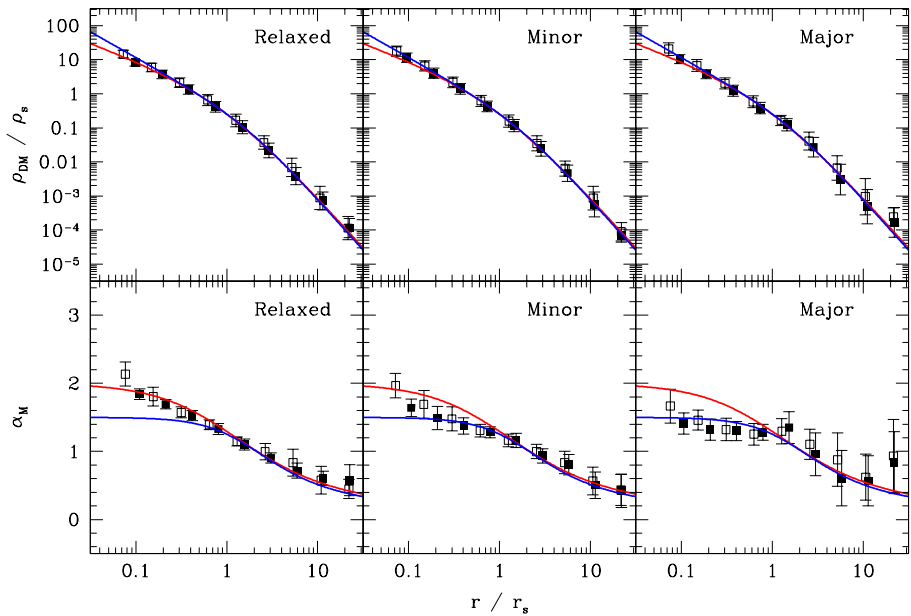}
  \caption[Dark matter density profiles]{Top panel: dark matter density profiles (squares) scaled by the characteristic density and radius of each halo. Red lines indicate NFW fit; blue is used for Moore et al. Bottom panel: logarithmic slope \am of the mass profile. Merging clusters display a pure power-law behaviour near the centre, while the slope of relaxed structures decreases gently with radius.}
  \label{figDM}
\end{figure}

We plot the radial density profiles of our clusters in the top panel of Figure~\ref{figDM}, rescaled by the best-fit values of $\rho_s$ and $r_s$ given in Table~\ref{tabDM}. The phenomenological formulae proposed by NFW (red) and Moore et al. (blue) are also shown for comparison. Since we have fitted our dark matter haloes to the NFW form, we have assumed $r_m=1.7r_s$ and $\rho_m=6\rho_s$ to represent the Moore et al. 'universal' profile in the figure.

%__________________________________
%__________________________________

We have plotted ART/\g data as empty/solid squares, respectively, and the innermost bin is set by the constraints explained in Section~\ref{secResolution}. The formal resolution is similar in both codes: the finest grid level in ART and the gravitational softening length in \g were set to $2\ h^{-1}$ kpc.

From the top panel of Figure~\ref{figDM}, we learn that most of the
clusters of galaxies in the present sample can be reasonably described
by either NFW or Moore et al. analytical fits, although some of them
(specially major mergers) display strong deviations from the
'universal' profile for $r\geq\Rv$ (a few times $r_s$; see
table~\ref{tabDM}). 

Nevertheless, a log-log plot of the density profile offers very limited information about the subtleties of the radial distribution (in fact, any deviation smaller than a factor of 2 is not easily perceived, given the scale of the figure). In order to study the shape of the mass distribution in the innermost regions, we plot the local value of the logarithmic slope of the cumulative mass profile \mbox{$\Am\equiv\dd\log M/\dd\log r$} as a function of radius (bottom panel of Figure~\ref{figDM}). The difference in the logarithmic slopes predicted by NFW and Moore et al. is mostly evident for $r\leq r_s$.

We see that relaxed and merging clusters show a remarkably different behaviour near the origin. Focusing only on the relaxed subset of the sample, we find that NFW provides a good fit to our data. The steeper slope predicted by Moore et al. formula can be confidently ruled out at radii as large as one order of magnitude above our resolution limit.

However, in agreement with the results of \citet{power}, we find that \am does not tend to any asymptotic value at all, but steadily increases towards the centre. Even more resolution is required in order to reach smaller radii (at the kpc scale) and test whether the logarithmic slope stabilises at a constant value of 2 (as in NFW fit) or, on the contrary, it still increases until $M\propto r^3$ (i.e. a constant density core).

This seems indeed to be the case when the simulated profiles are plotted slightly beyond the resolution limit. Yet, we would like to stress that the increasing slope at $r\leq0.1r_s$ may well be a numerical artifact induced by poor resolution. A this point, we prefer to be conservative and wait for higher-resolution data before reaching any conclusion.

%----------------------------------

\subsubsection*{Dependence on environment}

Relaxed clusters seem to follow indeed a universal profile dependent on
two scale parameters only (albeit a small scatter exists for
$r>\Rv$). This profile can be described with reasonable accuracy by NFW
expression up to our resolution limit, being the innermost values of the
logarithmic slope \am  clearly inconsistent with the proposal of
Moore et al. 

However, the situation is quite different for merging clusters. Minor mergers feature logarithmic slopes systematically steeper than those of relaxed systems. The shape of their radial density profiles is not exactly NFW or Moore et al.-like, but lies somewhere in between both fits, in many occasions closer to the latter ($\Am=1.5$) and sometimes even steeper.

Major mergers are an extreme case. Although they can be roughly approximated by NFW or Moore et al. formulae, significant deviations occur throughout the whole cluster. Nevertheless, our results hint that these recent mergers share some similar characteristics, the most remarkable being a power-law profile for more than one decade in radius. At the inner part, the logarithmic slope remains approximately constant. The exact value of \am shows some scatter, but the average seems to be slightly lower than 1.5.

A conjecture that deserves further investigation is that violent relaxation and subsequent equipartition of energy drive the inner part of the system into an isothermal state, and thus $\Am=1$ immediately after the merger. The cluster would then slowly relax, and the density profile would become increasingly shallow with time until virial equilibrium is reached. This would be consistent with the scenario proposed by \citet{Salvador98}, in which the halo formation time is defined by the last major merger.

%__________________________________
\begin{figure}
  \centering \includegraphics[width=12cm]{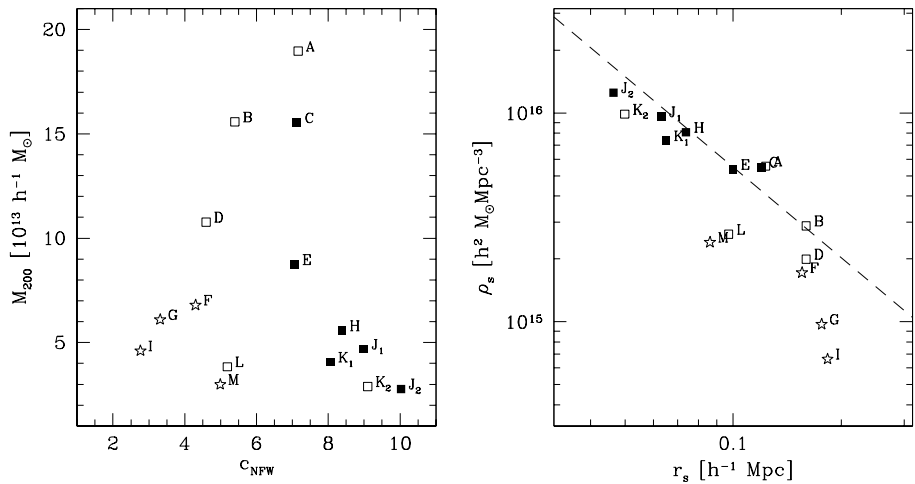}
  \caption[Relation between NFW parameters]{Relation between \mv and concentration parameter $c$ (left panel) and characteristic density and radius (right panel). Black squares represent relaxed systems, empty squares show minor mergers and stars are used for major mergers. Observational data from \citet{Sato00} is indicated by the dashed line.}
  \label{figMc}
\end{figure}
%__________________________________

The fact that major mergers rearrange the inner structure of dark matter haloes should leave some common imprint on the radial mass distribution. Apart from the steeper slope of the density profile, one of the features shared by all the merging systems in our sample (already shown in Table~\ref{tabDM}) was that merging haloes tend to be much less concentrated than relaxed ones. This is in agreement with the results of \citet{Jing00}, who concluded that the less virialised haloes (but still fitted by a NFW profile, similar to our 'minor mergers'), have a mean concentration about $15\%$ smaller than systems in virial equilibrium.

In Figure~\ref{figMc}, the $c-\Mv$ (or, equivalently, $r_s-\rho_s$)
relation is shown for our cluster sample. We see a clear dichotomy
between the behaviour of the $c-\Mv$ relation of relaxed and merging
systems. While relaxed clusters follow the usual trend of bigger   
concentration for smaller mass, major mergers are so extended that their
concentrations are much lower than those of the most massive systems.  

Due to the large variations that dark matter haloes experience
during a merging event, it seems unlikely that a $c-\Mv$ relation can
be established during these transient stages of their evolution. A
similar situation arises in minor mergers, where the effects on the
concentration (on the whole density profile, indeed) depend strongly on
the specifics of the encounter, such as the mass ratio of the CDM
haloes, their relative velocity, their former concentrations, etc. 

The relation seems to be a little tighter in the $r_s-\rho_s$ plane, shown on the right panel of Figure~\ref{figMc}. Although major mergers still show a different behaviour than relaxed clusters, their profiles have both larger values of the characteristic radius and lower central densities. This trend agrees well with recent X-ray observational estimates of the mass profiles of objects between $10^{12}$ and $10^{15}$ \msun \citep[][dashed line in Figure~\ref{figMc}]{Sato00}. 

%----------------------------------

\subsubsection*{Accuracy of NFW fit}

One way to quantify to what extent do clusters follow a given 'universal' profile is plotting the deviations with respect to the proposed form. Since we have fitted our CDM haloes by a NFW function in order to scale both axis in Figure~\ref{figDM}, we will present our results taking this profile as a reference.

%__________________________________
\begin{figure}
  \centering \includegraphics[width=12cm]{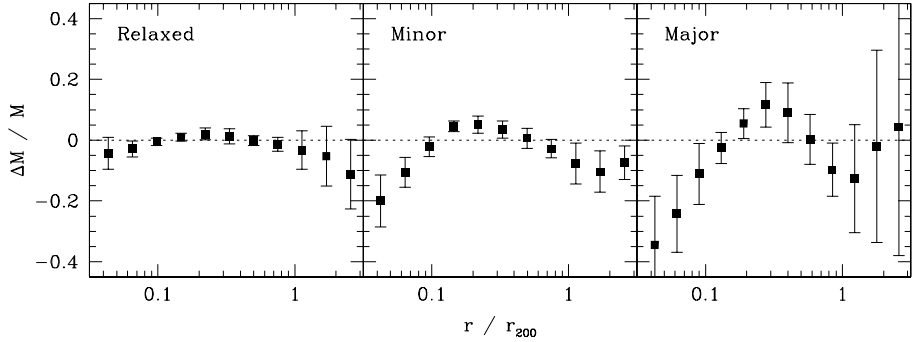}
  \caption[Accuracy of NFW fit]{Accuracy of NFW fit for the simulated
    clusters}
  \label{figNFWaccu}
\end{figure}
%__________________________________

We saw in Table~\ref{tabDM} that the r.m.s  values obtained for the NFW
fit indicated accuracies of the order of $5-10\%$. However, we recall
that this only applies for the cluster region that has been fitted
(i.e. from $0.05~\Rv$ to \RV). Outside this area, the profile of our
dark matter haloes is not so well described by the NFW form. 

The r.m.s deviation for the whole fit offers only a limited test of the
ability of NFW profile to trace the mass distribution found in
numerical experiments. We can obtain more information from
Figure~\ref{figNFWaccu}, where the relative difference between this
profile and the cumulative mass of our simulated haloes is plotted
against clustercentric radius. Excluding major mergers, our clusters do
not deviate appreciably from the NFW form throughout most of the radial
range $(0.1-1)\Rv$. Even in major mergers, the deviation
is seldom larger than $20\%$ in this region, although the formula
proposed by NFW offers a poor description of the shape (i.e. the
logarithmic slope) of the mass distribution in these systems. 

However, as pointed out by many authors \citep[e.g.][see
Section~\ref{secNFWmoo}]{Moore99}, departures from the NFW formula seem
to be similar for all clusters, displaying a characteristic 'S'
shape. Nonetheless, the mass excess detected at very small radii does
not necessarily imply a steeper inner slope of the radial density
profile. For example, the form proposed by \citet{TN01} (see Section~\ref{secPhaseTh})
 results in more concentrated CDM distributions,
 but  the asymptotic slope $\alpha (r\to 0)\sim 0.7$. Yet, the fact that our 
 simulated dark matter haloes show \emph{systematic} deviations with 
respect to the NFW formula hints  that it is not a good approximation to
 the shape of their density profile at the innermost regions. 

%___________________________________________________

\subsection{Phase-space structure}
\label{secJsims}

In addition to the radial mass distribution, the kinetic structure of
simulated CDM haloes can offer interesting insights into the formation
of galaxies and galaxy clusters. On cluster scales, rotation is
expected to give a negligible amount of support against gravitational
collapse because the angular momentum of the system is dominated by the
contribution of random motions.

In this section, we will discuss the phase-density structure of our
clusters, separating the global and random components and comparing our
results with previous work by \citet{Bullock01} and \citet{TN01},
respectively (see Section~\ref{secPhaseTh}). The details concerning the
computation of the angular momentum and velocity dispersion profiles
can be found in Section~\ref{secRprof}. 

%----------------------------------

\subsubsection*{Velocity Dispersion}

\begin{figure}
  \centering \includegraphics[width=8cm]{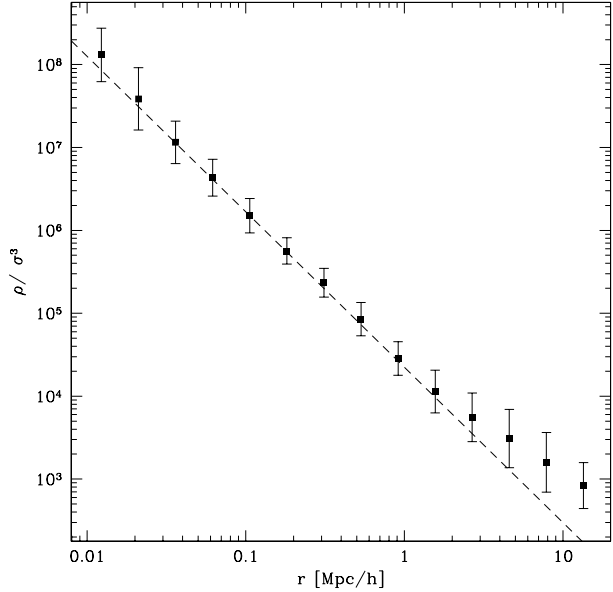}
  \caption[Phase space density profiles]{Phase space density profiles for  our haloes (dots) compared to the fit
  proposed by \citet{TN01} ($\rho/ \sigma^3 \propto r^{-1.879}$)}
  \label{figTN01}
\end{figure}

As was discussed in Section~\ref{secPhaseTh}, the velocity dispersion profile of a collissionless dark matter halo is related to the total mass distribution by virtue of the Jeans equation (\ref{ecJeans}). The average radial profile of the spherically-averaged phase-space density $\rho/\sigma^3$ has been computed for the whole sample of galaxy clusters, and it is plotted in Figure~\ref{figTN01}. Individual profiles display a remarkably low scatter, particularly if we take into account that 
\begin{enumerate}
  \item Major mergers have been included in the average.
  \item The profiles have not been normalised.
\end{enumerate}

%__________________________________
%__________________________________

The phase-space density of our simulated clusters is very well fitted
by the simple power-law
\be
\rho/\sigma^3=2.24\times10^4~r^{-1.875}
\label{tnfit}
\ee
where $\rho$, $\sigma$ and $r$ are expressed in units of $h^2$ \msun Mpc$^{-3}$, km s$^{-1}$ and $h^{-1}$ Mpc, respectively. The slope of the power law (-1.875) corresponds to the self-similar solution derived by \citet{Bertschinger85} for secondary infall onto a spherical perturbation in an otherwise homogeneous Einstein-de Sitter universe, and it has been found to describe also extremely well the phase-space density structure of galactic-size CDM haloes \citep{TN01}. An interesting issue which would be worth to investigate is whether the normalisation depends somehow on the scale of the halo, the assumed cosmological model, or any other factor. Unfortunately, the phase-space profiles computed by \citet{TN01} have been scaled arbitrarily, so it is impossible to compare with our result (\ref{tnfit}).

This correlation between density and velocity dispersion profiles has been used by \citet{TN01} to infer the radial mass distribution of the dark matter from the Jeans equation. \citet{hiotelis} has generalised this solution for the case of anisotropic velocity tensors, using an approximation for the anisotropy parameter
\be
  \beta(r)=\beta_1+2\beta_2\frac{r/r_*}{1+(r/r_*)^2}
  \label{ecBetaHiotelis}
\ee
where $\beta_1$ and $\beta_2$ are free parameters, and $r_*$ is twice the virial radius \citep{Carlberg97,Colin00vel}.

%__________________________________
\begin{figure}
  \centering \includegraphics[width=12cm]{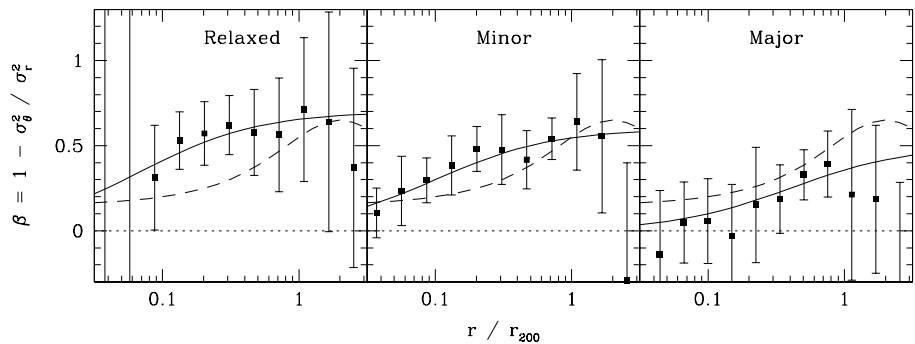}
  \caption[Anisotropy parameter]{ Anisotropy parameter $\beta$. 
See text for description of curves}
  \label{figAnisot}
\end{figure}
%__________________________________

In Figure~\ref{figAnisot}, the anisotropy profile of our simulated
clusters is compared with expression (\ref{ecBetaHiotelis}), using the
best-fit values found by \citet{Colin00vel}, $\beta_1=0.15$, and
$\beta_2=0.5$ (dashed lines). Although the average profile of the anisotropy parameter
seems to follow a similar trend in all haloes, the assumption of a
'universal' anisotropy profile seems hardly justifiable given the
scatter of each individual profile around the mean value\footnote{We note,
  however, that the numerical derivation of $\beta$ is not straightforward,
  since this parameter is extremely sensitive to inaccuracies in the
  velocity profiles, particularly near the centre.}. 

In any case, the simple formula (\ref{ecBetaHiotelis}) proposed by \citet{Carlberg97} does not provide a good description of the shape of the anisotropy profiles of our cluster sample. The exception could be the clusters with major mergers, which are much closer to isotropy in velocities than the rest, but we would need to modify the values of $\beta_1$ and $\beta_2$ and change $r_*$ by \RV. Nevertheless, expression (\ref{ecBetaHiotelis}) is a much better approximation to our results than purely isotropic profiles ($\beta=0$, dotted line), which can be ruled out in relaxed clusters and minor mergers at a several sigma confidence level over a broad radial range. 

Our numerical results can be better fit by the function (solid lines in Figure~\ref{figAnisot})
\be
  \beta(r)=\beta_{\rm max}\frac{r/r_*}{1+r/r_*}
\ee
where $\beta_{\rm max}=(0.7,0.6,0.5)$ and $r_*=(0.07,0.1,0.4)\Rv$ for relaxed clusters, minor and major mergers respectively. We insist, nonetheless, that this fit is a poor approximation to the radial dependence of $\beta$ in any individual halo, and it could only be applied in a statistical sense, such as in the determination of the \emph{average} mass profile attempted by \citet{hiotelis}.

%----------------------------------

\subsubsection*{Angular momentum}

The origin of angular momentum in dark matter haloes can be understood in terms of lineal tidal torque theory, in which protohaloes acquire their angular momentum because of the surrounding shear \citep{Peebles69,Doroshkevich70,White84,Porciani02a,Porciani02b}.

\begin{figure}
  \centering \includegraphics[width=12cm]{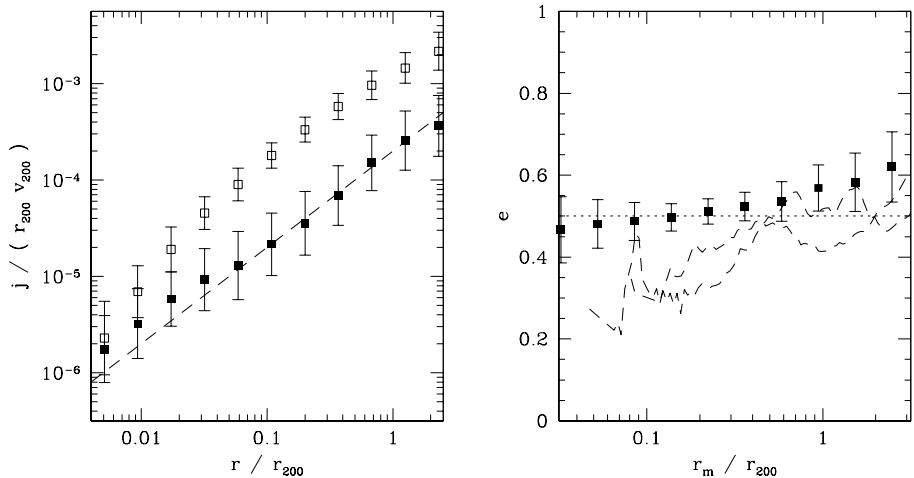}
  \caption[Angular momentum in the simulated haloes]{Angular momentum
    in the simulated haloes. Left panel: Contribution of bulk rotation
    (solid squares) and random motions (empty squares). $j\propto r$ is
    drawn as a dashed line. Right panel: Eccentricity of the individual
    orbits (haloes I and J$_2$ are shown separately).} 
  \label{figJ}
\end{figure}

In the left panel of Figure~\ref{figJ}, we plot the angular momentum of
 our haloes  computed in spherical shells around the centre of
mass. Solid squares indicate the mean value of $j$ in each shell
(i.e. rigid body rotation). The amount of angular momentum due to
random motions (i.e. tangential velocity dispersion) is represented by
the open symbols.

%__________________________________
%__________________________________

The first thing that can be appreciated in Figure~\ref{figJ} is that velocity dispersion is the dominant source of kinetic energy in the system. The amount of angular momentum contributed by random motions is higher by approximately one order of magnitude than that corresponding to bulk rotation.

The relative scatter in the radial profiles (when normalised on both axes as indicated in the figure) is larger for the mean angular momentum than for the r.m.s. value. Near the centre, the situation is less  clear, since at very small radii the low number of particles (and possible misidentifications of the centre of mass and velocities) makes difficult to disentangle ordered from random motions, as well as radial and tangential directions.

Building upon the pioneering work of \citet{BarnesEfstathiou87}, the question of a universal angular momentum profile has been recently investigated by several authors \citep[e.g.][]{Bullock01,bosch,chen}. The spin profiles found in numerical simulations grow approximately linearly with radius \citep[or as $j\propto r^{1.1}$,][]{Bullock01}. Our results are consistent with this behaviour (dashed line in the left panel of Figure~\ref{figJ}), but they deviate from this simple relation near the centre. However, the discrepancy ($\sim2$ over $j\propto r$ at $r\sim0.01\Rv$) is comparable to the scatter in the profiles. We must take into account that, at these radii, the lower number of particles may well lead to an overestimate of the mean angular momentum with respect to the random component.

Once the angular momentum distribution of the dark matter is known, it
is interesting to study the structure of the orbits of individual
particles around the centre of mass. More precisely, the eccentricity
of these orbits has important dynamical consequences, since it controls
directly the maximum depth that a dark matter particle is able to reach
within the gravitational potential of the dark matter halo. The ideal case $e=0$
corresponds to a configuration in which all particles would orbit at
constant radius with no mixing between concentric shells, while $e=1$
(often assumed in spherical infall models) implies that every dark
matter particle goes through the very centre of mass of the cluster. In
general, large values of the eccentricity lead to a situation in which
the core is mainly composed of particles coming from the outer shells
of the halo that happen to be near the pericenter of their orbits. 

We plot the the eccentricity profiles of our haloes as a function of the apocentric radius $r_m$ in the right panel of Figure~\ref{figJ}. Both quantities have been computed according to the prescriptions given in Section~\ref{secRprof}. Haloes I and J$_2$ are plotted separately as dashed line, since they feature much lower eccentricities than the rest of the sample. Except for these clusters, the rest of the sample exhibits a similar profile (as can be seen from the small scatter). For large apocentres, the eccentricities of different clusters show different behaviours, although the trend is that orbits are in general more radial as we move out to the turn-around radius.

Despite the small dependence of $e$ on $r$ seen in Figure~\ref{figJ}, we will make the approximation $e\simeq0.5$ (indicated by the horizontal dotted line) in order to model the orbits of dark matter particles. With this value, the minimum radius reached in the orbit is one third of the apocentric distance, while it would be $\sim0.43~r_m$ for $e=0.4$ and $0.25~r_m$ for $e=0.6$. Since the error we make is not large, we will use this simple approximation in our treatment of angular momentum in the spherical collapse formalism (Section~\ref{secSCj}).

%-----------------------------------------------------------------------------

\section{On the physical origin of dark matter profiles}

From the theoretical point of view, a number of plausible arguments have been advanced to try to explain the dynamical structure of dark matter haloes. The controversy regarding the 'universal' density profile and the asymptotic value of its logarithmic slope at the centre has stimulated a great deal of analytical work attempting to find the physical reasons behind this similarity in the mass distribution, as well as possible correlations with  halo size, environment, underlying power spectrum or even the nature of dark matter particles. 

The structure of collapsed and virialised objects such as galaxies and clusters poses a real challenge to our understanding of structure formation in the universe. The basic problem of the collisionless collapse of a spherical perturbation in an expanding background was addressed first by the two seminal papers of \citet{GG72} and \citet{Gunn77}, where the cosmological expansion and the role of adiabatic invariance were first introduced in the context of the formation of individual objects.

The next step was accomplished by \citet{FG84} and \citet{Bertschinger85}, who found analytical predictions for the density of collapsed objects seeded by scale-free primordial perturbations in a flat universe. \citet{HS85} generalised these solutions to realistic initial conditions in flat as well as open Friedmann models. Modifications of  the self-similar collapse model to include more realistic dynamics of the growth process have been attempted \citep[e.g.][]{AFH98,HenriksenWidrow99,Lokas00,Kull99,SCO00}. Several authors \citep[e.g.][]{syerWhite98, Salvador98, NusserSheth99} argue that the central density profile is linked to the accretion and merging history of dark matter substructure.

In this section, we will focus on the predictions of spherical collapse theory, showing that this simple model is able to explain the general features of the mass distribution  of our simulated haloes described in previous sections. First we explain our numerical implementation of the spherical collapse model and the primordial initial conditions. We then compare the resulting analytical profiles with our simulations. Finally, a brief discussion on the role of hierarchical merging is also included. 

%___________________________________________________

\subsection{Spherical collapse}
\label{secSC}

The most simple way of addressing the problem of structure formation is to assume spherical symmetry. For a homogeneous and isotropic universe, General Relativity leads to the Friedmann equation (\ref{ecFried}). Since we are interested on scales much smaller than the horizon, we can restrict our treatment to Newtonian dynamics. In this case, the equation of motion for a Lagrangian shell of mass $M$ can be derived from energy conservation
\be
\epsilon(r)=\frac{E(r)}{m}=\frac{\dot r^2}{2}-G\frac{M(r)+\frac{4\pi}{3}\rho_\Lambda r^3}{r}=\epsilon_i(r)
\label{ecConservE}
\ee
where $M(r)=M_i(r_i)$ is the mass contained within the shell, $\rho_\Lambda$ is the vacuum energy density, and the subscript 'i' denotes initial conditions. If we assume the universe to be homogeneous,
\be
M_i =\frac{4\pi}{3}\Om^i\rho_c^i r_i^3
\label{ecMi}
\ee

Substituting in (\ref{ecConservE}),
\be
\frac{\dot r^2}{2}   -G\frac{4\pi}{3}\rho_c^i~\frac{\Om^i r_i^3+\OL^i r^3}{r}=
\frac{(H_ir_i)^2}{2} -G\frac{4\pi}{3}\rho_c^i~\frac{\Om^i r_i^3+\OL^i r_i^3}{r_i}
\label{ecSCaux}
\ee

Since $\rho_c^i=\frac{3H_i^2}{8\pi G}$, we can make a coordinate transformation
\be
r(t)\equiv r_i\alpha(t)
\ee
and recover the Friedmann equation simply by multiplying both sides of (\ref{ecSCaux}) by a factor $\frac{2}{(H_ir_i)^2}$ :
\be
\frac{\dot \alpha^2}{H_i^2}-\Om^i \alpha^{-1}-\OL^i \alpha^2 = 1-\Om^i -\OL^i
\ee

Now, we introduce a small mass perturbation
\be
\Delta_i(r_i)=a_i\Delta(x_i)=\frac{M(r_i)}{\frac{4\pi}{3}\Om^i\rho_c^i r_i^3}-1
\ee
at an early epoch $a_i$, so that the growth is still in the linear phase. The initial positions and velocities of the spherical shells of matter are slightly modified, keeping only terms to first order in $\Delta_i(r_i)$:
\be
r_i'\simeq r_i\left(1-\frac{1}{3}\Delta_i(r_i)\right)~~;~~
v_i'\simeq H_ir_i\left(1-\frac{2}{3}\Delta_i(r_i)\right)
\ee
and the initial specific energy of each shell is thus
\bea
\nonumber\epsilon_i'(r_i) &\simeq & \frac{(H_ir_i)^2}{2}
\left[
  \left(1-\frac{2\Delta_i}{3}\right)^2 -
  \Om^i\left(1-\frac{\Delta_i}{3}\right)^{-1} -
  \OL^i\left(1-\frac{\Delta_i}{3}\right)^2
\right] \\
  &\simeq & \frac{(H_ir_i)^2}{2}
\left[
  \left(1-\Om^i-\OL^i\right)-\frac{\Delta_i}{3}\left(4+\Om^i-2\OL^i\right)
\right]
\eea

Since $a_i\ll1$,
\be
\Om^i=\frac{\Om a_i^{-3}}{\Om a_i^{-3}+\OL+\Ok a_i^{-2}}\simeq 1~~;~~
\OL^i=\frac{\OL}         {\Om a_i^{-3}+\OL+\Ok a_i^{-2}}\simeq 0
\ee
and energy conservation (\ref{ecConservE}) leads to the equation of motion
\be
\frac{2\epsilon_i}{(H_ir_i)^2}\simeq-\frac{5}{3}\Delta_i=
\frac{\dot \alpha^2}{H_i^2}-\Om^i \alpha^{-1}-\OL^i \alpha^2
\label{ecSC}
\ee
valid for a spherical shell enclosing a mass $M_i$.

%----------------------------------

\subsubsection*{Before turn-around}

According to (\ref{ecSC}), the evolution of a single spherical shell would thus be similar to that of a Friedmann universe. For high enough values of $\Delta_i$, the shell reaches a maximum radius $r_m$ at a {\em turn-around} time $t_m$ and re-collapses again to a point singularity. For an Einstein-de Sitter universe, equation (\ref{ecSC}) can be solved analytically, giving 
\be
r_m=\frac{3r_i}{5\Delta_i} ~~;~~ t_m=\frac{\pi}{2H_i(5\Delta_i/3)^{3/2}}
\label{ecRmTm}
\ee

This is also a valid approximation for shells reaching the turn-around radius $r_m$ before the cosmological constant term starts to dominate the expansion. Since the shells need at least another $t_m$ to virialise, in the \lcdm model discussed in Section~\ref{secIntroCosmo}, expression (\ref{ecRmTm}) can be applied to estimate the maximum expansion radius and time for the inner part of a virialised cluster. For the outer shells, however, the equation of motion (\ref{ecSC}) must be integrated numerically in order to find the trajectory up to the maximum radius.

In absence of shell-crossing, gravity would make the shell collapse back in a time-symmetric motion, so that the shell will reach the origin at $T=2t_m$. Since shells are assumed to be composed of collisionless cold dark matter particles, they will simply pass through the centre, describing an oscillatory motion with amplitude $r_m$ and period $T$.

%----------------------------------

\subsubsection*{After turn-around}

Equation (\ref{ecSC}) holds as long as the enclosed mass remains constant. However, as a shell re-collapses, it will cross the orbits of the inner shells that are bouncing  around the centre. When this happens, the assumption of constant mass breaks down because
\begin{enumerate}
\item The mass enclosed by the outer shell decreases
\item The mass enclosed by the inner shells increases
\end{enumerate}

To take into account the first point above, our model assumes that the final mass distribution can be approximated \emph{locally} by a pure power law. Since dark matter particles are expected to spend most of the time in the outermost regions of their orbits (see also Section~\ref{secSCj} on angular momentum), we only need this approximation to be valid over a relatively narrow range below the maximum radius. Near $r_m$, the enclosed mass is taken to vary with $r$ as 
\be
M(r)=M_i\left(\frac{r}{r_m}\right)^{\Am(r_m)}
\label{ecSCansatz}
\ee
where $\Am(r_m)$ is the {\em local} value of the logarithmic slope.

At first sight, this might seem similar to the classical approach followed in early works based on self-similar spherical collapse \citep[e.g.][]{Bertschinger85}, but in those cases the dark matter distribution was indeed assumed to be a power law, whereas in our model this ansatz is only an approximation to compute the orbits of particles in a given shell. The final mass profile is computed self-consistently as a sum of contributions from all shells, which have different values of $\Am(r_m)$. 

After turn-around, the particles belonging to a shell will oscillate (or, more generally, {\em orbit}) in the gravitational potential created by the dark matter halo. Taking the ansatz (\ref{ecSCansatz}), the gravitational potential is approximated as
\bea
\nonumber \Phi(r)-\Phi(r_m) &=& \int_{r_m}^r\frac{GM(r)}{r^2}~\dd r\\
  &\simeq& \frac{GM_i}{r_m^2[\Am(r_m)-1]}\left[\left(\frac{r}{r_m}\right)^{\Am(r_m)-1}-1\right]
\label{ecAnsatzM}
\eea

The probability of finding a particle inside radius $r$ is just proportional to the fraction of time that it spends within $r$:
\be
P(r,r_m)=\frac{1}{t_m} \int_0^r\frac{\dd x}{v_r(x)}=\frac{1}{t_m} \int_0^r\frac{\dd x}{\sqrt{\Phi(r_m)-\Phi(x)}}
\label{ecPr}
\ee

Note that, in this form, expression (\ref{ecPr}) is valid regardless of the assumed ansatz for $M(r)$. Taking different prescriptions to compute the potential, such as an isothermal profile ($\Am(r_m)=1$ for every shell, so $\Phi(r)$ is actually logarithmic) or considering $M$ constant ($\Am=0$, Keplerian potential $\Phi(r)=-GM/r$) does not lead to significant variations in the probability $P(r,r_m)$, and hence in the resulting mass distribution. 

If phase mixing is considered to be efficient, particles initially on the same shell will be spread out from $r=0$ to $r=r_m$ a short time after $t_m$. For the sake of simplicity, we will consider that phase mixing is instantaneous, so immediately after turn-around the shell is transformed into a density distribution whose cumulative mass is proportional to $P(r,r_m)$.

Effect (2) listed above has dramatic influence on the final structure of the dark matter halo. If the recently accreted shells have a non-zero contribution to the mass within $r$ (i.e. $P(r,r_m)\dd M$), the mass inside the inner shells whose maximum radius was $r$ changes from $M_i(r)$ to 
\be
M(r)=M_i(r)+\Madd(r)\equiv K(r)M_i(r)
\ee
where $\Madd(r)$ accounts for particles belonging to outer shells. To compute $\Madd(r)$ \citep[see][]{ZH93}, we must integrate the contribution of every shell whose maximum radius is larger than $r$, up to the current turn-around radius $R_m$:
\be
\Madd(r)=\int_r^{R_m}{\frac{\dd M_i(x)}{\dd x}~P(r,x)~\dd x}
\ee

\citet{Gunn77} was the first to apply the concept of adiabatic invariance to the secondary infall problem. If the orbital period of the inner particles is much shorter than the collapse timescale of the outer shells, the dynamics of the inner shell admits an adiabatic invariant
\be
J_r=\frac{1}{2\pi}\oint v_r(x)~\dd x
\ee
also known as the radial action. For a potential of the form (\ref{ecAnsatzM}), the radial action $J_r$ is proportional to $\sqrt{r_mM(r_m)}$. Therefore, the maximum radius $r_m$ of the inner shell decreases to a final value given by the implicit equation $r=F(r)r_m$, where
\be
F(r)=\frac{1}{K(r)}=\frac{M_i}{M_i+\Madd(r)}=\frac{M_i}{M_i+\Madd(F(r)r_m)}
\label{ecFr}
\ee
and whose solution must be obtained numerically for each shell.

%----------------------------------

\subsubsection*{Summary}

To summarise, the numerical procedure to compute the final radius of a Lagrangian shell of matter involves the following steps:
\begin{enumerate}
  \item Set the initial mass $M_i$ according to (\ref{ecMi})
  \item Integrate the equation of motion (\ref{ecSC}) up to the maximum radius $r_m$
  \item If $t_m>t_0$
    \begin{itemize}
      \item The shell is still expanding, $r=r(t_0)$
    \end{itemize}
  \item If $t_m<t_0$
    \begin{itemize}
      \item Solve (\ref{ecFr}) to compute the new apocenter $r=F(r)r_m$
      \item Add the contribution $P(r',r)\dd M$ to $\Madd(r')$
    \end{itemize}
  \item Repeat the process for the next shell (closer to the centre).
\end{enumerate}

%___________________________________________________

\subsection{Initial conditions}

The spherical collapse model allows us to compute the mass distribution
arising from a primordial fluctuation $\Delta_i(r_i)$, but does not say
anything about the shape of this function or its physical
origin. Nevertheless, it is important to note that the final density
profile is entirely determined by this initial condition. In the
spherical collapse paradigm, the case for universality in the radial
profiles of dark matter haloes can be reformulated in terms of
universality in the primordial density fluctuations that fix  $\Delta_i(r_i)$.

\citet{HS85} suggested that, according to the hierarchical scenario of
structure formation, haloes should collapse around local maxima of the
{\em smoothed} density field. The statistics of peaks in a Gaussian
random field has been extensively studied in the classical paper by
\citet[hereafter BBKS]{BBKS86}. One of the best known results of BBKS
is the expression for the radial density profile of a fluctuation 
centred on a  primordial peak of arbitrary height: 
\be
\frac{\left<\delta_f(r)\right>}{\sigma_0} = \nu\psi(r)-
\frac{\theta(\gamma,\gamma\nu)}{\gamma(1-\gamma^2)}
\left[ \gamma^2\psi(r) + \frac{R_*^2}{3}\nabla^2\psi(r) \right]
\label{ecBBKS}
\ee
where $\sigma_0\equiv\xi(0)^{1/2}$ is the rms density fluctuation, $\psi(r)\equiv\xi(r)/\sigma_0$ is the normalised two-point correlation function, $\nu\sigma_0$ is the height of the peak, and $\gamma\equiv\sigma_1^2/(\sigma_2\sigma_0)$ and $R_*\equiv\sqrt{3}\sigma_1/\sigma_2$ are related to the moments of the power spectrum
\be
\sigma_j^2 \equiv \frac{1}{2\pi^2}\int_0^\infty{P(k)~k^{2(j+1)}~\dd k}
\ee

The function $\theta(\gamma,\gamma\nu)$ parameterises the second derivative of the density fluctuation near the peak. BBKS suggest the approximate formula
\be
\theta(\gamma,\gamma\nu) \equiv -\frac{\left<\nabla^2\delta\right>}{\sigma_2}-\gamma\nu
\simeq
\frac{3(1-\gamma^2)+(1.216-0.9\gamma^4)\exp[-\frac{\gamma}{2}(\frac{\gamma\nu}{2})^2]}
     { \left[ 3(1-\gamma^2)+0.45+\frac{\gamma\nu}{2} \right]^{1/2} + \frac{\gamma\nu}{2}}
\ee
valid to $1\%$ accuracy in the range $0.4<\gamma<0.7$ and $1<\gamma\nu<3$, which is the scale relevant for both galaxies and galaxy clusters.

Expression (\ref{ecBBKS}) is often quoted in the literature\footnote{Some authors (Hoffman, private communication) use indeed the prescription described below. Yet (\ref{ecBBKS}) is stated in their paper to determine the initial conditions.}
\citep[e.g.][]{Hoffman88,LokasHoffman00,Popolo00,Hiotelis02} as the initial condition $\Delta_i(r_i)$, even though BBKS {\em explicitly} warn in their paper 
\begin{quote}
``Unless the filter is physical (as in pancake models), these profiles cannot be used for hydrodynamic or N-body studies of collapse since substructure must be included''.
\end{quote}

Although BBKS referred to a different problem (see our discussion in Section~\ref{secSCcomp} regarding the effects of merging and substructure), the point we want to stress is that (\ref{ecBBKS}) describes the {\em smoothed} density profile around a local maximum of the smoothed density field,
\be
\delta_f(\vec r)=\int{ W_f(\vec r-\vec x)\delta(\vec x) \dd^3\vec x}
\label{ecSmooth}
\ee
where the function $W_f(\vec r-\vec x)$ is a smoothing kernel that depends on a certain filtering scale $R_f$.

However, the smoothed density profile $\left<\delta_f(r)\right>$, given by equation (\ref{ecBBKS}), is {\em never} equal to the mean value of the local overdensity $\left<\delta(r)\right>$ that we need to integrate in order to compute $\Delta_i(r_i)$:
\be
\Delta_i(r_i)=4\pi\int_0^{r_i}\left<\delta(\vec x)\right>x^2\dd x
\label{ecDelta0}
\ee

Comparing this expression with (\ref{ecSmooth}), we see that $\Delta_i(r_i)$ is equivalent to $\delta_f(0)$ as long as $W_f$ is taken to be a spherical top hat function of radius $r_i$. In phase space, the Fourier transform of this function results in significant oscillations near the cut-off scale, so it is better to choose a Gaussian filter
\be
W_f(\vec r-\vec x)=\frac{1}{{(2\pi R_f^2)}^{-3/2}}~\exp{\left(-\frac{|\vec r-\vec x|^2}{2R_f^2}\right)}
\label{ecGauss}
\ee
with $R_f=0.64r_i$, such that the enclosed mass is approximately the same as in (\ref{ecDelta0}).

We are interested in the {\em physical} density profile around a local maximum of the {\em smoothed} density field. First, we fix the scale of the fluctuation $R_f$. Then, according to (\ref{ecDelta0}), $\Delta_i(r_i)=\delta_r(0)$, where $\delta_r(0)$ is the central density smoothed on a scale $r_0=0.64r_i$. Furthermore, we want to impose the condition that $r=0$ is a maximum of the density field $\delta_f(\vec r)$ smoothed on a scale $R_f$. BBKS show that the probability distribution of $\delta_r(0)$ is a Gaussian with mean
\be
\left<\delta_r(0)\right>= \nu_f\frac{\sigma_{0h}^2}{\sigma_{0f}} -
  \frac{\gamma_f\theta(\gamma_f,\nu_f)}{1-\gamma_f^2}\frac{\sigma_{0h}^2}{\sigma_{0f}}\left(1-\frac{\sigma_{1h}^2~\sigma_{0f}^2}{\sigma_{0h}^2~\sigma_{1f}^2}\right)
  \label{ecIC1}
\ee
for $r_0>R_f$ and
\be
  \left<\delta_0(0)\right>= \nu_f\frac{\sigma_{0h}^2}{\sigma_{0f}}
  \label{ecIC2}
\ee
for $r_0<R_f$. The moments
\be
\sigma_{jx}^2\equiv \frac{1}{2\pi^2}\int_0^\infty{P_x(k)~k^{2(j+1)}~\dd k}
\ee
are computed from
\be
P_f(k)\equiv P_{\Lambda{\rm CDM}}(k)~\exp\left[-(kR_f)^2\right]
\ee
and
\be
P_h(k)\equiv P_{\Lambda{\rm CDM}}(k)~\exp\left[-k^2~\frac{R_f^2+r_0^2}{2}~\right]
\ee

Thus, instead of (\ref{ecBBKS}), we have used (\ref{ecIC1}) and (\ref{ecIC2}) to set the initial condition $\Delta_i(r_i)$, assuming that the cluster collapses from a $\nu_f\sigma_{0f}$ peak on scale $R_f$.

%___________________________________________________

\subsection{Angular momentum}
\label{secSCj}

Two decades after the seminal paper by \citet{GG72}, \citet{WhiteZaritsky92} introduced the idea \citep[although see\footnote{If you can.} also the pioneering work of][]{GurevichZybin88} that angular momentum prevents dark matter particles from reaching the origin.

For pure radial orbits, the mass in the centre is dominated by the contribution $\Madd$ from the outer shells when the profile at turn-around is shallower than isothermal \citep{FG84}. The final density distribution found by these authors was $\rho(r)\propto r^{-2}$, independent on the initial logarithmic slope. More recently, \citet{Lokas00} and \citet{LokasHoffman00} found a similar result considering non-self-similar primordial fluctuations based on the peak formalism. \citet{Popolo00} obtains slightly shallower profiles, but
they rely on \citet{WhiteZaritsky92} to override the 'crowding effect' associated to radial orbits.

Angular momentum arises in linear theory \citep{White84} from the misalignment between the asymmetry of the collapsing region (i.e. the inertia tensor) and the tidal forces it experiences during the lineal regime. Also, violent relaxation \citep{LB67} transforms the radial infall energy into random velocity dispersion, due to two-body encounters taking place during shell crossing.

The total amount of angular momentum acquired by the system is, however, an open question. The usual approach \citep{AFH98,Nusser01,Hiotelis02} consists in assigning a specific angular momentum at turn-around
\be
j\propto\sqrt{GMr_m}
\ee

With this prescription, the orbit eccentricity $e$ is the same for all particles in the halo \citep{Nusser01}. As can be seen in Figure~\ref{figJ}, this is a fair approximation for all but two of our simulated clusters. Therefore, we will assume hereafter a constant eccentricity $e=0.5$ for the orbits of every dark matter particle after turn-around. For this value of the eccentricity, the pericenter radius of the orbit will be given by
\be 
r_{\rm min}=\frac{1-e}{1+e}~r_{\rm max}=\frac{r_{\rm max}}{3}
\label{ecRmin}
\ee
where $r_{\rm max}$ is computed from adiabatic invariance (\ref{ecFr}), following the procedure explained in Section~\ref{secSC}. In this case, angular momentum adds the usual effective term $j^2/(2r^2)$ to the gravitational potential $\Phi(r)$ to be substituted in equation (\ref{ecPr}). Although this changes the actual value of the radial action, $J_r$ is still proportional to $\sqrt{r_mM(r_m)}$. Therefore (\ref{ecFr}) still can be used to compute the collapse factor $F(r)$. The assumption of spherical symmetry leads to angular momentum conservation, which implies constant $e$ during the contraction. 

Moreover, since the radial coordinate of the particle changes only a factor of three (\ref{ecRmin}), the approximation $M(r)\propto r^{\Am}$ provides a reasonable description of the mass profile as long as the local value of the logarithmic derivative of the mass is used to set $\Am(r)$.

%__________________________________
\begin{figure}
  \centering \includegraphics[width=12cm]{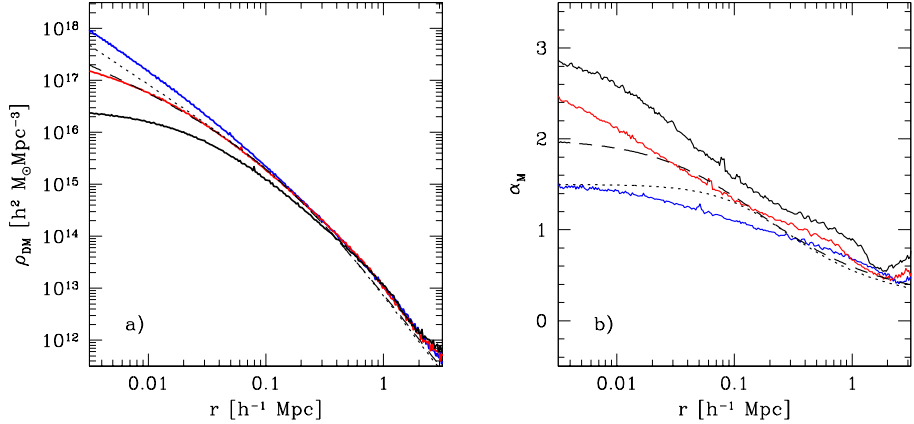}
  \caption[Mass distribution expected from spherical collapse]{Mass distribution resulting from a $3\sigma$ fluctuation on $1\ h^{-1}$ Mpc (solid lines). Black colour is used for $e=0.25$, red for $e=0.5$ and blue for pure radial orbits ($e=1$). NFW (dashed lines) and Moore et al (dotted lines) profiles are shown for comparison. Left panel: Radial density profile. Right panel: Logarithmic slope \am of the mass profile.}
  \label{figSCj}
\end{figure}
%__________________________________

In Figure~\ref{figSCj}, the final profile obtained for a $3\sigma$ peak in the primordial density field (smoothed on $1\ h^{-1}$ Mpc scales) is plotted for different values of the orbit eccentricity. As pointed out by \citet{WhiteZaritsky92} and \citet{Hiotelis02}, angular momentum plays a key role during secondary infall, preventing dark matter particles from reaching the centre of the halo and thus decreasing the amount of mass in the inner regions contributed by recently accreted shells.

As the amount of angular momentum is increased ($e$ is lowered), we
find that the predicted density profile becomes increasingly shallow,
in agreement with the results of \citet{Hiotelis02}. Pure radial orbits
give rise to a profile somewhat similar to the form proposed by
\citet{Moore99}, although the exact shape is slightly different (see
the right panel of Figure~\ref{figSCj}, where the logarithmic
derivative of the cumulative mass is plotted as a function of
radius). When the eccentricity is too low (as for example $e=0.25$,
represented by solid black lines in the figure), the halo develops a
constant density core inconsistent with the results of numerical
simulations. A value $e=0.5$, typical of our clusters (see
Section~\ref{secJsims}), yields a mass distribution almost
indistinguishable from the NFW formula up to $0.02\ h^{-1}$ Mpc, where
the logarithmic slope predicted by spherical collapse keeps increasing
towards the centre while NFW tends to an asymptotic value $\Am=2$.

This is precisely the kind of behaviour we see in the innermost part of
our relaxed clusters (see Figure~\ref{figDM}).  We cannot be totally 
sure that this is not a spurious effect caused by our finite numerical
resolution. Therefore, we  suggest  that a smoothing length of $\sim1\
h^{-1}$ kpc  would be needed to  confidently distinguish between both
profiles. We note, however, that this is only marginally beyond the
resolution of the numerical experiments presented here. We plan to test
this result with simulations from same initial conditions but with
another mass refinement level (i.e. effective  $1024^3$ particles
and $\epsilon \sim 200 pc$.)

%___________________________________________________

\subsection{Comparison with simulations}
\label{secSCcomp}

In principle, it is difficult to understand why the spherical collapse
picture should be able to predict the density distribution of dark
matter haloes, since the assembly of structures in the universe is not
expected to proceed by the infall of spherical shells, but rather
through a hierarchical series of merging events. Nothing of the
symmetric and ordered process envisaged by this simple formalism is
actually confirmed by the numerical simulations, in which the collapse
of haloes seems to take place in an extremely unordered and random
way. 

However, \citet{Zaroubi96} point out that the very complicated coalescence process looks very regular in energy space, where there is a high correlation between the initial an final energies of individual particles. In particular, the ranking in energy space is roughly preserved, an idea already proposed by \citet{QuinnZurek88} and \citet{Hoffman88}.

Furthermore, \citet{Moore99} simulated the collapse of a galaxy
cluster truncating the power spectrum of primordial fluctuations at a
scale $\sim4$ Mpc. The lack of small scale power causes the halo to form
via a single monolithic collapse rather than through mergers and
accretion of smaller haloes. Since the final density profile was very
similar to that of the same halo with the standard power spectrum, they
concluded that details of the merging history do not affect the final
density profiles. 

Predictions of spherical infall model have been confronted with numerical simulation in several occasions. Early work by \citet{QSZ86} found that in general the simple model correctly predicts the structure of bound objects, at least in a statistical sense. \citet{Zaroubi96} accomplished a series of 'simulations' in which forces were computed assuming spherical symmetry, obtaining the same results as a standard Tree N-body code (but, on the other hand, their analytic predictions did not agree so well with the numerical results). More recently, \citet{LokasHoffman00}, \citet{Popolo00} and \citet{Hiotelis02} obtained density profiles comparable to the NFW form using different implementations of the spherical collapse model. 

\begin{table}
\begin{center}
\begin{tabular}{lcrr}
{\sc Cluster} & {\sc State} & $\nu$~ & $R_f$~ \\ \hline\\[-2mm]
~~   A   &  Minor  & 3.5 & 1.3 \\
~~   B   &  Minor  & 3.4 & 1.2 \\
~~   C   & Relaxed & 3.4 & 1.3 \\
~~   D   &  Minor  & 3.2 & 0.9 \\
~~   E   & Relaxed & 2.8 & 1.4 \\
~~   H   & Relaxed & 2.5 & 1.2 \\
~~ J$_1$ & Relaxed & 2.7 & 1.0 \\
~~ K$_1$ & Relaxed & 2.7 & 0.9 \\
~~   L   &  Minor  & 2.7 & 0.7 \\
~~ K$_2$ &  Minor  & 2.5 & 0.9 \\
~~ J$_2$ & Relaxed & 2.8 & 0.8 \\
\end{tabular}
\end{center}
\caption[Spherical collapse parameters]{Spherical collapse parameters (peak height $\nu$ and smoothing scale $R_f$).}
\label{tabSCcomp}
\end{table}

In order to test the validity of the formalism outlined above, we now try to reproduce the individual density profiles of our relaxed and minor merging clusters. Assuming a fixed value of the orbit eccentricity, the model described in this section has two free parameters, which in this case have a clear physical meaning: namely, the smoothing scale $R_f$ and height $\nu$ (in units of sigma) of the primordial density peak from which the cluster is formed.

In Table~\ref{tabSCcomp}, we report  the best-fit values found for our clusters. Most of them can be described as structures formed from $\sim3\sigma$ peaks of the primordial density field, smoothed on $\sim1\ h^{-1}$ Mpc scales. It is interesting that there exists a certain degeneracy between the parameters $R_f$ and $\nu$ for a given mass. However, a higher peak on a smaller smoothed scale corresponds to an earlier formation time. The inner density profile of such peak would be much steeper than that of a similar object formed (at a later redshift) from a lower primordial peak on a larger smoothing scale. 

%__________________________________
\begin{figure}
  \centering \includegraphics[width=8cm]{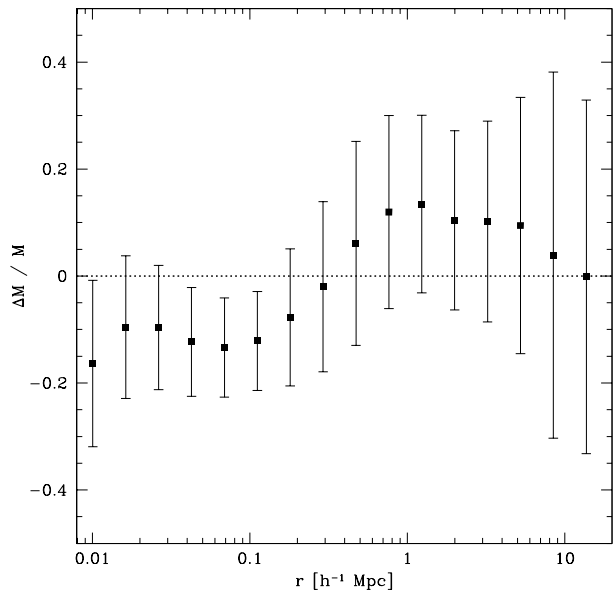}
  \caption{Accuracy of the spherical collapse prediction}
  \label{figAccuSC}
\end{figure}
%__________________________________

The average accuracy of the mass profiles predicted by our model based on spherical infall is shown in Figure~\ref{figAccuSC}. Individual profiles for each one of the clusters can be found in Appendix~\ref{apIndiv}. Although the overall quality  of the fit 
 is only slightly lower than that of the NFW fit (approximately $10\%$, compared to $5-10\%$ reported in Section~\ref{secDMprof}), there exists a large scatter between different dark matter haloes. This additional uncertainty rises considerably the error in the mass profiles of individual clusters, which can be as large as $\sim25\%$. 

Nonetheless, the spherical collapse formalism is able to \emph{predict} the mass distribution from a physically-motivated description, whereas the formulas proposed by NFW or Moore et al. are mere phenomenological fits (valid for a much
narrower radial range). Therefore, it constitutes a promising tool to understand the origin of the so-called 'universal' density profile despite the fact that it does not include \emph{explicitly} the merging effects on structure formation. 

We conclude that a possible explanation of why the density profiles predicted by the spherical infall model might describe the real (or numerical) dark matter haloes could be that merging is taken into account \emph{implicitly} through the smoothing scale $R_f$. Contrary to the common view (see e.g. BBKS), we argue that $R_f$ has a very precise physical interpretation: dark matter haloes do not form in \emph{local} maxima of the primordial density field. These are the seeds for the very first objects in the universe, which collide amongst themselves and give birth to heavier dark matter clumps. These clumps, in turn, correspond to peaks associated to a very small smoothing scale, below which all the information about the substructure of the primordial density field is lost. 

As time goes on, these protohaloes evolve through hierarchical clustering, and the smoothing length corresponding to the resulting objects increases (i.e. all the substructure below this scale is completely erased due to violent relaxation). Our point of view is that the smoothing length associated to a dark matter halo can be interpreted as a measure of the halo formation time or, equivalently, as the scale above which the system still remembers its initial conditions. 

%-----------------------------------------------------------------------------

%-----------------------------------------------------------------------------
\chapter{Gas}
\label{chapGas}
%-----------------------------------------------------------------------------

%\begin{quote}{\em
%I said pressure drop, oh pressure!, oh yeah!\\
%Pressures gonna drop on you\\
%I said when you drop, you've gotta feel it,\\
%All that you're doing is wrong}

%-- The Specials : {\em Pressure Drop} (19...) --\\
%\end{quote}

\begin{quote}{\em
Too much pressure! Too much pressure!\\
The pressure can't stop...\\
Too much pressure!}

-- The Selecter : {\em Too Much Pressure} (1980) --\\
\end{quote}
%-----------------------------------------------------------------------------

{\gothfamily {\Huge I}{\large n}} the hierarchical galaxy formation scenario, cosmic structure grows via gravitational instabilities from small perturbations seeded in an early inflationary epoch. Since most of the mass is assumed to be in the form of collisionless cold dark matter, this yet unidentified component determines the dynamics of baryons on large scales, where hydrodynamic forces are unimportant compared to gravity.

However, if we want our simulations to describe the observable properties of cosmological objects, we must obtain information about the physical state of the baryons, either by resorting to phenomenological relations between the dark and luminous components or by direct self-consistent numerical modelling. The most basic baryonic process to be included in a numerical simulation is gas hydrodynamics, i.e. the presence of a collisional term in the Boltzmann equation representing particle interactions.

In principle, hot gas in clusters of galaxies should be easy to understand. Because of the relatively low ratio of baryons to dark matter, the potential of a cluster should be dark-matter dominated. The dynamical time within a cluster potential is shorter than a Hubble time, so most clusters should be relaxed. Also, the cooling time of vast majority of intracluster gas is longer than a Hubble time. It would appear that cluster structure ought to be scale-free, as long as the shape of a cluster's potential well does not depend systematically on its mass.

If that were the case, then the global properties of clusters, such as halo mass, emission-weighted temperature, and X-ray luminosity, would scale self-similarly \citep{Kaiser86}. Indeed, numerical simulations that include adiabatic gasdynamics produce clusters of galaxies that obey these scaling laws \citep[e.g.][]{EMN96,BN98}.

Nonetheless, real clusters are not so simple. Even on cluster scales, radiative and feedback processes may not be completely negligible. This is expressed most clearly in the X-ray luminosity-temperature relation, probably due to an increase in entropy of the innermost gas. Observations indicate a much steeper temperature dependence than that expected from pure bremsstrahlung emission, especially for low mass systems \cite[e.g.][]{David93,Ponman96}. The amount of energy injection required depends on the gas density at the time when the heating occurred, either before, during or after the collapse.

In this chapter, we address the reliability of the current numerical implementations of gasdynamics, and then present our results for the sample of galaxy clusters described in Section~\ref{secSample}. The ICM of dynamically relaxed clusters is found to be in hydrostatic equilibrium and can be described by a polytropic equation of state \citep{clusters02}. The radial profiles of gas density and temperature are also investigated, as well as the scaling relations between the global properties of our clusters. In most cases, the physical characteristics observed in the simulations show an encouraging agreement with relatively simple analytical predictions.

%-----------------------------------------------------------------------------

\section{Numerical gasdynamics}
\label{secEulerLagrange}

Since the late 1980s a variety of techniques have been developed to simulate gasdynamics in a cosmological context. In part inspired by the success of the N-body scheme, the first gasdynamical techniques were based on a particle representation of Lagrangian gas elements using the smoothed particle hydrodynamics (SPH) technique \citep{Lucy77,GingoldMonaghan77}.

Soon thereafter, fixed-mesh Eulerian methods were adapted \citep{Cen90,Cen92} and, more recently, Eulerian methods with sub-meshing \citep{BryanNorman95}, deformable moving mesh \citep{Gnedin95,Pen95,Pen98} or adaptive mesh refinement \citep[AMR,][]{Bryan95,ARThydro02} have been developed, as well as extensions of the SPH technique \citep{Shapiro96,gadgetEntro02,deva}.

These codes are actively being applied to a variety of cosmological problems, ranging to the formation of individual galaxies and galaxy clusters to the evolution of Ly$\alpha$ forest clouds and the large-scale galaxy distribution. Because the inherent complexity of gasdynamics in a cosmological context, such simulations are more difficult to validate than N-body experiments. Standard test cases with known analytical solutions (such as shock tubes) are far from the conditions prevailing in cosmological situations where the gas is coupled to dark matter, and this, in turn, evolves through a hierarchy of mergers.

The Santa Barbara cluster comparison project \citep{SB99} attempted to assess the extent to which existing modelling techniques gave consistent and reproducible results in a realistic astrophysical application, simulating the formation of a galaxy cluster in a SCDM universe with 12 completely independent codes, seven of them based on the SPH technique, while the other five employed grid methods to implement gasdynamics.

The properties of the CDM component in all modes were encouragingly similar, most of the discrepancies being due to small differences in the timing of the final major merger, which happened to be uncomfortably close to $z=0$. There was less agreement in the gas-related quantities, although in most cases they were relatively similar, usually within $10-20\%$. Only the X-ray luminosity showed a strong dependence on the resolution of the different codes, with a spread as high as a factor of 10.

The most remarkable finding was the hint of a systematic trend in the
temperature profiles obtained for the inner regions ($r\leq100$ kpc). Near
the centre, SPH codes generated a flat or slightly declining (inwards)
temperature profile, while  grid codes produced temperature profiles
that are  still rising at the resolution limit.
 The entropy profile in  SPH codes decreased continuously 
towards the centre,  while grid codes develop 
an isentropic core at small radii. 

Several authors
\citep[e.g.][]{Hernquist93,NelsonPapaloizou93,NelsonPapaloizou94,Serna96,gadgetEntro02,deva}
have pointed out that an important shortcoming of conventional SPH
formulations is the poor conservation of entropy when a low number of
particles is used. In this section we intend to test the accuracy of
both Eulerian and Lagrangian numerical approaches for integrating 
the equations of gasdynamics. To this end,  we compare  results of the 
adaptive mesh code ART \citep{ARThydro02}, (see Section \ref{secART}
 for a  description) with two different
implementations of the Lagrangian code \G; one of them is the
publicly-available version of the code \citep{gadget01}, which
implements standard SPH gasdynamics, and the other is the explicit
entropy-conserving scheme described in \citet[ see also
Section \ref{secGadget}]{gadgetEntro02}. 

%___________________________________________________

\subsection{Santa Barbara cluster}

As a  first step  to test our codes, we have chosen the Santa Barbara
cluster initial conditions, corresponding to a $3\sigma$ peak in the
primordial density field smoothed on a 10 Mpc scale. Details concerning
the numerical experiments accomplished at different mass and spatial
resolutions can be found in Section~\ref{secExperSB}. 

We have plotted in Figure~\ref{figGasSB} the results of all these
experiments, using the criterion of 100 gas particles, 200 dark matter
particles and $r>3\epsilon$ to define the minimum radius in each case. In
particular, we compare the results of the Eulerian code ART \citep[][
triangles]{ARThydro02} with the standard and entropy-conserving
implementations of the SPH algorithm available in \g (red and blue
lines, respectively). The average of the simulations presented in
\citet{SB99} is shown by the solid squares. The results obtained by the
adaptive-mesh code of \citet{Bryan95} are plotted separately
(circles). 

%----------------------------------

\subsubsection*{Resolution}

\begin{figure}
  \centering \includegraphics[width=12cm]{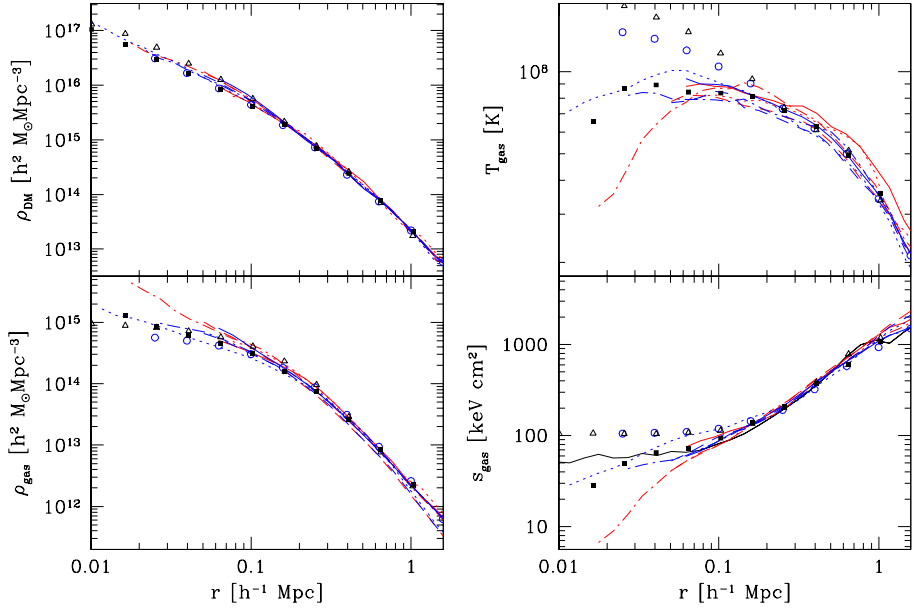}
  \caption[Simulations of the Santa Barbara Cluster]{Simulations of the
    Santa Barbara Cluster. Top left: Dark matter density. Bottom left:
    Gas density. Top right: Gas temperature. Bottom right: Gas
    entropy. Solid squares denote the average of the Santa Barbara
    experiment \citep{SB99}. Lines represent \g results for simulations
    with different mass resolutions and smoothing (see text),
    triangles are used for ART and circles for \citep{Bryan95}.
}
  \label{figGasSB}
\end{figure}

The profiles for the different numerical experiments of the Santa Barbara cluster reported in Table~\ref{tabSBsims} can be seen in Figure~\ref{figGasSB}. Blue lines are used to represent the simulations with $256^3$ particles and $\epsilon=2\ h^{-1}$ kpc (dotted), $128^3-5\ h^{-1}$ kpc (dotted-dashed), and $128^3-20\ h^{-1}$ kpc (solid). Red lines (standard SPH) correspond to $128^3-20\ h^{-1}$ kpc (solid), $128^3-5\ h^{-1}$ kpc (dotted-dashed), and $64^3-20\ h^{-1}$ kpc (dotted). The black line on the bottom right panel indicates the entropy profile obtained in the low mass resolution simulation with $64^3$ particles run with the standard SPH implementation of \G, ignoring the resolution limit.

%__________________________________
%__________________________________

In most cases, all the results are consistent when plotted according to the restrictions prescribed by \citet{Klypin01}. Nonetheless, the two experiments with the highest nominal spatial resolution (5 $h^{-1}$ kpc and $256^3$ particles with standard and entropy-SPH) develop a steep decrease in the central temperature. Since we have checked that the centre of mass of both gas and dark matter distributions are approximately coincident (see e.g. the high gas densities obtained for the standard SPH implementation), this effect is probably 
caused by an artificially low smoothing length, which leads to very
compact groups of cold particles that decouple from the surrounding
medium. Although Eulerian simulations are not affected by this problem,
the unphysical temperature drop described here is often seen in
clusters simulated with SPH-based codes at low and medium resolutions
\citep[e.g.][]{Eke98,Yoshikawa00,ME01}, and in some of the codes
reported in  \citet{SB99}.  

If the smoothing scale is not set excessively low, but a small number of particles is used, the resulting cluster displays an extended core of constant density and temperature up to the minimum radius we have used for our plot. This softening-dominated core yields a flat entropy profile at the inner regions (black line in Figure~\ref{figGasSB}), pointing to the misleading conclusion that this feature of clusters and groups of galaxies canç be reproduced with a fairly low resolution. The same effect has been shown in cluster simulations with variable resolution by \citet{Borgani02}.

%----------------------------------

\subsubsection*{Code comparison}

A different issue concerning the entropy profile of galaxy clusters is
whether the computations based on different integration schemes give
consistent results, for the same  numerical resolution. This
topic has been previously discussed by \citet{Kang94}, \citet{SB99},
\citet{gadgetEntro02} or \citet{deva}, among others.

The maximum number of particles allowed by the Santa Barbara initial conditions was $256^3$, which seems  insufficient to discern the presence of a physical flattening of the entropy from the effects of numerical smoothing. However, we confirm the trends shown in \citet{SB99}: Eulerian codes predict a systematically rising temperature profile near the centre, while SPH algorithms are more consistent with isothermality. 

This conclusion is strengthened by the fact that the two completely independent grid-based methods used in this comparison \citep[i.e. ART and][]{Bryan95} give very similar results, while both implementations of \g show profiles consistent  with the average value given in \citet{SB99}, which was dominated by SPH experiments. 

Nevertheless, it seems that the highest resolution simulations using the entropy-conserving version of \g develop a certain tendency towards higher central temperatures, displaying an increasing profile up to $\sim60$ $h^{-1}$ kpc. At this point, there is a sharp break and the temperature drops again in the inner part of the cluster. Quite interestingly, the entropy profile at that radius shows remarkable signs of flattening \citep[a fact also noted by][ who try to overcome the entropy problem by including the so-called $\nabla h$ terms arising from the spatial dependence of the SPH smoothing length]{deva}. In any case, more resolution is needed in order to study the structure of the ICM gas at even lower scales and make a reliable assessment on whether the entropy profile produced by these improved formulations of SPH is consistent with the results of Eulerian codes.

%___________________________________________________

\subsection{Cluster A}

%__________________________________
\begin{figure}
  \centering \includegraphics[width=12cm]{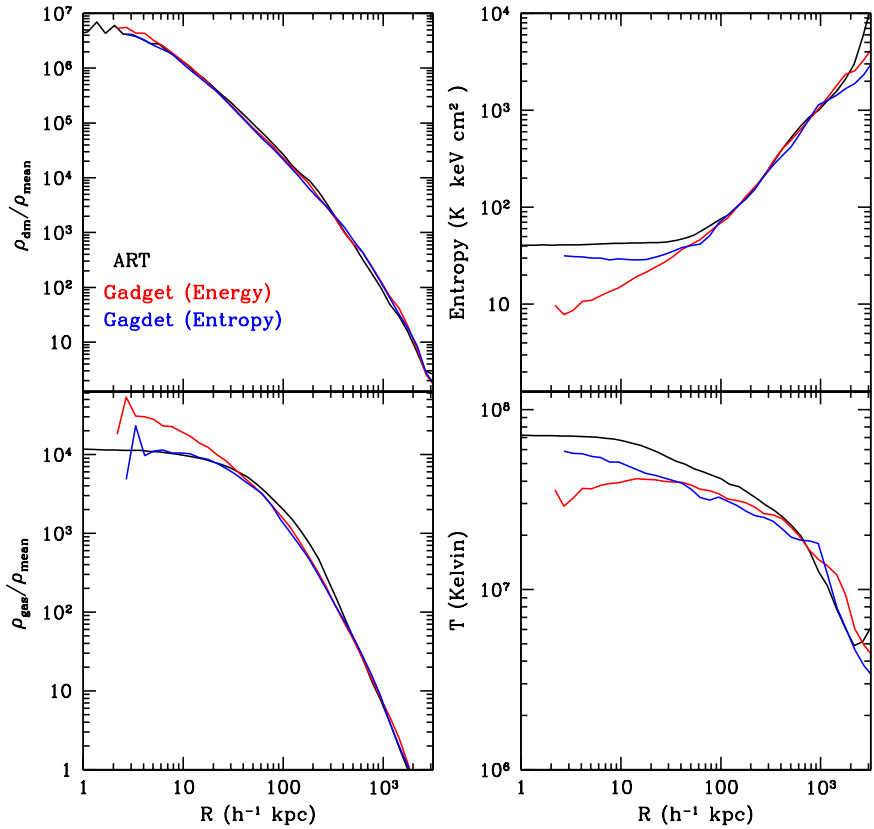}
  \caption{Simulations of Cluster A}
  \label{figGasClusterA}
\end{figure}
%__________________________________

Following the spirit of the Santa Barbara Cluster Comparison Project,
we have run several numerical experiments from the initial conditions
that  produced   the most massive  galaxy cluster of our catalogue, 
as described in Section~\ref{secSample}. 
In this case, the number of particles within
the virial radius is approximately a factor of 8  larger than in the
Santa Barbara cluster. Hence we have been 
 able to resolve scales much deeper into the cluster's potential. 

Figure~\ref{figGasClusterA} shows the results obtained with ART
\citep{cluster6} and both versions of \g for the radial profiles of
Cluster A. Here, the difference between energy and entropy conservation
is mostly evident at low radii (below $\sim20\ h^{-1}$ kpc). As noted in
previous studies \citep[e.g.][]{deva}, the density profile is
steeper in the standard formulation of SPH, whereas the temperature is
systematically higher when entropy conservation in SPH  is enforced. 

Consequently, the entropy profile is dramatically dependent on the
chosen implementation. In top right panel of
Figure~\ref{figGasClusterA}, it can be clearly seen that the entropy
profile of the explicit entropy-conserving scheme by
\citet{gadgetEntro02} unambiguously flattens in the central regions of
the simulated cluster, while the standard SPH decreases uninterruptedly
down  to the resolution limit. 

Since this effect is already manifest at clustercentric distances that
are confidently resolved, we claim that there is a fundamental
difference in the predictions based on Eulerian and Lagrangian codes
due to the unphysical entropy loss in the latter. The fact that the
entropy-conserving version of \g produces an entropy floor comparable
to that observed in ART gives reassuring support to the idea that this
numerical artifact is corrected in this formulation of the
SPH algorithm. 

%-----------------------------------------------------------------------------

\section{Structure of the ICM gas}

It has been twenty years since the realization that the extended X-ray emission from clusters \citep{Kellogg72} is thermal and arises from optically thin plasma filling the clusters \citep{Mitchell76}. If this plasma is in hydrostatic equilibrium within the gravitational potential created by the dark matter, the spatial distribution of the ICM density and temperature constitutes an invaluable source of information about the process of structure formation on large scales.

Most importantly, since clusters of galaxies are the largest
gravitationally bound structures in the universe, their physical
properties  have profound cosmological implications, and have
often been proposed as a probe to test the values of the cosmological
density parameters as well as the validity of the hierarchical CDM
scenario \citep[e.g.][]{White93}. If well calibrated, the slope and
evolution of cluster scaling relations, such as size, gas mass or
luminosity versus temperature, can be used to constrain cosmological
and structure formation models. Temperature profiles are also a
fundamental ingredient in the determination of the total mass, as well
as the gas entropy distribution, which is a powerful tool to  explore
non-gravitational processes that could alter the specific thermal
energy in the ICM. 

It is therefore important to investigate in detail the structure of the
intra-cluster medium. Throughout this section, we will assume spherical
symmetry and focus on the radial profiles of several quantities of
physical interest: first, we will address the validity of the usual
assumptions of hydrostatic equilibrium, isothermality, and polytropic
equation of state. Then, the average density and temperature profiles
of our  galaxy cluster catalogue  (whose main properties are described
in Section~\ref{secSample}) will be compared with  predictions based
on the existence of a universal density profile for the dark matter
component. 

%___________________________________________________

\subsection{Physical state}
\label{secHydroPolyt}

One of the hypothesis that are often used in the study of galaxy clusters is that gas is in hydrostatic equilibrium with the mass distribution dominated by the dark matter. \citet{Allen98} suggested that this assumption is only valid for cooling-flow clusters, which has been corroborated by recent observations with \emph{Chandra} and \emph{XMM-Newton}. 

Most analytical studies on the radial profiles of the ICM gas rely on
hydrostatic equilibrium, in many occasions supplemented by a polytropic
or isothermal equation of state for the gas. As we briefly discussed in
Section~\ref{secCentralDens}, these prescriptions are used, for
example, to estimate the gravitating mass from the observed X-ray
emission. It is therefore of crucial importance that the accuracy of
these assumptions is investigated by means of direct numerical
simulation. 

The main difficulty in studying the properties of gas in clusters
formed in numerical experiments is the of lack of resolution
\citep[e.g.][]{ThackerCouchman01}. Recently, \citet{Loken02} have  made a
detailed analysis of the gas temperature profile for a sample of 20
clusters simulated with a resolution ($<15h^{-1}$  kpc) which is 
 comparable to the present work, using an AMR gasdynamical code. 

A completely different problem is the implementation of all the
relevant physics involved in cluster formation. Radiative cooling and
stellar feedback play a major role on galactic scales, as well as in
the dense central parts of  galaxy clusters. Outside the core,
though, cooling times are longer than a Hubble time, and the effect of
cooling on the ICM structure is not expected to be significant, at
least on a first approximation. 

On the other hand, there are still many open questions regarding the assembly of galaxy clusters even when only adiabatic physics is taken into account (see e.g. the previous section). Before including any radiative process, it is necessary to make sure that the physics associated to gravothermal collapse and shock-wave heating are accurately modelled. This also applies to the many theoretical predictions that can be derived with this 'simple' physics, which are often used as a benchmark to be compared with the observed properties of real clusters of galaxies.

%----------------------------------

\subsubsection*{Hydrostatic equilibrium}

Assuming that the gravitational potential is counter-balanced by thermal support only (i.e. bulk -- infall and/or rotation -- and turbulent motions of the gas are negligible with respect to its internal energy), the hydrostatic equilibrium equation for a spherically symmetric system reduces to the simple form
\be
\frac{1}{\rho_g}\frac{\dd P}{\dd r}= -\frac{GM(r)}{r^2}
\label{ecGasEqHydro}
\ee
where $\rho_g(r)$ is the gas density, $P(r)=\frac{\rho_g(r)kT(r)}{\mu m_p}$ is the pressure and $M(r)$ is the total (gas + dark matter) mass enclosed within radius $r$. Since the dark component dominates the mass of the cluster at all radii, the cumulative mass can be approximated by any of the 'universal' profiles described in Section~\ref{secNFWmoo}.

%__________________________________
\begin{figure}
  \centering \includegraphics[width=12cm]{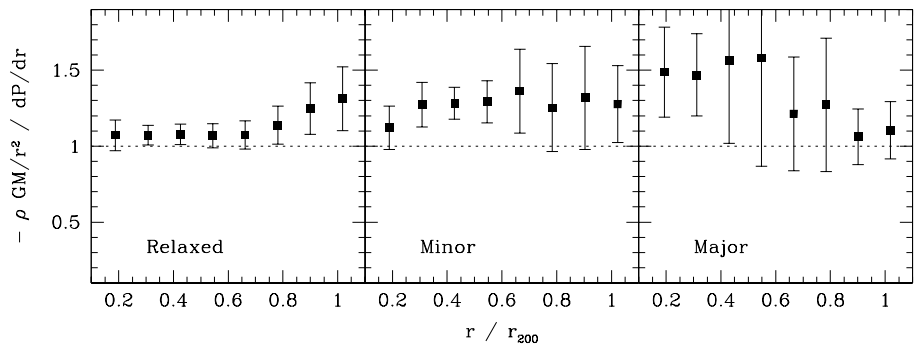}
  \caption{Test of the hydrostatic equilibrium assumption}
  \label{figEqHydro}
\end{figure}
%__________________________________

We have checked the reliability of the usual assumption of hydrostatic equilibrium in Figure~\ref{figEqHydro}, where the ratio of the pressure gradient to the gravitational term ($\rho GMr^{-2}$) is plotted as a function of clustercentric radius. If the gas was in thermally-supported hydrostatic equilibrium, this quotient should be equal to unity by virtue of equation (\ref{ecGasEqHydro}).

For clusters that have been classified as \emph{Relaxed} systems on dynamical grounds (see Section~\ref{secSample}), hydrostatic equilibrium can be applied with a reasonable degree of accuracy ($\sim10\%$). Deviations from the behaviour predicted by (\ref{ecGasEqHydro}) are due to the oversimplifying assumption that angular momentum and turbulent motions do not contribute significantly to support the gas against gravity. Besides, near the virial radius ($r\geq0.8\Rv$), radial infall cannot be completely neglected.

As can be seen in the middle and right panels of Figure~\ref{figEqHydro}, the assumption of hydrostatic equilibrium holds only marginally for minor mergers, and not at all for clusters that are undergoing a major merging event. At this point, it is important to recall that non-relaxed systems constitute 60\% of our sample. Even if the error is not extremely large (particularly, in the case of minor mergers), caution must be kept in mind whenever equation (\ref{ecGasEqHydro}) is applied to a real cluster ensemble.

%----------------------------------

\subsubsection*{Polytropic equation of state}

In many occasions, hydrostatic equilibrium is used in combination with
a simple equation of state. Usually, the gas is assumed to follow a
polytropic law
\be
P=\frac{\rho_gk_BT}{\mu m_p}=P_0\rho_g^\gamma
\label{ecPolyt}
\ee
where $\gamma$ is the polytropic index and the normalisation $P_0$ is an arbitrary constant. The value $\gamma=1$ corresponds to the special case of an isothermal gas, and is often seen in the literature, both in analytical and observational studies.

If the ICM gas could be described by the polytropic form (\ref{ecPolyt}), the radial temperature and the density profiles are related by
\be
\frac{T}{T_0}=\left(\frac{\rho_g}{\rho_0}\right)^{\gamma-1}
\label{ecTrho}
\ee
where $\rho_0$ and $T_0$ are the central values of both magnitudes.

In real clusters, there is no strong physical reason to expect that the gas should be described by a polytropic equation of state. However, the appealing simplicity of equation (\ref{ecPolyt}) makes it extremely useful in analytical work, and hence it would be interesting to test whether it can be used to interpret the results on a first-order approximation.

The most straightforward way to check whether the ICM follows a polytropic relation is to compute the \emph{local} value of the polytropic index 
\be
 \gamma(r)=1+\frac{{\rm d}\log(T)}{{\rm d}\log(\rho)}
\ee

%__________________________________
\begin{figure}
  \centering \includegraphics[width=12cm]{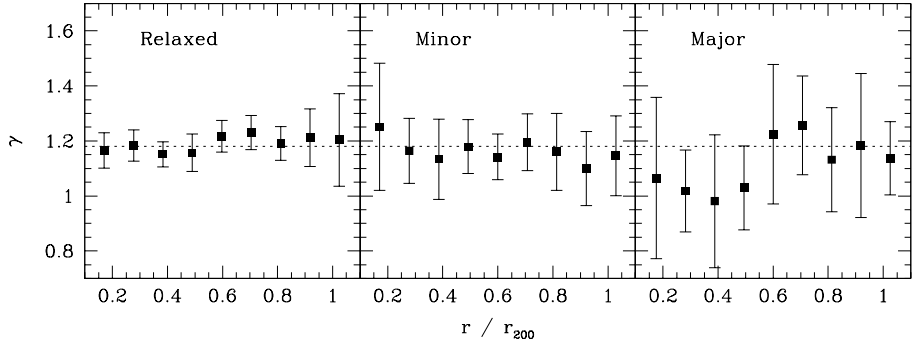}
  \caption{Polytropic index as a function of $r$}
  \label{figPolyt}
\end{figure}
%__________________________________

Figure~\ref{figPolyt} displays the radial dependence of $\gamma$ for our sample of simulated galaxy clusters. Albeit some scatter, the radial profiles of the diffuse gas component in relaxed systems and minor mergers seem to be consistent with a polytropic equation of state with $\gamma\sim1.18$ (shown by a horizontal dotted line in the figure).

Another important issue is that an isothermal profile ($\gamma=1$) offers a very poor approximation to our results. In the section below, where the temperature structure of our clusters will be described in more detail, we will see that the constant temperature approximation can be ruled out at a significant level of several sigmas.

%___________________________________________________

\subsection{Radial profiles}
\label{secGasRprof}

In the previous chapter, we thoroughly discussed the possibility that the radial mass distribution of dark matter haloes followed a 'universal' form, finding that relaxed haloes (and even minor mergers, albeit with a lower degree of accuracy) could be described by the simple analytical fits proposed in the literature.

Once we know the dark matter distribution, the hydrostatic equilibrium equation (\ref{ecGasEqHydro}) relates the gas density and temperature profiles to the cumulative mass, which can be approximated by the dark matter mass.

Several further assumptions must be made in order to compute the gas
and temperature profiles. By far, the most common prescription is the
so-called \BM, introduced by \citet{CavaliereFusco76} to describe the
X-ray surface brightness profiles of galaxy clusters. Assuming  that
gas is isothermal and in hydrostatic equilibrium with a  dark matter
potential that follows an analytical King profile \citep{king62}), 
\bea
\rho(r)= \frac{\rho_0}{(1+x^2)^{3/2}}~~;~~x\equiv \frac{r}{r_c}\\
  M(r)=4\pi\rho_0r_c^3\left[\ln(x+\sqrt{1+x^2})-\frac{x}{\sqrt{1+x^2}}\right]\\
\phi(r)=- 4\pi G \rho_0 r_c^2 \frac{\ln[x+(1+x^2)^{1/2}]}{x}
\eea
one can derive the  gas density profile
\be
\rho_g(r)=\frac{\rho_0}{\left[1+(r/r_c)^2\right]^{3\beta/2}}, 
\label{ecBM}
\ee
where $\rho_0$ is the central density, $r_c$ is a core radius and $\beta\sim
kT/(\mu m_p \sigma_{DM}^2)$ relates the thermal energy of the gas to the
velocity dispersion of the dark matter component. This form for the
density profile $\rho_g$ has been widely used in the literature \citep[see
e.g.][for a review]{Rosati02}. 
More recently, a generalisation of the \bm to account for temperature gradients in the surface brightness profile of a polytropic gas has been derived by \citet{Ettori00}. 

As we have shown in the previous chapter, the King profile is not the
model that best fits the cluster gravitational potential of dark matter
halos in numerical simulations.    

\citet{Suto98} applied the
hydrostatic equilibrium equation to a variety of dark matter density 
distributions. Assuming a polytropic
equation of state for the gas and a NFW profile for the dark component,
they find 
\be
\frac{T(r)}{T_0}=1-B\left(1-\frac{\ln(1+\rrs)}{\rrs}\right)
\ee
and a similar (though more complicated) expression for a Moore et
al. formula. The  $B$ parameter  gives the normalisation of the 
 temperature profile, $T_0$,  as a function of  cluster dark mass,
 parametrised  in terms of the characteristic density $\rho_s$ 
and radius $r_s$  of  the NFW profile:
\be
B=\frac{\gamma-1}{\gamma}\frac{4\pi G\mu m_p\rho_s r_s^2}{k_BT_0}
\ee

Needless to say, the corresponding gas density profiles are given by (\ref{ecTrho}). $B$ fixes not only the central value $T_0$, but also the asymptotic behaviour of both gas and temperature profiles ($T(\infty)\to1-B$). If we want the baryon fraction to remain approximately constant at large radii, we must choose $B=1$ and $\gamma\sim1.2$ in order that $\rho_g\sim r^{-3}$. 

A similar  result for $B$ and $\gamma$ has been found by \citet{KS01} from the same assumptions (i.e. NFW profiles, hydrostatic equilibrium, polytropic equation of state and constant baryon fraction at large radii). The normalisation of their temperature profiles is defined as
\be
\eta(0)\equiv \frac{3kT_0R_{200}}{GM_{200}\mu m_p}= 3\frac{\gamma -1}{\gamma}\frac{c g(c)}{B} 
\ee
where $g(c)\equiv[\ln(1+c)-\frac{c}{1+c}]^{-1}$. However, these authors did not require that $B=1$ to have a constant baryon fraction. Instead, they propose a relation between the normalisation $\eta(0)$, the polytropic index $\gamma$ and the concentration parameter $c$, which can be fitted by 
\be
\eta^{-1}(0)=0.00676\tilde c^2 +0.206\tilde c +2.48
\label{ecKSeta0}
\ee
and
\be
\gamma=1.15+0.01\tilde c
\label{ecKSgamma}
\ee
with $\tilde c\equiv c_{\rm NFW}-6.5$. This fit works well for $c>3$, as can be seen in Figure~3 of \citet{KS01}, but cannot reproduce the normalisation for less concentrated halos.

%__________________________________
\begin{figure}
  \centering \includegraphics[width=12cm]{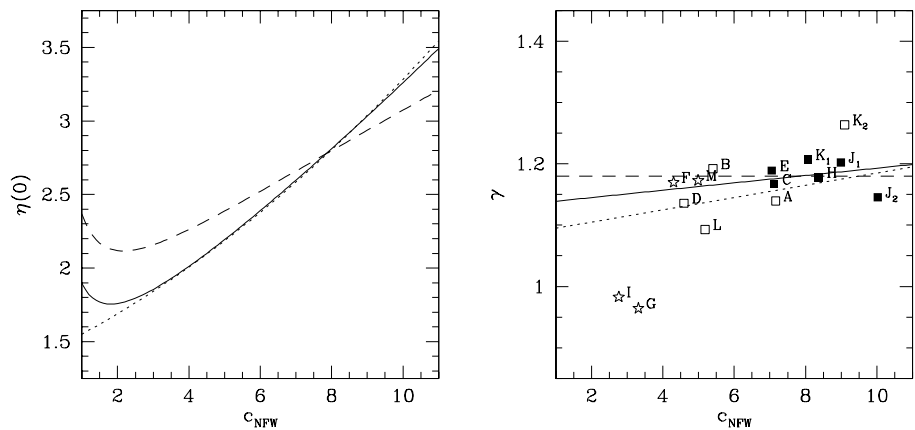}
  \caption[Temperature normalisation]{Left panel: Temperature normalisation $\eta(0)$. Right panel: Concentration dependence of the polytropic index. In both cases, solid lines are used for (\ref{ecGasGammaC}) and dashed lines for $\gamma=1.18$. Dotted lines represent the results of \citet{KS01}.}
  \label{figGasKS01}
\end{figure}
%__________________________________

In Figure~\ref{figGasKS01}, we compare (\ref{ecKSeta0}) with our prescription $B=1$, equivalent to $\eta(0)\equiv3\frac{\gamma-1}{\gamma}cg(c)$. Assuming a constant polytropic index, $\gamma\sim 1.18$, as our simulations seem to indicate (see Figure~\ref{figPolyt}), we obtain a slightly different normalisation. In order to obtain a similar prediction for $\eta(0)$, we should use the following dependence of $\gamma $ on the concentration parameter:
\be
\gamma=1.172+0.006\tilde c
\label{ecGasGammaC}
\ee

As can be seen in the figure, the match with the analytical fit (\ref{ecKSeta0}) is perfect for $c>3$. For lower values of $c$, our prescription follows the numerical results of \citet{KS01} (see Figure 3 of their paper).

In any case, both models are compatible with results from our simulated clusters. The most important conclusion drawn from these analyses is that physical properties of the gas are intimately related with the dark matter distribution. As we will show below, the radial temperature and density profiles of the ICM can be unambiguously determined from the parameters of the dark matter halo. Therefore, it is expected that they can also be characterised by 'universal' formulae.

%----------------------------------

\subsubsection*{Temperature}

Mass determinations from X-ray emission in clusters usually assume that
relaxed clusters (i.e. morphologically symmetric) are
isothermal. However, this assumption has been questioned recently by
\citet{Markevitch98}, who found evidence of a decreasing temperature
profile for nearby clusters observed with ASCA. A large subset of their
sample showed a remarkably self-similar temperature profile when
properly normalised and rescaled to the virial radius. On the contrary,
\citet{White00} and  \citet{ib00}, using data from {\em Beppo}SAX and
ASCA satellites, do not find any decrease of the temperature in a large
collection of clusters. More recently, the analysis of {\em Beppo}SAX
observations accomplished by \citet{GrandiMolendi02} concluded that
temperature profiles of galaxy clusters can be described by an
isothermal core followed by a rapid decrease. 

In this context, we would like to investigate what can we learn from the temperature profiles obtained in our numerical simulations. Particularly, we will focus on the shape of the radial temperature profiles, compared with both observational data and the simple analytical predictions based on hydrostatic equilibrium and a polytropic equation of state.

%__________________________________
\begin{figure}
  \centering \includegraphics[width=8cm]{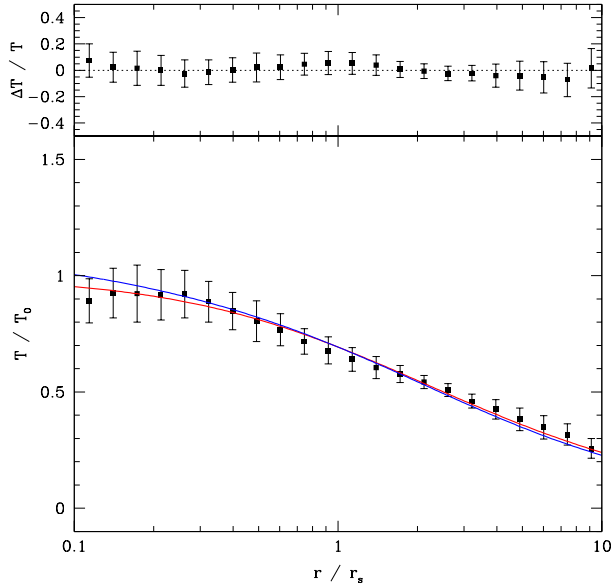}
  \caption[Temperature profile]{Average temperature profile (dots) compared to the theoretical predictions based on NFW (red line) and Moore et al. (blue line) profiles for the dark matter, assuming hydrostatic equilibrium and a polytropic equation of state. Accuracy of the NFW-based fit is shown in the top panel.}
  \label{figTgas}
\end{figure}
%__________________________________

The mass-weighted temperature of our sample of galaxy clusters (excluding major mergers) is shown in Figure~\ref{figTgas} as a function of radius. We see that, with this normalisation, the scatter is low enough ($\sim15\%$) to hint that the temperature profile can be described by a 'universal' form.

The scale factors in both axis of Figure~\ref{figTgas} have been chosen to compare with the predictions based on a NFW profile, hydrostatic equilibrium, polytropic equation of state and approximately constant baryon fraction at large radii (i.e. $B=1$, $T(\infty)\to0$):
\be
\frac{T(r)}{T_0}=\frac{\ln(1+\rrs)}{\rrs}~~;~~k_BT_0=\frac{\gamma-1}{\gamma}4\pi G\mu m_p\rho_s r_s^2
\label{ecTemp}
\ee
which is shown as a red line in Figure~\ref{figTgas}. In the top panel,
we plot deviations from this 'universal' profile. The temperature
profile derived from Moore et al. density profile is also shown for
comparison (blue line). As can be easily appreciated in the figure, the
difference is very small even near the centre. 

We would like to remark that equation (\ref{ecTemp}) {\em is not a fit} to the radial temperature profiles of our clusters, but a \emph{prediction} based on the existence of a universal dark matter distribution (in this case, given by NFW expression). The central temperatures $T_0$ have been computed assuming a polytropic index $\gamma=1.18$ for every cluster, so there are no free parameters related to the gas distribution. When only relaxed clusters are considered, the
scatter around the predicted profile (determined by the values of $\rho_s$ and $r_s$) is considerably reduced.

In spite of its physical significance, the radial temperature profile
has the disadvantage of not being a direct  measurable quantity.
 In order to make a more reliable  comparison with observations, the
projected emission-weighted temperature profiles must be computed from our
numerical data. 
More elaborate methods \citep[e.g.][]{ME01} could  be used
to mimic  the instrument-dependent observational
procedure, simulating annular spectra and estimating  the temperature
profiles in the same way  that is done in real clusters. 

%__________________________________
\begin{figure}
  \centering \includegraphics[width=8cm]{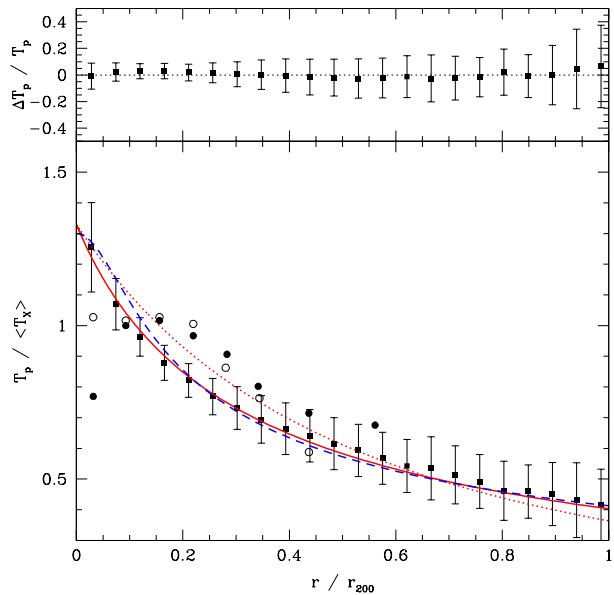}
  \caption[Projected temperature]{Averaged projected temperature profile (black squares with error bars) compared to the observational results of \citet{GrandiMolendi02} (solid circles for cooling-flow clusters, empty cicles for non-cooling flow) and \citet{Markevitch98} (dashed blue line). Red lines show the analytical fit proposed by \citet{Loken02} (see text).}
  \label{figGas2D}
\end{figure}
%__________________________________

Recent results from high resolution AMR gasdynamical simulations of
galaxy clusters \citep{Loken02} seem to indicate that the projected
X-ray temperature profiles  have  indeed an universal form which decreases
significantly with radius. These authors propose, as a best fit to
their numerical simulations, the simple   analytical formula 
\be
 T(r)=\frac{T_0}{{(1+r/a_x)}^\delta}
\label{ecLoken}
\ee
where $T_0$ is the central temperature and $a_x$ is a core radius. In
order to confront this exciting result with our numerical experiments,
we compare, Figure~\ref{figGas2D}, the averaged temperature profile 
(once again, excluding clusters with major mergers) 
projected along the line of sight normalised to the averaged
 X-ray temperature at the virial radius\footnote{
 For a detailed description on the computation of
these quantities, the reader is referred to Chapter~\ref{chapExper}.}.

Solid red line shows the best fit we obtain for our data using
expression (\ref{ecLoken}), whereas the dotted red line indicates the
radial dependence found by \citet{Loken02}. Although the results seem
to be consistent within the error bars, and the normalisation is
similar in both cases ($T_0\simeq1.33\left<\Tx\right>$), the best-fit values
of the core radius and the asymptotic exponent are extremely sensitive
to the particular details of the profile. While \citet{Loken02} quote
$a_x=\Rv/1.5$ and $\delta=1.6$ in their study, our clusters are better fit
by  $a_x=\Rv/4.5$ and $\delta=0.7$. 

As can be seen in Figure~\ref{figGas2D}, our  results are almost
identical to the observational estimates of \citet{Markevitch98}, who
 claim that they are  well represented by a polytropic \BM. But,  as 
noted by \citet{Loken02}, this is not entirely self-consistent, since
gas in the \bm is assumed to be isothermal and the data show a
pronounce temperature gradient. Nevertheless, because gas in our
cluster do follow a polytropic  relation between density and
temperature, and we expect the density of gas to be close to the
observed one, it is  not surprising that  the projected profiles agree
so well with observations.
 
However, the agreement is not so good with  the temperature profile
obtained by \citet{GrandiMolendi02} for a selection of clusters
observed with BeppoSax.
 Our clusters are marginally
consistent with their data, but we do not find any evidence of the
temperature flattening observed by these authors at  the inner regions.
 This could reflect the need of additional non-adiabatic physics in the
simulations, but it could also be  a consequence of the observational
difficulties involved in the spectral determination of the gas
temperature \citep[see the discussion in][]{ME01}. 

%----------------------------------

\subsubsection*{Density}

Assuming that the ICM gas can be described by a polytrope, the gas
density profile in hydrostatic equilibrium with a NFW dark matter
distribution is given by the formula 
\be
\frac{\rho_g(r)}{\rho_0}=\left(\frac{\ln(1+\rrs)}{\rrs}\right)^\frac{1}{\gamma-1}
\label{ecDens}
\ee
where we have taken $B=1$, as we did in (\ref{ecTemp}) for the
temperature. With this prescription, the gas density vanishes as $r\to\infty$
and the  logarithmic slope is
\be
\alpha(r)=\frac{1}{\gamma-1}\left[-1+\frac{\rrs}{(1+\rrs)\ln(1+\rrs)}\right]
\ee

Given the asymptotic behaviour at large radii,
$\alpha\sim\frac{1}{\gamma-1}\left[-1+\frac{1}{\ln(\rrs)}\right]$, it is impossible
that the gas density does \emph{exactly} follow the dark matter profile
in order to maintain a constant baryon fraction \citep[a fact already
noted by][]{KS01}. However, we should recall that NFW is not intended
(nor able) to fit the dark matter density much beyond the virial
radius, and the assumption of hydrostatic equilibrium is no longer
valid in the outer parts of the cluster (see Figure~\ref{figEqHydro}). Therefore, the only, and necessary, 
 requirement for (\ref{ecDens}) to be a good approximation to the
 ICM density is that it
had a \emph{similar} shape  to the NFW profile for a broad  radial
range around \RV. 

This led \citet{KS01} to the constraint (\ref{ecKSgamma}) on the polytropic
index of the gas, but no attempt was made to obtain a density
normalisation, arguing that the gas mass fraction does not appear in
the hydrostatic equilibrium equation, and  that there is
observational evidence suggesting  its dependence on cluster mass
\citep[e.g.][]{ArnaudEvrard99}. This makes the gas fraction
determination uncertain and hence indicates that it should be 
treated as an additional free parameter. 

Although the \emph{global} baryon fraction is not a constant among the
different clusters, we propose that a first-guess approximation to the
central density $\rho_0$ may come from the \emph{local} ratio of the gas
and dark matter densities. 

%__________________________________
\begin{figure}
  \centering \includegraphics[width=8cm]{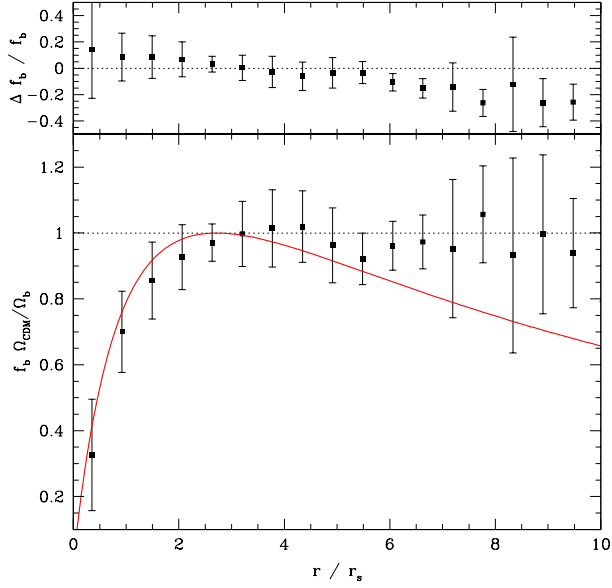}
  \caption{Radial dependence of the local baryon fraction }
  \label{figBaryonF}
\end{figure}
%__________________________________

This quantity is plotted in Figure~\ref{figBaryonF} as a function of
clustercentric radius, normalised to the cosmic baryon fraction
$\Ob/\Omega_{\rm dm}$. Not surprisingly, the radial profiles of both dark
and gaseous components are very similar throughout most of the
cluster\footnote{In fact, this was one of the basic assumptions in the
  work of \citet{KS01}.}, differing only near the origin, where
pressure is responsible for the formation of a shallow gas core. A flat
baryon fraction profile has also been observed by \citet{ASF02omega},
who applied this result to the estimation of the mean matter density of
the universe. 

The baryon fraction expected from our prescription (\ref{ecDens}), plotted as a red line in Figure~\ref{figBaryonF}, is simply
\be
\frac{\rho_g(r)}{\rho_{DM}(r)}=\frac{\rho_0}{\rho_s}\left(\frac{\ln(1+\rrs)}{\rrs}\right)^\frac{1}{\gamma-1}\rrs(1+\rrs)^2
\label{ecNormRho0}
\ee

Unfortunately, this expression does not tend to a definite asymptotic
value, but it varies slowly enough at large radii for the range of
polytropic indexes in which we are interested (as $\gamma=1.18$ in the
figure). Actually, the shape of the gas density profile would be better
modelled by slightly higher values ($\gamma\sim1.2$), resulting in a more
constant baryon fraction in the outer parts of the clusters, but we
will continue to use same value of $\gamma$ as in the temperature profiles
in order to maintain self-consistency. 

%__________________________________
\begin{figure}
  \centering \includegraphics[width=8cm]{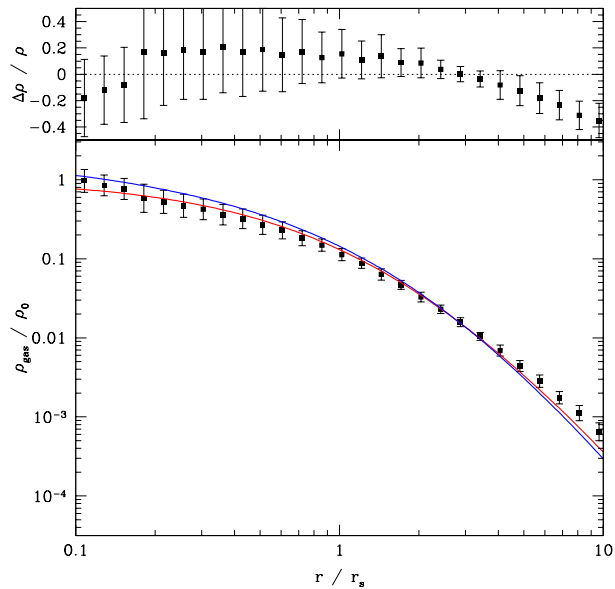}
  \caption{Average gas density profile}
  \label{figDenGas}
\end{figure}
%__________________________________

The averaged radial density profile of the ICM gas is shown in
Figure~\ref{figDenGas}, rescaled by $r_s$ on one axis and the value of
$\rho_0$ inferred from (\ref{ecNormRho0}) in the other. The normalisation
$\rho_0$ has been obtained by imposing the condition that $\rho_g/\rho_{\rm
  dm}=\Ob/\Omega_{DM}$ at the radius  where (\ref{ecNormRho0}) has a
maximum. For $\gamma=1.18$, that is 
\be
\rho_0\simeq 1.514\frac{\Ob}{\Omega_{\rm dm}}\rho_s
\ee

The quality of this approximation, given in the top panel for the
density profile derived from NFW (red line) is significantly poorer
($\sim20-30\%$ accuracy) than the phenomenological fits for the dark
matter. In this plot, we also show the gas density expected from a
Moore et al. profile (blue line). As in the temperature, the difference
with NFW is not very large (usually, of the same order as the scatter
around the NFW-based prediction). Part of this scatter can be
attributed to departures from hydrostatic equilibrium in merging
clusters, as well as deviations from purely thermal support that take
place even in the outer parts of relaxed clusters. 

Apart from self-consistency, another advantage of parametrisation
(\ref{ecDens}) over the conventional \bm is due to the instability of
the $\beta$ parameter with respect to changes in the outer radius used to
infer its best-fit value. Observational data restricted to radii much
smaller than \rv tend to find values close to $\beta\sim2/3$, while wider
field cluster images are usually better described by $\beta\sim1$ \citep[see
e.g.][]{NFW95}. Since the profile given in expression (\ref{ecDens})
depends only on the dark matter parameters, and these are much more
stable than those of the \BM, we consider this formula to be more
reliable in order to extrapolate the gas density up to the virial
radius. Moreover, it allows a direct estimate of the total mass if the
dark matter halo can be described by a NFW profile, although we have
already shown that the accuracy is somewhat limited. 

%-----------------------------------------------------------------------------

\section{Scaling relations}
\label{secScalingRelations}

Before the structure of clusters of galaxies could be resolved by the
available observations, they were already realised to exhibit a number
of simple correlations between their global properties. These scaling
relations could lend  insight into the physical nature of clusters when
compared with the analytical predictions expected to hold under the
assumption of self-similarity in the formation and evolution of large
scale structure \citep{Kaiser86}. 

However, deviations from self-similarity are expected under the effects
of, for example, merging \citep{JingSuto00} and any additional
physics acting on the intracluster gas over the simplistic infall in
the  dark matter potential  \citep[e.g.][ and
references therein]{EvrardHenry91,BN98,Bialek01}. The latter
case is particularly relevant to cool systems, where the extra
energy estimated from observed properties  is comparable
to their thermal energy \citep[e.g.][]{Ponman96,PCN99,TozziNorman01}. 

In the past 10 years, the advent of space-based X-ray observations of
galaxy clusters have revealed that there exist tight correlations
between the total gravitating mass, the X-ray luminosity and the
temperature of the intracluster medium \citep[e.g.][]{Rosati02}. A
close comparison of theoretical results and observational data can
therefore provide valuable constraints on the prevailing cosmological
models and even on the nature of the dark matter that dominates the
mass distribution and the dynamical evolution of clusters. 

%___________________________________________________

\subsection{Mass-temperature}
\label{secMT}

The number density of galaxy clusters as a function of their mass can
provide a powerful probe of models of large-scale structure. Since the
mass of a cluster is not a directly observable quantity, the abundance
of rich clusters of galaxies has been historically measured in terms of
some other parameter which is used as a approximation for mass. Several options
exist, but much attention has been focused recently on the X-ray
temperature. 

Cosmological simulations suggest that this magnitude is strongly
correlated with mass, showing little scatter
\citep[e.g.][]{EMN96,BN98}. How well simulations agree with
observational results is still far from clear, and several issues need
to be resolved. Numerical resolution and difficulties to include all
the relevant physics are the main sources of uncertainty invariably
coupled to cosmological simulations, while instrumental effects and
the lack of a reliable method to estimate the mass are the most
worrying aspects related to the observational determination of the
$M-T$ relation. 

%----------------------------------

\subsubsection*{Analytical prediction}

Within the framework of pure gravitational infall
\citep[e.g.][]{Lilje92}, the cluster mass and the gas temperature
should scale simply as $M\propto T^\alpha$, where the exponent $\alpha\simeq3/2$ is almost
unrelated to the particular  settings of cosmological parameters. This
self-similarity has been later justified by numerical simulations
\citep[e.g.][]{EMN96,BN98} and is often used as a reference against
which observational results are compared. 

There are several ways to obtain this relation. The simplest of them 
relates  the thermal energy of the gas to the potential energy due to
self-gravitation of the cluster. By virtue of the virial theorem: 
\be
\frac{kT_\Delta}{\mu m_p}\simeq\frac{1}{2}\frac{GM_\Delta}{r_\Delta}
\ee
where $T$ and $M$ refer to average temperature and cumulative mass. The
subscript $\Delta$ indicates the overdensity threshold at which these
quantities are measured. Substituting
%\footnote{We recall that we follow
%  the convention of defining overdensities with respect to the critical
%  density. In terms of the mean density, $\Delta\to\Omega_M\Delta$.} 
$M_\Delta\equiv\frac{4\pi}{3}\Delta\rho_cr_\Delta^3$ and $\rho_c\equiv\frac{3H_0^2}{8\pi G}$ above we get
\be
M_\Delta\simeq\frac{4}{GH_0\Delta^{1/2}}\left(\frac{kT_\Delta}{\mu m_p}\right)^{3/2}
\equiv M_0^{\rm vir}\Delta^{-1/2}\left(\frac{T_\Delta}{1\ {\rm keV}}\right)^{3/2}.
\label{ecMTvir}
\ee
Therefore, the virial theorem predicts not only the relation
between $M$ and $T^{1.5}$, but also the  normalisation $M_0^{\rm
  vir}\equiv\frac{4}{(\mu m_p)^{3/2}GH_0}\simeq6.12×10^{14}h^{-1}$ \Msun. 

A similar scaling law can be derived from our assumption  of a polytropic
equation of state for the gas, in hydrostatic equilibrium with a NFW
potential. Using the density profile (\ref{ecDens}), the averaged
temperature at overdensity $\Delta$ is just 
\be
T_\Delta=T_0\tau_m(\gamma,r_\Delta/r_s)~~;~~\tau_m(\gamma,r_\Delta/r_s)\equiv 
\frac{\int_0^{r_\Delta/r_s}{\left[\frac{\ln(1+x)}{x}\right]^\frac{\gamma}{\gamma-1}\ x^2\ \dd x} }
                 {\int_0^{r_\Delta/r_s}{\left[\frac{\ln(1+x)}{x}\right]^\frac{1}{\gamma-1}\ x^2\ \dd x} }
\ee
and the emission-weighted temperature
\be
\Tx^\Delta=T_0\tau_x(\gamma,r_\Delta/r_s)~~;~~\tau_x(\gamma,r_\Delta/r_s)\equiv
\frac{\int_0^{r_\Delta/r_s}{\left[\frac{\ln(1+x)}{x}\right]^{\frac{2}{\gamma-1}+\frac{3}{2}}\ x^2\ \dd x} }
                   {\int_0^{r_\Delta/r_s}{\left[\frac{\ln(1+x)}{x}\right]^{\frac{2}{\gamma-1}+\frac{1}{2}}\ x^2\ \dd x} }
\ee

The assumption $B=1$ led to the central temperature normalisation given
in equation (\ref{ecTemp}). Using the definition of $M_\Delta$, we recover
the scaling relation $M_\Delta=M_0\Delta^{-1/2}T_\Delta^{3/2}$, with $M_0$ given by  
\be
M_0=M_0^{\rm vir}\left[\frac{1}{2\tau_{m/x}}\frac{\gamma}{\gamma-1}\frac{\ln(1+r_\Delta/r_s)-\frac{r_\Delta/r_s}{1+r_\Delta/r_s}}{r_\Delta/r_s}\right]^{3/2}
\label{ecNormMT}
\ee
where the term in brackets  depends on the particular  values of the
polytropic index and 'concentration' $r_\Delta/r_s$ of each dark matter
halo. As we discussed in the previous section, the polytropic index can
be taken to be a constant value $\gamma\simeq1.18$ for the range of masses
covered by our simulated clusters. The validity of this
approximation outside this  mass range  must 
be tested by independent numerical simulations. 

%__________________________________
\begin{figure}
  \centering \includegraphics[width=12cm]{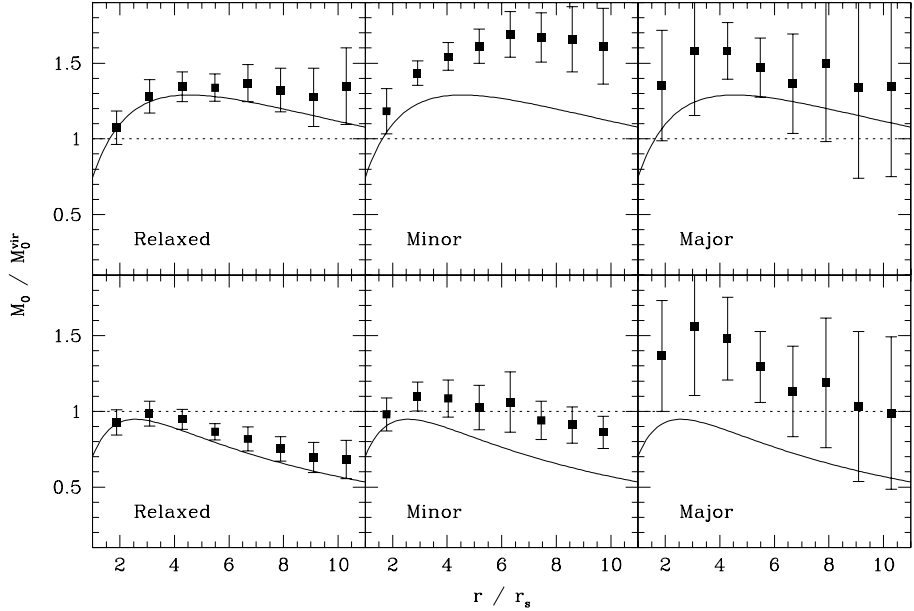}
  \caption[Analytical $M-T$ normalisation]{Analytical $M-T$ normalisation. Top panel: Mass-weighted temperature. Bottom panel: Emission weighted temperature $\Tx$.}
  \label{figGasM0}
\end{figure}
%__________________________________

The dependence of $M_0$ on $r_\Delta/r_s$ is shown  in
Figure~\ref{figGasM0}, which can be interpreted as a 'universal
profile' of the ratio between thermal and gravitational energy. The
normalisation $M_0\equiv M_\Delta\Delta^{1/2}T_\Delta^{-3/2}$ is plotted as a function of
radius, in units of the characteristic scale $r_s$. For a value of the
NFW concentration parameter $c=8$, the radius at which the
overdensity is $\Delta=2500$ approximately coincides with $r_s/3$,
$r_{1000}\sim r_s/2$ and $r_{500}\sim2r_s/3$. By definition, $r_{200}=cr_s$. 

As can be seen in Figure~\ref{figGasM0}, the normalisation of the $M-T$
relation predicted by equation (\ref{ecNormMT}) depends on whether $T$
or \tx is used, the difference (which can be  up to a factor of 2 near
the virial radius) is  due to the factor $\tau_{M/x}^{-3/2}$. Our formula
gives a fair approximation to the mass-temperature relation in relaxed
clusters, but it severely overestimates the mean temperature in merging
systems (by $\geq50\%$ in the outer parts). At high overdensities, though,
expression (\ref{ecNormMT}) is consistent with the results of our
simulations within the scatter from halo to halo. The effect of dark
halo concentration (which varies between 7 and 10 for our relaxed
clusters, 5 and 7 for minor mergers, and $3-5$ for the major merging
subset, see Table~\ref{tabDM}) is always of the order of 10\% at all radii. 

According to these results, dispersion around the predicted scaling
relation between mass and temperature arises mainly due to the excess
of gravitational energy present in merging events with respect to the
final virialised state. This agrees, at least at a qualitative level,
with the deviation from thermally-supported hydrostatic equilibrium
that was seen in Figure~\ref{figEqHydro}. 

%----------------------------------

\subsubsection*{Observations}

From an observational point of view, several studies have been devoted
to test the mass-temperature relation in clusters of
galaxies. \citet{Horner99} claim that the $M-T$ relation is steeper
than the traditional scaling, following a $M\propto T^{1.8-2.8}$ law when the
mass is estimated according to the \BM. Slopes steeper than 1.5 are
also observed in high-redshift clusters \citep[e.g.]{Schindler99} and
highly-luminous clusters \citep{EttoriFabian99}, where isothermality
has been assumed. \citet{NeumannArnaud99}, also using the \bm to
estimate the gravitating mass, obtain a $M-T$ relation consistent with
the classical scaling. \citet{ASF01}, using spatially resolved
spectroscopy with \emph{Chandra}, found consistency with the scaling
law prediction, but the normalisation was $\sim40\%$ lower with respect to
numerical simulations. 

However, it should be noted that in the above observational
analyses, only high-temperature ($kT>4-5$ keV) clusters are
involved. As has been realized in  recent studies
\citep[e.g.][]{Mohr99,PCN99}, the input of  energy feedback into the
ICM can break the self-similarity in the low-temperature clusters and
groups. This has been confirmed by \citet{Nevalainen00}, who inferred a
slope $\alpha=1.79\pm0.14$ and a normalisation significantly lower than the
one observed in numerical simulations, although the classical relation
$M\propto T^{3/2}$ was recovered when  the low-temperature groups are excluded. 

\citet{Finoguenov01} also found that the slope of the $M-T$ relation
for more massive clusters (i.e. $M_{500}>5\times10^{13}$ \Msun) was
considerably shallower ($1.58\pm0.07$) than that obtained for their whole
sample ($1.78\pm0.09$), the normalisation being more than $50\%$ lower
than the value quoted in \citet{EMN96}. These authors also point out
that mass estimates based on isothermality like the \bm result in a
steeper scaling relation \citep[a fact already noted by][]{Horner99}
due to the implicit assumption that the dark matter density scales as
$r^{-2}$, thus underestimating the mass at low radii and overestimating
it at large radii. An additional bias in introduced because the value
of $\beta$, and hence the asymptotic slope of the gas profile, was a
function of temperature. 

\citet{Xu01} compared the mass-temperature relation obtained from the
\bm mass estimates and an isothermal distribution in hydrostatic
equilibrium with a NFW potential. They found both models to be
indistinguishable, giving a slope steeper than the self-similarity
prediction unless the clusters at less than 3.5 keV are excluded, in
agreement with \citet{Finoguenov01}. 

\citet{Ettori02} have not only studied the slope of the $M-T$ relation,
but they have also extensively investigated the dependency of the
scaling relations with the dynamical state of the cluster as well as
the outer radius reached by the observational data. The slope at
overdensity 2500 is consistent with 1.5, the presence of non-cooling
flow clusters significantly increasing the scatter. However, the slope
at $\Delta=500$ steepened up to $M\propto T^{2.17\pm0.37}$. Regarding the
normalisation, these authors claim that the previously reported lower
values \citep[e.g.][]{Horner99,Nevalainen00} were due to the steeper
slopes found in those studies. 

\begin{table}
\begin{center}
\begin{tabular}{lrcll}
{\sc Reference} & $\Delta$~ & $M_0/M_0^{\rm vir}$ & ~$\alpha$ & \\ \cline{1-4}\\[-2mm]
\citet{Horner99}       &  200 & 0.92 & 1.53 & {\small velocity dispersion} \\
                       &  200 & 0.63 & 1.48 & {\small temperature profiles} \\
                       &  200 & 0.51 & 1.78 & {\small \BM} \\
                       &  200 & 0.47 & 2.06 & {\small $S_{\rm X}$ deprojection}\\
\citet{EttoriFabian99} &  500 & 0.40(0.97) & 1.93 & {\small \BM} \\
\citet{NeumannArnaud99}& 1750 & 0.75 & 1.5 & {\small \BM} \\
\citet{Mohr99}         & 1000 & 1.15 & 1.5 & {\small \BM} \\
\citet{Nevalainen00}   & 2000 & 0.58(0.75) & 1.77 & {\small temperature profiles} \\
                       & 1500 & 0.57(0.73) & 1.77 \\
                       & 1000 & 0.52(0.73) & 1.79 \\
                       &  500 & 0.41(0.74) & 1.84 \\
\citet{Finoguenov01}   &  500 & 0.65 & 1.68 & {\small \BM} \\
                       &  500 & 0.77 & 1.48 & {\small polyt. \BM, $T\geq3$} \\
                       &  500 & 0.45 & 1.87 & {\small idem. $T\leq4.5$ keV} \\
\citet{Xu01}           &  200 & 0.67 & 1.60 & {\small \BM} \\
                       &  200 & 0.40 & 1.81 & {\small isothermal NFW} \\
\citet{ASF01}          & 2500 & 0.97 & 1.47 & {\small $T_{\rm mw}$ NFW SCDM} \\
                       & 2500 & 0.93 & 1.52 & {\small $T_{\rm mw}$ NFW \LCDM} \\
\citet{Ettori02}       & 2500 & 0.82(1.03) & 1.54 & {\small $T_{\rm mw}$ (NFW or King)}\\
                       & 1000 & 0.78(1.18) & 1.76 \\
                       & 2500 & 0.88(1.03) & 1.51 & {\small $\Tx$}\\
                       & 1000 & 0.48(1.06) & 1.94 \\
\end{tabular}
\end{center}
\caption{Observed $M-T$ relation}
\label{tabMTobs}
\end{table}

Observational data are summarised in Table~\ref{tabMTobs}, where the
reference and overdensity of each study are listed in the first two
columns, followed by the best-fit values of the normalisation (rescaled
to the self-similar prediction) and the logarithmic slope $\alpha$. The
number in parentheses corresponds to the best-fit $M_0$ when the
slope is set to $\alpha=1.5$. Particularities of each work, such as the
assumptions employed to derive the gravitating mass, are noted on the
right margin of the table. 

Given the remarkable success of the \lcdm model in explaining
most of the relevant cosmological observations\footnote{As
  well as some others... The \lcdm model is almost as wonderful as SCDM
  was ten years ago! \citep[see e.g.][]{Cole94}}, it is somewhat
amazing that observers do show such a tenacious reluctance to accept
it as a fact. Except for  \citet{ASF01}, all the
mass estimates in Table~\ref{tabMTobs} have been made in a $\Omega=1$,
$\Lambda=0$, $h=0.5$ universe. Fortunately, the scaling relation between mass
and temperature does not depend on cosmology but through the usual
factor $h^{-1}$ in the mass, which is implicitly taken into account in
the units we used for $M_0^{\rm vir}$. 

A more worrying concern is raised by the discrepancy found between the
masses inferred from different methods and models, as well as that
arising from different analysis prescriptions. As can be inferred from
the compilation given in Table~\ref{tabMTobs}, we are far from reaching
an acceptable convergence in the observational value neither of the
normalisation nor the slope of the scaling relation. 

%----------------------------------

\subsubsection*{Simulations}

The mass-temperature relation has also been extensively studied by
means of cosmological numerical simulations. In most cases, from the
early work of \citet{NFW95} or \citet{EMN96} to the recent
high-resolution simulations of \citet{Muanwong02}, slopes consistent
with the self-similar value $\alpha=1.5$ are found when only adiabatic
physics is considered. 

The normalisation, however, has a little bit more scatter, although not
so much as that present in the different observational estimates. In
both cases, this scatter is partly due to the precise value found for
the logarithmic slope, since the normalisation of the scaling law is
closely related to this parameter (for a given sample, the higher the
best-value of $\alpha$, the lower the corresponding normalisation $M_0$). 

A compilation of different mass-temperature relations found in
numerical simulations is given in Table~\ref{tabMTsims}. The structure
of the table is the same of Table~\ref{tabMTobs}, in which the
observational data were presented. Unless otherwise noted, the
cumulative mass is compared to the emission-weighted temperature
\TX. \citet{Yoshikawa00} gives the normalisation referred to both
quantities, and \citet{ME01} also computed a spectral temperature $T_s$
based on a more detailed model of the line emission of the ICM gas in
the soft X-ray band as well as a subsequent fitting procedure closer to
the analysis of observational data. 

\begin{table}
\begin{center}
\begin{tabular}{lrcll}
{\sc Reference} & $\Delta$~ & $M_0/M_0^{\rm vir}$ & ~$\alpha$ & \\ \cline{1-4}\\[-2mm]
\citet{NFW95}       &  200 & 1.01 & 1.5 \\
\citet{EMN96}       & 2500 & 1.30 & 1.5 \\
                    & 1000 & 1.38 & 1.5 \\
                    &  500 & 1.30 & 1.5 \\
                    &  250 & 1.15 & 1.5 \\
                    &  100 & 0.89 & 1.5 \\
\citet{BN98}        &  200 & 1.42 & 1.5 \\
\citet{Pen98}       &  200 & 1.19 & 1.5 \\
\citet{Eke98}       &  100 &   1  & 1.5 \\
\citet{Yoshikawa00} &  100 & 1.40 & 1.5 & {\small $T_{\rm mw}$}\\
                    &      & 0.91 & 1.5 & {\small $\Tx$}\\
\citet{ME01}        &  500 & 1.34 & 1.52 & {\small $T_{\rm mw}$}\\
                    &  200 & 1.25 & 1.54 \\
                    &  500 & 1.85 & 1.38 & {\small $\Tx$}\\
                    &  200 & 1.65 & 1.39 \\
                    &  500 & 1.81 & 1.48 & {\small $T_{\rm s}$ (0.5-9.5)}\\
                    &  200 & 1.69 & 1.51 \\
                    &  500 & 1.34 & 1.62 & {\small $T_{\rm s}$ (2.0-9.5)}\\
                    &  200 & 1.20 & 1.64 \\
\citet{Muanwong02}  &  200 & 2.28 & 1.5 \\
\end{tabular}
\end{center}
\caption{$M-T$ relation in previous simulations}
\label{tabMTsims}
\end{table}

These authors find that the spectral fit temperature is generally lower
than the mass or emission-weighted average\footnote{Rather
  surprisingly, their values of \tx are similar, although slightly
  lower, than $T_{\rm mw}$, which is reflected in the normalisations
  quoted in Table~\ref{tabMTsims}. This is probably due to the presence
  of cold gas in the most inner part of their clusters.} due to the
influence of  cooler gas being accreted as part of the hierarchical
clustering process. Despite significant departures from isothermality,
single-temperature models produce acceptable fits for the spectral
temperature $T_s$. The unusual coincidence that a realistic spectrum
has nearly the same shape as an isothermal one was explained by these
authors in terms of a lack of spectral resolution. 

Although several cosmologies have been assumed in all these works, the
results do not seem to be very sensitive to this. As pointed out by
\citet{Muanwong02}, the numerical resolution of the simulation is much
more important, particularly in those cases where cooling was
implemented. Increasing the resolution lowers the emission-weighted
temperature due to the presence of cold, dense gas cores that survive
in sub-clumps. This has been argued by \citet{ME01} to explain the
lower spectral temperature detected in minor mergers, proposing the
deviations from a canonical $M-T$ relation to distinguish these systems
from relaxed clusters. 

\begin{table}
\begin{center}
\begin{tabular}{crcc|cc}
& & \multicolumn{2}{c|}{\sc Relaxed clusters} &  \multicolumn{2}{c}{\sc Whole sample} \\
& $\Delta$~ & $M_0/M_0^{\rm vir}$ & $\alpha$ & $M_0/M_0^{\rm vir}$ & $\alpha$ \\ \hline & & & & \\[-2mm]
{\small $T_{\rm mw}$}
& 2500 & 1.20 & 1.56 & 1.08 & 1.71 \\
& 1000 & 1.31 & 1.59 & 1.30 & 1.65 \\
&  500 & 1.34 & 1.53 & 1.45 & 1.54 \\
&  200 & 1.29 & 1.46 & 1.57 & 1.33 \\
{\small $\Tx$}
& 2500 & 0.93 & 1.61 & 0.91 & 1.68 \\
& 1000 & 0.90 & 1.62 & 1.04 & 1.57 \\
&  500 & 0.84 & 1.57 & 1.12 & 1.40 \\
&  200 & 0.71 & 1.51 & 1.15 & 1.17 \\
\end{tabular}
\end{center}
\caption{$M-T$ relation for our sample}
\label{tabMT}
\end{table}

Table~\ref{tabMT} summarises our results concerning the mass-temperature relation, where a separate fit has been performed for relaxed clusters (left columns). Our values are consistent with the self-similar prediction in almost every case. Clusters undergoing major merging events are usually responsible for the discrepancies.

The normalisation derived in the present study seems to be in agreement
with previous numerical work when the mass-weighted temperature is
considered, but is significantly lower when expressed in terms of the
emission-weighted temperature. We interpret this result as a 
resolution effect, maybe related to the numerical implementation of
gasdynamics (see Section~\ref{secEulerLagrange}). Since \tx is biased
towards the highest-density regions of the cluster, and both our
simulated temperature and density profiles seem to increase continuously
towards the centre (as expected from the analytical predictions based
on hydrostatic equilibrium), the emission-weighted temperature grows as
smaller scales are resolved in the cluster cores. 

Indeed, our $M-T$ relation agrees relatively well with the results of
\citet{Yoshikawa00}, as well as those of \citet{ME01} for the
mass-weighted temperature. As noted earlier, the values of \tx found by
these authors are  lower than $T_{\rm mw}$. This is only possible if
the densest gas is colder than the average. This seems to be the case
in their Figure 4, where they plot the phase-space diagram of the gas
particles of one of their clusters. As noted in
Section~\ref{secEulerLagrange}, we attribute   the presence of such cold
clumps to a  numerical artifact associated to SPH softening. 

A possible way to overcome the problem posed by the sensitivity of the
X-ray related quantities to the detailed structure of the ICM in the
central regions is to compute the  \emph{local} (instead of average)
values at $r_\Delta$. Observationally, this has been possible only very
recently, thanks to the deprojection of high-resolution maps obtained
by space-based observatories \citep{Ettori02}. Although it could be in
principle a good way to compare numerical and observational results, it
is still subject  to many uncertainties on the observational side. 

%__________________________________
\begin{figure}
  \centering \includegraphics[width=10cm]{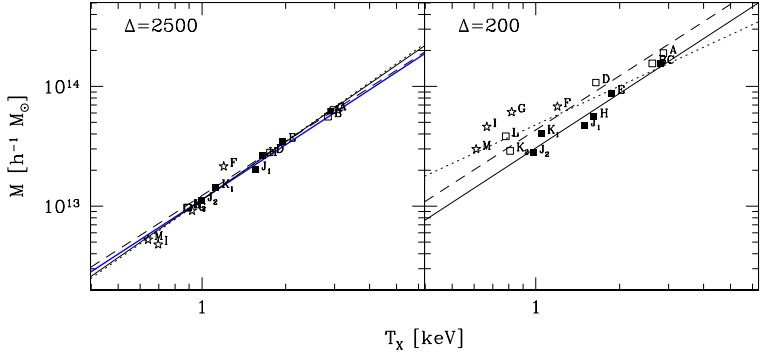}
  \caption[$M-T$ relation for our sample]{$M-T$ relation for our sample at two different overdensities. Solid lines are least-square fits to the relaxed population, dotted lines are used for the whole sample. The analytical prediction is shown by dashed lines and observational data from \citet{ASF01} by a blue line.}
  \label{figGasMT}
\end{figure}
%__________________________________

Our results at overdensities $\Delta=2500$ and $\Delta=200$ are compared with
both observational data and analytical estimates  in
Figure~\ref{figGasMT}. We see that the virial prediction based on
equation (\ref{ecMTvir}) accurately reproduces our simulated $M-T$
relation when only relaxed clusters are taken into account. As could be
seen in Figure~\ref{figGasM0}, merging clusters deviate from the
expected normalisation at large radii, being more than $50\%$ heavier
for a given temperature. This biases the best-fit slope towards
shallower values at \RV, as indicated in Table~\ref{tabMT}, but has an
almost negligible effect at higher overdensities. 

The best-fit mass-temperature relation found by \citet{ASF01} is drawn
as a blue line in the left panel of Figure~\ref{figGasMT}. Some authors
\citep{Ettori02,Mohr99} find similar normalisations at high
overdensities, whereas others \citep{NeumannArnaud99,Nevalainen00}
quote slightly lower values. The situation is much less clear near the
virial radius, but there is a general trend in the sense that
observational estimates seem to indicate hotter temperatures for a
given virial mass. 

%___________________________________________________

\subsection{Luminosity-temperature}
\label{secLT}

A scaling law that is more directly observable is the correlation
between bolometric X-ray luminosity and global emission-weighted
temperature. It has been known long ago that the slope of the relation
appears to fit $\Lx\propto \Tx^3$ 
\citep[e.g.][]{Mitchell79,EdgeStewart91,David93,Fabian94}
rather than $\Lx\propto \Tx^2$, as would be expected for a simple
self-similar model, and much theoretical work has been devoted to find
the reasons for such a discrepancy. 

One of the proposed solution involves some \emph{preheating} of the
intracluster gas by an agent other than gravity, such as supernova
explosions at high redshift \citep{Kaiser91,EvrardHenry91}. This
additional heating would decrease the central density and hence the
X-ray luminosity. This view was strengthened by the discovery of an
apparent entropy 'floor' in the centres of groups and clusters
\citep{PCN99}. 

However, the amount of heating required is substantial. Although
estimates vary, it seems likely that about 1 keV per particle is needed
, which may be challenging to explain from supernova
heating alone due to an excessive enrichment of the ICM
\citep{ky00,vs99}. Another difficulty is that observations 
of the Ly$\alpha$ forest indicate a much lower temperature for the majority 
of the intergalactic medium at $z\sim2-3$
\citep{BryanMachacek00},
 a condition that may
extend to even low redshifts. Although hardly
conclusive, these concerns may be pointing toward another explanation
for these  observations. 

Early on, it was suggested that the steepening of the $L-T$ relation
could be due to systematic variations of the cluster baryon fraction
with X-ray temperature \citep{David93}. \citet{Fabian94} and
\citet{Markevitch98} also note that the large scatter in the $\Lx-\Tx$
plane was mostly due to the strong cooling flows present in more than
half the low-redshift clusters \citep[e.g.][]{Edge92}. It has been
argued \citep{AllenFabian98} that these cooling flows, which can
account for up to $70\%$ of the total X-ray luminosity of a cluster
\citep{Allen98}, could be responsible for the observed
discrepancy in the slope of this scaling relation. 

%----------------------------------

\subsubsection*{Analytical prediction}

Simple arguments based on self-similarity predict that the bolometric
X-ray luminosity of galaxy clusters should scale with the temperature
of the gas according to $\Lx\propto \Tx^2$. Nevertheless, the normalisation
of the $L-T$ relation is often treated as a free parameter. In order to
attempt a theoretical prediction, we will calculate the X-ray
luminosity of a spherically symmetric cluster of galaxies, assuming
that it is dominated by thermal bremsstrahlung emission. With this
prescription, the cumulative bolometric luminosity can be computed as 
\be
 \Lx(r)=\int_0^r \Lambda_{\rm X} \left(\frac{\rho_g(x)}{\mu m_p}\right)^2 [kT(x)]^{1/2}\ 4\pi x^2\ \dd x
\label{ecGasLx}
\ee
where $\Lambda_{\rm X}=1.2\times10^{-24}$ erg s$^{-1}$ cm$^3$ keV$^{-1/2}$ \citep{NFW95}. As explained in Section~\ref{secRprof}, this is precisely the definition we use to compute the bolometric luminosity of our numerical clusters.

Although it has already been shown that clusters are definitely
\emph{not} isothermal, the term $T^{1/2}(r)$ varies very slowly
compared to $\rho_g^2$(r). Therefore, at a  first approximation we can take
its mean value out of the integral in (\ref{ecGasLx}). Since this
expression is biased towards the highest-density regions, it would be
more accurate to use the emission-weighted average. 

After some calculations, we obtain the expression
\be
    \Lx(r)\simeq \frac{\Lambda_{\rm X}}{(\mu m_p)^2} [k\Tx(r)]^{1/2}
\frac{\int_0^r \rho_g^2(x)\ 4\pi x^2\ \dd x/(\frac{4\pi}{3}r^3)}
     {\left[M_g(r)/(\frac{4\pi}{3}r^3)\right]^2}
\frac{M_g^2(r)}{\frac{4\pi}{3}r^3}
\ee
which can be simplified using the definitions of cumulative baryon
fraction, overdensity and the structure parameter
$Q(r)\equiv\left<\rho_g^2\right>/\left<\rho_g\right>^2$. The final scaling relation is
reduced to the form 
\be
    \Lx^\Delta\simeq \frac{\Lambda_{\rm X}}{(\mu m_p)^2}\Delta\rho_c F_\Delta^2 Q_\Delta M_\Delta [k\Tx^\Delta]^{1/2}\equiv L_0\ \Delta^{1/2}\left(\frac{\Ob}{\Om}\right)^2\left(\frac{\Tx^\Delta}{1\ {\rm keV}}\right)^2
\label{ecL0}
\ee

Proceeding in an analogous manner as we did for the $M-T$ relation, we
express the normalisation $L_0$ in expression (\ref{ecL0}) in terms of
the  quantity $L_0^{\rm vir}\equiv\frac{\Lambda_{\rm X}\rho_c}{(\mu
  m_p)^2}M_0^{\rm vir}\simeq2.85\times10^{43}\ h$ erg s$^{-1}$. For the sake of
simplicity, we will assume a cosmic baryon fraction $\Ob/\Om=0.14$
irrespective of the cosmological model. This is approximately the ratio
expected for a \lcdm universe  and it is roughly
consistent with the value observed in galaxy clusters
\citep[e.g.][]{ASF02omega}, assuming a low $\Omega_M$.

Another difficulty in comparing observational and theoretical results
arises from the different scaling with the Hubble constant. Since the
observed luminosity is proportional to  $d_L(z)^2$, 
it scales as $h^{-2}$ instead of $h$. For nearby clusters, this effect
accounts for a factor of 6 in the normalisation depending on whether
the comparison is made in units of $h=1$, $h=0.5$ or $h=0.7$. We
take $h=1$ as the  prescription that is  more consistent with
 both observational and numerical methods.

%__________________________________
\begin{figure}
  \centering \includegraphics[width=8cm]{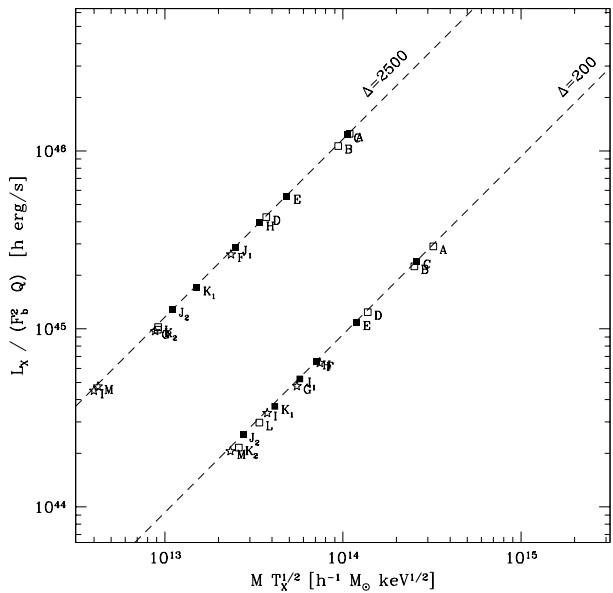}
  \caption[Normalisation of the $L-T$ relation]{Normalisation of the $L-T$ relation, corrected for the effects of baryon fraction and structure parameter.}
  \label{figLTfQ}
\end{figure}
%__________________________________

We test the validity of our analytic estimates  of the $L-T$ relation
by comparing the values of $\Lx/(f^2Q)$ and $M\Tx^{1/2}$. The values of
these quantities for all our simulated clusters are plotted in
Figure~\ref{figLTfQ}, as well as the results  of equation
(\ref{ecL0}) for two different overdensity values.
 The agreement is not surprising, since the only assumption we
made for  deriving the analytical prediction was to   consider
that $T^{1/2}$ varied much more slowly than $\rho_g^2$, which, as we
discussed in Section~\ref{secGasRprof}, is a very good approximation. 

%----------------------------------

\subsubsection*{Observations}

The luminosity-temperature relation of several cluster catalogs
observed with the first generation of X-ray satellites showed a
departure from the predicted slope $\alpha=2$
\citep[e.g.][]{EdgeStewart91,David93} with a measured value
of $\alpha\sim  3$ and a scatter along the mean relation that could be reduced
considerably once the effect of the cold  cores was taken into account
\citep{Fabian94}. 

When the impact of cooling flows on the bolometric luminosity and
emission-weighted temperature of the cluster is properly corrected, the
$L-T$ relation is shown to be tighter and closer to (yet not completely
consistent) the self-similar scaling prediction
\citep[e.g.][]{AllenFabian98,EAF01}. The scatter is significantly
reduced in analyses that either excise the core cooling flow regions
\citep{Markevitch98} or examine samples of clusters defined to possess
weak cooling cores \citep{ArnaudEvrard99}. 

The steeper $L-T$ dependence found in other studies
\citep[e.g.][]{David93,Ponman96,Cavaliere97,PCN99,XueWu00,TozziNorman01}
is often justified by invoking some sort of pre-heated of the
intracluster gas, which would rise the entropy level in the 
coolest systems. 

\begin{table}
\begin{center}
\begin{tabular}{lrcll}
{\sc Reference} & $\Delta$~ & $L_0/L_0^{\rm vir}$ & ~$\alpha$ & \\ \cline{1-4}\\[-2mm]
\citet{David93}        &  200 & 0.084 & 3.37 \\
\citet{Ponman96}       &  200 & 0.468 & 8.20 & {\small groups}\\
                       &  200 & 0.132 & 3.29 & {\small groups+clusters}\\
\citet{AllenFabian98}  &  200 & 0.287 & 3.22 & {\small cooling-flow (raw)}\\
                       &  200 & 1.698 & 2.13 & {\small CF (corrected)}\\
                       &  200 & 0.191 & 2.92 & {\small non-CF}\\
\citet{Markevitch98}   &  200 & 0.414 & 2.10 & {\small $0.1-2.4$ keV}\\
                       &  200 & 0.347 & 2.64 & {\small bolometric}\\
\citet{ArnaudEvrard99} &  200 & 0.209 & 2.88 \\
\citet{XueWu00}        &  200 & 0.017 & 5.57 & {\small groups}\\
                       &  200 & 0.029 & 2.79 & {\small clusters}\\
\citet{ASF01}          & 2500 & 0.296 & 2.17 & {\small $T_{\rm mw}$ SCDM} \\
                       & 2500 & 0.589 & 2.08 & {\small $T_{\rm mw}$ \LCDM} \\
\citet{Ettori02}       & 2500 & 0.055(0.258) & 2.79 & {\small $T_{\rm mw}$}\\
                       & 1000 & 0.347(0.677) & 2.37 \\
                       & 2500 & 0.076(0.258) & 2.64 & {\small $\Tx$}\\
                       & 1000 & 0.214(0.551) & 2.54 \\
\end{tabular}
\end{center}
\caption{Observed $L-T$ relation}
\label{tabLTobs}
\end{table}

A compilation on observational estimates of the $\Lx-\Tx$ relation is
given in Table~\ref{tabLTobs}. We see that both the normalisation and
the logarithmic slope differ considerably from author to author, being
extremely sensitive to the details of the data reduction and analysis
(e.g. instrumental specifics, treatment of cooling flows, fitting
procedure, etc.). 

Once again, we see that the parameters $L_0$ and $\alpha$ are tightly
correlated, reflecting the fact that the typical temperature of most
cluster samples is substantially larger than 1 keV. Thus, the
normalisations obtained for a self-similar slope $\alpha=2$ (quoted between
parentheses in Table~\ref{tabLTobs}) are much higher than those
obtained when the slope of the relation is not constrained. 

Contrary to the $M-T$ relation, in this case the assumed cosmological
model plays an important role in the exact value of $L_0$, while the
slope $\alpha$ is not so affected \citep[see][ the \emph{only} work that 
included an estimate based on a \lcdm cosmology]{ASF01}. 

The effect of differences in individual gas fractions and structure parameters of clusters on the observed logarithmic slope of the $L-T$ relation has been examined by \citet{ArnaudEvrard99}, who came to the conclusion that the higher gas concentration in hot clusters is responsible for at least half of the steepening of the $L-T$ relation. When cluster mass is estimated from a \bm, the temperature dependence of the structure factor $Q$ can explain the observed slope. But if virial theorem is used to derive masses, then the gas fraction increases with cluster temperature, and it contributes as much as $Q$ to the steepening of the logarithmic slope $\alpha$.

%----------------------------------

\subsubsection*{Simulations}

Despite the significant amount of observational work regarding
normalisation and slope of the $\Lx-\Tx$ relation, there are little
numerical work  done in this regard as compared to   works  done to
study the  mass-temperature relation (some of them listed in
Table~\ref{tabMTsims}). 

This is  a reflection of the difficulty in obtaining reliable estimates
of the bolometric X-ray luminosities. While it is reasonable to assume 
 that the results are sensitive to the ICM thermal evolution, 
it is well known \citep[e.g.][]{BN98,Yoshikawa00}
 that they are also affected by the numerical resolution.

 When equal mass particles are employed in a given
simulation,  the smallest clusters have
poorer resolution and hence their luminosities tend to be
systematically underestimated as compared with more massive ones.

This effect can steepen the slope of the simulated $L-T$ relation,
making it closer to the observed one. Although our numerical
experiments seem to have enough resolution to overcome this problem, it
is always necessary to keep in mind this consideration when
interpreting the results. 

\begin{table}
\begin{center}
\begin{tabular}{lrcll}
{\sc Reference} & $\Delta$~ & $L_0/L_0^{\rm vir}$ & ~$\alpha$ & \\ \cline{1-4}\\[-2mm]
\citet{NFW95}       &  200 & 1.46 & 2 \\
\citet{BN98}        &  200 & 1.27 & 2.59 \\
\citet{Eke98}       &  100 & 1.08 & 2 \\
\citet{Bialek01}    &  500 & 3.85 & 1.56 & {\small core included}\\
                    &      & 1.25 & 2.02 & {\small core excised}\\
\end{tabular}
\end{center}
\caption{$L-T$ relation in previous simulations}
\label{tabLTsims}
\end{table}

Many of the references cited in the present work did not study the
$L-T$ relation at all. Among them, only those reported in
Table~\ref{tabLTsims} quote the best-fit values found for the
parameters $L_0$ and $\alpha$. \citet{Bialek01} also includes a discussion
on the analytical derivation of the scaling laws, highlighting 
 the importance of  cluster structure, related to
$F_b$ and $Q$ parameters. 

These authors point out that the slope of the power-law fits can be
biased by a few points at one end. In their case, they had  two
low-temperature systems that appeared to be fortuitously observed
mergers.  In agreement  with our results presented in previous chapter
\citet{Bialek01} remark that these two objects featured steeper density
profiles compared to the rest of the sample, and used this fact to
justify that their luminosity was a factor of 3 higher than that
expected from their gas temperature. Exclusion of the core emission,
where the core is defined as a circular area of radius 0.13\RV, resulted
in a best-fit slope consistent with the analytical scaling (see
Table~\ref{tabLTsims}). 

\begin{table}
\begin{center}
\begin{tabular}{rcc|cc}
& \multicolumn{2}{c|}{\sc Relaxed clusters} &  \multicolumn{2}{c}{\sc Whole sample} \\
$\Delta$~ & $L_0/L_0^{\rm vir}$ & $\alpha$ & $L_0/L_0^{\rm vir}$ & $\alpha$ \\ \hline & & & & \\[-2mm]
 2500 & 1.34(1.11) & 1.89 & 0.55 & 2.86 \\
 1000 & 2.35(2.21) & 1.90 & 1.24 & 2.66 \\
  500 & 3.44(3.13) & 1.89 & 2.23 & 2.45 \\
  200 & 5.57(4.95) & 1.89 & 4.25 & 2.28 \\
\end{tabular}
\end{center}
\caption{$L-T$ relation for our sample}
\label{tabLT}
\end{table}

However, this is not what we find in our numerical experiments. In Table~\ref{tabLT}, we show
the values of $L_0$ and $\alpha$ that best describe the $\Lx-\Tx$ relation
of our galaxy cluster catalogue. The best-fit
parameters are quoted for different overdensity thresholds, making a
separate fit for the relaxed subset of the sample. The number in
parentheses is the normalisation $L_0$ when the logarithmic slope is
fixed to the canonical value $\alpha=2$. 

There is a striking difference between the behaviour of the relaxed
clusters and the merging systems in our simulations.
 Contrarily to the results reported by
\citet{Bialek01}, the latter exhibit significantly lower luminosities
for a given temperature, mainly due to their lower concentrations (and
thus, lower central densities). This is not surprising, since, despite
the steeper inner slope of the density profile, the normalisation $\rho_s$
is significantly reduced with respect to  relaxed systems (see
e.g. the right panel of Figure~\ref{figMc}). 

 We also find,  in agreement with \citet{Bialek01},  that the low-mass
 end of our sample is dominated by the presence of merging systems, but
 in our case this biases the X-ray luminosity towards \emph{lower}
 values, and the slope of the $L-T$ relation steepens considerably when
 low-mass mergers are included in a lest-square fit together with the
 more massive, relaxed clusters of galaxies. 

At high overdensities, merging systems raise the slope of the simulated
$L-T$ relation to values consistent (and sometimes even higher) than
those observed. However, since these are the least massive of our
clusters, caution must be exercised because of the lower relative
resolution (i.e. less particles within the virial radius) in these
objects. Nonetheless, a steepening of the luminosity-temperature
relation would be expected if 
\begin{enumerate}
  \item Clusters indeed followed a 'universal' baryon fraction profile (as it seems to be the case in Figure~\ref{figBaryonF}).
  \item Low-temperature systems were systematically less concentrated.
\end{enumerate}

In such a case, a dependence of the baryon fraction on temperature
would arise and hence the X-ray luminosity of the low-temperature
clusters would be significantly depressed with respect to that of the
more massive objects. This effect would be more evident at high
overdensities. Near the virial radius, the global baryon fraction is
expected to be only slightly lower than the cosmic value, since gas and
dark matter profiles are observed to be proportional everywhere except
in the core regions. 

The first hypothesis   above can be understood as a consequence of
hydrostatic equilibrium and  a 'universal' dark matter profile. The
second arises from the fact that low-mass systems are more prone to
merge(in agreement with the predictions based on the hierarchical
scenario). Therefore,  we think that this relation between baryon 
fraction and temperature is not an artifact due to a lack of
 resolution, but has a physical origin. However, this is not the 
only mechanism that can alter the observed $\Lx-\Tx$ relation. 
A more detailed discussion is given in the next section. 

%__________________________________
\begin{figure}
  \centering \includegraphics[width=10cm]{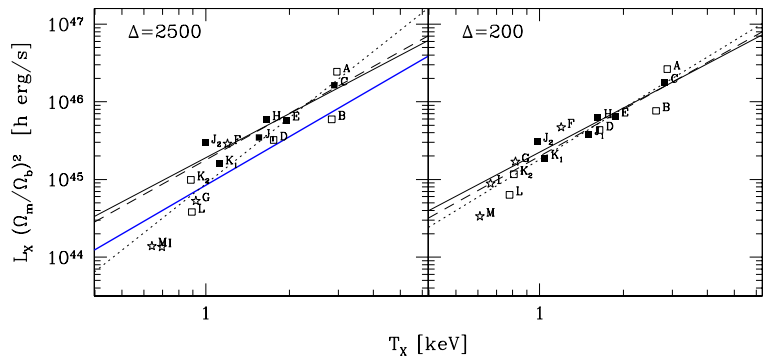}
  \caption[$L-T$ relation for our sample]{$L-T$ relation for our cluster sample. Symbols are as in
    previous Figures. Observational data from \citet{ASF01} are shown
    as solid blue. Dotted line is the self-similar relation. Dashed
    line is the unconstrained best fit to relaxed clusters and solid
    line is the best fit to relaxed clusters for $\alpha =2$ (see Table \ref{tabLT} }
  \label{figGasLT}
\end{figure}
%__________________________________

The steepening of the $L-T$ scaling law can be clearly seen in
Figure~\ref{figGasLT}, where the effects of observational 'aperture'
(i.e. overdensity) can be appreciated at first sight by comparing the
left and right panels. At $\Delta=2500$, our results for the whole sample
are perfectly compatible with a power law as steep as $\Lx\propto\Tx^3$. 

Observational data from \citet{ASF01} have been represented by a blue
line in the left panel of Figure~\ref{figGasLT}. It is not clear to
what extent our results are compatible with this fit, but the
disagreement is manifest if we compare with estimates based on a SCDM
cosmology. However, such a comparison is not straightforward. These authors, for instance, quote a normalisation which is a factor of 2 lower (as well as a steeper slope) for the same cluster sample if a SCDM cosmological scenario is assumed for the analysis. 

Given the many difficulties surrounding the reliable calibration of the $\Lx-\Tx$ scaling law from both theoretical and observational grounds, the mass-temperature relation seems to be much more convenient in order to use clusters of galaxies as cosmological probes. 

%----------------------------------

\subsubsection*{Departure from self-similarity}

We have already shown in Figure~\ref{figLTfQ} that our clusters follow
quite closely the analytical prediction of the $L-T$ relation given by
expression (\ref{ecL0}) when the effects of substructure are taken into
account. However, it is evident from both Figure~\ref{figGasLT} and the
slopes quoted in Table~\ref{tabLT} that the best-fit $\Lx-\Tx$ relation
deviates significantly from the self-similar scaling law $L\propto T^2$,
particularly at high overdensities. 

Since (\ref{ecL0}) holds, these deviations arise from the factor
$F_b^2QM_0$, where we recall that $F_b$ denotes the cumulative gas
fraction, $Q\equiv\left<\rho_g^2\right>/\left<\rho_g\right>^2$ is the structure
parameter and $M_0$ is the normalisation of the $M-T$ relation. It has
been often argued that self-similarity implies that these quantities
must be the same for all clusters, and hence the normalisation $L_0$
would be a constant multiple of $L_0^{\rm vir}$. 

A systematic dependence of $L_0$ on temperature would tilt the observed
$L-T$ relation, changing the best-fit value of the logarithmic slope
$\alpha$. Several authors \citep[e.g.][]{David93,ArnaudEvrard99} argued that
$F_b^2\sim T$ would reconcile the slope $\alpha\sim3$ derived from observational
data with the self-similar prediction. Some attention has been paid
\citep[e.g.][]{ArnaudEvrard99} to the effect of the structure parameter, and
almost none to possible variations in $M_0$. 

%__________________________________
\begin{figure}
  \centering \includegraphics[width=12cm]{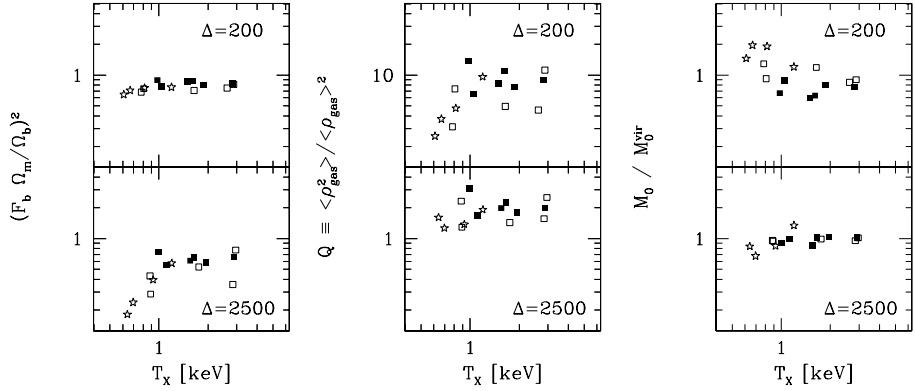}
  \caption[Deviations from the self-similar $L-T$ scaling law]{Deviations from the self-similar $L-T$ scaling law at different overdensities. Left panel: Baryon fraction. Middle panel: Structure parameter. Right panel: Mass-temperature normalisation.}
  \label{figGasFQ}
\end{figure}
%__________________________________

Figure~\ref{figGasFQ} displays the behaviour of these quantities as a
function of temperature and overdensity. In the top panel, we see that
the cumulative baryon fraction at \rv is slightly lower than the cosmic
value, and does not vary significantly even in
clusters undergoing a major merging event. However, these systems are
much less concentrated than relaxed structures, which manifests as a
factor of $\sim3-4$ difference in the structure parameter, $Q$. However, this
variation is anti-correlated with the normalisation of the $M-T$
relation, since $M_0$ tends to increase for lower values of $\Rv/r_s$
(see Figure~\ref{figGasM0}). These two effects compensate  each other 
and the total deviation from the self-similar scaling is  relatively
small for $\Delta=200$. 

As we restrict ourselves to inner radii, both the structure parameter and the normalisation of the $M-T$ relation tend  to more stable values. Now, differences in  central concentrations make that smaller systems, whose gas distribution is puffer, get lower baryon fractions  than relaxed clusters. Hence, the bolometric X-ray luminosity  is suppressed  by a factor of 3 with respect to the expected  behaviour. As can be seen  in Table~\ref{tabLT}, this effect is responsible for a dramatic steepening of the slope of the $\Lx-\Tx$ relation. 

%-----------------------------------------------------------------------------

%%% Local Variables: 
%%% mode: latex
%%% TeX-master: t
%%% End: 

%-----------------------------------------------------------------------------
\chapter{Stars}
   \label{chapStars}
%-----------------------------------------------------------------------------

\begin{quote}{\em
Oh let the sun beat down upon my face\\
stars to fill my dream\\
I am a traveler of both time and space\\
to be where I have been}

-- Led Zeppelin : {\em Kashmir} (1975) --
\end{quote}
%-----------------------------------------------------------------------------

{\gothfamily {\Huge S}{\large tar}} formation and feedback to the
interstellar medium (ISM) play a key role in the whole process of
structure formation. The cosmological model provides the space-time
framework in which the events take place and fixes several key
parameters, such as the expansion rate of the universe, the amount of
matter and the amplitude of primordial density fluctuations. The
large-scale structure of the universe, the clustering properties of
matter or the mass distribution in individual objects can be understood
in terms of hierarchical growth of dark matter
fluctuations. Hydrodynamic simulations have  shown to be a  valuable
tool to understand the physical conditions of the gas at low and
moderate densities. Detailed theoretical studies of,  e.g. the
Lyman-alpha forest \citep{Cen92,Hernquist96} 
 and the intergalactic medium could be done thanks to them.

At higher densities, radiative cooling and star formation become an essential ingredient in the description of baryons. Unfortunately, our knowledge of the complex physical processes involved in galaxy formation is much less certain. For instance, in the absence of stellar feedback most of the baryonic matter would have cooled into small mini-galaxies \citep{col91,bla92}. Photoionization and supernova explosions inhibit cooling and star formation in very low-mass objects, biasing star formation at early times towards high overdensity regions. Feedback has been proposed, as we discussed in the previous chapter, to solve the discrepancy between the observed and theoretical scaling relations of galaxy clusters, as well as the problems of angular momentum in discs \citep{Silk01} and the low number of observed dwarf galaxies \citep{Bullock00}.

Despite the numerous attempts to include star formation in cosmological simulations \citep[e.g.][and references therein]{shyk3}, progress has been comparatively low with respect to gravity or gasdynamics, since the physics involved in star formation and feedback into the interstellar medium is substantially more complex, and also because it acts on scales of a few pc and masses of $\sim 10^5$ M$_\odot$, which are well below the resolution limits.

The present chapter describes the results obtained from the model
implemented by \citet[see also Section \ref{secYK3}]{YK3}. The star formation history of the universe \citep{Ascasibar02} is discussed in detail, as well as the main differences between the star formation regimes observed in clusters and field galaxies.

%-----------------------------------------------------------------------------

\section{The cosmic star formation history}

The star formation history of the universe, i.e.  the global star formation rate (SFR) density as a function of redshift, is a crucial test for galaxy formation scenarios.  Many measurements of this quantity have been undertaken during the past few years, and the number and accuracy of available observations are still rapidly increasing.

Although there is a large scatter among different indicators at any given redshift, basically all studies find a significant increase (by about one decade) in the cosmic SFR density from the present day to $z=1$.  However, it is still a controversial issue \citep{hop01} whether it reaches a broad maximum there \citep[e.g.][]{mad96,gis00} or it declines gradually towards higher redshifts \citep[e.g.][]{saw97,pas98,ste99}.

From a theoretical point of view, star formation is a highly non-linear process, which precludes any simple analytical treatment. Most efforts towards modeling the star formation history of the universe resort to semi-analytical methods or numerical simulations to tackle the feedback mechanisms that self-regulate the SFR within the hierarchical clustering scenario of structure formation.

Semi-analytical models \citep[see e.g.][]{kau93,col94,Salvador98} construct a large sample of Monte Carlo realizations of halo merging histories, using the Press-Schechter formalism. Hydrodynamics, star formation and feedback are implemented through simple recipes, whose parameters are fixed in order to match the observed properties of real galaxies, specially the luminosity function and the Tully-Fisher relation \citep{SP99}. Their high computational efficiency allows a fast exploration of the parameter space in search of a viable model, but the number of free parameters asks for supplementary modeling based on more fundamental physics.

Numerical simulations are intended to solve directly the set of differential equations of gravitation and gas hydrodynamics, and hence require much fewer model assumptions than the semi-analytic approach. However, despite the staggering advances in computer technology and numerical algorithms, resolution in large-scale hydrodynamical simulations is severely restricted by computational and data storage requirements. Although gas can be treated self-consistently, some phenomenological recipes are still required in order to model star formation and feedback.

The cosmic SFR density is a direct outcome of semi-analytical methods \citep{som01} and hydrodynamical simulations \citep{Tissera00,nag01,Ascasibar02,SpringelHernquist}. Both of them predict a significant star formation activity at high $z$, favoring the hypothesis of a gradual evolution of the comoving SFR as a function of time.

%-----------------------------------------------------------------------------

\subsection{Observations}
\label{secMadau}

A variety of observations and observational methods have contributed during the past years to our understanding of the evolution of the star formation rate density on cosmological scales. Nonetheless, different wavelength regimes and estimation prescriptions yield rather discrepant results. In this section, the merits and flaws of each spectral range are reviewed, as well as the usual procedure to construct a consistent diagram of the SFR density as a function of redshift, also known as the 'Madau plot'.

%----------------------------------

\subsubsection*{Wavelength}

Photometric redshift estimates based on Lyman-limit systems \citep{st96a,st96b} stimulated the study of the rest frame UV continuum flux around 2000 {\AA} as a preferred wavelength range for objects in the far universe.  This emission traces the presence of massive (and therefore young) stars and can be related directly to the actual star formation rate.  It is relatively easy to detect at high $z$ as it is redshifted into the optical band.  However, older stellar populations and AGN can make a significant contribution as well, leading to an overestimation of the total SFR.  On the other hand, dust enshrouding star formation regions absorb very efficiently the UV light, re-emitting it in the far IR.  Unfortunately, there is still some uncertainty to account for dust extinction and its evolution with redshift \citep[e.g.][]{AdelbergerSteidel00}.

Recombination lines from {H}{II} regions can be in principle a more reliable estimator of the instantaneous SFR, since the ionizing radiation ($\lambda\leq912${\AA}) comes from more massive (younger) stars than the softer UV continuum.  These optical lines are less dramatically affected by dust extinction and can be mapped with higher resolution in the local universe, but they require infrared spectroscopy at moderate redshifts, which has not been possible until very recently due to the faintness of the sources involved \citep{pet98}.  For H$_{\rm \alpha}$, this happens at $z\sim0.5$, and for forbidden lines, such as [{O}{II}]3727{\AA}, at $z\sim1.5$.  There is also a considerable degree of uncertainty in the relation between recombination lines and star formation activity.

Finally, the far infrared (FIR) spectrum of a galaxy can be roughly separated into two components, namely the thermal emission of dust (heated by star formation bursts) around $\lambda\sim60\ \mu$m and an infrared cirrus powered by the global radiation field, which dominates for $\lambda\geq100\ \mu$m.  Neglecting cirrus and AGN contamination, the FIR luminosity would also be an excellent tracer, but instrumentation is not as developed in the IR and radio bands as it is in the optical, and the conversion factor between FIR luminosity and SFR is for the moment rather model dependent.

%----------------------------------

\subsubsection*{The Madau plot}

The first step to obtain an observational estimate of the cosmic SFR density is the selection of a {\em complete} sample of galaxies at a given redshift and wavelength. Some authors \citep[e.g.][]{ste99} claim that the HDF covers a very small area ($\sim5$ arcmin$^2$), and hence it is not statistically representative of the whole universe.  This cosmic variance is difficult to quantify, but the strong clustering of high redshift galaxies (and the fact that most estimates beyond $z=2$ are based on the HDF) seems to indicate that this effect could introduce an important systematic bias in the star formation rates inferred for galaxies in the high redshift universe.

Then, the sample must be corrected for incompleteness before
constructing the luminosity function, which can be done following the
$V_{\rm max}$ formalism \citep{sch68} or resorting to more elaborate
Monte Carlo algorithms \citep[such as that proposed by][]{ste99}. The
comoving luminosity density is easily obtained by integration over all
magnitudes, extrapolating the faint end as a Schechter function.  
The last step is to convert the LF to a star formation rate density with the help of a population synthesis model, taking into account the absorption of interstellar dust.

\begin{table*}
\begin{center}
\footnotesize
\begin{tabular}{lccccc}
 & & & & \multicolumn{2}{c}{$\log{\dot\rho_*}$} \\
 {\sc ~~ Survey} & {\sc Estimator} & $z$ & $\log{\rho_{\rm L}^{\rm SCDM}}$ & SCDM & \LCDM \\
\hline\\[-2mm]
HDF (\cite{pas98}) $^{\rm d}$ & 1500 {\AA}
  & $0.25 \pm 0.25$ & $26.47^{+0.31}_{-0.22}$ & -1.38 & -1.36\\
& & $0.75 \pm 0.25$ & $26.61^{+0.23}_{-0.14}$ & -1.24 & -1.26 \\
& & $1.25 \pm 0.25$ & $26.80^{+0.24}_{-0.12}$ & -1.05 & -1.10 \\
& & $1.75 \pm 0.25$ & $26.83^{+0.24}_{-0.12}$ & -1.02 & -1.08 \\
& & $2.50 \pm 0.50$ & $26.62^{+0.28}_{-0.21}$ & -1.23 & -1.30 \\
& & $3.50 \pm 0.50$ & $26.59^{+0.34}_{-0.27}$ & -1.26 & -1.34 \\
& & $4.50 \pm 0.50$ & $26.70^{+0.44}_{-0.37}$ & -1.15 & -1.23 \\
& & $5.50 \pm 0.50$ & $26.34^{+0.59}_{-0.38}$ & -1.51 & -1.59 \\
HDF (\cite{mad98}) $^{\rm d}$ & 1500 {\AA} & $2.75\pm0.75$ 
 & $26.90\pm0.15$ & -0.95 & -1.02 \\
& & $4.00\pm0.50$ & $26.50\pm0.20$ & -1.35 & -1.43\\
(\cite{ste99}) & 1700 {\AA} & $3.04\pm0.35$ & $27.04\pm0.07$ & -0.81 & -0.89\\
& & $4.13\pm0.35$ & $26.95\pm0.10$ & -0.90 & -0.98\\
(\cite{tre98}) $^{\rm d}$ & 2000 {\AA} 
& $0.15\pm0.15$ & $26.45\pm0.15$ & -1.40 & -1.35 \\
CFRS (\cite{lil96}) $^{\rm d}$ & 2800 {\AA} & $0.35 \pm0.15$ 
 & $26.14\pm0.07$ & -1.71 & -1.68\\
& & $0.625\pm0.125$& $26.46\pm0.08$ & -1.39 & -1.39\\
& & $0.875\pm0.125$& $26.78\pm0.15$ & -1.07 & -1.10\\
HDF (\cite{cow99}) $^{\rm d}$ & 2800 {\AA} & $0.35 \pm0.15$
 & $25.91\pm0.12$ & -1.69 & -1.66\\
& & $0.625\pm0.125$& $26.15\pm0.12$ & -1.45 & -1.45\\
& & $0.875\pm0.125$& $26.06\pm0.12$ & -1.54 & -1.57\\
& & $1.25\pm0.25$ &  $26.35\pm0.10$ & -1.25 & -1.30\\
HDF (\cite{con97}) $^{\rm d}$ & 2800 {\AA} & $0.75\pm0.25$ 
 & $26.77\pm0.15$ & -1.08 & -1.10 \\
& & $1.25\pm0.25$ & $26.94\pm0.15$ & -0.91 & -0.96\\
& & $1.75\pm0.25$ & $26.84\pm0.15$ & -1.26 & -1.32\\
HDF (\cite{saw97}) $^{\rm d}$ & 3000 {\AA} & $0.35\pm0.15$ 
 & $26.51\pm0.47$ & -1.34 & -1.31 \\
& & $0.75\pm0.25$ & $26.74\pm0.06$ & -1.11 & -1.13\\
& & $1.50\pm0.50$ & $26.93\pm0.05$ & -0.92 & -0.97\\
& & $2.50\pm0.50$ & $27.28\pm0.06$ & -0.57 & -0.64\\
& & $3.50\pm0.50$ & $26.91\pm0.10$ & -0.94 & -1.02\\
(\cite{gal95}) & H$_{\rm \alpha}$ & $0.025\pm0.025$&$39.09\pm0.04$ & -2.01 & -1.89 \\
CFRS (\cite{tm98}) & H$_{\rm \alpha}$ 
 & $0.2\pm0.1$ & $39.44\pm0.04$ & -1.66 & -1.60 \\
CFRS (\cite{gla99}) & H$_{\rm \alpha}$ 
 & $0.875\pm0.125$& $40.01\pm0.15$ & -0.91 & -0.94\\
(\cite{yan99}) & H$_{\rm \alpha}$ & $1.3\pm0.5$ & $40.21\pm0.13$ & -0.89 & -0.94 \\
CFRS (\cite{ham97}) $^{\rm d}$ & {O}{II} 3727 {\AA}
 & $0.35 \pm 0.15$ & $38.63^{+0.06}_{-0.08}$ & -1.97 & -1.94 \\
& & $0.625\pm 0.125$ & $39.16^{+0.11}_{-0.15}$ & -1.44 & -1.44\\
& & $0.875\pm 0.125$ & $39.56^{+0.20}_{-0.38}$ & -1.04 & -1.07\\
CFRS (\cite{flo99}) & 15 $\mu$m 
 & $0.35\pm0.15$ & $41.97\pm0.25$ & -1.46 & -1.43\\
& & $0.625\pm0.125$& $42.20\pm0.25$ & -1.15 & -1.15\\
& & $0.875\pm0.125$& $42.53\pm0.25$ & -0.82 & -0.85\\
HDF (\cite{hug98}) & 60 $\mu$m & $3.0 \pm 1.0$
 & $\leq 42.26$ & -0.98 & -1.05 \\
(\cite{haa00}) & 1.4 GHz & $0.28\pm0.12$
 & $26.75^{+0.14}_{-0.21}$ & -1.17 & -1.13\\
& & $0.46\pm0.05$ & $27.03^{+0.24}_{-0.21}$ & -0.89 & -0.87\\
& & $0.60\pm0.05$ & $27.12^{+0.16}_{-0.26}$ & -0.80 & -0.80\\
& & $0.81\pm0.08$ & $27.39^{+0.13}_{-0.18}$ & -0.53 & -0.55\\
& & $1.60\pm0.64$ & $27.54^{+0.13}_{-0.18}$ & -0.38 & -0.44\\
\end{tabular}
\end{center}
\caption[Observational estimates of the cosmic SFR]{Observational estimates of the cosmic SFR density at different epochs. $\rho_{\rm L}$ refers to the luminosity density at the appropriate wavelength, expressed in erg s$^{-1}$Hz$^{-1}$ Mpc$^{-3}$. $\dot\rho_*$ represents the comoving SFR density (in M$_\odot$ yr$^{-1}$ Mpc$^{-3}$) for our SCDM and $\Lambda$CDM cosmologies. $^{\rm d}$ Original data has been corrected for dust extinction.}
\label{tableSFR}
\end{table*}

A compilation of recent results is given in Table~\ref{tableSFR}, where the first column indicates the bibliographic reference, as well as the survey from which the galaxy sample was extracted. Next columns correspond to the chosen wavelength and redshifts of each analysis, and finally the comoving luminosity and SFR densities.

Conversion factors between the last two quantities differ significantly from author to author, and in some cases only the luminosity density was provided. If there is a SFR estimate in the original paper, then it is quoted in Table~\ref{tableSFR} {\em corrected by dust extinction}; else, we follow \citet{Kennicutt98} prescription for an exponential burst and a Salpeter IMF between 0.1 and 100 M$_\odot$:
\begin{equation}
 \dot\rho_*\ [\rm{M}_\odot\ yr^{-1}]=\cases{
 1.4\times10^{-28}L_{\rm UV}\ [\rm{erg\ s^{-1}Hz^{-1}}] \cr
 1.4\times10^{-41}L_{3727}\ [\rm{erg\ s^{-1}Hz^{-1}}]}
\end{equation}

For dust extinction, we have applied a correction of A(1500-2000{\AA})=1.2 mag and A(2880{\AA},{O}{II})=0.625 mag, which correspond to factors of 3.02 and 1.78 respectively. All modifications to the original data are noted by a footnote mark on the bibliographic reference in Table~\ref{tableSFR}.

%----------------------------------

\subsubsection*{Cosmology}

Most observational values given in Table~\ref{tableSFR} were calculated assuming a standard cold dark matter (SCDM) cosmology with dimensionless Hubble parameter $h=0.5$ and mean matter density $\Omega_{\rm m} = 1$.  Since the conversion factor from luminosity to SFR and the correction for dust extinction are completely independent of the cosmological scenario, the computation of the comoving SFR density in a $\Lambda$CDM ($h=0.7$) model only involves the transformation of the luminosity densities.

In a flat universe, the volume enclosed by a solid angle $\Delta\omega$ between
$z-\Delta z$ and $z+\Delta z$ is
\begin{equation}
 \displaystyle V(z,\Delta z)=\frac{4\pi}{3}\Delta\omega\left( d_{\rm m}^3(z+\Delta
z)-d_{\rm m}^3(z-\Delta z)\right),
\end{equation}
where $d_{\rm m}(z)$ is the comoving distance. Luminosity scales with luminosity distance, $d_{\rm L}(z)=(1+z)d_{\rm m}(z)$, as $d_{\rm L}^2$. Then, assuming that all galaxies are located at the center of the interval in $z$, the luminosity density (and hence the total SFR density) will be proportional to
\begin{equation}
 \dot{\rho}_*(z)\propto\frac{L(z)}{V(z,\Delta z)} \propto  \frac{d_{\rm m}^2(z)}{d_{\rm m}^3(z+\Delta
z)-d_{\rm m}^3(z-\Delta z)}
\label{ecLambda}
\end{equation}

The conversion between SFR densities in $\Lambda$CDM and variants of the SCDM
($h=0.5$) cosmologies can be approximated, to first order in $\Delta z$, 
 as
\begin{equation}
 \frac{\dot\rho_*^{\rm \Lambda CDM}}{\dot\rho_*^{\rm SCDM}}\sim\frac{h^{\rm \Lambda
 CDM}}{h^{\rm SCDM}}\ \sqrt{\Omega_{\rm m}+\Omega_\Lambda(1+z)^{-3}}
\label{ecVolker}
\end{equation}

%________________________________________
\begin{figure}
  \centering \includegraphics[width=8cm]{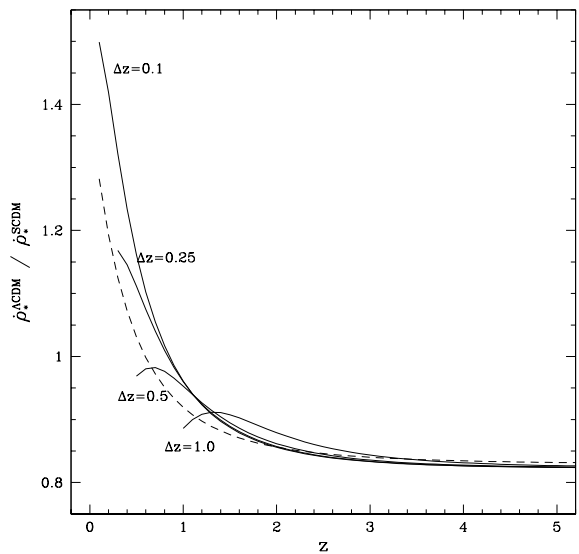}
  \caption[Conversion between SCDM and \lcdm star formation rates]{Conversion factor between star formation densities in SCDM and \lcdm cosmological models, for several redshift intervals $\Delta z$. Dashed line corresponds to the approximation (\ref{ecVolker}).}
 \label{volker}
\end{figure}
%________________________________________

Figure~\ref{volker} shows this factor, as well as the exact expression for different values of $\Delta z$, which requires a numerical evaluation of the luminosity distance in the $\Lambda$CDM model. Although the results are very similar, we have chosen to use the exact expression (\ref{ecLambda}) to compute the SFRs in Table~\ref{tableSFR} in order to minimize errors.

At low redshifts, the data are consistent for all the independent
tracers of the star formation activity once the effect of dust
obscuration is taken into account. The steep increase in the cosmic SFR
density from the present day to $z\sim1$ is firmly established. However,
there is not yet a full agreement on  whether there is a characteristic
epoch of star formation in the universe around $z\sim1.5$ or, on the
contrary,   the SFR declines very slowly towards higher $z$, as more recent observations seem to point out. The results of ongoing deep surveys will be extremely helpful in order to reach a definitive conclusion from the observational point of view.

%-----------------------------------------------------------------------------

\subsection{Simulations}
\label{sims}

In order to study the cosmic star formation history, several
simulations have been accomplished using the Eulerian gasdynamical code
\YK3 (succintly described in Section~\ref{secYK3}), which incorporates
a simple physically motivated model to implement star formation and
feedback into the ISM. Our aim was to investigate the separate effects
of cosmology, supernova explosions and photoionisation on the
self-regulation of the star formation activity. For details on the
numerical experiments, the reader is referred to Section~\ref{simyk3}.

%----------------------------------

\subsubsection*{Cosmology}

\begin{figure}
  \centering \includegraphics[width=12cm]{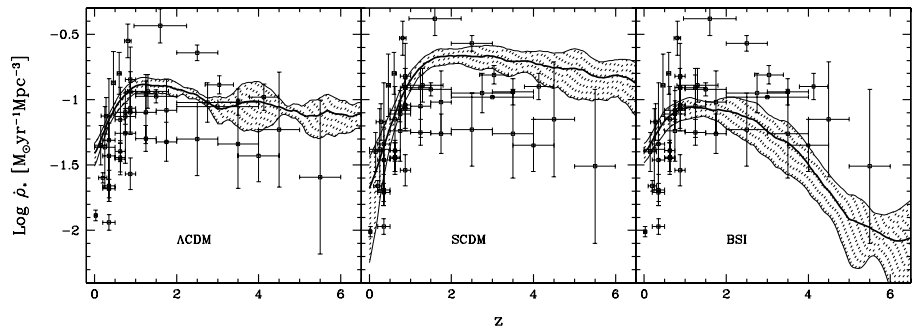}
  \caption[Evolution of the SFR density in different cosmologies]{Evolution of the comoving SFR density in different cosmologies (experiments $\Lambda$CDM5, SCDM and BSI). The average value over different realizations is plotted as a thick solid line, while the shaded area represents the $1\sigma$ deviation from the mean. Dots correspond to the observational data points listed in \citet{Ascasibar02}}
  \label{figCosmo}
\end{figure}

Three different cosmological scenarios have been considered in the
present study: the currently-favoured \lcdm universe, the
'old-fashioned'  standard CDM model, and a modified version of the
latter: The Broken Scale Invariance  (BSI) model which is based on a
double inflation scenario \citep[see e.g.][for a description]{bsi95}. The post-recombination power spectrum 
 of the BSI model  is similar to that of the $\tau$CDM scenario, in which 
the dark matter consists of a decaying neutrino.  The global star
formation histories found in our simulations of each one of those three
cosmologies are compared in Figure~\ref{figCosmo}.  A crucial factor is
the initial normalisation of the power spectrum, since this parameter
sets the amount and evolution of substructure, and therefore represents
one of the key ingredients in determining the rate of cooling and
subsequent conversion of gas  into stars.

%________________________________________
%________________________________________

Cosmological scenarios with CDM alone cannot explain structure formation on both small and very large scales. However, we have included SCDM as a reference model in our comparison. Due to its large power at small scales, this scenario produces many structures at high redshift, and so it features a flat star formation rate even beyond $z\sim6$, with a plateau about one decade over the observations (Figure~\ref{figCosmo}, middle panel). It does not seem very likely that this difference can be explained by dust absorption alone. In this model, too many baryons are transformed into stars, and it happens too early with respect to observations. We claim that the cosmic star formation history poses an additional problem for the SCDM model.

On the contrary, the BSI model leads to a clear maximum in the SFR at $z\sim2$, but the absolute value is too small and the decrease at higher redshifts is clearly too steep (Figure~\ref{figCosmo}, right). This model has not enough power at small scales for producing enough stars at high redshift. As a side effect, the cosmic star formation rate density remains too high at low redshifts, since a large part of cold gas remains in high density regions triggering further star formation activity.

A cosmological model with non-zero cosmological constant offers the best agreement with observations (Figure~\ref{figCosmo}, left). Both the overall trend and the height of the plateau fit reasonably well the observational data points. At present, the $\Lambda$CDM model fits best all known observational data \citep[see e.g.][]{Bahcall99}. Our results confirm that this model can also reproduce the observed star formation history of the universe when dust extinction is taken into account, in which case a significant amount of star formation is expected to take place at high redshift.

A further feature, common to all cosmologies, is the smooth redshift
dependence of the mean SFR and the small range of scatter. This is in
sharp contrast to the evolution of individual baryonic clumps or the
behavior of different regions in a single time-step of the
simulation. There, we observe huge bursts of star formation (with an
increase in activity by a factor of 10 and higher with respect to the
average value), as well as a quiescent regime in which the star
formation rate remains  approximately constant.

In these simulations, the first stars are created around $z\sim10$, which seems to be not too unrealistic, but it may depend on resolution. This was not a primary issue of the present study.

%----------------------------------

\subsubsection*{Feedback}

\begin{figure}
  \centering \includegraphics[width=10cm]{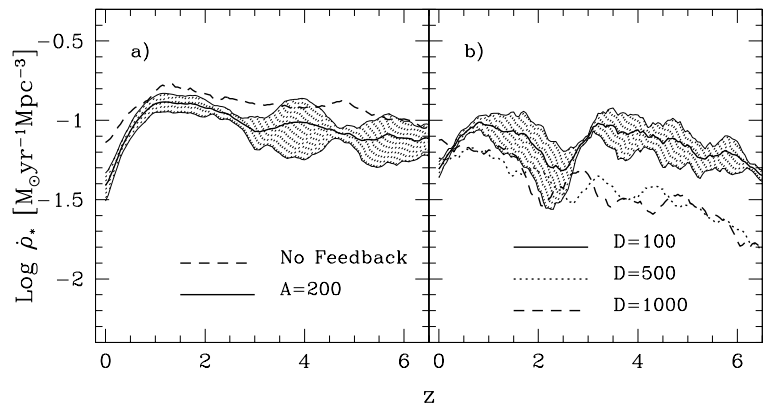}
  \caption[Effects of stellar feedback]{Effects of stellar feedback on the cosmic SFR density. a) $\Lambda$CDM5 simulations with supernova feedback parameter $A=200$ (solid line) within $1\sigma $ error (shaded area) and $A=0$ (dashed line). b) Same as (a) but for $\Lambda$CDM4 simulations run with different values of the overdensity threshold $\mathcal D$ imposed by photoionization.}
  \label{figFeedback}
\end{figure}

Figure~\ref{figFeedback} shows the results of experiments $\Lambda$CDM5 and $\Lambda$CDM4, from where the effects of supernova feedback and photoionization were studied. The same trends described in this section were also seen in SCDM and BSI simulations.

%________________________________________
%________________________________________

Supernovae (left panel of Figure~\ref{figFeedback}) play a key role in
the slope of the SFR density at low $z$. Gas heating and evaporation
act as a self-regulating mechanism that inhibits subsequent gas infall
when many new stars are formed. This is specially important in the most
massive objects, which show a much higher star formation activity when
feedback is not implemented. 

In addition, supernova explosions can also blow away the gas reservoirs
of dwarf galaxies. The effect on the global star formation rate is more
noticeable at high redshift, when there are still very few galaxies
massive enough to retain the heated gas within their gravitational
potentials. However, photoionization (right panel) is the dominant
effect at very early epochs, preventing the gas from cooling and
forming stars in low density regions \citep{efs2}. As we have discussed
in Section~\ref{secYK3}, the model implemented in \YK3 takes into account this effect through the overdensity threshold parameter ${\mathcal D}$ for star formation. Only in regions denser than ${\mathcal D}$, molecular clouds will be able to form, and the onset of cosmic star formation activity is thus retarded until objects become dense enough to screen the photoionizing background.

We found a SFR density consistent with observational data for ${\mathcal D}\sim100$. This is in roughly agreement with results from a more detailed theoretical estimate of the same parameter \citep{muc97}, although in that case the actual value of ${\mathcal D}$, as well as the temperature range in which thermal instability is an efficient process, depend on the local gas density and the intensity of the ionizing flux.

%----------------------------------

\subsubsection*{Resolution and volume}
\label{secSFRresol}

Star formation and feedback processes are incorporated in gasdynamical simulations (either SPH or Eulerian) by means of rather simple recipes which pretend to {\em extrapolate} the effects of local physics acting on scales several orders of magnitude below the resolution limit (often at kpc or even higher, see Table~\ref{tableSFRsims} for these simulations). In the algorithm implemented in \YK3, part of this extrapolation is hidden in two parameters: The supernova feedback parameter $A$ and the overdensity threshold for star formation ${\mathcal D}$. Whether this parameterization is a reasonable extrapolation of the underlying subgrid physics depends on the stability of the results against changes in spatial resolution of the simulations.

%________________________________________
\begin{figure}
  \centering \includegraphics[width=8cm]{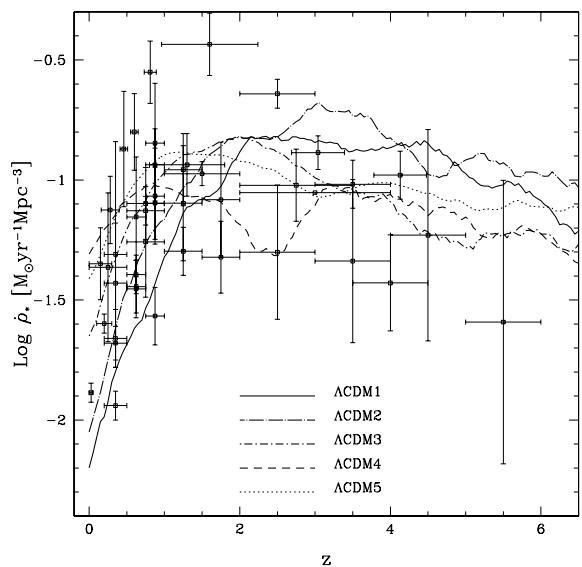}
  \caption[Effects of resolution and volume]{Comoving SFR in our 5 $\Lambda$CDM experiments, all with $A=200$ and ${\mathcal D}=100$. A significant star formation activity is found at high $z$ regardless of resolution and volume. This is consistent with the full set of observational data points (Table \ref{tableSFR}).}
  \label{figCSFR}
\end{figure}
%________________________________________

On the other hand, when one tries to estimate cosmological averaged
quantities, like the SFR density, it is necessary to check for
deviations of these quantities due to small number statistics. In the
case that matters here, a volume averaged quantity, the {\em cosmic
  variance} can have a relatively large contribution in the particular
determination of the comoving SFR density. Thus, we have simulated
different volumes with similar resolutions, depending on
the computational facilities at our disposal.

To check for these two effects we compare in Figure~\ref{figCSFR} our results for $\Lambda$CDM simulations with different resolutions and volumes. The most noticeable feature that can be observed in this figure is the steeper slope of the local SFR for larger volume simulations. As we increase the volume, larger structures are being simulated (i.e.  clusters), in which the star formation activity is significantly reduced during recent epochs. In simulations with smaller volumes, most of the galaxies that form are either isolated or in small groups, and hence have a completely different star formation activity. The effects of environment on the SFR are investigated in more detail in the next section.

If this trend we found in our simulations is indeed real, we can conclude that the effect of cosmic variance might be quite important in the quantitative estimation of the SFR in the nearby universe, and hence it seems very likely that it must be taken into account in the determiniation (either numerical or observational) of the evolution of the cosmic star formation rate density from $z\geq 1$ to the present.

Although we do not find significant differences in our estimates of the SFR density from simulations with different spatial resolutions, there is a general trend to have more SFR density at early times in the highest resolution experiments. This is a consequence of the overcooling problem \citep[see e.g.][for a recent revuew]{bal01}, and it is directly related to the increasing number of low-mass halos resolved in these simulations, particularly at high redshift. A detailed analysis of this question would require a more realistic treatment of photoionization than the simplistic prescription of an overdensity threshold for star formation.

Keeping all these considerations in mind, we should be extremely cautious when making {\em quantitative} estimates of the cosmic SFR density at any given $z$. However, this quantity shows a similar behavior in all $\Lambda$CDM experiments. This is a very promising result indicating that its {\em qualitative} evolution (i.e. the shape of the curve) is a robust prediction of our simulations. More precisely, we claim that no characteristic epoch of cosmic star formation is found for a $\Lambda$CDM universe. The comoving SFR density increases with lookback time until it reaches a nearly constant mean plateau beyond $z=2$. Stars probably formed gradually at high redshift, according to most recent observations and taking dust extinction into account. At some point around $z\sim1$, the global star formation activity would have declined sharply until the present day, although individual SFRs of active galaxies are kept approximately constant.

\begin{table}
\begin{center}
\begin{tabular}{ccc}
{\sc Experiment} & $\rho_*(z=0)$ & $\rho_*(z=3)/\rho_*(0)$ \\ \hline \\
 $\Lambda$CDM1 & 6.50 & 0.252 \\
 $\Lambda$CDM2 & 8.60 & 0.243 \\
 $\Lambda$CDM3 & 9.51 & 0.115 \\
 $\Lambda$CDM4 & 9.01 & 0.116 \\
 $\Lambda$CDM5 & 10.9 & 0.119 \\
 SCDM & 12.3 & 0.115 \\
 BSI & 7.37 & 0.041 \\
\end{tabular}
\end{center}
\caption[Cumulative star density]{Present star density (in $10^8\ h$ M$_\odot$ Mpc$^{-3}$) and fraction of stars older than $z=3$ for the different numerical experiments.}
\label{rhoSt}
\end{table}

The star density $\rho_*(z)$ shows a similar behavior for both CDM and $\Lambda$CDM models. The most remarkable difference is found in the present value of $\rho_*(0)$ (see Table \ref{rhoSt}), which can be observationally deduced from the local luminosity density, assuming a constant mass-to-light ratio \citep[e.g.][]{mad96,fuk98}:
\begin{equation}
 \rho_*^{\rm obs}(0)\sim7\times10^8\ h\ {\rm M}_\odot\ {\rm Mpc}^{-3}
\end{equation}

Based on the fraction of stars in bulges, ellipticals and S0s,
\citet{ren98} estimated that about $30\%$ of the stellar content of the
universe should have already been formed by $z=3$, regardless of the
cosmological model. The amount of stars produced in experiments $\Lambda$CDM1
and 2 (Table \ref{rhoSt}), both at $z=3$ and $z=0$, are in fair
agreement with the observationally derived values for these
quantities. For the other experiments, the fraction of stars older than
$z=3$ is significantly lower, not only beacuse of a low star formation
activity at high redshift, but also due to a higher SFR in the local
universe, as well as to the cosmic bias towards low-density regions
introduced by the small simulated volumes.

These results have been recently confirmed by \citet{SpringelHernquist} who estimated the cosmic star formation history in the $\Lambda $CDM model using a similar prescription for star formation and feedback in the \g code. Through a series of simulations with different resolutions and volumes, they claim to have found stable results from the numerical point of view. The general features of the cosmic star formation history found by these authors are similar to those reported in \citet{Ascasibar02}. Although sistematically lower, their SFR density is marginally consistent with observational estimates. The reason for this discrepacy could be atrributed to the extreme feedback model they have used in their simulations.

%-----------------------------------------------------------------------------

\section{Star formation in clusters of galaxies}
\label{secSFRclus}

In the hierarchical scenario of structure formation, the first
cosmological objects are expected to collapse around the highest
overdensity peaks, merging untily they eventually form the giant
ellipticals in the centers of clusters and the bulges of the most
massive isolated galaxies.  These objects contain the vast majority of
stars older than $z=3$, and hence their star formation history must be
completely different from that of galaxies living in less dense
environments. 

The strong morphology segregation observed in rich galaxy clusters \citep[e.g.][]{Dressler80} testifies the fundamental role played by the environment on the evolution of galaxies. Yet, the physical mechanisms responsible for such transformations are still a matter of debate. Several processes might alter the star formation conditions in cluster galaxies. Some are related to the interaction with the intracluster medium \citep[a fact already pointed out by][]{GG72} and others account for the effects of tidal forces due to the gravitational potential of the cluster \citep{Merritt83} or galaxy-galaxy interactions \citep{Moore96,Moore98h,Moore99h}. All these mechanisms can produce strong perturbations in the galaxy morphology, which include the formation of tidal tails, dynamical disturbances that appear as asymmetries in the rotation curves \citep{Dahle01} and significant removal of the diffuse gas \citep{GiovanelliHaynes85,ValluriJog90,Quilis00}\footnote{In this regard, our simulations of clusters of galaxies described in
previous chapters give also an indication that gas
stripping is a very important phenomenon  to inhibit star formation in
cluster galaxies. We do observe that most of the small halos that orbit
the cluster potential do loose their gas  very fast. A visual
impression of this phenomenon can be seen  from a series of animations
done  from one of our  SPH cluster simulations. They can be accesed
from the following web page
{\tt http://pollux.ft.uam.es/gustavo/VIDEOS/LCDM80/clustervideo.html}
We plan to do a detailed quatitative  study of this effect in the near
future.}.

Most of these processes are expected to change significantly the star
formation activity of galaxies in clusters. Several studies have
addressed the issue of the influence of the cluster environment on the
SFR of disk galaxies, but no agreement has been established so far:
whereas some authors proposed similar or even enhanced star formation
in cluster spirals than in the field, some others
claim quenched SFRs in cluster spirals. This
discrepancy could arise from non-uniformity of the adopted methods and
wavelengths or from real differences in the studied clusters. 

If the star formation rates are reduced when galaxies are accreted into
larger groups or clusters, the cosmological implications could be
profound, since approximately $30-40\%$ of the present-day galaxies are
observed not to be isolated
\citep[e.g.][and references therein]{Carlberg01}, in agreement
with the results found in numerical models
\citep[e.g.][]{Gottloeber01}. As structure builds up in
the universe, more and more galaxies can be found in groups and, if
these environments serve to terminate star formation, it will be
clearly reflected in the evolution of the universe as a whole,
explaining part of the observed decline in the global SFR at recent
epochs. 
 
%___________________________________________________

\subsection{Observations}

The effect of local environment on galaxy evolution in general is not well understood. Many observational studies of environmental effects have been devoted to the study of galaxies in the cores of rich clusters, which differ so dramatically from more common galaxies \citep[e.g.][]{Dressler80,Dressler85,Couch87,Balogh97,Balogh99,Poggianti99,MossWhittle00,Couch01,Solanes01}. In particular, \citet{Balogh97,Balogh98} showed that the mean cluster galaxy star formation may be supressed as far as twice the virial radius from the cluster center, relative to a field sample with similar bulge-to-disk ratios, physical disk size, and luminosity, which suggests that the observed decrease in the SFR may not be fully explained by the morphology-density relation.

This conclusion has also been suggested by \citep{Hashimoto98}, who studied the relation between local galaxy density and star formation activity in the Las Campanas Redshift Survey, and similar results have been obtained in more recent studies based on the 2dF Galaxy Redshift Survey \citep{Lewis02} and the Early Data Release of the Sloan Digital Sky Survey \citep{Gomez02}.

From a wide field photometric survey using the SUBARU telescope, \citet{Kodama01} reported the detection of an abrupt change in the colours of galaxies at a 'critical' projected surface density of $\sim200\ h_{75}^{-2}$ gal. Mpc$^{-2}$. This corresponds to a radial distace of $\sim1\ h_{75}^{-1}$ Mpc from the cluster center. Such a 'break' has been confirmed by the SFRs derived from the SDSS \citet{Gomez02}, although it occurs at an order of magnitude lower surface density, i.e. at a larger clustercentric radius. Further observations are needed in order to discern whether this difference is physical (i.e. due to the differences in redshift or luminosity limits of both studies) or an artifact due to the specifics of the analysis techniques. 

In any case, the 'critical' density (or radius) where the star formation rate of galaxies is found to change from that of the field seems to be well beyond the virialised regions of clusters of galaxies. Since the mechanisms that alter the SFR in the cores of rich clusters, e.g. ram-pressure stripping \citep{GG72,Quilis00}, galaxy harassment \citep{Moore99h}, or tidal disruption \citep{ByrdValtonen90}, are unlikely to be important at low overdensities, these results point in the direction that the SFR is significantly reduced due to close interactions in less massive groups within the infall region of clusters \citep[e.g.]{Zabludoff96,ZabludoffMulchaey98,Gomez02}.

In this scenario, environmental effects associated to hierarchical
structure formation could be responsible for some of the observed
decline in the cosmic SFR from $z\sim1$ to the present. Results from the
SDSS \citep{Gomez02} seem to support this view, indicating that the
break in the SFR-density relation corresponds approximately to the
turn-around radius of the studied clusters (a few times $\Rv$), which
roughly marks the limit of the gravitational influence of the dark
matter halo, but is too large for the ICM gas to have any significant
influence on the infalling galaxies. 

Although \citet{Lewis02} found that the relation between SFR and local density was independent on the cluster velocity dispersion and presumably mass, a spectral analysis of galaxy groups in the 2dFGRS accomplished by \citet{Martinez02} concludes that even the environment of low mass-systems ($M\sim10^{13}$ \Msun) is effetive in diminishing the process of star formation in their member galaxies, but the fraction of spectral types associated to star forming galaxies decreases with group virial mass. Furthermore, \citet{Balogh02} have also found some evidence that the morphology-density relation is correlated with the cluster X-ray luminosity.

To evaluate the contribution of clusters of galaxies to the current
star formation rate density, \citet{Iglesias02} computed the H$\alpha$
luminosity function of the central regions of two nearby clusters,
obtaining the total SFR per unit volume of the member
galaxies. Multiplying by the cluster density in the universe
\citep{Bramel00}, these authors conclude that the SFR in
type 2 and type 1 clusters can account for approximately $0.25\%$ and
$10.8\%$ of the cosmic stellar production at $z=0$, respectively. 

%___________________________________________________

\subsection{Simulations}

The environmental effects on the star formation activity that can be
seen in numerical experiments are due to differences in merging history
and removal of the hot gas reservoir that surrounds every isolated
galaxy. In semi-analytical models
\citep[e.g.][]{Kauffmann93,SP99,Cole00}, it is assumed that this supply
of fresh fuel for star formation is lost when galaxies are accreted
into larger haloes. Therefore, in this simple scheme, star formation
rates begin to decline for any satellite galaxy, whether in a poor
group or in a rich cluster. 

Although these models are able to successfully match some of the observed trends, such as the radial gradients in SFR within the virial radius of clusters \citep{Balogh00,Diaferio01,OkamotoNagashima01}, they lack at present the capability to explore how galaxies lose their gas \citep[see e.g.][]{Bekki02}. One strength of numerical simulations is that they address the problem from first principles, making it possible to constrain the contribution of the proposed mechanisms acting on the SFR, as well as the role played by different environments in galaxy formation and evolution.

\citet{Gottloeber01} have studied the merging history of dark matter halos as a function of their environment, finding that haloes located inside clusters have formed earlier than isolated objects of the same mass. Moreover, they showed that at higher redshifts ($z\sim 1-4$), progenitors of cluster and group galaxies have 3--5 times higher merger rates than isolated galaxies. Mergers of galaxies are thought to play a crucial role in the evolution of the SFR.  In particular, the inflow of material may serve as a source of high-density gas and therefore increase the star formation activity.

Amongst all the numerical experiments we have accomplished with \YK3,
only $\Lambda$CDM 1 and 2 sample a volume large enough to contain clusters
and groups of galaxies. 

%________________________________________ 
\begin{figure}
  \centering \includegraphics[width=10cm]{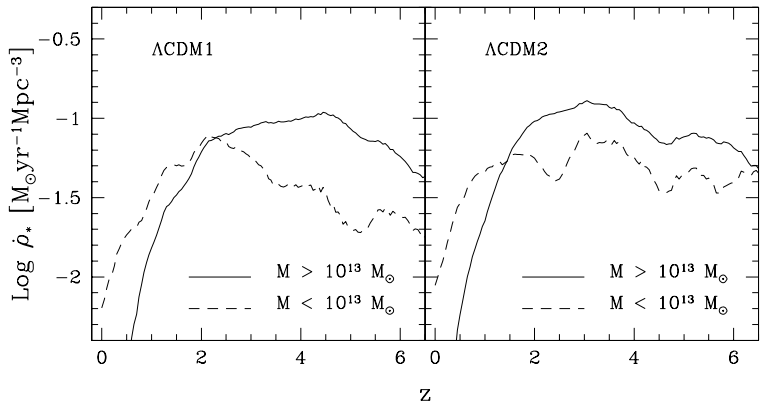}
  \caption[Star formation rate densities in galaxy clusters]{Star formation rate densities in models $\Lambda$CDM1 and $\Lambda$CDM2 due to halos more (solid lines) and less (dashed lines) massive than  $10^{13}$ M$_\odot$ at $z=0$.}
  \label{figBIG}
\end{figure}
%________________________________________ 

In Figure~\ref{figBIG} we plot the contribution to the cosmic SFR density due to the progenitors of present-day objects, according to the mass at $z=0$. The stellar production of objects that end up in haloes more massive than $10^{13}$ M$_\odot$ at present (clusters, large groups) are plotted as solid lines, whereas the SFR density of those that are in less massive objects at the end of the simulation (isolated galaxies or small groups) is shown by the dashed lines.

The most massive objects have their peaks of star formation activity
around $z\sim3$, undergoing passive evolution since $z\sim2$. This is in
excellent agreement with the merging history found by
\citet{Gottloeber01} from N-body simulations.
 The progenitors of these objects were responsible
for the bulk of star formation at high redshift, and could be very likely
associated with the Lyman Break Galaxy population. 

On the other hand, galaxies that are isolated at $z=0$ or galaxies
residing in small groups tend to be much younger. In agreement with the
results of \citet{Iglesias02}, these objects are responsible for most of
the current stellar production. In these low-density areas, halos
massive enough to retain their gas content display a nearly constant
SFR with sporadic burst episodes. 

Unfortunately, the simulated volume in these numerical experiments is
too small to be cosmologically representative. In fact, the larger
volume covered by $\Lambda$CDM1 contains only one galaxy cluster of
approximately $3\times10^{14}$ M$_\odot$, whereas the simulation $\Lambda$CDM2
contains four massive groups of about $5\times 10^{13}$ M$_\odot$. Comparing the
solid lines in the left and right panels of Figure~\ref{figBIG}, we
find qualitatively the expected scenario: the peak of star formation in
the progenitors of the more massive cluster happened earlier (at
$z\sim4.5$) than for the less massive groups ($z\sim3$). This lends some
support to the claim that the effect of environment on the SFR of
infalling galaxies is related to the group mass \citep{Martinez02} or
luminosity \citep{Balogh02}. 

As we have seen in Section~\ref{sims}, photoionization and supernova
explosions inhibit cooling and star formation in very low-mass objects,
providing a physical mechanism to suppress the over-cooling of gas at
very early epochs. Thus, the formation of the first stars is biased
towards high overdensity regions that will collapse later to form
the cores of the present-day galaxy clusters. As their hierarchical
assembly proceeds, the gas component of these objects is heated by
shock waves produced during mergers, and the cooling times become too
long for efficient star formation to take place. Consequently, the SFR
drops drastically once the initial reservoir of cold gas has been
exhausted. 

At the same time, new galaxies will form and produce stars. Some of
them will remain in rather isolated areas until present (dashed lines
in Figure~\ref{figBIG}) and some others will fall recently into the
potential wells of the clusters. It seems very likely that tidal
interactions trigger intense star formation bursts in these infalling
galaxies \citep[e.g.][]{lav88}.

Observations and simulations coincide in pointing out that star formation activity in galaxies located in dense cluster environments decays (or even stops). Thus, their colours become redder, giving rise to the \citet{but78} effect. Moreover, these objects are expected to experience many close encounters and sometimes mergers with neighpouring cluster galaxies, which could explain their morphological transformation from an initial spiral Hubble type into the S0 population that is observed to dominate the inner regions of clusters \citep{dre97}. In fact, our high resolution simulations of cluster formation  done with \g show that dark matter sub-haloes orbit many times in the cluster potential, loosing most of their gas envelope, but with keeping most of their dark matter content\footnote{see {\tt http://pollux.ft.uam.es/gustavo/VIDEOS/LCDM80/clustervideo.html} for detailed animation of dark subhalo orbiting within the  cluster potential}.

%-----------------------------------------------------------------------------

%%% Local Variables: 
%%% mode: latex
%%% TeX-master: t
%%% End: 

%-----------------------------------------------------------------------------
   \chapter{Conclusions and future prospects}
   \label{chapConclus}
%-----------------------------------------------------------------------------

\begin{quote}{\em
So here we are again to experience the bitter, scalding end\\
and we're the only ones who can perceive it\\
But others sing of beauty and the story that's unfolded\\
as one that deserves praise and ritual}

-- Bad Religion : {\em Pessimistic Lines} (1988) --\\
\end{quote}
%-----------------------------------------------------------------------------

% {\gothfamily {\Huge V}{\large aya}},

%-----------------------------------------------------------------------------

\section{Conclusions}

{\gothfamily {\Huge F}{\large rom}} a set of numerical experiments, using different codes and initial conditions, we have been able to compile a database of numerical galaxy clusters. High resolution (spatial and mass) has been achieved in selected areas of the computational box by means of refinement techniques. The simulations were carried out within the framework of the \lcdm cosmological scenario, which is the most favoured by current observational data. The initial realisation was equivalent to a set-up with $1024^3$ homogeneous particles.

From our cluster database, we have been able to study the physical properties of both the dark and baryonic components. Most of the work accomplished for the present thesis focuses on the analysis of the simulated clusters at $z=0$. Below we summarise the major conclusions that can be derived from the results of this analysis.

%___________________________________________________

\subsection{Dark matter}

\begin{enumerate}

 \item Numerical predictions about the dark matter distribution are robust within the resolution limits. ART and \g give consistent results. Perturbations to the dark matter potential due to the presence of baryons are negligible on cluster scales when only adiabatic physics is taken into account.

  \item From a sample of 15 clusters, 6 have been catalogued as relaxed systems, 4 as minor mergers and 5 as major mergers. The amount of dynamical activity is significantly higher in the low-mass end of the sample, in agreement with the expectation from hierarchical assembly.

  \item There are systematic differences between the inner profiles of our numerical clusters according to their dynamical state. Relaxed clusters are better described by the profile proposed by \citet{NFW97}, while minor mergers are closer to the form advocated by \citet{Moore99}. Major mergers follow a pure power-law distribution for more than one order of magnitude in radius; their profiles are even steeper than $\rho\propto r^{-1.5}$.

  \item The logarithmic slope of the mass profile of relaxed haloes does not reach any asymptotic slope at our resolution limit. Indeed, it seems to increase above the NFW value for small enough radii. More resolution is needed in order to verify this conclusion.

  \item Merging systems are systematically more extended than relaxed clusters. This effect completely disrupts the usual $c-\Mv$ relation, although some correlation is still found in the $\rho_s-r_s$ plane. The best-fit values of these parameters are consistent with those inferred from observations.

  \item The phase-space density of our clusters is well fitted a power-law $\rho/\sigma^3=2.24\times10^4~r^{-1.875}$, similar to that found by \citet{TN01} for galaxy-size haloes.

  \item We find no clear evidence of a 'universal' profile of the anisotropy parameter $\beta\equiv1-\sigma_\theta^2/\sigma_r^2$. Although there are hints of some average trend, the scatter shown by individual profiles is extremely high. An isotropic velocity field ($\beta=0$) can be confidently ruled out in relaxed clusters and minor mergers.

  \item Angular momentum grows roughly proportional to $r$ is our clusters, in agreement with previous numerical work. Rotational support from bulk rotation is negligible with respect to the contribution of random motions. The eccentricity of individual particle orbits is approximately $e=0.5$ throughout the whole cluster, but two of our objects deviate from this behaviour.

  \item The inclusion of angular momentum and the choice of the initial conditions are the most delicate issues concerning the spherical infall model. Our clusters are very well described by $\sim3\sigma$ peaks of the density field, smoothed on $\sim1\ h^{-1}$ Mpc scales, when the eccentricity of the orbits is fixed to $e=0.5$.

\end{enumerate}

%___________________________________________________

\subsection{Gas}

\begin{enumerate}

  \item We find additional evidence of non-physical entropy losses in the standard implementation of the SPH algorithm. This effect can be of critical importance in the computation of the entropy profiles in groups and clusters of galaxies. The new formulation of SPH proposed by \citet{gadgetEntro02} seems to overcome this problem; high-resolution simulations accomplished with this code are consistent with the results of the Eulerian code ART. An excessively low number of particles can produce an isentropic core due to numerical effects.

  \item The assumption of hydrostatic equilibrium is justified for the clusters classified as 'relaxed' on dynamical grounds, although some deviations occur in the outer parts of the cluster. It holds only marginally for minor mergers, and not at all for systems undergoing a major merging event.

  \item A polytropic equation of state with $\gamma\sim1.18$ provides a good description of the ICM gas. Only major mergers deviate from a polytropic relation.

  \item Under the assumptions of hydrostatic equilibrium, polytropic equation of state, a universal profile for the dark matter, and constant baryon fraction at large radii, the temperature profile is {\em determined} by the underlying dark matter distribution. Our results are consistent with \citet{KS01}, who follow a slightly different prescription. The temperature profile derived from a NFW profile provides an excellent fit to the simulation data.

  \item Projected X-ray temperature profiles are very similar to those observed in real clusters. Although the situation is less clear in the inner regions, probably in connection with non-adiabatic phenomena, an isothermal temperature profile can be ruled out both on observational and theoretical grounds. Our data is very well fitted by the formula proposed by \citet{Loken02}, but no agreement is found regarding the best-fit values of the free parameters.

  \item The density profile of the ICM gas can also be derived by the procedure outlined above (4). Again, it is {\em entirely} determined once the parameters describing the dark matter halo are known. Nevertheless, the agreement with numerical data is considerably poorer than for the temperature profile.

  \item The relationship between dark and baryonic profiles is reflected also in the scaling relations between global quantities. The $M-\Tx$ relation is predicted to follow the self-similar scaling relation, particularly at high overdensities. As we go to larger radii, the density of merging systems is significantly higher. This adds to the fact that our prescription severely underestimates the gas density at large radii.

  \item Our mass-temperature relation agrees relatively well with previous independent numerical experiments. However, some discrepancies are found in the emission-weighted temperature due to differences in the inner shape of the profile. We claim that this is probably a resolution effect. The agreement with observational estimates based on the \lcdm model and a NFW profile is good; other observational methods tend to predict slightly lower normalisations, but the scatter is rather large.

  \item The $\Lx-\Tx$ relation can be derived from the definition of X-ray luminosity, substituting the appropriate profiles for the gas density and temperature. The scaling law so obtained is {\em not} expected to follow the self-similar prediction $\Lx\propto\Tx^2$, since it depends on three additional factors: the square of the baryon fraction, $F^2$, the structure parameter, $Q$, and the exact normalisation of the $M-T$ relation, $M_0$. When these factors are taken into account, the agreement between predicted and simulated scaling laws is extremely good.

  \item At low overdensities (e.g. $\Delta=200$), the baryon fraction is close to the cosmic value for all clusters, the structure parameter $Q$ increases dramatically with $\Tx$ (i.e. low-mass haloes are much less concentrated due to merging), but $M_0$ is a decreasing function of temperature. The last two effects roughly compensate, so they simply introduce some scatter in the $\Lx-\Tx$ relation observed at the virial radius.

  \item At high overdensities (e.g. $\Delta=2500$), both the structure factor and the normalisation of the $M-T$ relation are almost the same for every cluster, but the baryon fraction is lower in the less-concentrated low-temperature clusters. Hence, the X-ray luminosity of these systems is suppressed with respect to that of the massive clusters, and the observed $\Lx-\Tx$ relation steepens considerably.

  \item Our normalisation of the $\Lx-\Tx$ relation is significantly higher than that found in previous numerical studies. This could well be a numerical artifact due to our higher resolution. Low-resolution studies would not be able to solve the high-density regions near the centre of the cluster, and hence the bolometric luminosity would be underestimated. This issue must be investigated in more detail.

  \item Observational estimates also yield lower luminosities for a given temperature. The discrepancy amounts to a factor of 2 for the only estimate based on a \lcdm cosmology, and it is much worse for SCDM estimates. However, the comparison between different cosmologies is not straightforward.

\end{enumerate}

%___________________________________________________

\subsection{Stars}

\begin{enumerate}

\item A $\Lambda$CDM cosmology shows the best agreement with observations of the cosmic star formation history. Standard CDM tends to overpredict the cosmic star formation rate density, while BSI fails to form enough stars at high $z$ due to its reduced power at small scales.

\item Photoionisation fixes the onset of star formation in the universe, suppressing star formation in low-mass objects at high $z$. Supernovae feedback is very efficient to self-regulate the process of converting gas into stars, preventing the most massive halos to form too many stars at later times.

\item In contrast to the first observational estimates \citep{mad96}, but in agreement with more recent data \citep{ste99}, the $\Lambda$CDM model predicts almost no drop in the cosmic SFR for $2 < z < 5$. Star formation seems to be a gradual process with no characteristic epoch.

\item The comoving simulated volumes in our experiments are comparable to those covered by observations. Thus, a non-negligible error is expected in the observational measurements due to cosmic variance associated with small volume statistics. This is most important when one wants to determine the steep slope of the SFR density from $z\sim 1$ to the present time.

\item Star formation histories are very different for objects that end up in the cores of clusters and isolated galaxies. Galaxies in cluster cores have much older stellar populations, and their activity is drastically reduced when they loose their cold gas reservoirs.

\item In agreement with recent observations, we find that the mechanism quenching the star formation in infalling galaxies is efficient in low-mass groups ($M\sim10^{13}$ \Msun) as well as in clusters.

\end{enumerate}

%-----------------------------------------------------------------------------

\section{Future prospects}

Our database still has a lot of {\em juice} to be squeezed from it. In particular, we have restricted our analysis to the present epoch, and there is a lot of work that can be done regarding the study of the evolutionary trends of both dark matter and gas properties.

These numerical simulations are also very interesting to investigate realistic cluster mergers. All our objects have experienced a major merging event since z=0.6. With an appropriate time resolution, we will be able to study in detail the physics of mergers and put constrains in the relaxation time-scale as a function of the merger characteristics. Moreover, we can simultaneously compare the X-ray properties of the system with the dark matter distribution before, during and after the merger. Then, an unbiased tracer of the dynamical state of a cluster could be applied to real systems in order to understand the combined observations in X-ray and optical wavelengths.

On the other hand, new high-resolution simulations are being carried out at the present time. Their aim is to add more clusters to the present sample, not only to improve statistics, but also to enlarge the mass coverage. This is particularly important for the study of the scaling relations found between dark matter and gas properties, as well as to test the validity of our prescription relating the structure of both components.

Finally, the implementation of cooling and star formation would offer new insights on the process of galaxy formation and evolution. Some steps are also being taken in this direction. In the long term, we would like to self-consistently model photoionisation and chemical enrichment, computing the intensity of the ionising background at each timestep, as well as metal production and advection.

From the point of view of analytical modelling, we would like to extend our treatment of spherical collapse in order to predict the evolution of the dark matter profiles. It would be also extremely interesting to study the multiplicity function of dark matter haloes in combination with the statistics of Gaussian random peaks. Regarding the gas, taking into account infall and rotational motions into the hydrostatic equilibrium equation will surely improve our theoretical estimations of the ICM density and temperature profiles, as well as the normalisation of the scaling relations.

%-----------------------------------------------------------------------------

%%% Local Variables: 
%%% mode: latex
%%% TeX-master: t
%%% End: 

\section{Conclusiones y perspectivas futuras}

\subsection{Materia oscura}

\begin{enumerate}
\item
Las predicciones num\'ericas sobre la distribuci\'on de materia oscura son robustas dentro de los l\'\i mites de resoluci\'on. ART y GADGET dan resultados consistentes. Las perturbaciones del potencial de materia oscura debido a la presencia de bariones son despreciables a la escala de los c\'umulos, al menos cuando s\'olo se considera la f\'\i sica adiab\'atica.

\item
De una muestra de 15 c\'umulos, 6 han sido catalogados como sistemas relajados, 4 como fusiones' menores y 5 como fusiones mayores. La magnitud de la actividad din\'amica es significativamente mayor en el extremo de baja masa de la muestra, como se podia esperar debido a la organizaci\'on jer\'arquica.

\item
Hay diferencias sistem\'aticas entre los perfiles internos de nuestros c\'umulos num\'ericos dependiendo de su estado din\'amico. Los c\'umulos relajados se describen mejor con el perfil propuesto por Navarro et al. (1997), mientras que las fusiones menores est\'an m\'as pr\'oximos a la forma defendida por Moore et al. (1999). Los fusiones mayores siguen una ley de potencias pura para m\'as de un orden de magnitud en el radio; sus perfiles tienen una pendiente incluso mayor que $\rho \propto r^{-1.5}$.

\item
La pendiente logar\'\i tmica del perfil de masas de los halos relajados no alcanza ninguan pendiente asint\'otica para nuestro l\'\i mite de resoluci\'on. De hecho, parece aumentar por encima del valor de ``NFW'' para radios suficientemente peque\~nos. Se necesita mayor resoluci\'on para verificar esta conclusi\'on.

\item
La extensi\'on de los sistemas fusionados es sistem\'aticamente mayor que la de los c\'umulos relajados. Este efecto distorsiona completamente la relaci\'on $c-M_{200}$ usual, aunque todavia se puede encontrar algo de correlaci\'on en el plano $\rho_s-r_s$. Los valores que mejor ajustan estos par\'ametros son consistentes con los inferidos a partir de las observaciones.

\item
La densidad del espacio de fase de nuestros c\'umulos se ajusta bien con la ley de potencias $\rho/\sigma^3=2.24\times10 r^{-1.875}$, similar a la encontrada por Taylor y Navarro (2001) para halos del tama\~no de galaxias.

\item
No encontramos una evidencia clara de un perfil universal en el par\'a-metro de anisotrop\'\i a, $\beta \equiv 1-<\sigma_{\theta}^2/\sigma_r^2$. Aunque hay indicaciones de una tendencia promedio, la variabilidad de los perfiles individuales es extremadamente alta. Podemos descartar con seguridad la existencia de un campo de velocidades is\'otropo ($\beta=0$) en los c\'umulos relajados y en los fusiones menores.

\item
El momento angular crece aproximadamente de forma proporcional a $r$, consistentemente con trabajo num\'erico previo.El soporte rotacional es despreciable con respecto a las contribuciones debidas al movimiento aleatorio. La excentricidad de las orbitas de part\'\i culas individuales es aproximadamente $e=0.5$ para todo el c\'umulo. Dos c\'umulos se desvian de este comportamiento.

\item
La inclusi\'on del momento angular y la elecci\'on de las condiciones iniciales son los elementos m\'as delicados en lo que respecta al modelo de caida esf\'erico. Nuestros c\'umulos se describen muy bien con picos $\approx 3\sigma$ del campo de densidades, suavizados en las escalas $\approx 1 h^{-1}$ Mpc, cuando se fija la excentricidad de las \'orbitas a $e=0.5$.

\end{enumerate}

\subsection{Gas}

\begin{enumerate}
\item
Encontramos evidencia adicional de p\'erdidas de entrop\'\i a no f\'\i
sicas en la implementaci\'on estandar del algoritmo SPH. Este efecto
puede tener una importancia cr\'\i tica en el c\'alculo del perfil de
entrop\'\i a en grupos y c\'umulos de galaxias. La nueva formulaci\'on
de SPH propuesta parece solucionar este problema; Las simulaciones de
alta resoluci\'on realizadas con este c\'odigo son consistentes con
los resultados del c\'odigo Euleriano ART. Un n\'umero excesivamente
bajo de part\'\i culas puede producir un n\'ucleo ``isentr\'opico''
debido a efectos num\'ericos.

\item
La hip\'otesis de equilibrio hidrost\'atico esta justificada para los c\'umulos relajados desde un punto de vista din\'amico, aunque se tienen algunas desviaciones en las partes externas del c\'umulo. Es v\'alida s\'olo de forma marginal para fusiones menores y falla completamente para sistemas en proceso claro de fusi\'on.

\item
La ecuaci\'on de estado del gas intracumular se puede describir bien mediante un politr\'opo de \'\i ndice $\gamma \approx 1.18$. S\'olo las fusiones mayores se desvian de una relaci\'on politr\'opica.

\item
Bajo las hip\'otesis de equilibrio hidrost\'atico, una ecuaci\'on de estado politr\'opica, un perfil universal para la materia oscura y una fraci\'on bari\'onica constante a radios grandes, el perfil de temperatura queda {\it determinado} por la distribuci\'on de materia oscura subyacente. Nuestros resultados son consistentes con los de Komatsu y Seljak (2001), que tienen un modelo ligeramente diferente. El perfil de temperatura obtenido a partir de un perfil de NFW proporciona un ajuste excelente de los resultados de la simulaci\'on.

\item
Los perfiles de temperatura de rayos X proyectados son muy similares a los observados en c\'umulos reales. Aunque la situaci\'on es menos clara en las regiones internas, lo que est\'a probablemente relacionado con fenomenos no adiab\'aticos, se puede descartar un perfil de temperaturas isot\'ermico por razones tanto te\'oricas como observacionales. Nuestros datos se ajustan muy bien con la f\'ormula propuesta por Loken et al. (2002), pero no se encuentra concordancia seg\'un los valores de mejor ajuste de los par\'ametros libres.

\item
El perfil de densidad del gas ICM se puede obtener tambien con el procedimiento descrito m\'as arriba. De nuevo, queda {\it completamente} determinado una vez que se conocen los par\'ametros que describen el halo de materia oscura. De todas formas, la concordancia con los datos num\'ericos es considerablemente peor que con el perfil de temperatura.

\item
La relaci\'on entre los perfiles oscuro y bari\'onico se refleja tambien en las relaciones de escala entre los promedios globales. Se predice que la relaci\'on $M-T_x$ sigue una relaci\'on de escala de autosimilaridad, en particular, para sobredensidades altas. Cuando incrementamos el radio, la densidad de los sistemas que se fusionan se incrementa significativamente por encima del valor esperado por su temperatura. Esto se acumula con el hecho de que nuestro modelo infraestima severamente la densidad del gas para radios grandes.

\item
Nuestra relaci\'on temperatura-masa concuerda relativamente bien con experimentos num\'ericos previos independientes. Sin embargo, se encuentran algunas discrepancias en la temperatura pesada con las emisiones debido a diferencias en la forma interna del perfil. Pensamos que es probablemente un efecto de la resoluci\'on. La concordancia con estimaciones observacionales basadas en el modelo $\Lambda$CDM y el perfil NFW es buena; otros m\'etodos observacionales tienden a predecir normalizaciones ligeramente menores, aunque con una variabilidad significativa.

\item
La relaci\'on $L_x-T_x$ puede obtenerse de la definici\'on de la luminosidad de rayos X, substituyendo los perfiles adecuados de densidad de gas y temperatura. No se espera que la ley de potencias que se obtiene asi siga la predicci\'on autosimilar $L_x \alpha {T_x}^2 $, porque depende de tres factores adicionales: el cuadrado de la fracci\'on bari\'onica, $F^2$, el par\'ametro de estructura, Q, y la normalizaci\'on exacta de la relaci\'on $M-T$, $M_0$. Cuando estos factores se toman en cuenta, la concordancia entre las leyes de escala predecidas y simuladas es extremadamente buena.

\item
Para sobredensidades bajas (e.j. $\Delta=200$), la fracci\'on bari\'onica es cercana al valor c\'osmico para todos los c\'umulos, el par\'ametro Q de estructura crece dram\'aticamente con $T_X$ (es decir, los halos poco masivos estan mucho menos concentrados debido a la fusi\'on), pero $M_0$ es una funci\'on decreciente de la temperatura. Los dos \'ultimos efectos se compensan aproximadamente, asi que introducen alguna variabilidad en la relaci\'on $L_x-T_x$ observada en el radio virial.

\item
A grandes sobredensidades (e.j. $\Delta=2500$) el factor de estructura y la normalizaci\'on de la relaci\'on $M-T$ son la misma para todos los c\'umulos pero la fraci\'on bari\'onica es m\'as baja en los c\'umulos menos concentrados y con baja temperatura. Por lo tanto, la luminosidad en rayos X de estos sistemas es menor respecto de la de los c\'umulos masivos, y la relaci\'on $L_x-T_x$ observada aumenta su pendiente considerablemente. 

\item
Nuestra normalizaci\'on de la relaci\'on  $L_x-T_x$ es significativamente mayor que la encontrada en estudios num\'ericos previos. Esto se podr\'\i a deber a un artefacto num\'erico de nuestra mayor resoluci\'on. Los estudios a baja resoluci\'on podr\'\i an ser incapaces de resolver las regiones de alta densidad cerca del centro del c\'umulo, y por lo tanto la luminosidad volum\'etrica podria estar infravalorada. Esta cuesti\'on debe ser investigada en mayor detalle.

\item
Estimaciones observacionales tambi\'en dan luminosidades menores a temperatura dada. La discrepancia es de un factor $2$ para la estimaci\'on basada s\'olo en la cosmolog\'\i a $\Lambda$CDM, y es mucho peor para las estimaciones SCDM. Sin embargo, la comparaci\'on entre las diferentes cosmolog\'\i as no es trivial.

\end{enumerate}

\subsection{Estrellas}

Hemos realizado 66 simulaciones num\'ericas agrupadas en siete experimentos diferentes, con el fin de obtener una predicci\'on te\'orica de la historia de la formaci\'on estelar c\'osmica. Comparamos nuestras predicciones con estimaciones observacionales publicadas. Resumimos nuestros resultados:
\begin{enumerate}

\item $\Lambda CDM$ muestra el mejor acuerdo con las observaciones. El modelo $CDM$ Est\'andar tiende a sobreestimar la densidad c\'osmica $SFR$, mientras que la $BSI$ falla al forma suficientes estrellas a gran corrimiento hacia el rojo debido a su fuerza reducida a escalas peque\~nas.

\item La fotoionizaci\'on fija el comienzo de la formaci\'on estelar en el unviverso, suprimiendo la formaci\'on estelar de objetos masivos a alto $z$. La retroalimentaci\'on de las supernovas es muy eficiente para autoregular el proceso de convertir gas en estrellas, previniendo as\'\i\  que los halos mas masivos formen demasiadas estrellas en \'epocas tard\'\i as.

\item Al contrario de las primeras estimaciones observacionales, pero de acuerdo con los datos m\'as recientes, el modelo $\Lambda CDM$ predice que no existe decaimiento en el $SFR$ c\'osmico para $2<z<5$. La formaci\'on estelar parece ser un proceso gradual sin ninguna \'epoca caracter\'\i stica.

\item Los vol\'umenes com\'oviles en los experimentos son comparables con las observaciones. Por lo tanto se puede esperar un error apreciable en las medidas observacionales debidas a la variancia c\'osmica asociada a la estad\'\i stica de vol\'umenes peque\~nos. Esto es lo m\'as importante cuando
se quiere determinar el salto de pendiente de la densidad $SFR$ para $z \sim 1$ en el momento actual.

\item La historia de la formaci\'on estelar es muy diferente en objetos que acaban en los centros de los c\'umulos de las galaxias de campo. Las galaxias de los centros de los c\'umulos tienen una poblaci\'on estelar mucho m\'as vieja y su actividad se reduce dr\'asticamente cuando pierden la reserva de gas fr\'\i o a partir de la que se forman las estrellas.

\item En concordancia con las observaciones m\'as recientes, encontramos que los mecanismos que inhiben la formaci\'on estelar en c\'umulos de galaxias son tambi\'en eficientes en entornos correspondientes a grupos m\'as peque\~nos ($M\sim 10^{13}M_\odot$).

\end{enumerate}

Los efectos de seleci\'on, extinci\'on por polvo y conversi\'on de luminosidades a tasas de formaci\'on estelar son las principales fuentes de incertidumbre de las estimaciones observacionales. as\'\i\  mismo el peque\~no volumen estudiado por el $HDF$ tambi\'en podr\'\i a introducir alg\'un tipo de sesgo estad\'\i stico en las observaciones a alto redshift. 

Las incertidumbres en los experimentos num\'ericos provienen fundamentalmente de la resoluci\'on espacial y de la modelizaci\'on de la formaci\'on estelar y de los procesos de realimentaci\'on. El uso de c\'odigos eulerianos con malla adaptativa ser\'a muy \'util para conseguir un aumento significativo de la resoluci\'on espacial en vol\'umenes cosmol\'ogicos que sean estad\'\i sticamente significativos.

El enriquecimiento qu\'\i mico y los procesos de fotoionizaci\'on han sido tenidos en cuenta de forma fenomenol\'ogica en el c\'odigo que hemos utilizado. Ser\'\i a extremadamente interesante realizar un modelo autoconsistente de ambos procesos simult\'aneamente, calculando la intensidad del fondo ionizante ultravioleta, no s\'olo proveniente de qu\'asares y $AGN$, sino tambi\'en de las estrellas generadas en la simulaci\'on en cada paso de tiempo, as\'\i\  como la producci\'on de metales en las mismas y su tratamiento hidrodin\'amico como otra variable m\'as.

%------------------------------------------- 

\appendix
%\include{Text/Algor}
%-----------------------------------------------------------------------------
\chapter{Cluster Atlas}
\label{apIndiv}
%-----------------------------------------------------------------------------

\begin{quote}{\em
Is this the real life? Is this just fantasy?\\
Caught in a landslide, no escape from reality\\
Open your eyes, look up to the skies and see!}

-- Queen : {\em Bohemian Rhapsody} (1975) --\\
\end{quote}
%-----------------------------------------------------------------------------

{\gothfamily {\Huge T}{\large hroughout}} this work, the sample of galaxy clusters described in Section~\ref{secSample} has been analysed in a statistical way. The present appendix is devoted to test whether the theoretical considerations that have been discussed in the previous chapters can be applied to {\em predict} the structure of individual clusters.

%-----------------------------------------------------------------------------

\section{Figure description}

We have been plotted two figures per cluster. The first one is divided into two columns, which in turn have been sub-divided into three panels. The second one contains a contour plot overlaid on a colour map.

The left column of the first figure is devoted to the dark matter distribution, while the right column is used to display the radial properties of the ICM gas. In both cases, the numerical results are indicated by dots, while solid lines are used to represent the theoretical predictions.

{\sc Dark matter:} Dark matter density is plotted on the top panel, normalised to the mean cosmic value. In the middle panel, the cumulative mass is shown, and the lower panel is used to display its logarithmic slope \AM. The solid lines show the predictions based on the spherical collapse formalism, as discussed in Section~\ref{secSCcomp}. The values of the peak height and smoothing scale appropriate for each cluster can be found in Table~\ref{tabSCcomp}. We recall that major mergers have not been fit by the spherical infall model.

{\sc Gas:} Top panel shows the gas density, also normalised to mean cosmic value. Solid line corresponds to the prediction (\ref{ecDens}) based on hydrostatic equilibrium with a NFW-like potential, a polytropic equation of state and constant baryon fraction at large radii. Middle panel is used for the temperature, which has been compared with our prescription (\ref{ecTemp}). The lower panel shows the entropy profile of the cluster, computed as $S=kTn_e^{-2/3}$ both for the experimental and theoretical profiles. As has been shown in Section~\ref{secEulerLagrange}, the entropy-conserving version of \g is able to produce isentropic cores in many of our clusters. For details on the computation of the theoretical predictions of the gas profiles, the reader is referred to Section~\ref{secGasRprof}.

{\sc Colour plot:} This image represents the a map of the dark matter density, projected along the three axes. As can be appreciated in some figures, the cluster might look more or less 'relaxed' (where, in this context, we understand 'relaxed' as 'spherically symmetric') depending on the projection observed. If we assume that light traces mass, this colour plots can be thought of as a coarse approximation to an observational image of the cluster in an optical band.

{\sc Contour plot:} These contours correspond to the projected X-ray luminosity of the clusters. Although they approximately trace the observed shape of the dark mater distribution, it is evident that the gas has much less substructure than the dark matter. This is due to the gas stripping that galaxies experience as they orbit through the cluster. Most of the infalling haloes lose their gas reservoir during the first pericentre passage.

%-----------------------------------------------------------------------------

\newpage
\section{Individual clusters}

\subsection*{Cluster A}

\vspace{1.1cm}
\begin{figure}[h]
  \centering \includegraphics[width=12cm]{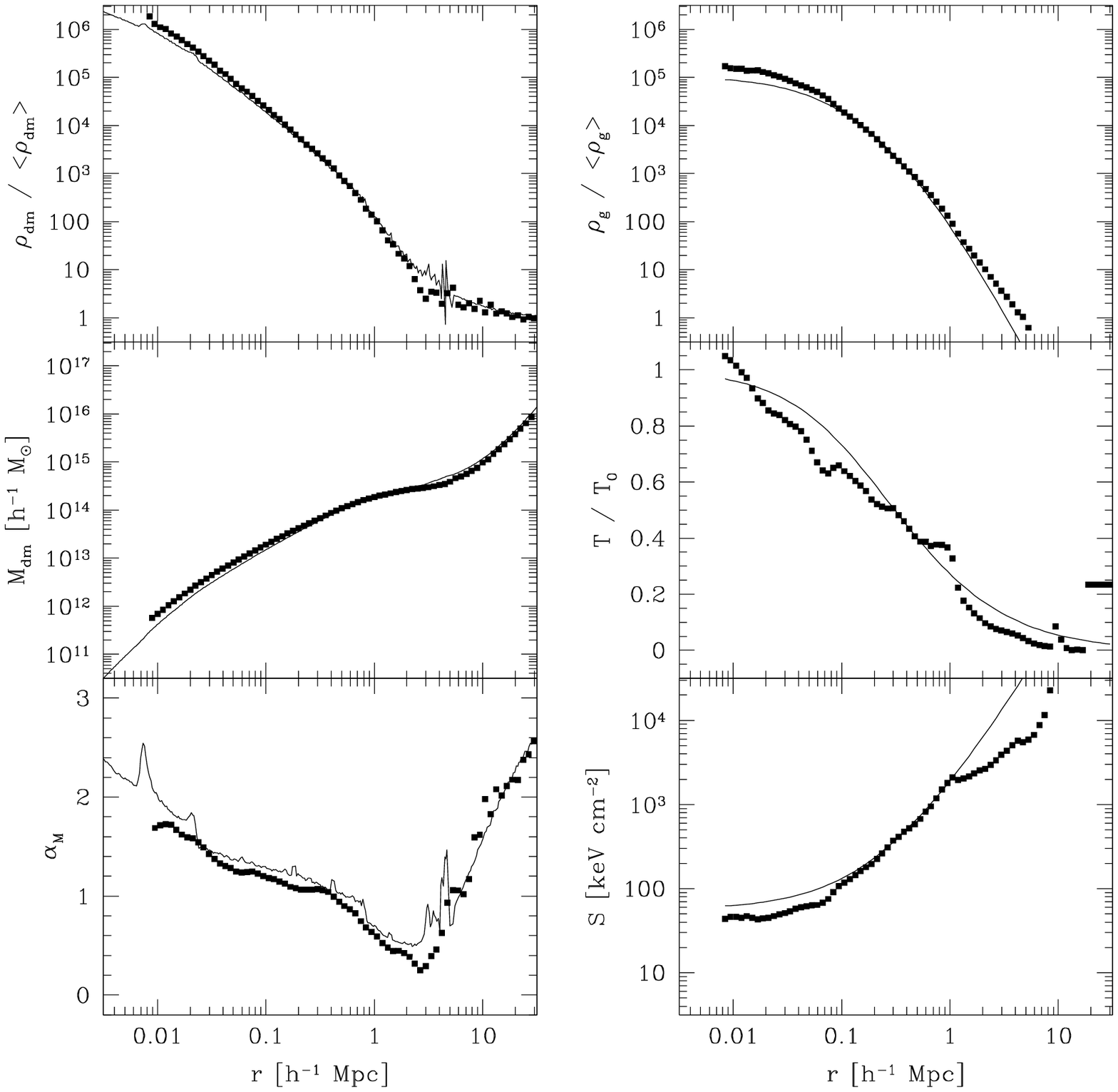}
  \caption{Cluster A}
\end{figure}
\newpage\ \newpage
%__________________________________

\subsection*{Cluster B}

\vspace{2cm}
\begin{figure}[h]
  \centering \includegraphics[width=12cm]{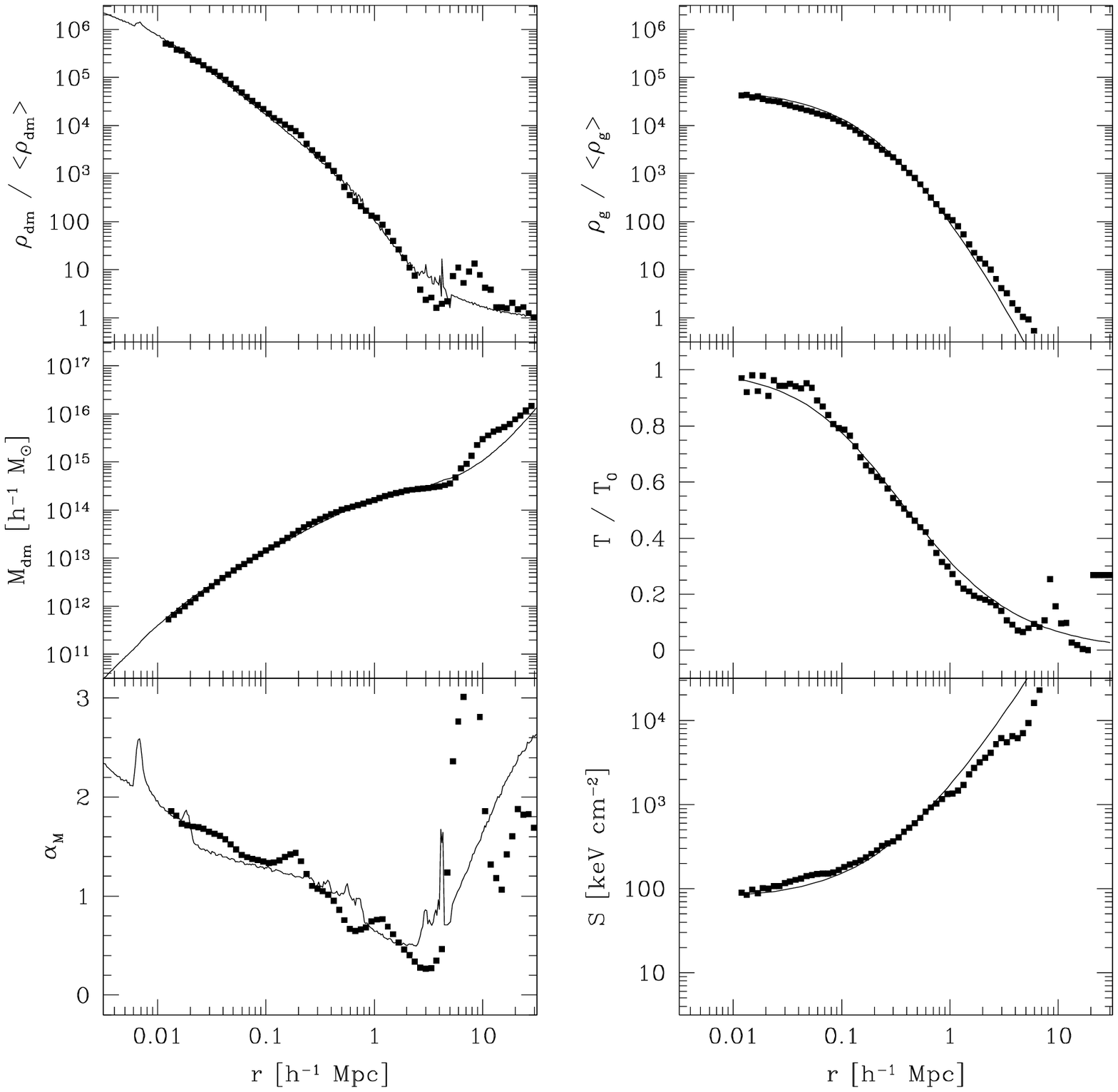}
  \caption{Cluster B}
\end{figure}
\newpage\ \newpage
%__________________________________

\subsection*{Cluster C}

\vspace{2cm}
\begin{figure}[h]
  \centering \includegraphics[width=12cm]{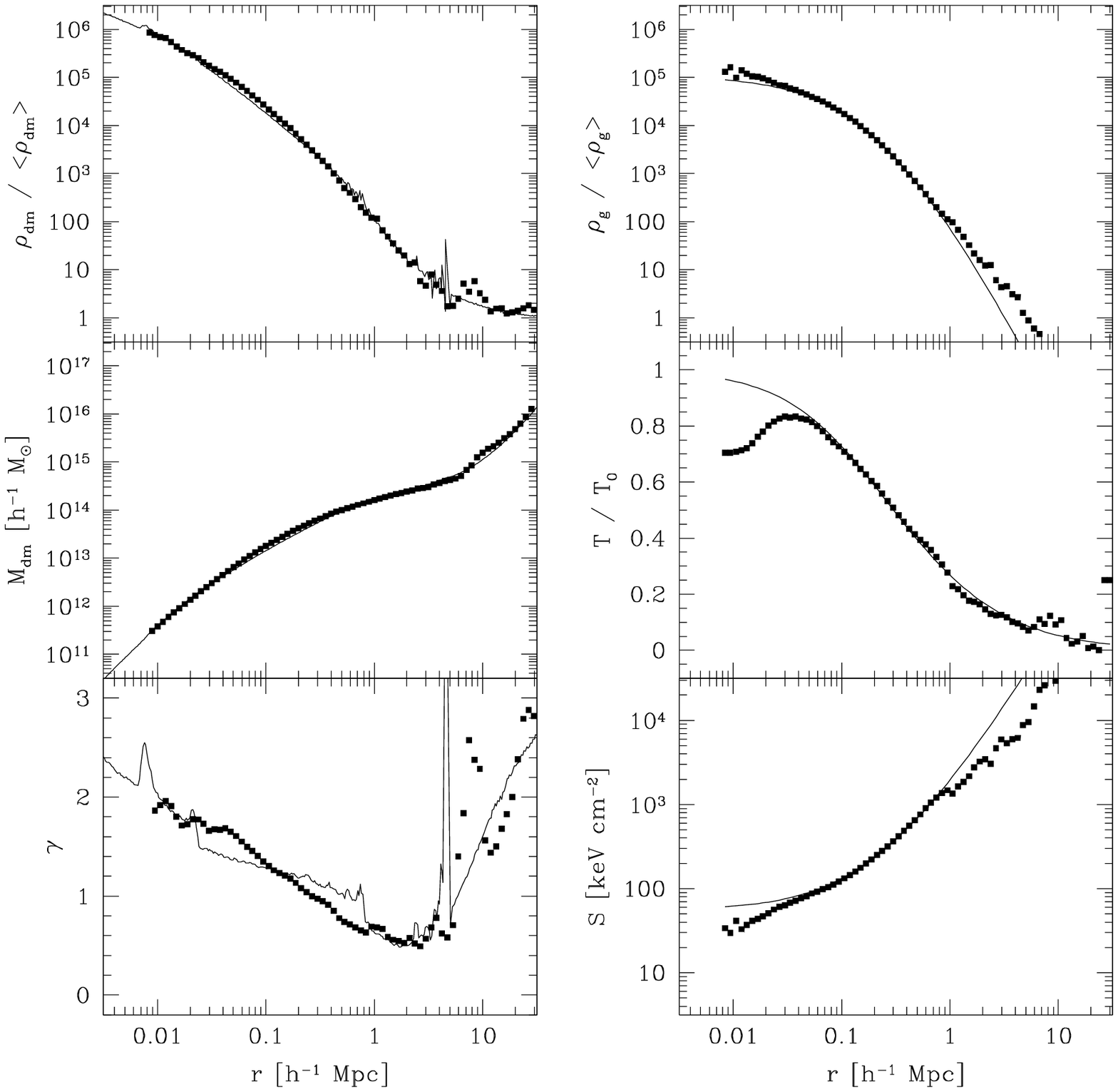}
  \caption{Cluster C}
\end{figure}
\newpage\ \newpage
%__________________________________

\subsection*{Cluster D}

\vspace{2cm}
\begin{figure}[h]
  \centering \includegraphics[width=12cm]{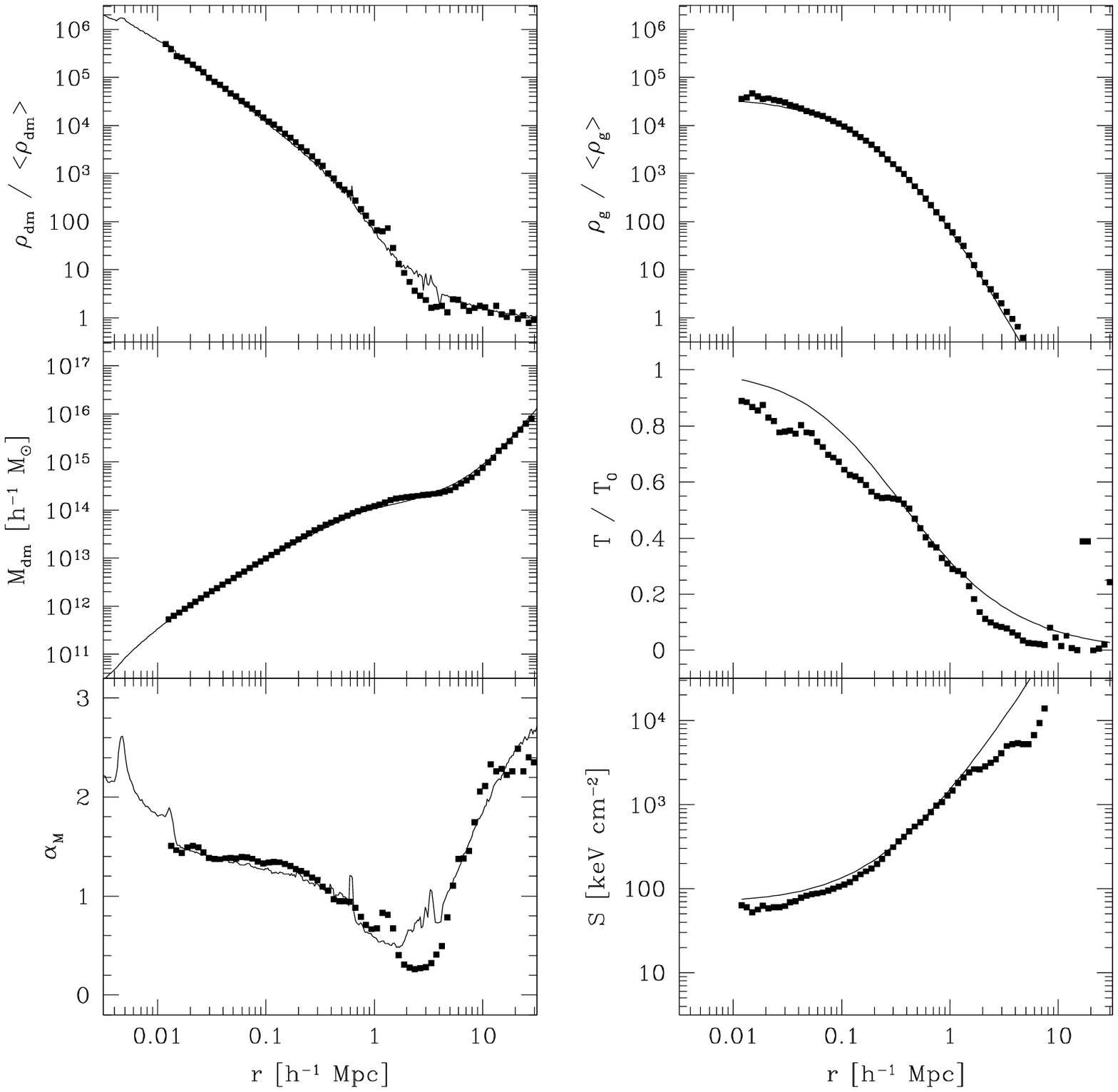}
  \caption{Cluster D}
\end{figure}
\newpage\ \newpage
%__________________________________

\subsection*{Cluster E}

\vspace{2cm}
\begin{figure}[h]
  \centering \includegraphics[width=12cm]{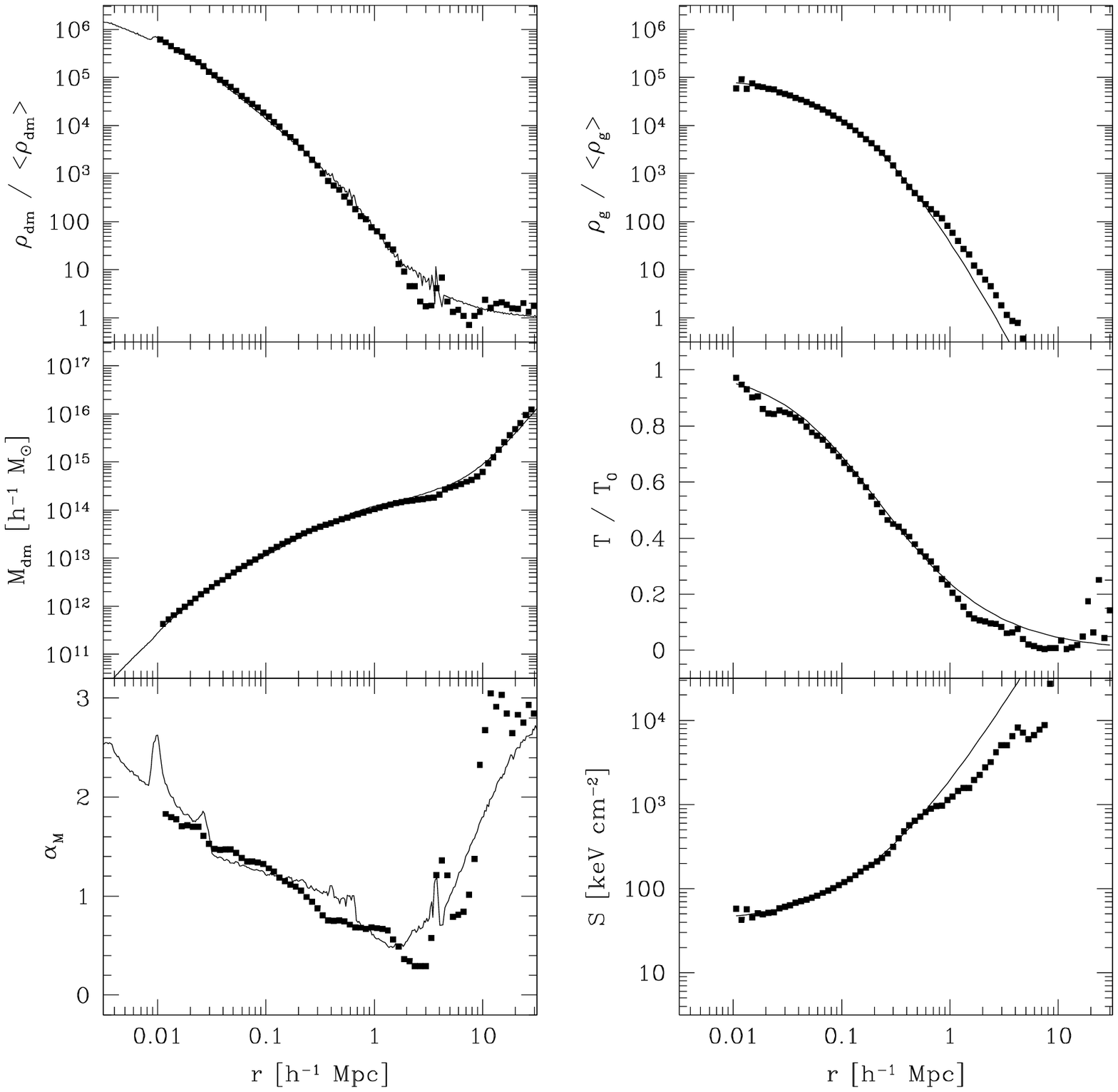}
  \caption{Cluster E}
\end{figure}
\newpage\ \newpage
%__________________________________

\subsection*{Cluster F}

\vspace{2cm}
\begin{figure}[h]
  \centering \includegraphics[width=12cm]{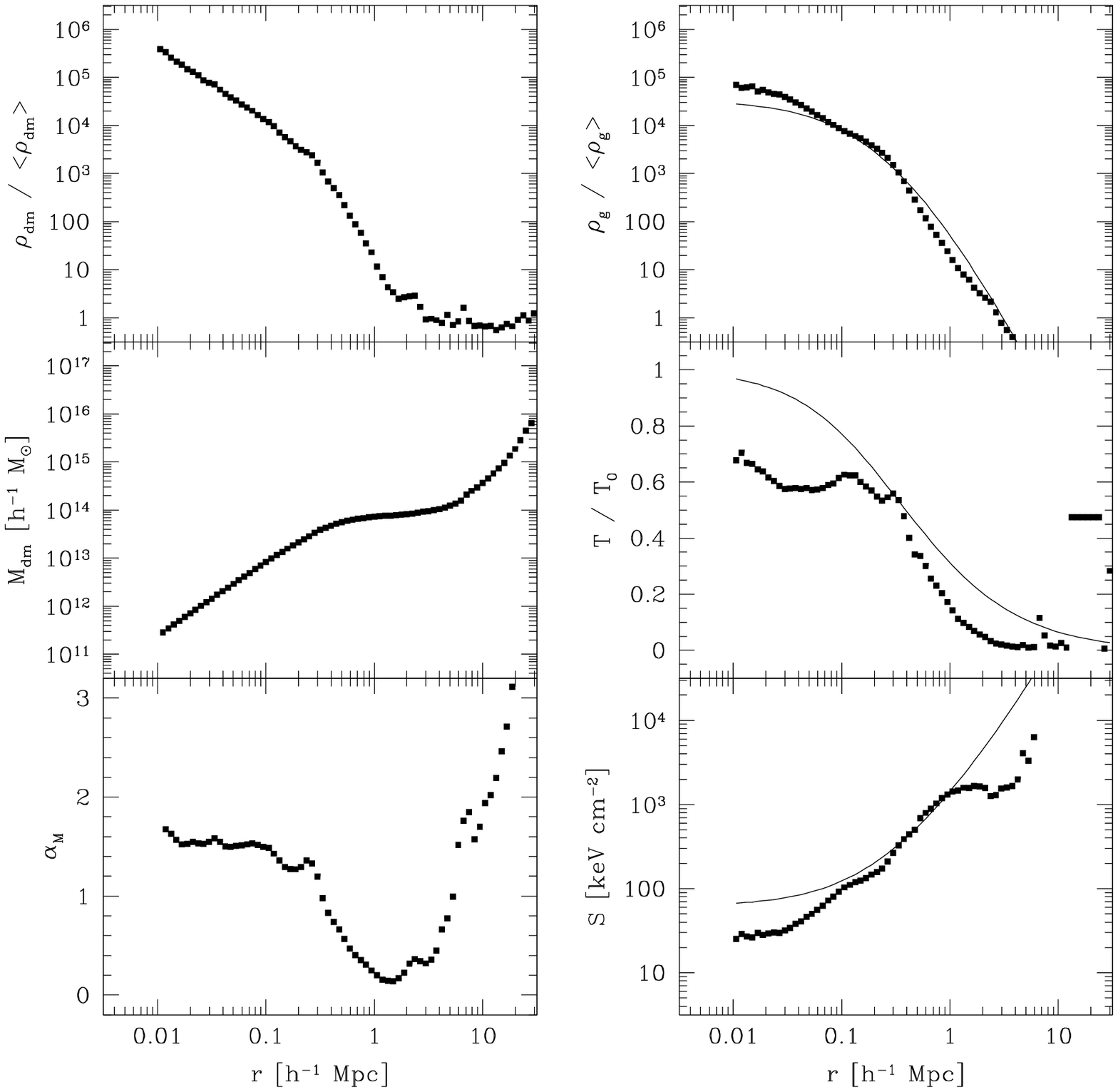}
  \caption{Cluster F}
\end{figure}
\newpage\ \newpage
%__________________________________

\subsection*{Cluster G}

\vspace{2cm}
\begin{figure}[h]
  \centering \includegraphics[width=12cm]{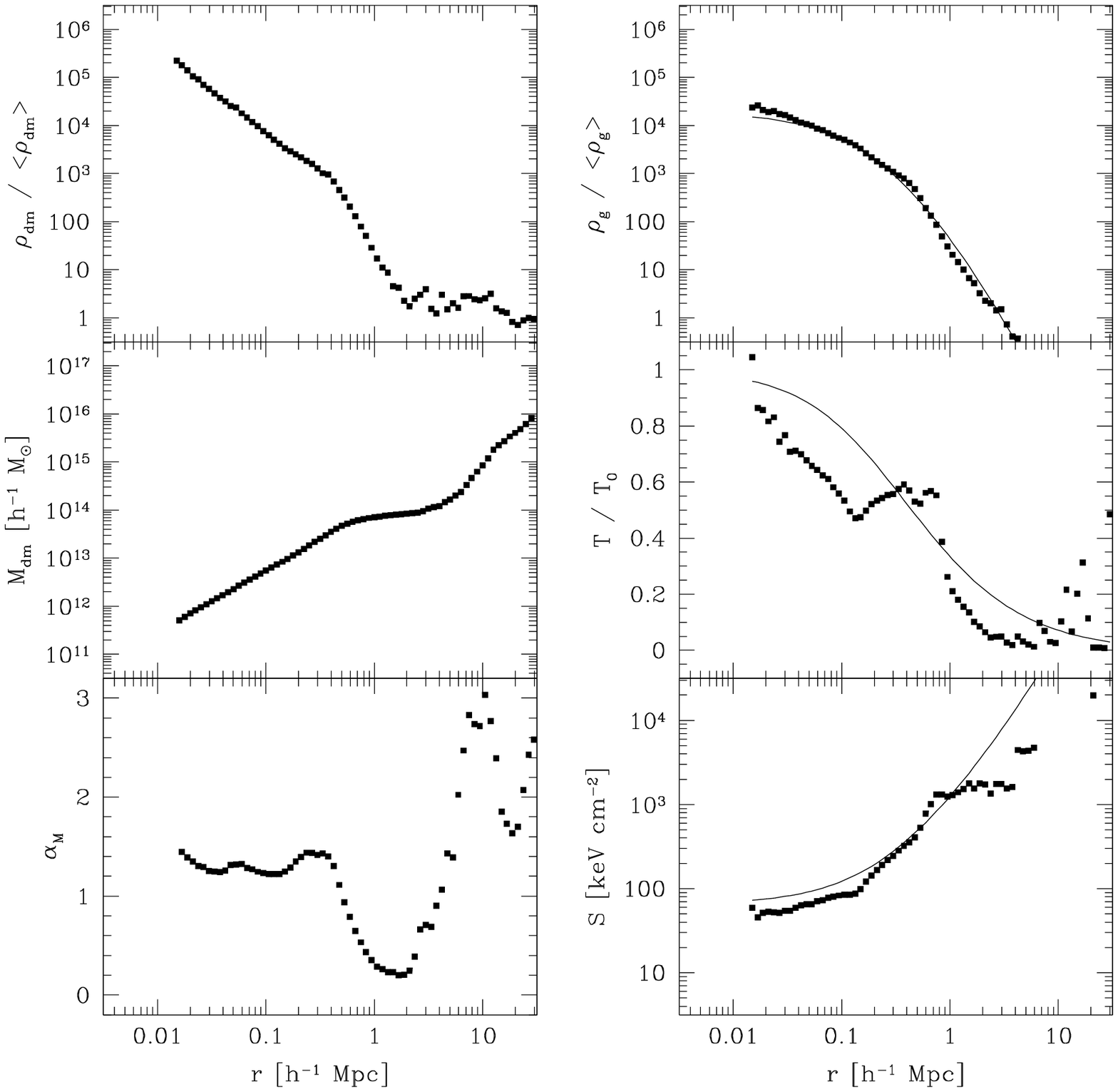}
  \caption{Cluster G}
\end{figure}
\newpage\ \newpage
%__________________________________

\subsection*{Cluster H}

\vspace{2cm}
\begin{figure}[h]
  \centering \includegraphics[width=12cm]{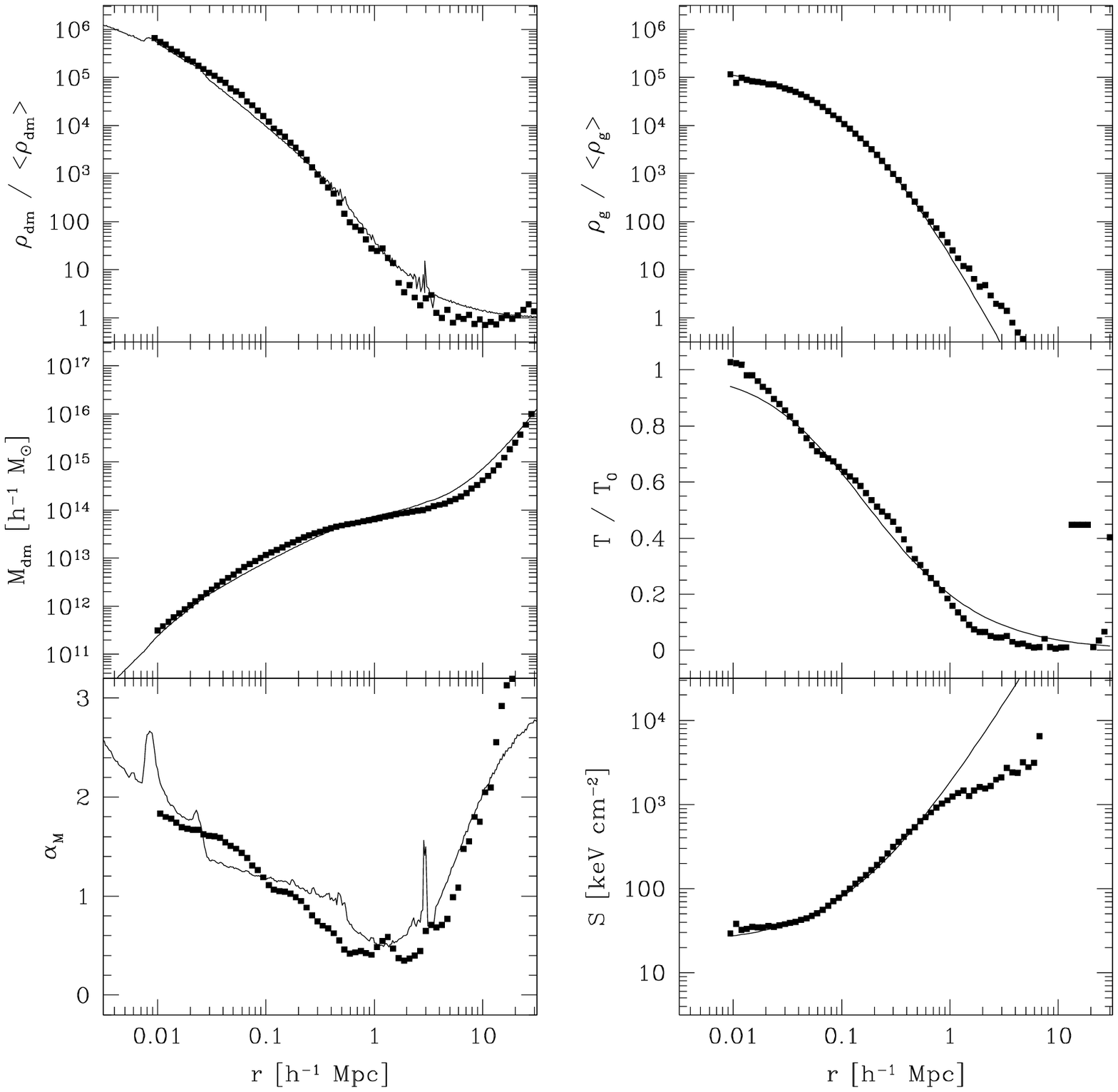}
  \caption{Cluster H}
\end{figure}
\newpage\ \newpage
%__________________________________

\subsection*{Cluster I}

\vspace{2cm}
\begin{figure}[h]
  \centering \includegraphics[width=12cm]{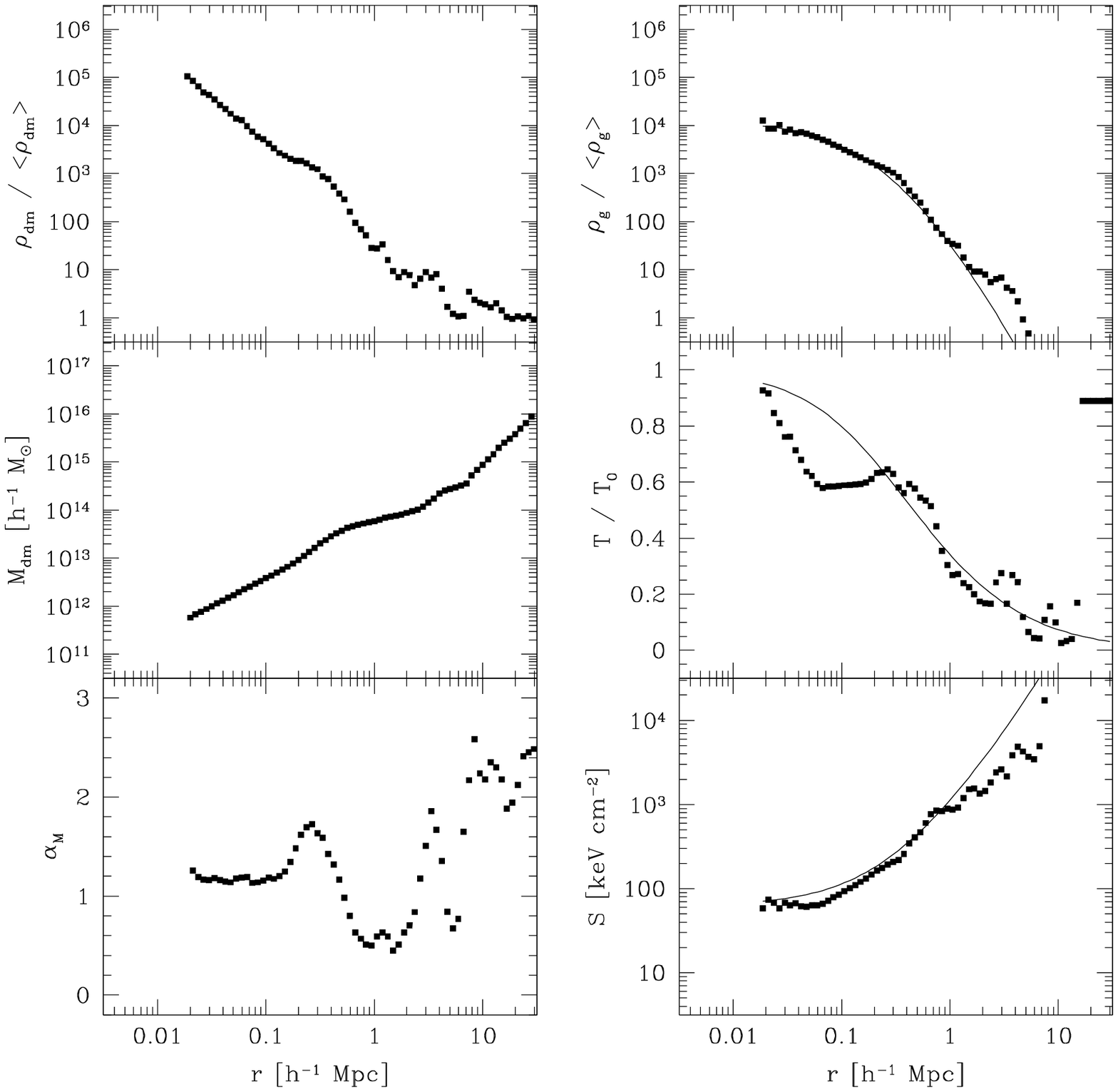}
  \caption{Cluster I}
\end{figure}
\newpage\ \newpage
%__________________________________

\subsection*{Cluster J$_1$}

\vspace{2cm}
\begin{figure}[h]
  \centering \includegraphics[width=12cm]{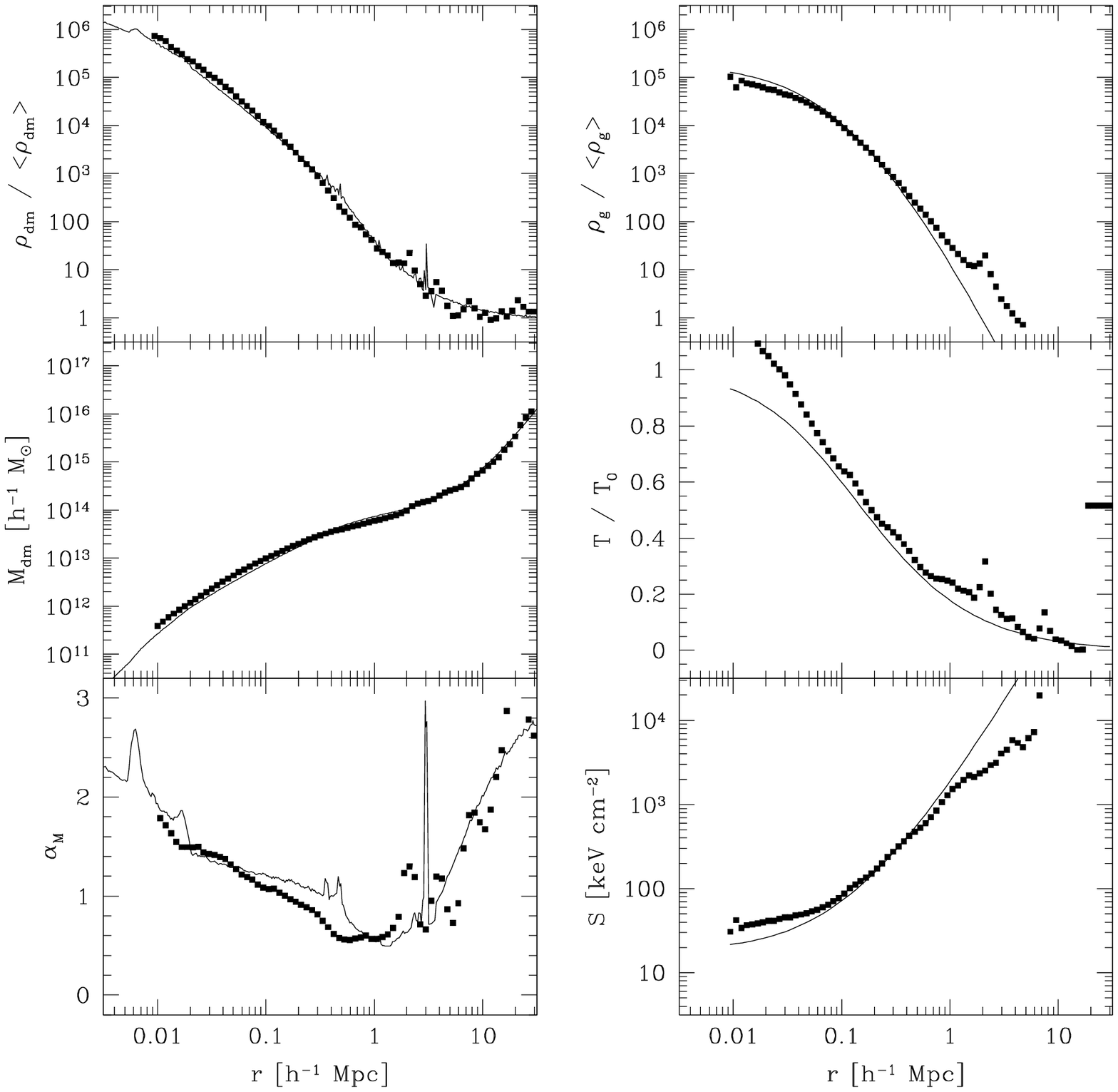}
  \caption{Cluster J$_1$}
\end{figure}
\newpage\ \newpage
%__________________________________

\subsection*{Cluster K$_1$}

\vspace{2cm}
\begin{figure}[h]
  \centering \includegraphics[width=12cm]{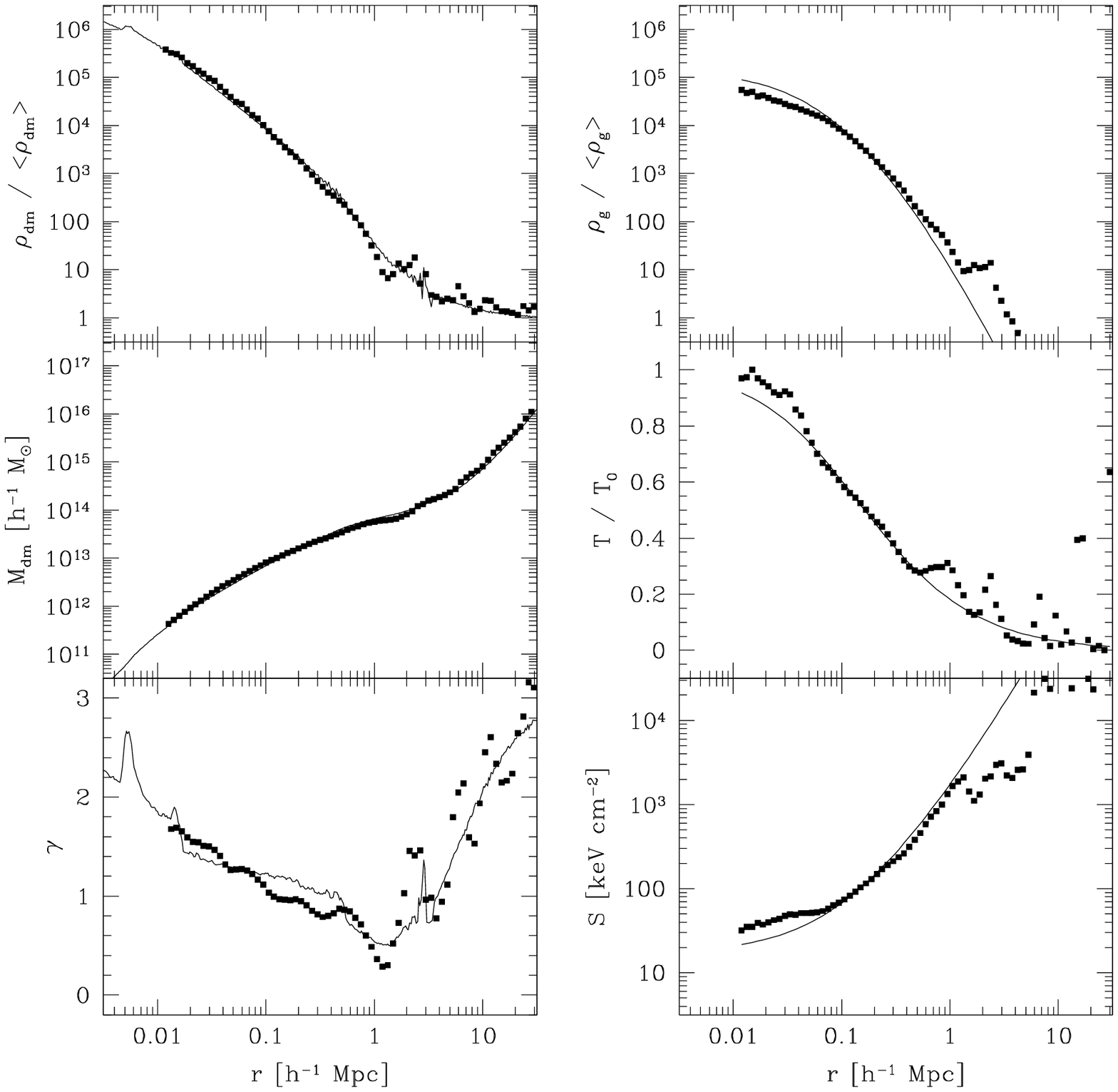}
  \caption{Cluster K$_1$}
\end{figure}
\newpage\ \newpage
%__________________________________

\subsection*{Cluster L}

\vspace{2cm}
\begin{figure}[h]
  \centering \includegraphics[width=12cm]{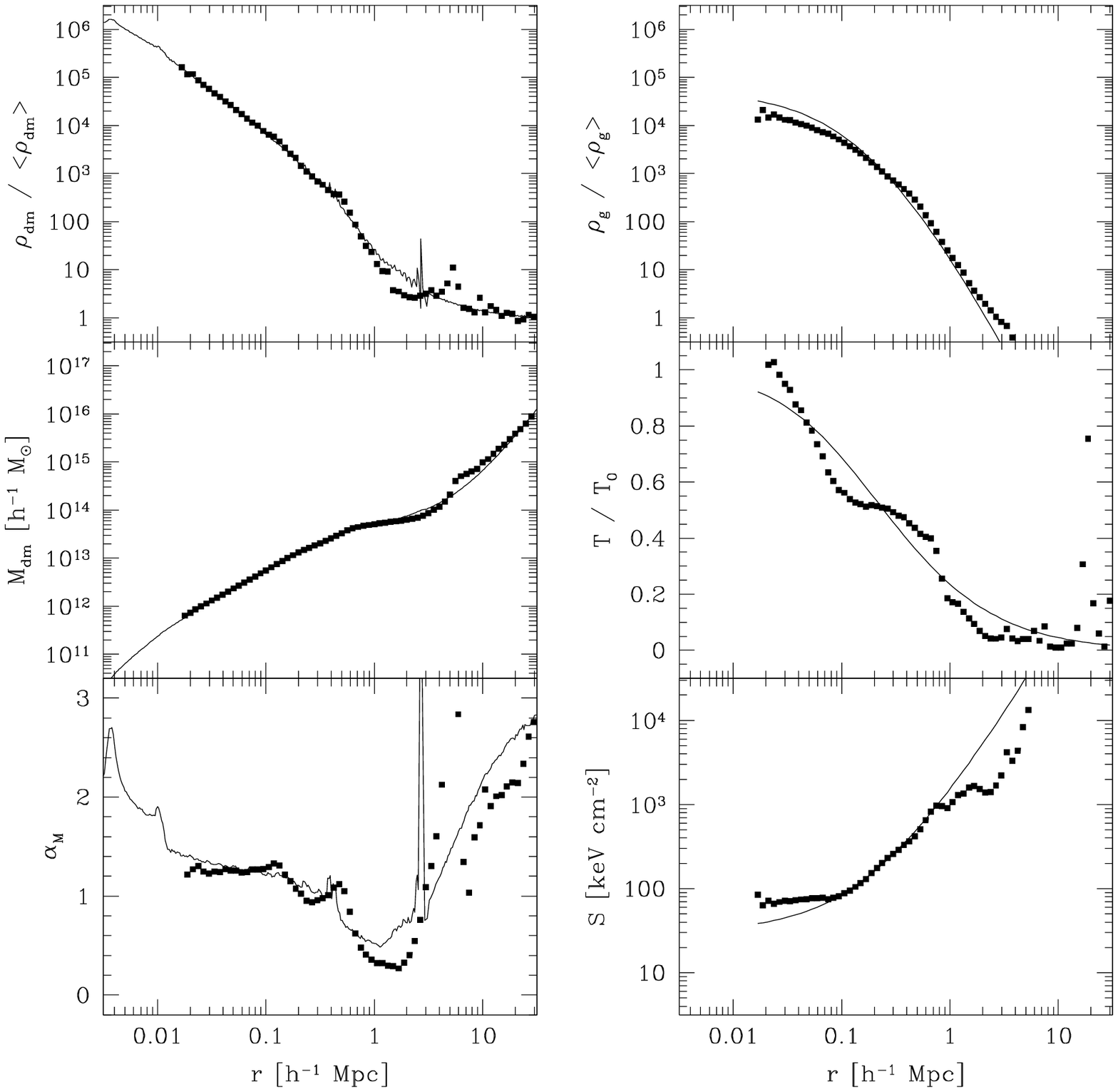}
  \caption{Cluster L}
\end{figure}
\newpage\ \newpage
%__________________________________

\subsection*{Cluster M}

\vspace{2cm}
\begin{figure}[h]
  \centering \includegraphics[width=12cm]{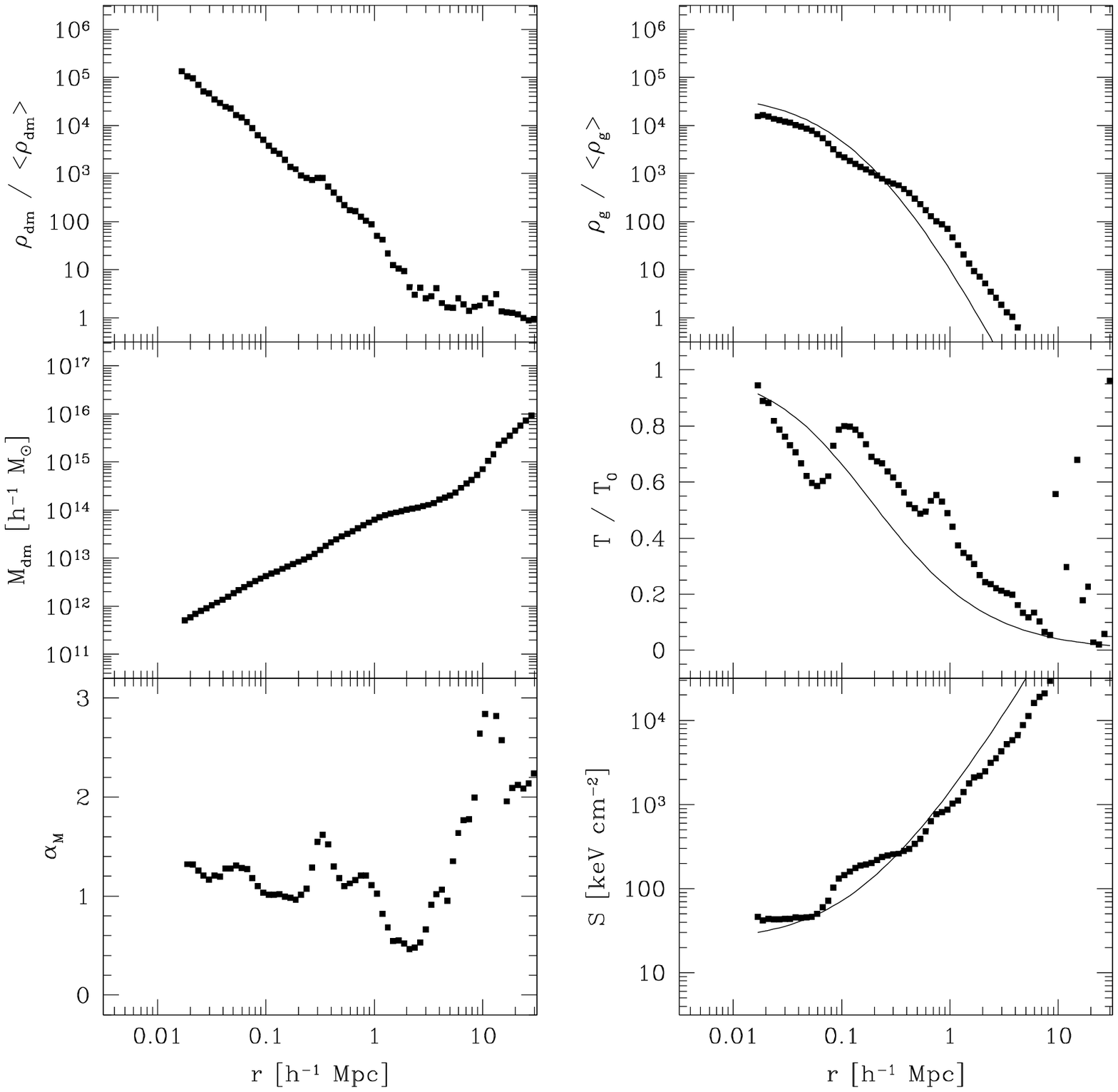}
  \caption{Cluster M}
\end{figure}
\newpage\ \newpage
%__________________________________

\subsection*{Cluster K$_2$}

\vspace{2cm}
\begin{figure}[h]
  \centering \includegraphics[width=12cm]{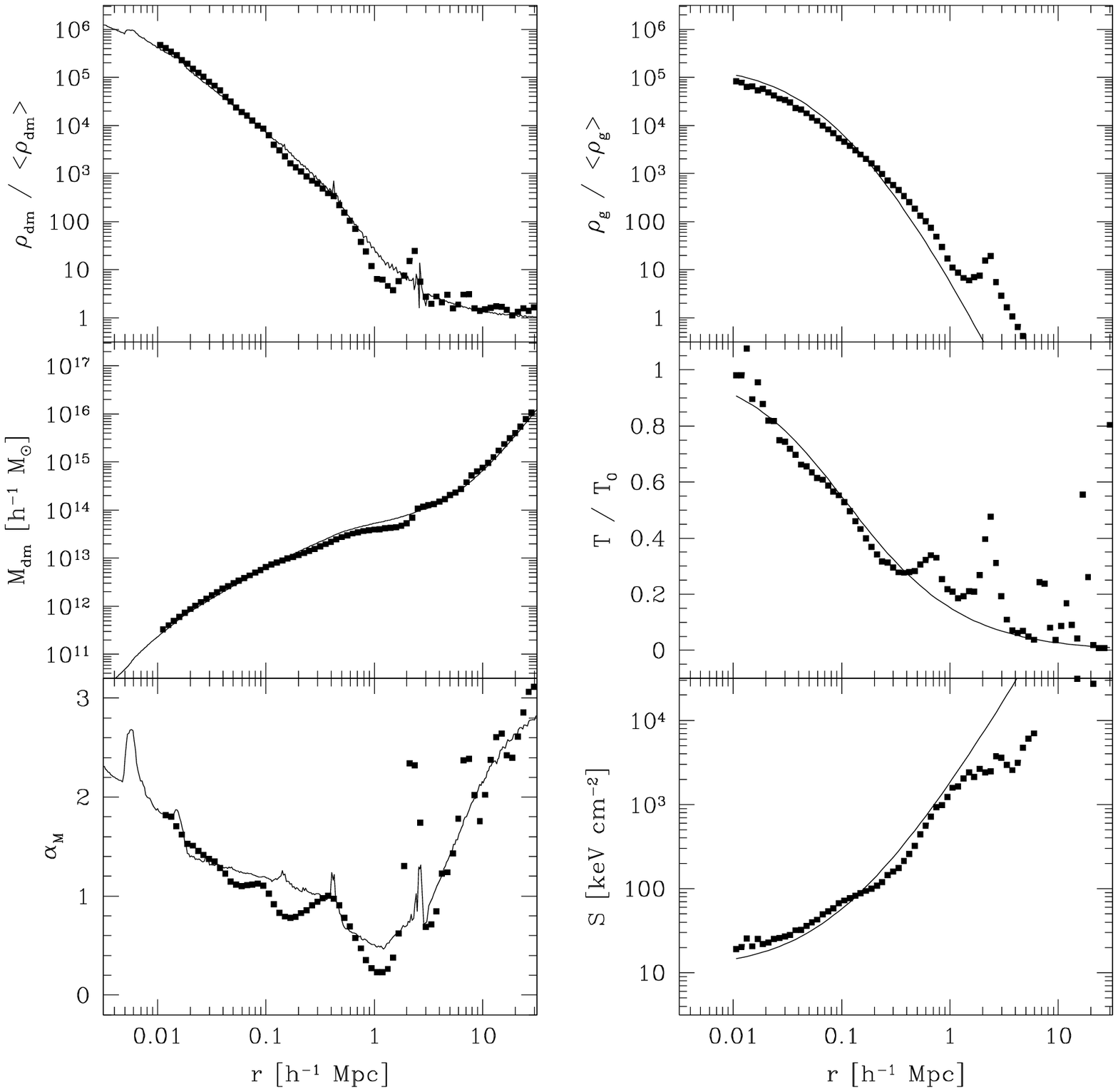}
  \caption{Cluster K$_2$}
\end{figure}
\newpage\ \newpage
%__________________________________

\subsection*{Cluster J$_2$}

\vspace{2cm}
\begin{figure}[h]
  \centering \includegraphics[width=12cm]{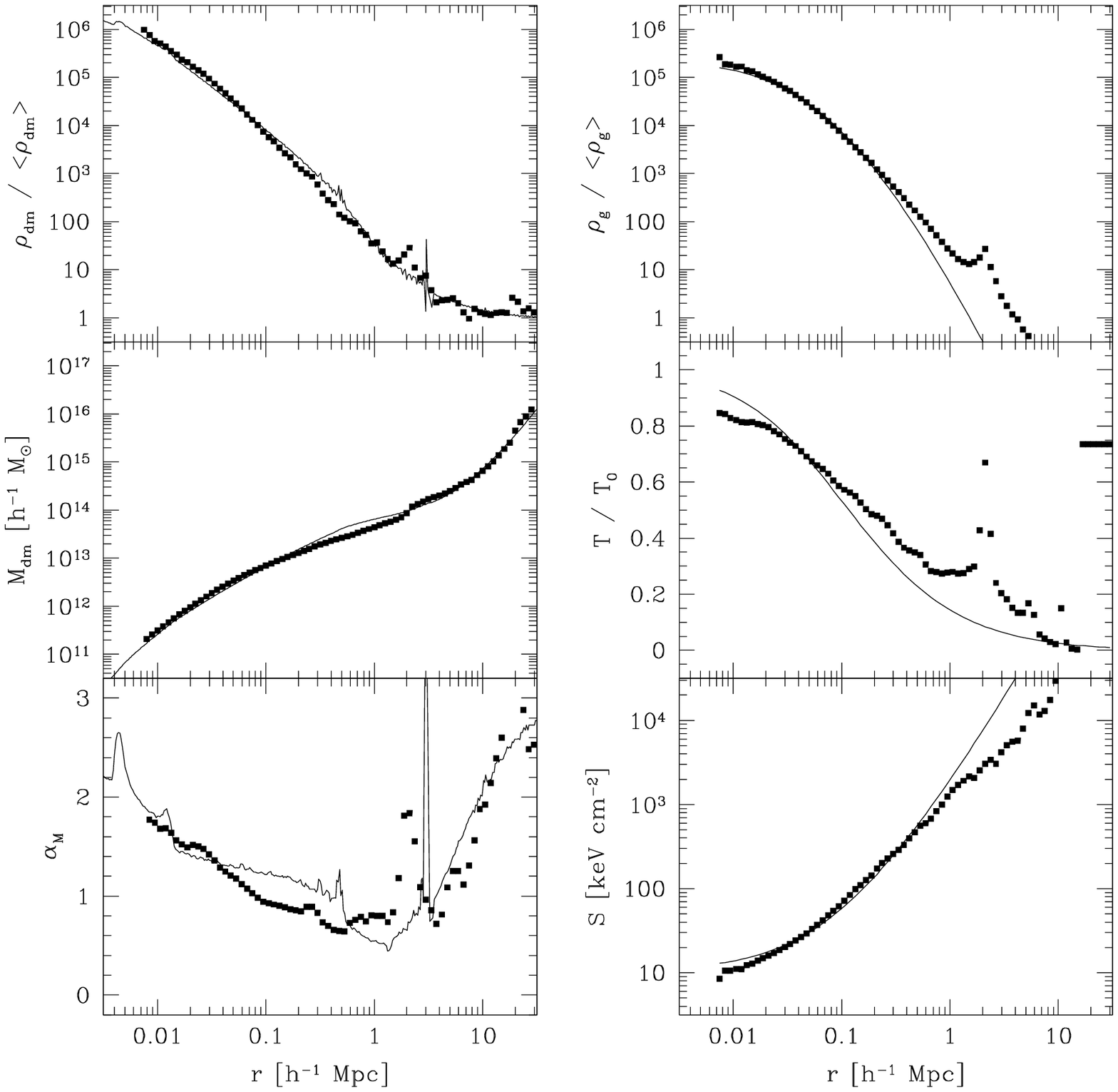}
  \caption{Cluster J$_2$}
\end{figure}
\newpage\ \newpage

%-----------------------------------------------------------------------------

%------------------------------------------- 

 \bibliographystyle{apalike}
 \bibliography{tesis}

\begin{thebibliography}{}

 \addcontentsline{toc}{chapter}{\ ~~ Bibliography}
 
 \bibitem[{Jethro Tull}, 1995]{.}
 \begin{quote}{\em\ \\
 Words get written. Words get twisted.\\
 Old meanings move in the drift of time.\\
 Lift the flickering torches. See gentle shadows change\\
 the features of the faces cut in unmoving stone.}
   
 -- Jethro Tull : {\em Roots to Branches} (1995) --\\
 \end{quote}

\bibitem[{Aarseth} et~al., 1979]{Aarseth79}
{Aarseth}, S.~J., {Turner}, E.~L., and {Gott}, J.~R. (1979).
\newblock {N-body simulations of galaxy clustering. I - Initial conditions and
  galaxy collapse times}.
\newblock {\em \apj}, 228:664--683.

\bibitem[{Adams}, 1976]{Adams76}
{Adams}, T.~F. (1976).
\newblock {The detectability of deuterium Lyman alpha in QSOs}.
\newblock {\em \aap}, 50:461.

\bibitem[{Adelberger} and {Steidel}, 2000]{AdelbergerSteidel00}
{Adelberger}, K.~L. and {Steidel}, C.~C. (2000).
\newblock {Multiwavelength Observations of Dusty Star Formation at Low and High
  Redshift}.
\newblock {\em \apj}, 544:218--241.

\bibitem[{Allen}, 1998]{Allen98}
{Allen}, S.~W. (1998).
\newblock {Resolving the discrepancy between X-ray and gravitational lensing
  mass measurements for clusters of galaxies}.
\newblock {\em \mnras}, 296:392--406.

\bibitem[{Allen} et~al., 2001a]{AEF01}
{Allen}, S.~W., {Ettori}, S., and {Fabian}, A.~C. (2001a).
\newblock {Chandra measurements of the distribution of mass in the luminous
  lensing cluster Abell 2390}.
\newblock {\em \mnras}, 324:877--890.

\bibitem[{Allen} and {Fabian}, 1998]{AllenFabian98}
{Allen}, S.~W. and {Fabian}, A.~C. (1998).
\newblock {The impact of cooling flows on the T$_X$-L$_{\rm Bol}$ relation for
  the most luminous clusters}.
\newblock {\em \mnras}, 297:L57--L62.

\bibitem[{Allen} et~al., 2001b]{ASF01}
{Allen}, S.~W., {Schmidt}, R.~W., and {Fabian}, A.~C. (2001b).
\newblock {The X-ray virial relations for relaxed lensing clusters observed
  with Chandra}.
\newblock {\em \mnras}, 328:L37--L41.

\bibitem[{Allen} et~al., 2002]{ASF02omega}
{Allen}, S.~W., {Schmidt}, R.~W., and {Fabian}, A.~C. (2002).
\newblock {Cosmological constraints from the X-ray gas mass fraction in relaxed
  lensing clusters observed with Chandra}.
\newblock {\em \mnras}, 334:L11--L15.

\bibitem[{Arabadjis} et~al., 2002]{Arabadjis02}
{Arabadjis}, J.~S., {Bautz}, M.~W., and {Garmire}, G.~P. (2002).
\newblock {Chandra Observations of the Lensing Cluster EMSS 1358+6245:
  Implications for Self-interacting Dark Matter}.
\newblock {\em \apj}, 572:66--78.

\bibitem[{Arnaud} and {Evrard}, 1999]{ArnaudEvrard99}
{Arnaud}, M. and {Evrard}, A.~E. (1999).
\newblock {The L\_X-T relation and intracluster gas fractions of X-ray
  clusters}.
\newblock {\em \mnras}, 305:631--640.

\bibitem[{Ascasibar} et~al., 2002a]{Ascasibar02}
{Ascasibar}, Y., {Yepes}, G., {Gottl{\" o}ber}, S., and {M{\" u}ller}, V.
  (2002a).
\newblock {Numerical simulations of the cosmic star formation history}.
\newblock {\em \aap}, 387:396--405.

\bibitem[{Ascasibar} et~al., 2002b]{sc02}
{Ascasibar}, Y., {Yepes}, G., {Gottl{\" o}ber}, S., and {M{\" u}ller}, V.
  (2002b).
\newblock {The structure of cold dark matter haloes}.
\newblock {\em in preparation}.

\bibitem[{Ascasibar} et~al., 2002c]{clusters02}
{Ascasibar}, Y., {Yepes}, G., {M{\" u}ller}, V., and {Gottl{\" o}ber}, S.
  (2002c).
\newblock {The Universal Density and Temperature Profile of Galaxy Clusters}.
\newblock {\em \apjl, in preparation}.

\bibitem[{Avila-Reese} et~al., 1998]{AFH98}
{Avila-Reese}, V., {Firmani}, C., and {Hern{\' a}ndez}, X. (1998).
\newblock {On the Formation and Evolution of Disk Galaxies: Cosmological
  Initial Conditions and the Gravitational Collapse}.
\newblock {\em \apj}, 505:37--49.

\bibitem[{Bahcall} et~al., 1999]{Bahcall99}
{Bahcall}, N.~A., {Ostriker}, J.~P., {Perlmutter}, S., and {Steinhardt}, P.~J.
  (1999).
\newblock {The Cosmic Triangle: Revealing the State of the Universe}.
\newblock {\em Science}, 284:1481.

\bibitem[{Balbi} et~al., 2000]{Balbi00}
{Balbi}, A., {Ade}, P., {Bock}, J., {Borrill}, J., {Boscaleri}, A., {De
  Bernardis}, P., {Ferreira}, P.~G., {Hanany}, S., {Hristov}, V., {Jaffe},
  A.~H., {Lee}, A.~T., {Oh}, S., {Pascale}, E., {Rabii}, B., {Richards}, P.~L.,
  {Smoot}, G.~F., {Stompor}, R., {Winant}, C.~D., and {Wu}, J.~H.~P. (2000).
\newblock {Constraints on Cosmological Parameters from MAXIMA-1}.
\newblock {\em \apjl}, 545:L1--L4.

\bibitem[{Balogh} et~al., 1997]{Balogh97}
{Balogh}, M.~L., {Morris}, S.~L., {Yee}, H.~K.~C., {Carlberg}, R.~G., and
  {Ellingson}, E. (1997).
\newblock {Star Formation in Cluster Galaxies at $0.2<z<0.55$}.
\newblock {\em \apjl}, 488:L75.

\bibitem[{Balogh} et~al., 1999]{Balogh99}
{Balogh}, M.~L., {Morris}, S.~L., {Yee}, H.~K.~C., {Carlberg}, R.~G., and
  {Ellingson}, E. (1999).
\newblock {Differential Galaxy Evolution in Cluster and Field Galaxies at
  z\~{}0.3}.
\newblock {\em \apj}, 527:54--79.

\bibitem[{Balogh} et~al., 2000]{Balogh00}
{Balogh}, M.~L., {Navarro}, J.~F., and {Morris}, S.~L. (2000).
\newblock {The Origin of Star Formation Gradients in Rich Galaxy Clusters}.
\newblock {\em \apj}, 540:113--121.

\bibitem[{Balogh} et~al., 2001]{bal01}
{Balogh}, M.~L., {Pearce}, F.~R., {Bower}, R.~G., and {Kay}, S.~T. (2001).
\newblock {Revisiting the cosmic cooling crisis}.
\newblock {\em \mnras}, 326:1228--1234.

\bibitem[{Balogh} et~al., 1998]{Balogh98}
{Balogh}, M.~L., {Schade}, D., {Morris}, S.~L., {Yee}, H.~K.~C., {Carlberg},
  R.~G., and {Ellingson}, E. (1998).
\newblock {The Dependence of Cluster Galaxy Star Formation Rates on the Global
  Environment}.
\newblock {\em \apjl}, 504:L75.

\bibitem[{Balogh} et~al., 2002]{Balogh02}
{Balogh}, M.~L., {Smail}, I., {Bower}, R.~G., {Ziegler}, B.~L., {Smith}, G.~P.,
  {Davies}, R.~L., {Gaztelu}, A., {Kneib}, J.-P., and {Ebeling}, H. (2002).
\newblock {Distinguishing Local and Global Influences on Galaxy Morphology: A
  Hubble Space Telescope Comparison of High and Low X-Ray Luminosity Clusters}.
\newblock {\em \apj}, 566:123--136.

\bibitem[{Bardeen} et~al., 1986]{BBKS86}
{Bardeen}, J.~M., {Bond}, J.~R., {Kaiser}, N., and {Szalay}, A.~S. (1986).
\newblock {The statistics of peaks of Gaussian random fields}.
\newblock {\em \apj}, 304:15--61.

\bibitem[{Barnes} and {Efstathiou}, 1987]{BarnesEfstathiou87}
{Barnes}, J. and {Efstathiou}, G. (1987).
\newblock {Angular momentum from tidal torques}.
\newblock {\em \apj}, 319:575--600.

\bibitem[{Barnes} and {Hut}, 1986]{BarnesHut86}
{Barnes}, J. and {Hut}, P. (1986).
\newblock {A Hierarchical O(NlogN) Force-Calculation Algorithm}.
\newblock {\em \nat}, 324:446--449.

\bibitem[{Bekki} et~al., 2002]{Bekki02}
{Bekki}, K., {Couch}, W.~J., and {Shioya}, Y. (2002).
\newblock {Passive Spiral Formation from Halo Gas Starvation: Gradual
  Transformation into S0s}.
\newblock {\em \apj}, 577:651--657.

\bibitem[{Benoit} et~al., 2002]{benoit}
{Benoit}, A., {Ade}, P., {Amblard}, A., {Ansari}, R., {Aubourg}, E., {Bargot},
  S., {Bartlett}, J.~G., {Bernard}, J.-P., {Bhatia}, R.~S., {Blanchard}, A.,
  {Bock}, J.~J., {Boscaleri}, A., {Bouchet}, F.~R., {Bourrachot}, A., {Camus},
  P., {Couchot}, F., {de Bernardis}, P., {Delabrouille}, J., {Desert}, F.~X.,
  {Dore}, O., {Douspis}, M., {Dumoulin}, L., {Dupac}, X., {Filliatre}, P.,
  {Fosalba}, P., {Ganga}, K., {Gannaway}, F., {Gautier}, B., {Giard}, M.,
  {Giraud-Heraud}, Y., {Gispert}, R., {Guglielmi}, L., {Hamilton}, J.-C.,
  {Hanany}, S., {Henrot-Versille}, S., {Kaplan}, J., G., L., {Lamarre}, J.-M.,
  {Lange}, A.~E., {Macias-Perez}, J.~F., {Madet}, K., {Maffei}, B.,
  {Magneville}, C., {Marrone}, D.~P., {Masi}, S., {Mayet}, F., {Murphy}, A.,
  {Naraghi}, F., {Nati}, F., {Patanchon}, G., {Perrin}, G., {Piat}, M.,
  {Ponthieu}, N., {Prunet}, S., {Puget}, J.-L., {Renault}, C., {Rosset}, C.,
  {Santos}, D., {Starobinsky}, A., {Strukov}, I., {Sudiwala}, R.~V.,
  {Teyssier}, R., {Tristram}, M., {Tucker}, C.,  (2002).
\newblock {Cosmological constraints from Archeops}.
\newblock {\em \tt(astro-ph/0210306)}.

\bibitem[{Bertschinger}, 1985]{Bertschinger85}
{Bertschinger}, E. (1985).
\newblock {Self-similar secondary infall and accretion in an Einstein-de Sitter
  universe}.
\newblock {\em \apjs}, 58:39--65.

\bibitem[{Bialek} et~al., 2001]{Bialek01}
{Bialek}, J.~J., {Evrard}, A.~E., and {Mohr}, J.~J. (2001).
\newblock {Effects of Preheating on X-Ray Scaling Relations in Galaxy
  Clusters}.
\newblock {\em \apj}, 555:597--612.

\bibitem[{Binney} and {Tremaine}, 1987]{BT}
{Binney}, J. and {Tremaine}, S. (1987).
\newblock {\em {Galactic dynamics}}.
\newblock Princeton, NJ, Princeton University Press, 1987, 747 p.

\bibitem[{Blais-Ouellette} et~al., 2001]{blais01}
{Blais-Ouellette}, S., {Carignan}, C., and {Amram}, P. (2001).
\newblock {Mass Distribution in Spiral Galaxies: Multiwavelength Rotation
  Curves to Test Dark Halos}.
\newblock In {\em Galaxies: the Third Dimension}.

\bibitem[{Blanchard} et~al., 1992]{bla92}
{Blanchard}, A., {Valls-Gabaud}, D., and {Mamon}, G.~A. (1992).
\newblock {The origin of the galaxy luminosity function and the thermal
  evolution of the intergalactic medium}.
\newblock {\em \aap}, 264:365--378.

\bibitem[{Blumenthal} et~al., 1984]{Blumenthal84}
{Blumenthal}, G.~R., {Faber}, S.~M., {Primack}, J.~R., and {Rees}, M.~J.
  (1984).
\newblock {Formation of galaxies and large-scale structure with cold dark
  matter}.
\newblock {\em \nat}, 311:517--525.

\bibitem[{Borgani} et~al., 1999]{Borgani99}
{Borgani}, S., {Girardi}, M., {Carlberg}, R.~G., {Yee}, H.~K.~C., and
  {Ellingson}, E. (1999).
\newblock {Velocity Dispersions of CNOC Clusters and the Evolution of the
  Cluster Abundance}.
\newblock {\em \apj}, 527:561--572.

\bibitem[{Borgani} et~al., 2002]{Borgani02}
{Borgani}, S., {Governato}, F., {Wadsley}, J., {Menci}, N., {Tozzi}, P.,
  {Quinn}, T., {Stadel}, J., and {Lake}, G. (2002).
\newblock {The effect of non-gravitational gas heating in groups and clusters
  of galaxies}.
\newblock {\em \mnras}, 336:409--424.

\bibitem[{Borgani} and {Guzzo}, 2001]{BorganiGuzzo01}
{Borgani}, S. and {Guzzo}, L. (2001).
\newblock {X-ray clusters of galaxies as tracers of structure in the Universe}.
\newblock {\em \nat}, 409:39--45.

\bibitem[{Borriello} and {Salucci}, 2001]{BorrielloSalucci01}
{Borriello}, A. and {Salucci}, P. (2001).
\newblock {The dark matter distribution in disc galaxies}.
\newblock {\em \mnras}, 323:285--292.

\bibitem[{Bramel} et~al., 2000]{Bramel00}
{Bramel}, D.~A., {Nichol}, R.~C., and {Pope}, A.~C. (2000).
\newblock {The Local Space Density of Optically Selected Clusters of Galaxies}.
\newblock {\em \apj}, 533:601--610.

\bibitem[{Bryan} and {Machacek}, 2000]{BryanMachacek00}
{Bryan}, G.~L. and {Machacek}, M.~E. (2000).
\newblock {The B Distribution of the Ly{$\alpha$} Forest: Probing Cosmology and
  the Intergalactic Medium}.
\newblock {\em \apj}, 534:57--68.

\bibitem[{Bryan} and {Norman}, 1995]{BryanNorman95}
{Bryan}, G.~L. and {Norman}, M.~L. (1995).
\newblock {Simulating X-ray Clusters with Adaptive Mesh Refinement}.
\newblock {\em Bulletin of the American Astronomical Society}, 27:1421.

\bibitem[{Bryan} and {Norman}, 1998]{BN98}
{Bryan}, G.~L. and {Norman}, M.~L. (1998).
\newblock {Statistical Properties of X-Ray Clusters: Analytic and Numerical
  Comparisons}.
\newblock {\em \apj}, 495:80.

\bibitem[{Bryan} et~al., 1995]{Bryan95}
{Bryan}, G.~L., {Norman}, M.~L., {Stone}, J.~M., {Cen}, R., and {Ostriker},
  J.~P. (1995).
\newblock {\em Comput. Phys. Comm.}, 89:149.

\bibitem[{Bullock} et~al., 2001]{Bullock01}
{Bullock}, J.~S., {Dekel}, A., {Kolatt}, T.~S., {Kravtsov}, A.~V., {Klypin},
  A.~A., {Porciani}, C., and {Primack}, J.~R. (2001).
\newblock {A Universal Angular Momentum Profile for Galactic Halos}.
\newblock {\em \apj}, 555:240--257.

\bibitem[{Bullock} et~al., 2000]{Bullock00}
{Bullock}, J.~S., {Kravtsov}, A.~V., and {Weinberg}, D.~H. (2000).
\newblock {Reionization and the Abundance of Galactic Satellites}.
\newblock {\em \apj}, 539:517--521.

\bibitem[{Burkert}, 1995]{Burkert95}
{Burkert}, A. (1995).
\newblock {The Structure of Dark Matter Halos in Dwarf Galaxies}.
\newblock {\em \apjl}, 447:L25.

\bibitem[{Burles} and {Tytler}, 1998a]{BurlesTytler98a}
{Burles}, S. and {Tytler}, D. (1998a).
\newblock {The Deuterium Abundance toward Q1937-1009}.
\newblock {\em \apj}, 499:699.

\bibitem[{Burles} and {Tytler}, 1998b]{BurlesTytler98b}
{Burles}, S. and {Tytler}, D. (1998b).
\newblock {The Deuterium Abundance toward QSO 1009+2956}.
\newblock {\em \apj}, 507:732--744.

\bibitem[{Butcher} and {Oemler}, 1978]{but78}
{Butcher}, H. and {Oemler}, A. (1978).
\newblock {The evolution of galaxies in clusters. II - The galaxy content of
  nearby clusters}.
\newblock {\em \apj}, 226:559--565.

\bibitem[{Byrd} and {Valtonen}, 1990]{ByrdValtonen90}
{Byrd}, G. and {Valtonen}, M. (1990).
\newblock {Tidal generation of active spirals and S0 galaxies by rich
  clusters}.
\newblock {\em \apj}, 350:89--94.

\bibitem[{Carlberg} et~al., 1996]{Carlberg96}
{Carlberg}, R.~G., {Yee}, H.~K.~C., {Ellingson}, E., {Abraham}, R., {Gravel},
  P., {Morris}, S., and {Pritchet}, C.~J. (1996).
\newblock {Galaxy Cluster Virial Masses and Omega}.
\newblock {\em \apj}, 462:32.

\bibitem[{Carlberg} et~al., 1997]{Carlberg97}
{Carlberg}, R.~G., {Yee}, H.~K.~C., {Ellingson}, E., {Morris}, S.~L.,
  {Abraham}, R., {Gravel}, P., {Pritchet}, C.~J., {Smecker-Hane}, T.,
  {Hartwick}, F.~D.~A., {Hesser}, J.~E., {Hutchings}, J.~B., and {Oke}, J.~B.
  (1997).
\newblock {The Average Mass Profile of Galaxy Clusters}.
\newblock {\em \apjl}, 485:L13.

\bibitem[{Carlberg} et~al., 2001]{Carlberg01}
{Carlberg}, R.~G., {Yee}, H.~K.~C., {Morris}, S.~L., {Lin}, H., {Hall}, P.~B.,
  {Patton}, D.~R., {Sawicki}, M., and {Shepherd}, C.~W. (2001).
\newblock {Galaxy Groups at Intermediate Redshift}.
\newblock {\em \apj}, 552:427--444.

\bibitem[{Cavaliere} and {Fusco-Femiano}, 1976]{CavaliereFusco76}
{Cavaliere}, A. and {Fusco-Femiano}, R. (1976).
\newblock {X-rays from hot plasma in clusters of galaxies}.
\newblock {\em \aap}, 49:137--144.

\bibitem[{Cavaliere} et~al., 1997]{Cavaliere97}
{Cavaliere}, A., {Menci}, N., and {Tozzi}, P. (1997).
\newblock {The Luminosity-Temperature Relation for Groups and Clusters of
  Galaxies}.
\newblock {\em \apjl}, 484:L21.

\bibitem[{Cen}, 1992]{Cen92}
{Cen}, R. (1992).
\newblock {A hydrodynamic approach to cosmology - Methodology}.
\newblock {\em \apjs}, 78:341--364.

\bibitem[{Cen}, 2001]{Cen01}
{Cen}, R. (2001).
\newblock {Decaying Cold Dark Matter Model and Small-Scale Power}.
\newblock {\em \apjl}, 546:L77--L80.

\bibitem[{Cen} et~al., 1990]{Cen90}
{Cen}, R.~Y., {Ostriker}, J.~P., {Jameson}, A., and {Liu}, F. (1990).
\newblock {The universe in a box - Thermal effects in the standard cold dark
  matter scenario}.
\newblock {\em \apjl}, 362:L41--L45.

\bibitem[{Chen} and {Jing}, 2002]{chen}
{Chen}, D.~N. and {Jing}, Y.~P. (2002).
\newblock {The Angular Momentum Distribution within Halos in Different Dark
  Matter Models}.
\newblock {\em \tt(astro-ph/0201520)}.

\bibitem[{Cole}, 1991]{col91}
{Cole}, S. (1991).
\newblock {Modeling galaxy formation in evolving dark matter halos}.
\newblock {\em \apj}, 367:45--53.

\bibitem[{Cole} et~al., 1994a]{Cole94}
{Cole}, S., {Aragon-Salamanca}, A., {Frenk}, C.~S., {Navarro}, J.~F., and
  {Zepf}, S.~E. (1994a).
\newblock {A Recipe for Galaxy Formation}.
\newblock {\em \mnras}, 271:781.

\bibitem[{Cole} et~al., 1994b]{col94}
{Cole}, S., {Aragon-Salamanca}, A., {Frenk}, C.~S., {Navarro}, J.~F., and
  {Zepf}, S.~E. (1994b).
\newblock {A Recipe for Galaxy Formation}.
\newblock {\em \mnras}, 271:781.

\bibitem[{Cole} et~al., 2000]{Cole00}
{Cole}, S., {Lacey}, C.~G., {Baugh}, C.~M., and {Frenk}, C.~S. (2000).
\newblock {Hierarchical galaxy formation}.
\newblock {\em \mnras}, 319:168--204.

\bibitem[{Colella} and {Woodward}, 1984]{ColellaWoodward84}
{Colella}, P. and {Woodward}, P.~R. (1984).
\newblock {Hydrodynamical simulations of galaxy formation: effects of supernova
  feedback}.
\newblock {\em J.Comp.Phys}, 54:174--.

\bibitem[{Colella} and {Woodward}, 1985]{App85}
{Colella}, P. and {Woodward}, P.~R. (1985).
\newblock {\em J.Sci. Stat. Comp.}, 6:85.

\bibitem[{Col{\'\i }n} et~al., 2000a]{Colin00}
{Col{\'\i }n}, P., {Avila-Reese}, V., and {Valenzuela}, O. (2000a).
\newblock {Substructure and Halo Density Profiles in a Warm Dark Matter
  Cosmology}.
\newblock {\em \apj}, 542:622--630.

\bibitem[{Col{\'\i }n} et~al., 2000b]{Colin00vel}
{Col{\'\i }n}, P., {Klypin}, A.~A., and {Kravtsov}, A.~V. (2000b).
\newblock {Velocity Bias in a {$\Lambda$} Cold Dark Matter Model}.
\newblock {\em \apj}, 539:561--569.

\bibitem[{Col{\'{\i}}n} et~al., 1999]{Colin99}
{Col{\'{\i}}n}, P., {Klypin}, A.~A., {Kravtsov}, A.~V., and {Khokhlov}, A.~M.
  (1999).
\newblock {Evolution of Bias in Different Cosmological Models}.
\newblock {\em \apj}, 523:32--53.

\bibitem[{Connolly} et~al., 1997]{con97}
{Connolly}, A.~J., {Szalay}, A.~S., {Dickinson}, M., {Subbarao}, M.~U., and
  {Brunner}, R.~J. (1997).
\newblock {The Evolution of the Global Star Formation History as Measured from
  the Hubble Deep Field}.
\newblock {\em \apjl}, 486:L11.

\bibitem[{Couch} et~al., 2001]{Couch01}
{Couch}, W.~J., {Balogh}, M.~L., {Bower}, R.~G., {Smail}, I., {Glazebrook}, K.,
  and {Taylor}, M. (2001).
\newblock {A Low Global Star Formation Rate in the Rich Galaxy Cluster AC 114
  at z=0.32}.
\newblock {\em \apj}, 549:820--831.

\bibitem[{Couch} and {Sharples}, 1987]{Couch87}
{Couch}, W.~J. and {Sharples}, R.~M. (1987).
\newblock {A spectroscopic study of three rich galaxy clusters at Z = 0.31}.
\newblock {\em \mnras}, 229:423--456.

\bibitem[{Couchman}, 1991]{hydra91}
{Couchman}, H.~M.~P. (1991).
\newblock {Mesh-refined P3M - A fast adaptive N-body algorithm}.
\newblock {\em \apjl}, 368:L23--L26.

\bibitem[{Cowie} et~al., 1999]{cow99}
{Cowie}, L.~L., {Songaila}, A., and {Barger}, A.~J. (1999).
\newblock {Evidence for a Gradual Decline in the Universal Rest-Frame
  Ultraviolet Luminosity Density for $z<1$}.
\newblock {\em \aj}, 118:603--612.

\bibitem[{Crone} et~al., 1994]{Crone94}
{Crone}, M.~M., {Evrard}, A.~E., and {Richstone}, D.~O. (1994).
\newblock {The cosmological dependence of cluster density profiles}.
\newblock {\em \apj}, 434:402--416.

\bibitem[{Dahle} et~al., 2002]{dahle}
{Dahle}, H., {Hannestad}, S., and J., S. (2002).
\newblock {The Density Profile of Cluster-scale Dark Matter Halos}.
\newblock {\em \apj, submitted {\tt(astro-ph/0206455)}}.

\bibitem[{Dahle} et~al., 2001]{Dahle01}
{Dahle}, H., {Kaiser}, N., {Squires}, G., and {Broadhurst}, T.~J. (2001).
\newblock {Weak Lensing Analysis of a WFPC2 Mosaic of Abell 1689}.
\newblock In {\em ASP Conf. Ser. 237: Gravitational Lensing: Recent Progress
  and Future Go}, page 311.

\bibitem[{Dav{\' e}} et~al., 2001]{Dave01}
{Dav{\' e}}, R., {Spergel}, D.~N., {Steinhardt}, P.~J., and {Wandelt}, B.~D.
  (2001).
\newblock {Halo Properties in Cosmological Simulations of Self-interacting Cold
  Dark Matter}.
\newblock {\em \apj}, 547:574--589.

\bibitem[{David} et~al., 1993]{David93}
{David}, L.~P., {Slyz}, A., {Jones}, C., {Forman}, W., {Vrtilek}, S.~D., and
  {Arnaud}, K.~A. (1993).
\newblock {A catalog of intracluster gas temperatures}.
\newblock {\em \apj}, 412:479--488.

\bibitem[{Davis} et~al., 1985]{Davis85}
{Davis}, M., {Efstathiou}, G., {Frenk}, C.~S., and {White}, S.~D.~M. (1985).
\newblock {The evolution of large-scale structure in a universe dominated by
  cold dark matter}.
\newblock {\em \apj}, 292:371--394.

\bibitem[{de Bernardis} et~al., 2000]{deBernardis00}
{de Bernardis}, P., {Ade}, P.~A.~R., {Bock}, J.~J., {Bond}, J.~R., {Borrill},
  J., {Boscaleri}, A., {Coble}, K., {Crill}, B.~P., {De Gasperis}, G.,
  {Farese}, P.~C., {Ferreira}, P.~G., {Ganga}, K., {Giacometti}, M., {Hivon},
  E., {Hristov}, V.~V., {Iacoangeli}, A., {Jaffe}, A.~H., {Lange}, A.~E.,
  {Martinis}, L., {Masi}, S., {Mason}, P.~V., {Mauskopf}, P.~D., {Melchiorri},
  A., {Miglio}, L., {Montroy}, T., {Netterfield}, C.~B., {Pascale}, E.,
  {Piacentini}, F., {Pogosyan}, D., {Prunet}, S., {Rao}, S., {Romeo}, G.,
  {Ruhl}, J.~E., {Scaramuzzi}, F., {Sforna}, D., and {Vittorio}, N. (2000).
\newblock {A flat Universe from high-resolution maps of the cosmic microwave
  background radiation}.
\newblock {\em \nat}, 404:955--959.

\bibitem[{de Blok} and {Bosma}, 2002]{BlokBosma02}
{de Blok}, W.~J.~G. and {Bosma}, A. (2002).
\newblock {High-resolution rotation curves of low surface brightness galaxies}.
\newblock {\em \aap}, 385:816--846.

\bibitem[{de Blok} et~al., 2001]{Blok01}
{de Blok}, W.~J.~G., {McGaugh}, S.~S., {Bosma}, A., and {Rubin}, V.~C. (2001).
\newblock {Mass Density Profiles of Low Surface Brightness Galaxies}.
\newblock {\em \apjl}, 552:L23--L26.

\bibitem[{De Grandi} and {Molendi}, 2002]{GrandiMolendi02}
{De Grandi}, S. and {Molendi}, S. (2002).
\newblock {Temperature Profiles of Nearby Clusters of Galaxies}.
\newblock {\em \apj}, 567:163--177.

\bibitem[{Del Popolo} et~al., 2000]{Popolo00}
{Del Popolo}, A., {Gambera}, M., {Recami}, E., and {Spedicato}, E. (2000).
\newblock {Density profiles of dark matter halos in an improved secondary
  infall model}.
\newblock {\em \aap}, 353:427--434.

\bibitem[{den Hartog} and {Katgert}, 1996]{HartogKatgert96}
{den Hartog}, R. and {Katgert}, P. (1996).
\newblock {On the dynamics of the cores of galaxy clusters.}
\newblock {\em \mnras}, 279:349--388.

\bibitem[{Diaferio} et~al., 2001]{Diaferio01}
{Diaferio}, A., {Kauffmann}, G., {Balogh}, M.~L., {White}, S.~D.~M., {Schade},
  D., and {Ellingson}, E. (2001).
\newblock {The spatial and kinematic distributions of cluster galaxies in a
  {$\Lambda$}CDM universe: comparison with observations}.
\newblock {\em \mnras}, 323:999--1015.

\bibitem[{Doroshkevich}, 1970]{Doroshkevich70}
{Doroshkevich}, A.~G. (1970).
\newblock {\em Astrofisika}, 6:581.

\bibitem[{Dressler}, 1980]{Dressler80}
{Dressler}, A. (1980).
\newblock {Galaxy morphology in rich clusters - Implications for the formation
  and evolution of galaxies}.
\newblock {\em \apj}, 236:351--365.

\bibitem[{Dressler} et~al., 1997]{dre97}
{Dressler}, A., {Oemler}, A.~J., {Couch}, W.~J., {Smail}, I., {Ellis}, R.~S.,
  {Barger}, A., {Butcher}, H., {Poggianti}, B.~M., and {Sharples}, R.~M.
  (1997).
\newblock {Evolution since Z = 0.5 of the Morphology-Density Relation for
  Clusters of Galaxies}.
\newblock {\em \apj}, 490:577.

\bibitem[{Dressler} et~al., 1985]{Dressler85}
{Dressler}, A., {Thompson}, I.~B., and {Shectman}, S.~A. (1985).
\newblock {Statistics of emission-line galaxies in rich clusters}.
\newblock {\em \apj}, 288:481--486.

\bibitem[{Dubinski}, 1996]{Dubinski96}
{Dubinski}, J. (1996).
\newblock {A parallel tree code}.
\newblock {\em New Astronomy}, 1:133--147.

\bibitem[{Dubinski} and {Carlberg}, 1991]{DubinskiCarlberg91}
{Dubinski}, J. and {Carlberg}, R.~G. (1991).
\newblock {The structure of cold dark matter halos}.
\newblock {\em \apj}, 378:496--503.

\bibitem[{Edge} and {Stewart}, 1991]{EdgeStewart91}
{Edge}, A.~C. and {Stewart}, G.~C. (1991).
\newblock {EXOSAT Observations of Clusters of Galaxies - Part Two - X-Ray to
  Optical Correlations}.
\newblock {\em \mnras}, 252:428.

\bibitem[{Edge} et~al., 1992]{Edge92}
{Edge}, A.~C., {Stewart}, G.~C., and {Fabian}, A.~C. (1992).
\newblock {Properties of cooling flows in a flux-limited sample of clusters of
  galaxies}.
\newblock {\em \mnras}, 258:177--188.

\bibitem[{Efstathiou}, 1992]{efs2}
{Efstathiou}, G. (1992).
\newblock {Suppressing the formation of dwarf galaxies via photoionization}.
\newblock {\em \mnras}, 256:43P--47P.

\bibitem[{Efstathiou} et~al., 1992]{Efstathiou92}
{Efstathiou}, G., {Bond}, J.~R., and {White}, S.~D.~M. (1992).
\newblock {COBE background radiation anisotropies and large-scale structure in
  the universe}.
\newblock {\em \mnras}, 258:1P--6P.

\bibitem[{Efstathiou} et~al., 2002]{Efstathiou02}
{Efstathiou}, G., {Moody}, S., {Peacock}, J.~A., {Percival}, W.~J., {Baugh},
  C., {Bland-Hawthorn}, J., {Bridges}, T., {Cannon}, R., {Cole}, S., {Colless},
  M., {Collins}, C., {Couch}, W., {Dalton}, G., {de Propris}, R., {Driver},
  S.~P., {Ellis}, R.~S., {Frenk}, C.~S., {Glazebrook}, K., {Jackson}, C.,
  {Lahav}, O., {Lewis}, I., {Lumsden}, S., {Maddox}, S., {Norberg}, P.,
  {Peterson}, B.~A., {Sutherland}, W., and {Taylor}, K. (2002).
\newblock {Evidence for a non-zero {$\Lambda$} and a low matter density from a
  combined analysis of the 2dF Galaxy Redshift Survey and cosmic microwave
  background anisotropies}.
\newblock {\em \mnras}, 330:L29--L35.

\bibitem[{Eke} et~al., 1998]{Eke98}
{Eke}, V.~R., {Navarro}, J.~F., and {Frenk}, C.~S. (1998).
\newblock {The Evolution of X-Ray Clusters in a Low-Density Universe}.
\newblock {\em \apj}, 503:569.

\bibitem[{Ettori}, 2000]{Ettori00}
{Ettori}, S. (2000).
\newblock {Note on a polytropic {$\beta$}-model to fit the X-ray surface
  brightness of clusters of galaxies}.
\newblock {\em \mnras}, 311:313--316.

\bibitem[{Ettori} et~al., 2001]{EAF01}
{Ettori}, S., {Allen}, S.~W., and {Fabian}, A.~C. (2001).
\newblock {BeppoSAX observations of three distant, highly luminous clusters of
  galaxies: RXJ1347-1145, Zwicky 3146 and Abell 2390}.
\newblock {\em \mnras}, 322:187--194.

\bibitem[{Ettori} et~al., 2002a]{Ettori02}
{Ettori}, S., {De Grandi}, S., and {Molendi}, S. (2002a).
\newblock {Gravitating mass profiles of nearby galaxy clusters and relations
  with X-ray gas temperature, luminosity and mass}.
\newblock {\em \aap}, 391:841--855.

\bibitem[{Ettori} and {Fabian}, 1999]{EttoriFabian99}
{Ettori}, S. and {Fabian}, A.~C. (1999).
\newblock {ROSAT PSPC observations of 36 high-luminosity clusters of galaxies:
  constraints on the gas fraction}.
\newblock {\em \mnras}, 305:834--848.

\bibitem[{Ettori} et~al., 2002b]{EttoriFAJ02}
{Ettori}, S., {Fabian}, A.~C., {Allen}, S.~W., and {Johnstone}, R.~M. (2002b).
\newblock {Deep inside the core of Abell 1795: the Chandra view}.
\newblock {\em \mnras}, 331:635--648.

\bibitem[{Evrard} and {Henry}, 1991]{EvrardHenry91}
{Evrard}, A.~E. and {Henry}, J.~P. (1991).
\newblock {Expectations for X-ray cluster observations by the ROSAT satellite}.
\newblock {\em \apj}, 383:95--103.

\bibitem[{Evrard} et~al., 1996]{EMN96}
{Evrard}, A.~E., {Metzler}, C.~A., and {Navarro}, J.~F. (1996).
\newblock {Mass Estimates of X-Ray Clusters}.
\newblock {\em \apj}, 469:494.

\bibitem[{Fabian}, 1994]{Fabian94}
{Fabian}, A.~C. (1994).
\newblock {Cooling Flows in Clusters of Galaxies}.
\newblock {\em \araa}, 32:277--318.

\bibitem[{Fabian}, 2002]{fabian}
{Fabian}, A.~C. (2002).
\newblock {Cluster cores and cooling flows}.
\newblock {\em \tt(astro-ph/0210150)}.

\bibitem[{Fadda} et~al., 1996]{Fadda96}
{Fadda}, D., {Girardi}, M., {Giuricin}, G., {Mardirossian}, F., and {Mezzetti},
  M. (1996).
\newblock {The Observational Distribution of Internal Velocity Dispersions in
  Nearby Galaxy Clusters}.
\newblock {\em \apj}, 473:670.

\bibitem[{Fall}, 1978]{Fall78}
{Fall}, S.~M. (1978).
\newblock {On the evolution of galaxy clustering and cosmological N-body
  simulations}.
\newblock {\em \mnras}, 185:165--178.

\bibitem[{Fillmore} and {Goldreich}, 1984]{FG84}
{Fillmore}, J.~A. and {Goldreich}, P. (1984).
\newblock {Self-similar gravitational collapse in an expanding universe}.
\newblock {\em \apj}, 281:1--8.

\bibitem[{Finoguenov} et~al., 2001]{Finoguenov01}
{Finoguenov}, A., {Reiprich}, T.~H., and {B{\" o}hringer}, H. (2001).
\newblock {Details of the mass-temperature relation for clusters of galaxies}.
\newblock {\em \aap}, 368:749--759.

\bibitem[{Flores} et~al., 1999]{flo99}
{Flores}, H., {Hammer}, F., {Thuan}, T.~X., {C{\' e}sarsky}, C., {Desert},
  F.~X., {Omont}, A., {Lilly}, S.~J., {Eales}, S., {Crampton}, D., and {Le F{\`
  e}vre}, O. (1999).
\newblock {15 Micron Infrared Space Observatory Observations of the 1415+52
  Canada-France Redshift Survey Field: The Cosmic Star Formation Rate as
  Derived from Deep Ultraviolet, Optical, Mid-Infrared, and Radio Photometry}.
\newblock {\em \apj}, 517:148--167.

\bibitem[{Flores} and {Primack}, 1994]{FloresPrimack94}
{Flores}, R.~A. and {Primack}, J.~R. (1994).
\newblock {Observational and theoretical constraints on singular dark matter
  halos}.
\newblock {\em \apjl}, 427:L1--L4.

\bibitem[{Freedman} et~al., 2001]{Freedman01}
{Freedman}, W.~L., {Madore}, B.~F., {Gibson}, B.~K., {Ferrarese}, L., {Kelson},
  D.~D., {Sakai}, S., {Mould}, J.~R., {Kennicutt}, R.~C., {Ford}, H.~C.,
  {Graham}, J.~A., {Huchra}, J.~P., {Hughes}, S.~M.~G., {Illingworth}, G.~D.,
  {Macri}, L.~M., and {Stetson}, P.~B. (2001).
\newblock {Final Results from the Hubble Space Telescope Key Project to Measure
  the Hubble Constant}.
\newblock {\em \apj}, 553:47--72.

\bibitem[{Frenk} et~al., 1999]{SB99}
{Frenk}, C.~S., {White}, S.~D.~M., {Bode}, P., {Bond}, J.~R., {Bryan}, G.~L.,
  {Cen}, R., {Couchman}, H.~M.~P., {Evrard}, A.~E., {Gnedin}, N., {Jenkins},
  A., {Khokhlov}, A.~M., {Klypin}, A., {Navarro}, J.~F., {Norman}, M.~L.,
  {Ostriker}, J.~P., {Owen}, J.~M., {Pearce}, F.~R., {Pen}, U.-L., {Steinmetz},
  M., {Thomas}, P.~A., {Villumsen}, J.~V., {Wadsley}, J.~W., {Warren}, M.~S.,
  {Xu}, G., and {Yepes}, G. (1999).
\newblock {The Santa Barbara Cluster Comparison Project: A Comparison of
  Cosmological Hydrodynamics Solutions}.
\newblock {\em \apj}, 525:554--582.

\bibitem[{Frenk} et~al., 1988]{Frenk88}
{Frenk}, C.~S., {White}, S.~D.~M., {Davis}, M., and {Efstathiou}, G. (1988).
\newblock {The formation of dark halos in a universe dominated by cold dark
  matter}.
\newblock {\em \apj}, 327:507--525.

\bibitem[{Frenk} et~al., 1990]{Frenk90}
{Frenk}, C.~S., {White}, S.~D.~M., {Efstathiou}, G., and {Davis}, M. (1990).
\newblock {Galaxy clusters and the amplitude of primordial fluctuations}.
\newblock {\em \apj}, 351:10--21.

\bibitem[{Fukugita} et~al., 1998]{fuk98}
{Fukugita}, M., {Hogan}, C.~J., and {Peebles}, P.~J.~E. (1998).
\newblock {The Cosmic Baryon Budget}.
\newblock {\em \apj}, 503:518.

\bibitem[{Fukushige} and {Makino}, 1997]{FukushigeMakino97}
{Fukushige}, T. and {Makino}, J. (1997).
\newblock {On the Origin of Cusps in Dark Matter Halos}.
\newblock {\em \apjl}, 477:L9.

\bibitem[{Fukushige} and {Makino}, 2001]{FukushigeMakino01}
{Fukushige}, T. and {Makino}, J. (2001).
\newblock {Structure of Dark Matter Halos from Hierarchical Clustering}.
\newblock {\em \apj}, 557:533--545.

\bibitem[{Gallego} et~al., 1995]{gal95}
{Gallego}, J., {Zamorano}, J., {Aragon-Salamanca}, A., and {Rego}, M. (1995).
\newblock {The Current Star Formation Rate of the Local Universe}.
\newblock {\em \apjl}, 455:L1.

\bibitem[{Gamow}, 1946]{Gamow46}
{Gamow}, G. (1946).
\newblock {\em Phys. Rev.}, 70:572--573.

\bibitem[{Geller} et~al., 1999]{Geller99}
{Geller}, M.~J., {Diaferio}, A., and {Kurtz}, M.~J. (1999).
\newblock {The Mass Profile of the Coma Galaxy Cluster}.
\newblock {\em \apjl}, 517:L23--L26.

\bibitem[{Ghigna} et~al., 1998]{Ghigna98}
{Ghigna}, S., {Moore}, B., {Governato}, F., {Lake}, G., {Quinn}, T., and
  {Stadel}, J. (1998).
\newblock {Dark matter haloes within clusters}.
\newblock {\em \mnras}, 300:146--162.

\bibitem[{Ghigna} et~al., 2000]{Ghigna00}
{Ghigna}, S., {Moore}, B., {Governato}, F., {Lake}, G., {Quinn}, T., and
  {Stadel}, J. (2000).
\newblock {Density Profiles and Substructure of Dark Matter Halos: Converging
  Results at Ultra-High Numerical Resolution}.
\newblock {\em \apj}, 544:616--628.

\bibitem[{Gingold} and {Monaghan}, 1977]{GingoldMonaghan77}
{Gingold}, R.~A. and {Monaghan}, J.~J. (1977).
\newblock {Smoothed particle hydrodynamics - Theory and application to
  non-spherical stars}.
\newblock {\em \mnras}, 181:375--389.

\bibitem[{Giovanelli} and {Haynes}, 1985]{GiovanelliHaynes85}
{Giovanelli}, R. and {Haynes}, M.~P. (1985).
\newblock {Gas deficiency in cluster galaxies - A comparison of nine clusters}.
\newblock {\em \apj}, 292:404--425.

\bibitem[{Gispert} et~al., 2000]{gis00}
{Gispert}, R., {Lagache}, G., and {Puget}, J.~L. (2000).
\newblock {Implications of the cosmic infrared background for light production
  and the star formation history in the Universe}.
\newblock {\em \aap}, 360:1--9.

\bibitem[{Glazebrook} et~al., 1999]{gla99}
{Glazebrook}, K., {Blake}, C., {Economou}, F., {Lilly}, S., and {Colless}, M.
  (1999).
\newblock {Measurement of the star formation rate from Halpha in field galaxies
  at z=1}.
\newblock {\em \mnras}, 306:843--856.

\bibitem[{Gnedin}, 1995]{Gnedin95}
{Gnedin}, N.~Y. (1995).
\newblock {Softened Lagrangian hydrodynamics for cosmology}.
\newblock {\em \apjs}, 97:231--257.

\bibitem[{Gomez} et~al., 2002]{Gomez02}
{Gomez}, P., {Nichol}, R., {Miller}, C., {Balogh}, M., {Goto}, T., {Zabludoff},
  A., {Romer}, K., {Bernardi}, M., {Sheth}, R., {Hopkins}, A., {Castander}, F.,
  {Connolly}, A., {Schneider}, D., {Brinkmann}, J., {Lamb}, D., {SubbaRao}, M.,
  and {York}, D. (2002).
\newblock {Galaxy Star-Formation as a Function of Environment in the Early Data
  Release of the Sloan Digital Sky Survey}.
\newblock In {\em Accepted for publication in ApJ, 24 pages in emulateapj.sty},
  page 10193.

\bibitem[{Goodman}, 2000]{Goodman00}
{Goodman}, J. (2000).
\newblock {Repulsive dark matter}.
\newblock {\em New Astronomy}, 5:103--107.

\bibitem[{Gottl{\" o}ber} et~al., 2001]{Gottloeber01}
{Gottl{\" o}ber}, S., {Klypin}, A., and {Kravtsov}, A.~V. (2001).
\newblock {Merging History as a Function of Halo Environment}.
\newblock {\em \apj}, 546:223--233.

\bibitem[{Gottl{\" o}ber} et~al., 1991]{Gottloeber91}
{Gottl{\" o}ber}, S., {M{\" u}ller}, V., and {Starobinsky}, A.~A. (1991).
\newblock {Analysis of inflation driven by a scalar field and a
  curvature-squared term}.
\newblock {\em \prd}, 43:2510--2520.

\bibitem[{Greengard}, 1988]{Greengard88}
{Greengard}, L.~F. (1988).
\newblock {\em {The rapid evaluation of potential fields in particle systems}}.
\newblock MIT Press, 1988.

\bibitem[{Gunn}, 1977]{Gunn77}
{Gunn}, J.~E. (1977).
\newblock {Massive galactic halos. I - Formation and evolution}.
\newblock {\em \apj}, 218:592--598.

\bibitem[{Gunn} and {Gott}, 1972]{GG72}
{Gunn}, J.~E. and {Gott}, J.~R.~I. (1972).
\newblock {On the Infall of Matter Into Clusters of Galaxies and Some Effects
  on Their Evolution}.
\newblock {\em \apj}, 176:1.

\bibitem[{Gurevich} and {Zybin}, 1988]{GurevichZybin88}
{Gurevich}, A.~V. and {Zybin}, K.~P. (1988).
\newblock {Nondissipative gravitational turbulence}.
\newblock {\em Zhurnal Eksperimental noi i Teoreticheskoi Fiziki}, 94:3--25.

\bibitem[{Haarsma} et~al., 2000]{haa00}
{Haarsma}, D.~B., {Partridge}, R.~B., {Windhorst}, R.~A., and {Richards}, E.~A.
  (2000).
\newblock {Faint Radio Sources and Star Formation History}.
\newblock {\em \apj}, 544:641--658.

\bibitem[{Halverson} et~al., 2002]{Halverson02}
{Halverson}, N.~W., {Leitch}, E.~M., {Pryke}, C., {Kovac}, J., {Carlstrom},
  J.~E., {Holzapfel}, W.~L., {Dragovan}, M., {Cartwright}, J.~K., {Mason},
  B.~S., {Padin}, S., {Pearson}, T.~J., {Readhead}, A.~C.~S., and {Shepherd},
  M.~C. (2002).
\newblock {Degree Angular Scale Interferometer First Results: A Measurement of
  the Cosmic Microwave Background Angular Power Spectrum}.
\newblock {\em \apj}, 568:38--45.

\bibitem[{Hammer} et~al., 1997]{ham97}
{Hammer}, F., {Flores}, H., {Lilly}, S.~J., {Crampton}, D., {Le Fevre}, O.,
  {Rola}, C., {Mallen-Ornelas}, G., {Schade}, D., and {Tresse}, L. (1997).
\newblock {Canada-France Redshift Survey. XIV. Spectral Properties of Field
  Galaxies up to z=1}.
\newblock {\em \apj}, 481:49.

\bibitem[{Hanany} et~al., 2000]{Hanany00}
{Hanany}, S., {Ade}, P., {Balbi}, A., {Bock}, J., {Borrill}, J., {Boscaleri},
  A., {de Bernardis}, P., {Ferreira}, P.~G., {Hristov}, V.~V., {Jaffe}, A.~H.,
  {Lange}, A.~E., {Lee}, A.~T., {Mauskopf}, P.~D., {Netterfield}, C.~B., {Oh},
  S., {Pascale}, E., {Rabii}, B., {Richards}, P.~L., {Smoot}, G.~F., {Stompor},
  R., {Winant}, C.~D., and {Wu}, J.~H.~P. (2000).
\newblock {MAXIMA-1: A Measurement of the Cosmic Microwave Background
  Anisotropy on Angular Scales of $10'-5{\deg}$}.
\newblock {\em \apjl}, 545:L5--L9.

\bibitem[{Hancock} et~al., 1998]{Hancock98}
{Hancock}, S., {Rocha}, G., {Lasenby}, A.~N., and {Gutierrez}, C.~M. (1998).
\newblock {Constraints on cosmological parameters from recent measurements of
  cosmic microwave background anisotropy}.
\newblock {\em \mnras}, 294:L1--L6.

\bibitem[{Hashimoto} et~al., 1998]{Hashimoto98}
{Hashimoto}, Y., {Oemler}, A.~J., {Lin}, H., and {Tucker}, D.~L. (1998).
\newblock {The Influence of Environment on the Star Formation Rates of
  Galaxies}.
\newblock {\em \apj}, 499:589.

\bibitem[{Henriksen} and {Widrow}, 1999]{HenriksenWidrow99}
{Henriksen}, R.~N. and {Widrow}, L.~M. (1999).
\newblock {Relaxing and virializing a dark matter halo}.
\newblock {\em \mnras}, 302:321--336.

\bibitem[{Hernquist}, 1993]{Hernquist93}
{Hernquist}, L. (1993).
\newblock {Some cautionary remarks about smoothed particle hydrodynamics}.
\newblock {\em \apj}, 404:717--722.

\bibitem[{Hernquist} et~al., 1996]{Hernquist96}
{Hernquist}, L., {Katz}, N., {Weinberg}, D.~H., and {Jordi}, M. (1996).
\newblock {The Lyman-Alpha Forest in the Cold Dark Matter Model}.
\newblock {\em \apjl}, 457:L51.

\bibitem[{Hiotelis}, 2002a]{Hiotelis02}
{Hiotelis}, N. (2002a).
\newblock {Density profiles in a spherical infall model with non-radial
  motions}.
\newblock {\em \aap}, 382:84--91.

\bibitem[{Hiotelis}, 2002b]{hiotelis}
{Hiotelis}, N. (2002b).
\newblock {Density profiles of dark matter halos with anisotropic velocity
  tensors}.
\newblock {\em \tt(astro-ph/0207628)}.

\bibitem[{Hockney} and {Eastwood}, 1981]{HockneyEastwood81}
{Hockney}, R.~W. and {Eastwood}, J.~W. (1981).
\newblock {\em {Computer Simulation Using Particles}}.
\newblock Computer Simulation Using Particles, New York: McGraw-Hill, 1981.

\bibitem[{Hockney} et~al., 1974]{Hockney74}
{Hockney}, R.~W., {Goel}, S.~P.~., and {Eastwood}, J.~W. (1974).
\newblock {Quiet high-resolution computer simulations}.
\newblock {\em J.~Comp.~Phys}, 14:148.

\bibitem[{Hoffman}, 1988]{Hoffman88}
{Hoffman}, Y. (1988).
\newblock {On the formation and structure of galactic halos}.
\newblock {\em \apj}, 328:489--498.

\bibitem[{Hoffman} and {Shaham}, 1985]{HS85}
{Hoffman}, Y. and {Shaham}, J. (1985).
\newblock {Local density maxima - Progenitors of structure}.
\newblock {\em \apj}, 297:16--22.

\bibitem[{Hopkins} et~al., 2001]{hop01}
{Hopkins}, A.~M., {Connolly}, A.~J., {Haarsma}, D.~B., and {Cram}, L.~E.
  (2001).
\newblock {Toward a Resolution of the Discrepancy between Different Estimators
  of Star Formation Rate}.
\newblock {\em \aj}, 122:288--296.

\bibitem[{Horner} et~al., 1999]{Horner99}
{Horner}, D.~J., {Mushotzky}, R.~F., and {Scharf}, C.~A. (1999).
\newblock {Observational Tests of the Mass-Temperature Relation for Galaxy
  Clusters}.
\newblock {\em \apj}, 520:78--86.

\bibitem[{Hu} et~al., 2000]{Hu00}
{Hu}, W., {Barkana}, R., and {Gruzinov}, A. (2000).
\newblock {Observational Evidence for Self-Interacting Cold Dark Matter}.
\newblock {\em Physical Review Letters}, 85:1158--1161.

\bibitem[{Hubble}, 1929]{Hubble29}
{Hubble}, E. (1929).
\newblock {A Relation between Distance and Radial Velocity among Extra-Galactic
  Nebulae}.
\newblock {\em Proceedings of the National Academy of Science}, 15:168--173.

\bibitem[{Hughes} et~al., 1998]{hug98}
{Hughes}, D.~H., {Serjeant}, S., {Dunlop}, J., {Rowan-Robinson}, M., {Blain},
  A., {Mann}, R.~G., {Ivison}, R., {Peacock}, J., {Efstathiou}, A., {Gear}, W.,
  {Oliver}, S., {Lawrence}, A., {Longair}, M., {Goldschmidt}, P., and
  {Jenness}, T. (1998).
\newblock {High-redshift star formation in the Hubble Deep Field revealed by a
  submillimetre-wavelength survey.}
\newblock {\em \nat}, 394:241--247.

\bibitem[{Iglesias-P{\' a}ramo} et~al., 2002]{Iglesias02}
{Iglesias-P{\' a}ramo}, J., {Boselli}, A., {Cortese}, L., {V{\'{\i}}lchez},
  J.~M., and {Gavazzi}, G. (2002).
\newblock {A deep Halpha survey of galaxies in the two nearby clusters Abell
  1367 and Coma. The Halpha luminosity functions}.
\newblock {\em \aap}, 384:383--392.

\bibitem[{Irwin} and {Bregman}, 2000]{ib00}
{Irwin}, J.~A. and {Bregman}, J.~N. (2000).
\newblock {Radial Temperature Profiles of 11 Clusters of Galaxies Observed with
  BEPPOSAX}.
\newblock {\em \apj}, 538:543--554.

\bibitem[{Jansen} et~al., 2001]{Jansen01}
{Jansen}, F., {Lumb}, D., {Altieri}, B., {Clavel}, J., {Ehle}, M., {Erd}, C.,
  {Gabriel}, C., {Guainazzi}, M., {Gondoin}, P., {Much}, R., {Munoz}, R.,
  {Santos}, M., {Schartel}, N., {Texier}, D., and {Vacanti}, G. (2001).
\newblock {XMM-Newton observatory. I. The spacecraft and operations}.
\newblock {\em \aap}, 365:L1--L6.

\bibitem[{Jernigan} and {Porter}, 1989]{JerniganPorter89}
{Jernigan}, J.~G. and {Porter}, D.~H. (1989).
\newblock {A tree code with logarithmic reduction of force terms, hierarchical
  regularization of all variables, and explicit accuracy controls}.
\newblock {\em \apjs}, 71:871--893.

\bibitem[{Jimenez} et~al., 2002]{jimenez}
{Jimenez}, R., {Verde}, L., and {Oh}, S.~P. (2002).
\newblock {Dark halo properties from rotation curves}.
\newblock {\em \mnras, submitted {\tt(astro-ph/0201352)}}.

\bibitem[{Jing}, 2000]{Jing00}
{Jing}, Y.~P. (2000).
\newblock {The Density Profile of Equilibrium and Nonequilibrium Dark Matter
  Halos}.
\newblock {\em \apj}, 535:30--36.

\bibitem[{Jing} and {Suto}, 2000]{JingSuto00}
{Jing}, Y.~P. and {Suto}, Y. (2000).
\newblock {The Density Profiles of the Dark Matter Halo Are Not Universal}.
\newblock {\em \apjl}, 529:L69--L72.

\bibitem[{Kaiser}, 1986]{Kaiser86}
{Kaiser}, N. (1986).
\newblock {Evolution and clustering of rich clusters}.
\newblock {\em \mnras}, 222:323--345.

\bibitem[{Kaiser}, 1991]{Kaiser91}
{Kaiser}, N. (1991).
\newblock {Evolution of clusters of galaxies}.
\newblock {\em \apj}, 383:104--111.

\bibitem[{Kang} et~al., 1994]{Kang94}
{Kang}, H., {Ostriker}, J.~P., {Cen}, R., {Ryu}, D., {Hernquist}, L., {Evrard},
  A.~E., {Bryan}, G.~L., and {Norman}, M.~L. (1994).
\newblock {A comparison of cosmological hydrodynamic codes}.
\newblock {\em \apj}, 430:83--100.

\bibitem[{Kaplinghat} et~al., 2000]{Kaplinghat00}
{Kaplinghat}, M., {Knox}, L., and {Turner}, M.~S. (2000).
\newblock {Observational Evidence for Self-Interacting Cold Dark Matter}.
\newblock {\em Physical Review Letters}, 85:3335--3338.

\bibitem[{Kates} et~al., 1995]{bsi95}
{Kates}, R., {Muller}, V., {Gottlober}, S., {Mucket}, J.~P., and {Retzlaff}, J.
  (1995).
\newblock {Large-scale structure formation for power spectra with broken scale
  invariance}.
\newblock {\em \mnras}, 277:1254--1268.

\bibitem[{Kauffmann} et~al., 1993a]{kau93}
{Kauffmann}, G., {White}, S.~D.~M., and {Guiderdoni}, B. (1993a).
\newblock {The Formation and Evolution of Galaxies Within Merging Dark Matter
  Haloes}.
\newblock {\em \mnras}, 264:201.

\bibitem[{Kauffmann} et~al., 1993b]{Kauffmann93}
{Kauffmann}, G., {White}, S.~D.~M., and {Guiderdoni}, B. (1993b).
\newblock {The Formation and Evolution of Galaxies Within Merging Dark Matter
  Haloes}.
\newblock {\em \mnras}, 264:201.

\bibitem[{Kellogg} et~al., 1972]{Kellogg72}
{Kellogg}, E., {Gursky}, H., {Tananbaum}, H., {Giacconi}, R., and {Pounds}, K.
  (1972).
\newblock {The Extended X-Ray Source at M87}.
\newblock {\em \apjl}, 174:L65.

\bibitem[{Kennicutt}, 1998]{Kennicutt98}
{Kennicutt}, R.~C. (1998).
\newblock {Star Formation in Galaxies Along the Hubble Sequence}.
\newblock {\em \araa}, 36:189--232.

\bibitem[{Khokhlov}, 1998]{ftt}
{Khokhlov}, A.~M. (1998).
\newblock {Fully Threaded Tree Algorithms for Adaptive Refinement Fluid
  Dynamics Simulations}.
\newblock {\em J.Comp.Phys}, 143:239--267.

\bibitem[{King}, 1962]{king62}
{King}, I. (1962).
\newblock {The structure of star clusters. I. an empirical density law}.
\newblock {\em \aj}, 67:471.

\bibitem[{Klypin} et~al., 1999]{Klypin99}
{Klypin}, A., {Gottl{\" o}ber}, S., {Kravtsov}, A.~V., and {Khokhlov}, A.~M.
  (1999).
\newblock {Galaxies in N-Body Simulations: Overcoming the Overmerging Problem}.
\newblock {\em \apj}, 516:530--551.

\bibitem[{Klypin} et~al., 2001]{Klypin01}
{Klypin}, A., {Kravtsov}, A.~V., {Bullock}, J.~S., and {Primack}, J.~R. (2001).
\newblock {Resolving the Structure of Cold Dark Matter Halos}.
\newblock {\em \apj}, 554:903--915.

\bibitem[{Klypin} et~al., 1996]{Klypin96}
{Klypin}, A., {Primack}, J., and {Holtzman}, J. (1996).
\newblock {Small-Scale Power Spectrum and Correlations in Lambda + Cold Dark
  Matter Models}.
\newblock {\em \apj}, 466:13.

\bibitem[{Klypin} and {Shandarin}, 1983]{KlypinShandarin83}
{Klypin}, A.~A. and {Shandarin}, S.~F. (1983).
\newblock {Three-dimensional numerical model of the formation of large-scale
  structure in the Universe}.
\newblock {\em \mnras}, 204:891--907.

\bibitem[{Knebe} et~al., 2001]{mlapm01}
{Knebe}, A., {Green}, A., and {Binney}, J. (2001).
\newblock {Multi-level adaptive particle mesh (MLAPM): a c code for
  cosmological simulations}.
\newblock {\em \mnras}, 325:845--864.

\bibitem[{Knebe} et~al., 2000]{Knebe00}
{Knebe}, A., {Kravtsov}, A.~V., {Gottl{\" o}ber}, S., and {Klypin}, A.~A.
  (2000).
\newblock {On the effects of resolution in dissipationless cosmological
  simulations}.
\newblock {\em \mnras}, 317:630--648.

\bibitem[{Kodama} and {Smail}, 2001]{Kodama01}
{Kodama}, T. and {Smail}, I. (2001).
\newblock {Testing the hypothesis of the morphological transformation from
  field spiral to cluster S0}.
\newblock {\em \mnras}, 326:637--642.

\bibitem[{Kolb} and {Turner}, 1990]{KolbTurner90}
{Kolb}, E.~W. and {Turner}, M.~S. (1990).
\newblock {\em {The early universe}}.
\newblock Frontiers in Physics, Reading, MA: Addison-Wesley, 1988, 1990.

\bibitem[{Komatsu} and {Seljak}, 2001]{KS01}
{Komatsu}, E. and {Seljak}, U. (2001).
\newblock {Universal gas density and temperature profile}.
\newblock {\em \mnras}, 327:1353--1366.

\bibitem[{Koranyi} and {Geller}, 2000]{KoranyiGeller00}
{Koranyi}, D.~M. and {Geller}, M.~J. (2000).
\newblock {Kinematics and Mass Profile of AWM 7}.
\newblock {\em \aj}, 119:44--58.

\bibitem[{Kravtsov} et~al., 2002]{ARThydro02}
{Kravtsov}, A.~V., {Klypin}, A., and {Hoffman}, Y. (2002).
\newblock {Constrained Simulations of the Real Universe. II. Observational
  Signatures of Intergalactic Gas in the Local Supercluster Region}.
\newblock {\em \apj}, 571:563--575.

\bibitem[{Kravtsov} et~al., 1998]{KKBP98}
{Kravtsov}, A.~V., {Klypin}, A.~A., {Bullock}, J.~S., and {Primack}, J.~R.
  (1998).
\newblock {The Cores of Dark Matter-dominated Galaxies: Theory versus
  Observations}.
\newblock {\em \apj}, 502:48.

\bibitem[{Kravtsov} et~al., 1997]{ART97}
{Kravtsov}, A.~V., {Klypin}, A.~A., and {Khokhlov}, A.~M. (1997).
\newblock {Adaptive Refinement Tree: A New High-Resolution N-Body Code for
  Cosmological Simulations}.
\newblock {\em \apjs}, 111:73.

\bibitem[{Kravtsov} and {Yepes}, 2000]{ky00}
{Kravtsov}, A.~V. and {Yepes}, G. (2000).
\newblock {On the supernova heating of the intergalactic medium}.
\newblock {\em \mnras}, 318:227--238.

\bibitem[{Kull}, 1999]{Kull99}
{Kull}, A. (1999).
\newblock {A Model for the Density Distribution of Virialized Cold Dark Matter
  Halos}.
\newblock {\em \apjl}, 516:L5--L8.

\bibitem[{Lacey} and {Cole}, 1994]{LaceyCole94}
{Lacey}, C. and {Cole}, S. (1994).
\newblock {Merger Rates in Hierarchical Models of Galaxy Formation - Part Two -
  Comparison with N-Body Simulations}.
\newblock {\em \mnras}, 271:676.

\bibitem[{Lange} et~al., 2001]{Lange01}
{Lange}, A.~E., {Ade}, P.~A., {Bock}, J.~J., {Bond}, J.~R., {Borrill}, J.,
  {Boscaleri}, A., {Coble}, K., {Crill}, B.~P., {de Bernardis}, P., {Farese},
  P., {Ferreira}, P., {Ganga}, K., {Giacometti}, M., {Hivon}, E., {Hristov},
  V.~V., {Iacoangeli}, A., {Jaffe}, A.~H., {Martinis}, L., {Masi}, S.,
  {Mauskopf}, P.~D., {Melchiorri}, A., {Montroy}, T., {Netterfield}, C.~B.,
  {Pascale}, E., {Piacentini}, F., {Pogosyan}, D., {Prunet}, S., {Rao}, S.,
  {Romeo}, G., {Ruhl}, J.~E., {Scaramuzzi}, F., and {Sforna}, D. (2001).
\newblock {Cosmological parameters from the first results of Boomerang}.
\newblock {\em \prd}, 63:42001.

\bibitem[{Lavery} and {Henry}, 1988]{lav88}
{Lavery}, R.~J. and {Henry}, J.~P. (1988).
\newblock {Evidence for galaxy-galaxy interactions as an active agent of the
  'Butcher-Oemler' effect at a redshift of 0.2}.
\newblock {\em \apj}, 330:596--600.

\bibitem[{Lee} et~al., 2001]{Lee01}
{Lee}, A.~T., {Ade}, P., {Balbi}, A., {Bock}, J., {Borrill}, J., {Boscaleri},
  A., {de Bernardis}, P., {Ferreira}, P.~G., {Hanany}, S., {Hristov}, V.~V.,
  {Jaffe}, A.~H., {Mauskopf}, P.~D., {Netterfield}, C.~B., {Pascale}, E.,
  {Rabii}, B., {Richards}, P.~L., {Smoot}, G.~F., {Stompor}, R., {Winant},
  C.~D., and {Wu}, J.~H.~P. (2001).
\newblock {A High Spatial Resolution Analysis of the MAXIMA-1 Cosmic Microwave
  Background Anisotropy Data}.
\newblock {\em \apjl}, 561:L1--L5.

\bibitem[{Lewis} et~al., 2002a]{lewis}
{Lewis}, A.~D., {Buote}, D.~A., and {Stoke}, J.~T. (2002a).
\newblock {Chandra Observations of Abell 2029: The Dark Matter Profile at
  $<0.01 R_{vir}$ in an Unusually Relaxed Cluster}.
\newblock {\em \apj, submitted {\tt(astro-ph/0209205)}}.

\bibitem[{Lewis} et~al., 2002b]{Lewis02}
{Lewis}, I., {Balogh}, M., {De Propris}, R., {Couch}, W., {Bower}, R., {Offer},
  A., {Bland-Hawthorn}, J., {Baldry}, I.~K., {Baugh}, C., {Bridges}, T.,
  {Cannon}, R., {Cole}, S., {Colless}, M., {Collins}, C., {Cross}, N.,
  {Dalton}, G., {Driver}, S.~P., {Efstathiou}, G., {Ellis}, R.~S., {Frenk},
  C.~S., {Glazebrook}, K., {Hawkins}, E., {Jackson}, C., {Lahav}, O.,
  {Lumsden}, S., {Maddox}, S., {Madgwick}, D., {Norberg}, P., {Peacock}, J.~A.,
  {Percival}, W., {Peterson}, B.~A., {Sutherland}, W., and {Taylor}, K.
  (2002b).
\newblock {The 2dF Galaxy Redshift Survey: the environmental dependence of
  galaxy star formation rates near clusters}.
\newblock {\em \mnras}, 334:673--683.

\bibitem[{Lia} and {Carraro}, 2001]{LiaCarraro01}
{Lia}, C. and {Carraro}, G. (2001).
\newblock {Parallel Tree-SPH: A Tool for Galaxy Formation}.
\newblock {\em \apss}, 276:1049--1056.

\bibitem[{Lilje}, 1992]{Lilje92}
{Lilje}, P.~B. (1992).
\newblock {Abundance of rich clusters of galaxies - A test for cosmological
  parameters}.
\newblock {\em \apjl}, 386:L33--L36.

\bibitem[{Lilly} et~al., 1996]{lil96}
{Lilly}, S.~J., {Le Fevre}, O., {Hammer}, F., and {Crampton}, D. (1996).
\newblock {The Canada-France Redshift Survey: The Luminosity Density and Star
  Formation History of the Universe to $z$ approximately 1}.
\newblock {\em \apjl}, 460:L1.

\bibitem[{Lineweaver} et~al., 1997]{Lineweaver97}
{Lineweaver}, C.~H., {Barbosa}, D., {Blanchard}, A., and {Bartlett}, J.~G.
  (1997).
\newblock {Constraints on h, $\{$OMEGA$\}$\_b\_ and $\{$lambda$\}$\_o\_ from
  cosmic microwave background observations.}
\newblock {\em \aap}, 322:365--374.

\bibitem[{{\L}okas}, 2000]{Lokas00}
{{\L}okas}, E.~L. (2000).
\newblock {Universal profile of dark matter haloes and the spherical infall
  model}.
\newblock {\em \mnras}, 311:423--432.

\bibitem[{{\L}okas} and {Hoffman}, 2000]{LokasHoffman00}
{{\L}okas}, E.~L. and {Hoffman}, Y. (2000).
\newblock {Formation of Cuspy Density Profiles: A Generic Feature of
  Collisionless Gravitational Collapse}.
\newblock {\em \apjl}, 542:L139--L142.

\bibitem[{{\L}okas} and {Mamon}, 2001]{LokasMamon01}
{{\L}okas}, E.~L. and {Mamon}, G.~A. (2001).
\newblock {Properties of spherical galaxies and clusters with an NFW density
  profile}.
\newblock {\em \mnras}, 321:155--166.

\bibitem[{Loken} et~al., 2002]{Loken02}
{Loken}, C., {Norman}, M.~L., {Nelson}, E., {Burns}, J., {Bryan}, G.~L., and
  {Motl}, P. (2002).
\newblock {A Universal Temperature Profile for Galaxy Clusters}.
\newblock {\em \apj}, 579:571--576.

\bibitem[{Lucey}, 1983]{Lucey83}
{Lucey}, J.~R. (1983).
\newblock {An assessment of the completeness and correctness of the Abell
  catalogue}.
\newblock {\em \mnras}, 204:33--43.

\bibitem[{Lucy}, 1977]{Lucy77}
{Lucy}, L.~B. (1977).
\newblock {A numerical approach to the testing of the fission hypothesis}.
\newblock {\em \aj}, 82:1013--1024.

\bibitem[{Lynden-Bell}, 1967]{LB67}
{Lynden-Bell}, D. (1967).
\newblock {Statistical mechanics of violent relaxation in stellar systems}.
\newblock {\em \mnras}, 136:101.

\bibitem[{Macfarland} et~al., 1998]{p3mpar98}
{Macfarland}, T., {Couchman}, H.~M.~P., {Pearce}, F.~R., and {Pichlmeier}, J.
  (1998).
\newblock {A new parallel P\^{}3M code for very large-scale cosmological
  simulations}.
\newblock {\em New Astronomy}, 3:687--705.

\bibitem[{Madau} et~al., 1996]{mad96}
{Madau}, P., {Ferguson}, H.~C., {Dickinson}, M.~E., {Giavalisco}, M.,
  {Steidel}, C.~C., and {Fruchter}, A. (1996).
\newblock {High-redshift galaxies in the Hubble Deep Field: colour selection
  and star formation history to z\~{}4}.
\newblock {\em \mnras}, 283:1388--1404.

\bibitem[{Madau} et~al., 1998]{mad98}
{Madau}, P., {Pozzetti}, L., and {Dickinson}, M. (1998).
\newblock {The Star Formation History of Field Galaxies}.
\newblock {\em \apj}, 498:106.

\bibitem[{Makino}, 2002]{Makino02}
{Makino}, J. (2002).
\newblock {An efficient parallel algorithm for O(N$^{2}$) direct summation
  method and its variations on distributed-memory parallel mac}.
\newblock {\em New Astronomy}, 7:373--384.

\bibitem[{Marchesini} et~al., 2002]{Marchesini02}
{Marchesini}, D., {D'Onghia}, E., {Chincarini}, G., {Firmani}, C., {Conconi},
  P., {Molinari}, E., and {Zacchei}, A. (2002).
\newblock {H{$\alpha$} Rotation Curves: The Soft Core Question}.
\newblock {\em \apj}, 575:801--813.

\bibitem[{Markevitch} et~al., 1998]{Markevitch98}
{Markevitch}, M., {Forman}, W.~R., {Sarazin}, C.~L., and {Vikhlinin}, A.
  (1998).
\newblock {The Temperature Structure of 30 Nearby Clusters Observed with ASCA:
  Similarity of Temperature Profiles}.
\newblock {\em \apj}, 503:77.

\bibitem[{Martel} and {Shapiro}, 1998]{martelshapiro98}
{Martel}, H. and {Shapiro}, P.~R. (1998).
\newblock {A convenient set of comoving cosmological variables and their
  application}.
\newblock {\em \mnras}, 297:467--485.

\bibitem[{Mart{\'{\i}}nez} et~al., 2002]{Martinez02}
{Mart{\'{\i}}nez}, H.~J., {Zandivarez}, A., {Dom{\'{\i}}nguez}, M., {Merch{\'
  a}n}, M.~E., and {Lambas}, D.~G. (2002).
\newblock {Galaxy groups in the 2dF Galaxy Redshift Survey: effects of
  environment on star formation}.
\newblock {\em \mnras}, 333:L31--L34.

\bibitem[{Mathiesen} and {Evrard}, 2001]{ME01}
{Mathiesen}, B.~F. and {Evrard}, A.~E. (2001).
\newblock {Four Measures of the Intracluster Medium Temperature and Their
  Relation to a Cluster's Dynamical State}.
\newblock {\em \apj}, 546:100--116.

\bibitem[{Mazure} et~al., 1996]{Mazure96}
{Mazure}, A., {Katgert}, P., {den Hartog}, R., {Biviano}, A., {Dubath}, P.,
  {Escalera}, E., {Focardi}, P., {Gerbal}, D., {Giuricin}, G., {Jones}, B., {Le
  Fevre}, O., {Moles}, M., {Perea}, J., and {Rhee}, G. (1996).
\newblock {The ESO Nearby Abell Cluster Survey. II. The distribution of
  velocity dispersions of rich galaxy clusters.}
\newblock {\em \aap}, 310:31--48.

\bibitem[{Mellier}, 1999]{Mellier99}
{Mellier}, Y. (1999).
\newblock {Probing the Universe with Weak Lensing}.
\newblock {\em \araa}, 37:127--189.

\bibitem[{Merritt}, 1983]{Merritt83}
{Merritt}, D. (1983).
\newblock {Relaxation and tidal stripping in rich clusters of galaxies. I -
  Evolution of the mass distribution}.
\newblock {\em \apj}, 264:24--48.

\bibitem[{Miralda-Escude} and {Babul}, 1995]{MiraldaBabul95}
{Miralda-Escude}, J. and {Babul}, A. (1995).
\newblock {Gravitational Lensing in Clusters of Galaxies: New Clues Regarding
  the Dynamics of Intracluster Gas}.
\newblock {\em \apj}, 449:18.

\bibitem[{Mitchell} et~al., 1976]{Mitchell76}
{Mitchell}, R.~J., {Culhane}, J.~L., {Davison}, P.~J.~N., and {Ives}, J.~C.
  (1976).
\newblock {Ariel 5 observations of the X-ray spectrum of the Perseus Cluster}.
\newblock {\em \mnras}, 175:29P--34P.

\bibitem[{Mitchell} et~al., 1979]{Mitchell79}
{Mitchell}, R.~J., {Dickens}, R.~J., {Burnell}, S.~J.~B., and {Culhane}, J.~L.
  (1979).
\newblock {The X-ray spectra of clusters of galaxies and their relationship to
  other cluster properties}.
\newblock {\em \mnras}, 189:329--361.

\bibitem[{Mohr} et~al., 1999]{Mohr99}
{Mohr}, J.~J., {Mathiesen}, B., and {Evrard}, A.~E. (1999).
\newblock {Properties of the Intracluster Medium in an Ensemble of Nearby
  Galaxy Clusters}.
\newblock {\em \apj}, 517:627--649.

\bibitem[{Monaghan}, 1992]{Mo92}
{Monaghan}, J.~J. (1992).
\newblock {Smoothed particle hydrodynamics}.
\newblock {\em \araa}, 30:543--574.

\bibitem[{Moore}, 1994]{Moore94}
{Moore}, B. (1994).
\newblock {Evidence against Dissipationless Dark Matter from Observations of
  Galaxy Haloes}.
\newblock {\em \nat}, 370:629.

\bibitem[{Moore} et~al., 1998a]{Moore98}
{Moore}, B., {Governato}, F., {Quinn}, T., {Stadel}, J., and {Lake}, G.
  (1998a).
\newblock {Resolving the Structure of Cold Dark Matter Halos}.
\newblock {\em \apjl}, 499:L5.

\bibitem[{Moore} et~al., 1996]{Moore96}
{Moore}, B., {Katz}, N., {Lake}, G., {Dressler}, A., and {Oemler}, A. (1996).
\newblock {Galaxy harassment and the evolution of clusters of galaxies.}
\newblock {\em \nat}, 379:613--616.

\bibitem[{Moore} et~al., 1998b]{Moore98h}
{Moore}, B., {Lake}, G., and {Katz}, N. (1998b).
\newblock {Morphological Transformation from Galaxy Harassment}.
\newblock {\em \apj}, 495:139.

\bibitem[{Moore} et~al., 1999a]{Moore99h}
{Moore}, B., {Lake}, G., {Quinn}, T., and {Stadel}, J. (1999a).
\newblock {On the survival and destruction of spiral galaxies in clusters}.
\newblock {\em \mnras}, 304:465--474.

\bibitem[{Moore} et~al., 1999b]{Moore99}
{Moore}, B., {Quinn}, T., {Governato}, F., {Stadel}, J., and {Lake}, G.
  (1999b).
\newblock {Cold collapse and the core catastrophe}.
\newblock {\em \mnras}, 310:1147--1152.

\bibitem[{Moss} and {Whittle}, 2000]{MossWhittle00}
{Moss}, C. and {Whittle}, M. (2000).
\newblock {An H{$\alpha$} survey of eight Abell clusters: the dependence of
  tidally induced star formation on cluster density}.
\newblock {\em \mnras}, 317:667--686.

\bibitem[{Mould} et~al., 2000a]{Mould00}
{Mould}, J.~R., {Huchra}, J.~P., {Freedman}, W.~L., {Kennicutt}, R.~C.,
  {Ferrarese}, L., {Ford}, H.~C., {Gibson}, B.~K., {Graham}, J.~A., {Hughes},
  S.~M.~G., {Illingworth}, G.~D., {Kelson}, D.~D., {Macri}, L.~M., {Madore},
  B.~F., {Sakai}, S., {Sebo}, K.~M., {Silbermann}, N.~A., and {Stetson}, P.~B.
  (2000a).
\newblock {The Hubble Space Telescope Key Project on the Extragalactic Distance
  Scale. XXVIII. Combining the Constraints on the Hubble Constant}.
\newblock {\em \apj}, 529:786--794.

\bibitem[{Mould} et~al., 2000b]{Mould00err}
{Mould}, J.~R., {Huchra}, J.~P., {Freedman}, W.~L., {Kennicutt}, R.~C.,
  {Ferrarese}, L., {Ford}, H.~C., {Gibson}, B.~K., {Graham}, J.~A., {Hughes},
  S.~M.~G., {Illingworth}, G.~D., {Kelson}, D.~D., {Macri}, L.~M., {Madore},
  B.~F., {Sakai}, S., {Sebo}, K.~M., {Silbermann}, N.~A., and {Stetson}, P.~B.
  (2000b).
\newblock {Erratum: The Hubble Space Telescope Key Project on the Extragalactic
  Distance Scale. XXVIII. Combining the Constraints on the Hubble Constant}.
\newblock {\em \apj}, 545:547--547.

\bibitem[{Muanwong} et~al., 2002]{Muanwong02}
{Muanwong}, O., {Thomas}, P.~A., {Kay}, S.~T., and {Pearce}, F.~R. (2002).
\newblock {The effect of cooling and preheating on the X-ray properties of
  clusters of galaxies}.
\newblock {\em \mnras}, 336:527--540.

\bibitem[{Muecket} and {Kates}, 1997]{muc97}
{Muecket}, J.~P. and {Kates}, R.~E. (1997).
\newblock {Evolution of a hot primordial gas in the presence of an ionizing
  ultraviolet background: instabilities, bifurcations, and the formation of
  Lyman limit systems.}
\newblock {\em \aap}, 324:1--10.

\bibitem[{Nagai} and {Kravtsov}, 2002]{cluster6}
{Nagai}, D. and {Kravtsov}, A.~V. (2002).
\newblock {Cold Fronts in CDM clusters}.
\newblock {\em \apj, submitted {\tt(astro-ph/0206469)}}.

\bibitem[{Nagamine} et~al., 2001]{nag01}
{Nagamine}, K., {Fukugita}, M., {Cen}, R., and {Ostriker}, J.~P. (2001).
\newblock {Star Formation History and Stellar Metallicity Distribution in a
  Cold Dark Matter Universe}.
\newblock {\em \apj}, 558:497--504.

\bibitem[{Navarro} et~al., 1995]{NFW95}
{Navarro}, J.~F., {Frenk}, C.~S., and {White}, S.~D.~M. (1995).
\newblock {Simulations of X-ray clusters}.
\newblock {\em \mnras}, 275:720--740.

\bibitem[{Navarro} et~al., 1996]{NFW96}
{Navarro}, J.~F., {Frenk}, C.~S., and {White}, S.~D.~M. (1996).
\newblock {The Structure of Cold Dark Matter Halos}.
\newblock {\em \apj}, 462:563.

\bibitem[{Navarro} et~al., 1997]{NFW97}
{Navarro}, J.~F., {Frenk}, C.~S., and {White}, S.~D.~M. (1997).
\newblock {A Universal Density Profile from Hierarchical Clustering}.
\newblock {\em \apj}, 490:493.

\bibitem[{Nelson} and {Papaloizou}, 1993]{NelsonPapaloizou93}
{Nelson}, R.~P. and {Papaloizou}, J.~C.~B. (1993).
\newblock {Three-Dimensional Hydrodynamic Simulations of Collapsing Prolate
  Clouds}.
\newblock {\em \mnras}, 265:905.

\bibitem[{Nelson} and {Papaloizou}, 1994]{NelsonPapaloizou94}
{Nelson}, R.~P. and {Papaloizou}, J.~C.~B. (1994).
\newblock {Variable Smoothing Lengths and Energy Conservation in Smoothed
  Particle Hydrodynamics}.
\newblock {\em \mnras}, 270:1.

\bibitem[{Netterfield} et~al., 2002]{Netterfield02}
{Netterfield}, C.~B., {Ade}, P.~A.~R., {Bock}, J.~J., {Bond}, J.~R., {Borrill},
  J., {Boscaleri}, A., {Coble}, K., {Contaldi}, C.~R., {Crill}, B.~P., {de
  Bernardis}, P., {Farese}, P., {Ganga}, K., {Giacometti}, M., {Hivon}, E.,
  {Hristov}, V.~V., {Iacoangeli}, A., {Jaffe}, A.~H., {Jones}, W.~C., {Lange},
  A.~E., {Martinis}, L., {Masi}, S., {Mason}, P., {Mauskopf}, P.~D.,
  {Melchiorri}, A., {Montroy}, T., {Pascale}, E., {Piacentini}, F., {Pogosyan},
  D., {Pongetti}, F., {Prunet}, S., {Romeo}, G., {Ruhl}, J.~E., and
  {Scaramuzzi}, F. (2002).
\newblock {A Measurement by BOOMERANG of Multiple Peaks in the Angular Power
  Spectrum of the Cosmic Microwave Background}.
\newblock {\em \apj}, 571:604--614.

\bibitem[{Netterfield} et~al., 1997]{Netterfield97}
{Netterfield}, C.~B., {Devlin}, M.~J., {Jarolik}, N., {Page}, L., and
  {Wollack}, E.~J. (1997).
\newblock {A Measurement of the Angular Power Spectrum of the Anisotropy in the
  Cosmic Microwave Background}.
\newblock {\em \apj}, 474:47.

\bibitem[{Neumann} and {Arnaud}, 1999]{NeumannArnaud99}
{Neumann}, D.~M. and {Arnaud}, M. (1999).
\newblock {Regularity in the X-ray surface brightness profiles of galaxy
  clusters and the M-T relation}.
\newblock {\em \aap}, 348:711--727.

\bibitem[{Nevalainen} et~al., 2000]{Nevalainen00}
{Nevalainen}, J., {Markevitch}, M., and {Forman}, W. (2000).
\newblock {The Cluster M-T Relation from Temperature Profiles Observed with
  ASCA and ROSAT}.
\newblock {\em \apj}, 532:694--699.

\bibitem[{Norman} and {Bryan}, 1999]{NormanBryan99}
{Norman}, M.~L. and {Bryan}, G.~L. (1999).
\newblock {Cosmological Adaptive Mesh Refinement\^{}$\{$CD$\}$}.
\newblock In {\em ASSL Vol. 240: Numerical Astrophysics}, page~19.

\bibitem[{Nusser}, 2001]{Nusser01}
{Nusser}, A. (2001).
\newblock {Self-similar spherical collapse with non-radial motions}.
\newblock {\em \mnras}, 325:1397--1401.

\bibitem[{Nusser} and {Sheth}, 1999]{NusserSheth99}
{Nusser}, A. and {Sheth}, R.~K. (1999).
\newblock {Mass growth and density profiles of dark matter haloes in
  hierarchical clustering}.
\newblock {\em \mnras}, 303:685--695.

\bibitem[{Oguri} et~al., 2001]{Oguri01}
{Oguri}, M., {Taruya}, A., and {Suto}, Y. (2001).
\newblock {Probing the Core Structure of Dark Halos with Tangential and Radial
  Arc Statistics}.
\newblock {\em \apj}, 559:572--583.

\bibitem[{Okamoto} and {Nagashima}, 2001]{OkamotoNagashima01}
{Okamoto}, T. and {Nagashima}, M. (2001).
\newblock {Morphology-Density Relation for Simulated Clusters of Galaxies in
  Cold Dark Matter-dominated Universes}.
\newblock {\em \apj}, 547:109--116.

\bibitem[{O'Meara} et~al., 2001]{OMeara01}
{O'Meara}, J.~M., {Tytler}, D., {Kirkman}, D., {Suzuki}, N., {Prochaska},
  J.~X., {Lubin}, D., and {Wolfe}, A.~M. (2001).
\newblock {The Deuterium to Hydrogen Abundance Ratio toward a Fourth QSO: HS
  0105+1619}.
\newblock {\em \apj}, 552:718--730.

\bibitem[{Pascarelle} et~al., 1998]{pas98}
{Pascarelle}, S.~M., {Lanzetta}, K.~M., and {Fern{\' a}ndez-Soto}, A. (1998).
\newblock {The Ultraviolet Luminosity Density of the Universe from Photometric
  Redshifts of Galaxies in the Hubble Deep Field}.
\newblock {\em \apjl}, 508:L1--L4.

\bibitem[{Pearce} et~al., 2000]{Pearce00}
{Pearce}, F.~R., {Thomas}, P.~A., {Couchman}, H.~M.~P., and {Edge}, A.~C.
  (2000).
\newblock {The effect of radiative cooling on the X-ray properties of galaxy
  clusters}.
\newblock {\em \mnras}, 317:1029--1040.

\bibitem[{Pearson} et~al., 2002]{pearson}
{Pearson}, T.~J., {Mason}, B.~S., {Readhead}, A.~C.~S., {Shepherd}, M.~S.,
  {Sievers}, J., {Udomprasert}, P.~S., {Cartwright}, J.~K., {Farmer}, A.,
  {Padin}, S., {Myers}, S.~T., {Bond}, J.~R., {Contaldi}, C.~R., {Pen}, U.,
  {Prunet}, S., {Pogosyan}, D., {Carlstrom}, J.~E., {Kovac}, J., {Leitch},
  E.~M., {Pryke}, C., {Halverson}, N.~W., {Holzapfel}, W.~L., {Altamirano}, P.,
  {Bronfman}, L., {Casassus}, S., {May}, J., and {Joy}, M. (2002).
\newblock {The Anisotropy of the Microwave Background to l = 3500: Mosaic
  Observations with the Cosmic Background Imager}.
\newblock {\em \apj, submitted {\tt(astro-ph/0205388)}}.

\bibitem[{Peebles}, 1969]{Peebles69}
{Peebles}, P.~J.~E. (1969).
\newblock {Origin of the Angular Momentum of Galaxies}.
\newblock {\em \apj}, 155:393.

\bibitem[{Peebles}, 1982]{Peebles82}
{Peebles}, P.~J.~E. (1982).
\newblock {Large-scale background temperature and mass fluctuations due to
  scale-invariant primeval perturbations}.
\newblock {\em \apjl}, 263:L1--L5.

\bibitem[{Peebles}, 2000]{Peebles00}
{Peebles}, P.~J.~E. (2000).
\newblock {Fluid Dark Matter}.
\newblock {\em \apjl}, 534:L127--L129.

\bibitem[{Pen}, 1995]{Pen95}
{Pen}, U. (1995).
\newblock {A Linear Moving Adaptive Particle-Mesh N-Body Algorithm}.
\newblock {\em \apjs}, 100:269.

\bibitem[{Pen}, 1998]{Pen98}
{Pen}, U. (1998).
\newblock {A High-Resolution Adaptive Moving Mesh Hydrodynamic Algorithm}.
\newblock {\em \apjs}, 115:19.

\bibitem[{Penzias} and {Wilson}, 1965]{PenziasWilson65}
{Penzias}, A.~A. and {Wilson}, R.~W. (1965).
\newblock {A Measurement of Excess Antenna Temperature at 4080 Mc/s.}
\newblock {\em \apj}, 142:419--421.

\bibitem[{Perlmutter} et~al., 1999]{Perlmutter99}
{Perlmutter}, S., {Aldering}, G., {Goldhaber}, G., {Knop}, R.~A., {Nugent}, P.,
  {Castro}, P.~G., {Deustua}, S., {Fabbro}, S., {Goobar}, A., {Groom}, D.~E.,
  {Hook}, I.~M., {Kim}, A.~G., {Kim}, M.~Y., {Lee}, J.~C., {Nunes}, N.~J.,
  {Pain}, R., {Pennypacker}, C.~R., {Quimby}, R., {Lidman}, C., {Ellis}, R.~S.,
  {Irwin}, M., {McMahon}, R.~G., {Ruiz-Lapuente}, P., {Walton}, N., {Schaefer},
  B., {Boyle}, B.~J., {Filippenko}, A.~V., {Matheson}, T., {Fruchter}, A.~S.,
  {Panagia}, N., {Newberg}, H.~J.~M., {Couch}, W.~J., and {The Supernova
  Cosmology Project} (1999).
\newblock {Measurements of Omega and Lambda from 42 High-Redshift Supernovae}.
\newblock {\em \apj}, 517:565--586.

\bibitem[{Pettini} et~al., 1998]{pet98}
{Pettini}, M., {Kellogg}, M., {Steidel}, C.~C., {Dickinson}, M., {Adelberger},
  K.~L., and {Giavalisco}, M. (1998).
\newblock {Infrared Observations of Nebular Emission Lines from Galaxies at Z
  \~{}= 3}.
\newblock {\em \apj}, 508:539--550.

\bibitem[{Plionis} and {Basilakos}, 2002]{PlionisBasilakos02}
{Plionis}, M. and {Basilakos}, S. (2002).
\newblock {The cluster substructure-alignment connection}.
\newblock {\em \mnras}, 329:L47--L51.

\bibitem[{Poggianti} et~al., 1999]{Poggianti99}
{Poggianti}, B.~M., {Smail}, I., {Dressler}, A., {Couch}, W.~J., {Barger},
  A.~J., {Butcher}, H., {Ellis}, R.~S., and {Oemler}, A.~J. (1999).
\newblock {The Star Formation Histories of Galaxies in Distant Clusters}.
\newblock {\em \apj}, 518:576--593.

\bibitem[{Ponman} et~al., 1996]{Ponman96}
{Ponman}, T.~J., {Bourner}, P.~D.~J., {Ebeling}, H., and {Bohringer}, H.
  (1996).
\newblock {A ROSAT survey of Hickson's compact galaxy groups.}
\newblock {\em \mnras}, 283:690--708.

\bibitem[{Ponman} et~al., 1999]{PCN99}
{Ponman}, T.~J., {Cannon}, D.~B., and {Navarro}, J.~F. (1999).
\newblock {The thermal imprint of galaxy formation on X-ray clusters.}
\newblock {\em \nat}, 397:135--137.

\bibitem[{Porciani} et~al., 2002a]{Porciani02a}
{Porciani}, C., {Dekel}, A., and {Hoffman}, Y. (2002a).
\newblock {Testing tidal-torque theory - I. Spin amplitude and direction}.
\newblock {\em \mnras}, 332:325--338.

\bibitem[{Porciani} et~al., 2002b]{Porciani02b}
{Porciani}, C., {Dekel}, A., and {Hoffman}, Y. (2002b).
\newblock {Testing tidal-torque theory - II. Alignment of inertia and shear and
  the characteristics of protohaloes}.
\newblock {\em \mnras}, 332:339--351.

\bibitem[{Power} et~al., 2002]{power}
{Power}, C., {Navarro}, J.~F., {Jenkins}, A., {Frenk}, C.~S., {White},
  S.~D.~M., {Springel}, V., {Stadel}, J., and {Quinn}, T. (2002).
\newblock {The Inner Structure of $\Lambda$CDM Halos I: A Numerical Convergence
  Study}.
\newblock {\em \mnras, submitted {\tt(astro-ph/0201544)}}.

\bibitem[{Pratt} and {Arnaud}, 2002]{PrattArnaud02}
{Pratt}, G.~W. and {Arnaud}, M. (2002).
\newblock {The mass profile of <ASTROBJ>A1413</ASTROBJ> observed with
  XMM-Newton: Implications for the M-T relation}.
\newblock {\em \aap}, 394:375--393.

\bibitem[{Press} et~al., 1986]{Press86}
{Press}, W.~H., {Flannery}, B.~P., and {Teukolsky}, S.~A. (1986).
\newblock {\em {Numerical recipes. The art of scientific computing}}.
\newblock Cambridge: University Press, 1986.

\bibitem[{Quilis} et~al., 2000]{Quilis00}
{Quilis}, V., {Moore}, B., and {Bower}, R. (2000).
\newblock {Gone with the Wind: The Origin of S0 Galaxies in Clusters}.
\newblock {\em Science}, 288:1617--1620.

\bibitem[{Quinn} et~al., 1986]{QSZ86}
{Quinn}, P.~J., {Salmon}, J.~K., and {Zurek}, W.~H. (1986).
\newblock {Primordial density fluctuations and the structure of galactic
  haloes}.
\newblock {\em \nat}, 322:329--335.

\bibitem[{Quinn} and {Zurek}, 1988]{QuinnZurek88}
{Quinn}, P.~J. and {Zurek}, W.~H. (1988).
\newblock {The angular momentum distribution in galactic halos}.
\newblock {\em \apj}, 331:1--18.

\bibitem[{Renzini}, 1998]{ren98}
{Renzini}, A. (1998).
\newblock {The Main Epoch of Metal Production in the Universe}.
\newblock In {\em ASP Conf. Ser. 146: The Young Universe: Galaxy Formation and
  Evolution at Intermediate and High Redshift}, page 298.

\bibitem[{Ricker} et~al., 2000]{cosmos00}
{Ricker}, P.~M., {Dodelson}, S., and {Lamb}, D.~Q. (2000).
\newblock {COSMOS: A Hybrid N-Body/Hydrodynamics Code for Cosmological
  Problems}.
\newblock {\em \apj}, 536:122--143.

\bibitem[{Riess} et~al., 1998]{Riess98}
{Riess}, A.~G., {Filippenko}, A.~V., {Challis}, P., {Clocchiatti}, A.,
  {Diercks}, A., {Garnavich}, P.~M., {Gilliland}, R.~L., {Hogan}, C.~J., {Jha},
  S., {Kirshner}, R.~P., {Leibundgut}, B., {Phillips}, M.~M., {Reiss}, D.,
  {Schmidt}, B.~P., {Schommer}, R.~A., {Smith}, R.~C., {Spyromilio}, J.,
  {Stubbs}, C., {Suntzeff}, N.~B., and {Tonry}, J. (1998).
\newblock {Observational Evidence from Supernovae for an Accelerating Universe
  and a Cosmological Constant}.
\newblock {\em \aj}, 116:1009--1038.

\bibitem[{Rogstad} and {Shostak}, 1972]{RogstadShostak72}
{Rogstad}, D.~H. and {Shostak}, G.~S. (1972).
\newblock {Gross Properties of Five Scd Galaxies as Determined from 21 cm
  Observations}.
\newblock {\em \apj}, 176:315.

\bibitem[{Rosati} et~al., 2002]{Rosati02}
{Rosati}, P., {Borgani}, S., and {Norman}, C. (2002).
\newblock {The Evolution of X-ray Clusters of Galaxies}.
\newblock {\em \araa}, 40:539--577.

\bibitem[{Rubin} and {Ford}, 1970]{RubinFord70}
{Rubin}, V.~C. and {Ford}, W.~K.~J. (1970).
\newblock {Rotation of the Andromeda Nebula from a Spectroscopic Survey of
  Emission Regions}.
\newblock {\em \apj}, 159:379.

\bibitem[{Sadat} and {Blanchard}, 2001]{SadatBlanchard01}
{Sadat}, R. and {Blanchard}, A. (2001).
\newblock {New light on the baryon fraction in galaxy clusters}.
\newblock {\em \aap}, 371:19--24.

\bibitem[{Salvador-Sole} et~al., 1998]{Salvador98}
{Salvador-Sole}, E., {Solanes}, J.~M., and {Manrique}, A. (1998).
\newblock {Merger versus Accretion and the Structure of Dark Matter Halos}.
\newblock {\em \apj}, 499:542.

\bibitem[{Sand} et~al., 2002]{Sand02}
{Sand}, D.~J., {Treu}, T., and {Ellis}, R.~S. (2002).
\newblock {The Dark Matter Density Profile of the Lensing Cluster MS 2137-23: A
  Test of the Cold Dark Matter Paradigm}.
\newblock {\em \apjl}, 574:L129--L133.

\bibitem[{Sarazin}, 1988a]{Sarazin88}
{Sarazin}, C.~L. (1988a).
\newblock {\em {X-ray emission from clusters of galaxies}}.
\newblock Cambridge Astrophysics Series, Cambridge: Cambridge University Press,
  1988.

\bibitem[{Sarazin}, 1988b]{sar88}
{Sarazin}, C.~L. (1988b).
\newblock {\em {X-ray emission from clusters of galaxies}}.
\newblock Cambridge Astrophysics Series, Cambridge: Cambridge University Press,
  1988.

\bibitem[{Sato} et~al., 2000]{Sato00}
{Sato}, S., {Akimoto}, F., {Furuzawa}, A., {Tawara}, Y., {Watanabe}, M., and
  {Kumai}, Y. (2000).
\newblock {The Observed Mass Profiles of Dark Halos and the Formation Epoch of
  Galaxies}.
\newblock {\em \apjl}, 537:L73--L76.

\bibitem[{Sawicki} et~al., 1997]{saw97}
{Sawicki}, M.~J., {Lin}, H., and {Yee}, H.~K.~C. (1997).
\newblock {Evolution of the Galaxy Population Based on Photometric Redshifts in
  the Hubble Deep Field}.
\newblock {\em \aj}, 113:1--12.

\bibitem[{Schindler}, 1999]{Schindler99}
{Schindler}, S. (1999).
\newblock {Distant clusters of galaxies: X-ray properties and their relations}.
\newblock {\em \aap}, 349:435--447.

\bibitem[{Schmidt} et~al., 1998]{Schmidt98}
{Schmidt}, B.~P., {Suntzeff}, N.~B., {Phillips}, M.~M., {Schommer}, R.~A.,
  {Clocchiatti}, A., {Kirshner}, R.~P., {Garnavich}, P., {Challis}, P.,
  {Leibundgut}, B., {Spyromilio}, J., {Riess}, A.~G., {Filippenko}, A.~V.,
  {Hamuy}, M., {Smith}, R.~C., {Hogan}, C., {Stubbs}, C., {Diercks}, A.,
  {Reiss}, D., {Gilliland}, R., {Tonry}, J., {Maza}, J.~., {Dressler}, A.,
  {Walsh}, J., and {Ciardullo}, R. (1998).
\newblock {The High-Z Supernova Search: Measuring Cosmic Deceleration and
  Global Curvature of the Universe Using Type IA Supernovae}.
\newblock {\em \apj}, 507:46--63.

\bibitem[{Schmidt}, 1968]{sch68}
{Schmidt}, M. (1968).
\newblock {Space Distribution and Luminosity Functions of Quasi-Stellar Radio
  Sources}.
\newblock {\em \apj}, 151:393.

\bibitem[{Schmidt} et~al., 2001]{SAF01}
{Schmidt}, R.~W., {Allen}, S.~W., and {Fabian}, A.~C. (2001).
\newblock {Chandra observations of the galaxy cluster Abell 1835}.
\newblock {\em \mnras}, 327:1057--1070.

\bibitem[{Scott et al.}, 2002]{scott}
{Scott et al.} (2002).
\newblock {First results from the Very Small Array -- III. The CMB power
  spectrum}.
\newblock {\em \mnras, submitted {\tt(astro-ph/0205380)}}.

\bibitem[{Seljak} and {Zaldarriaga}, 1996]{SeljakZaldarriaga96}
{Seljak}, U. and {Zaldarriaga}, M. (1996).
\newblock {A Line-of-Sight Integration Approach to Cosmic Microwave Background
  Anisotropies}.
\newblock {\em \apj}, 469:437.

\bibitem[{Serna} et~al., 1996]{Serna96}
{Serna}, A., {Alimi}, J.-M., and {Chieze}, J.-P. (1996).
\newblock {Adaptive Smooth Particle Hydrodynamics and Particle-Particle Coupled
  Codes: Energy and Entropy Conservation}.
\newblock {\em \apj}, 461:884.

\bibitem[{Serna} et~al., 2002]{deva}
{Serna}, A., {Dom{\'\i}nguez-Tenreiro}, R., and {S{\'a}iz}, A. (2002).
\newblock {Conservation Laws in Smoothed Particle Hydrodynamics: the DEVA
  Code}.
\newblock {\em \mnras}, page submitted.

\bibitem[{Shapiro} et~al., 1996]{Shapiro96}
{Shapiro}, P.~R., {Martel}, H., {Villumsen}, J.~V., and {Owen}, J.~M. (1996).
\newblock {Adaptive Smoothed Particle Hydrodynamics, with Application to
  Cosmology: Methodology}.
\newblock {\em \apjs}, 103:269.

\bibitem[{Silk}, 2001]{Silk01}
{Silk}, J. (2001).
\newblock {The formation of galaxy discs}.
\newblock {\em \mnras}, 324:313--318.

\bibitem[{Smith}, 1936]{smith36}
{Smith}, S. (1936).
\newblock {The Mass of the Virgo Cluster}.
\newblock {\em \apj}, 83:23--30.

\bibitem[{Sofue} and {Rubin}, 2001]{SofueRubin01}
{Sofue}, Y. and {Rubin}, V. (2001).
\newblock {Rotation Curves of Spiral Galaxies}.
\newblock {\em \araa}, 39:137--174.

\bibitem[{Solanes} et~al., 2001]{Solanes01}
{Solanes}, J.~., {Manrique}, A., {Garc{\'{\i}}a-G{\' o}mez}, C., {Gonz{\'
  a}lez-Casado}, G., {Giovanelli}, R., and {Haynes}, M.~P. (2001).
\newblock {The H I Content of Spirals. II. Gas Deficiency in Cluster Galaxies}.
\newblock {\em \apj}, 548:97--113.

\bibitem[{Somerville} and {Primack}, 1999]{SP99}
{Somerville}, R.~S. and {Primack}, J.~R. (1999).
\newblock {Semi-analytic modelling of galaxy formation: the local Universe}.
\newblock {\em \mnras}, 310:1087--1110.

\bibitem[{Somerville} et~al., 2001]{som01}
{Somerville}, R.~S., {Primack}, J.~R., and {Faber}, S.~M. (2001).
\newblock {The nature of high-redshift galaxies}.
\newblock {\em \mnras}, 320:504.

\bibitem[{Sommer-Larsen} and {Dolgov}, 2001]{SommerLarsenDolgov01}
{Sommer-Larsen}, J. and {Dolgov}, A. (2001).
\newblock {Formation of Disk Galaxies: Warm Dark Matter and the Angular
  Momentum Problem}.
\newblock {\em \apj}, 551:608--623.

\bibitem[{Spergel} and {Steinhardt}, 2000]{SpergelSteinhardt00}
{Spergel}, D.~N. and {Steinhardt}, P.~J. (2000).
\newblock {Observational Evidence for Self-Interacting Cold Dark Matter}.
\newblock {\em Physical Review Letters}, 84:3760--3763.

\bibitem[{Springel} and {Hernquist}, 2002a]{gadgetEntro02}
{Springel}, V. and {Hernquist}, L. (2002a).
\newblock {Cosmological smoothed particle hydrodynamics simulations: the
  entropy equation}.
\newblock {\em \mnras}, 333:649--664.

\bibitem[{Springel} and {Hernquist}, 2002b]{shyk3}
{Springel}, V. and {Hernquist}, L. (2002b).
\newblock {Cosmological SPH simulations: A hybrid multi-phase model for star
  formation}.
\newblock In {\em submitted to MNRAS, 25 pages, version with high-resolution
  figures available at http://www.mpa-garching.mpg.de/
  volker/paper\_multiphase/}, page 6393.

\bibitem[{Springel} and {Hernquist}, 2002c]{SpringelHernquist}
{Springel}, V. and {Hernquist}, L. (2002c).
\newblock {The history of star formation in a LCDM universe}.
\newblock In {\em submitted to MNRAS, 25 pages, version with high-resolution
  figures available at http://www.mpa-garching.mpg.de/ volker/paper\_sfr/},
  page 6395.

\bibitem[{Springel} et~al., 2001a]{Springel01}
{Springel}, V., {White}, S.~D.~M., {Tormen}, G., and {Kauffmann}, G. (2001a).
\newblock {Populating a cluster of galaxies - I. Results at z=0}.
\newblock {\em \mnras}, 328:726--750.

\bibitem[{Springel} et~al., 2001b]{gadget01}
{Springel}, V., {Yoshida}, N., and {White}, S.~D.~M. (2001b).
\newblock {GADGET: a code for collisionless and gasdynamical cosmological
  simulations}.
\newblock {\em New Astronomy}, 6:79--117.

\bibitem[{Steidel} et~al., 1999]{ste99}
{Steidel}, C.~C., {Adelberger}, K.~L., {Giavalisco}, M., {Dickinson}, M., and
  {Pettini}, M. (1999).
\newblock {Lyman-Break Galaxies at $z>4$ and the Evolution of the Ultraviolet
  Luminosity Density at High Redshift}.
\newblock {\em \apj}, 519:1--17.

\bibitem[{Steidel} et~al., 1996a]{st96b}
{Steidel}, C.~C., {Giavalisco}, M., {Dickinson}, M., and {Adelberger}, K.~L.
  (1996a).
\newblock {Spectroscopy of Lyman Break Galaxies in the Hubble Deep Field}.
\newblock {\em \aj}, 112:352.

\bibitem[{Steidel} et~al., 1996b]{st96a}
{Steidel}, C.~C., {Giavalisco}, M., {Pettini}, M., {Dickinson}, M., and
  {Adelberger}, K.~L. (1996b).
\newblock {Spectroscopic Confirmation of a Population of Normal Star-forming
  Galaxies at Redshifts $z>3$}.
\newblock {\em \apjl}, 462:L17.

\bibitem[{Subramanian} et~al., 2000]{SCO00}
{Subramanian}, K., {Cen}, R., and {Ostriker}, J.~P. (2000).
\newblock {The Structure of Dark Matter Halos in Hierarchical Clustering
  Theories}.
\newblock {\em \apj}, 538:528--542.

\bibitem[{Suto} et~al., 1998]{Suto98}
{Suto}, Y., {Sasaki}, S., and {Makino}, N. (1998).
\newblock {Gas Density and X-Ray Surface Brightness Profiles of Clusters of
  Galaxies from Dark Matter Halo Potentials: Beyond the Isothermal beta-Model}.
\newblock {\em \apj}, 509:544--550.

\bibitem[{Swaters} et~al., 2002]{swaters}
{Swaters}, R.~A., {Madore}, B.~F., {van den Bosch}, F.~C., and {Balcells}, M.
  (2002).
\newblock {The Central Mass Distribution in Dwarf and Low Surface Brightness
  Galaxies}.
\newblock {\em \apj, in press {\tt(astro-ph/0210152)}}.

\bibitem[{Syer} and {White}, 1998]{syerWhite98}
{Syer}, D. and {White}, S.~D.~M. (1998).
\newblock {Dark halo mergers and the formation of a universal profile}.
\newblock {\em \mnras}, 293:337.

\bibitem[{Taylor} and {Navarro}, 2001]{TN01}
{Taylor}, J.~E. and {Navarro}, J.~F. (2001).
\newblock {The Phase-Space Density Profiles of Cold Dark Matter Halos}.
\newblock {\em \apj}, 563:483--488.

\bibitem[{Tegmark}, 1996]{Tegmark96}
{Tegmark}, M. (1996).
\newblock {The Angular Power Spectrum of the Four-Year COBE Data}.
\newblock {\em \apjl}, 464:L35.

\bibitem[{Teyssier}, 2002]{ramses02}
{Teyssier}, R. (2002).
\newblock {Cosmological hydrodynamics with adaptive mesh refinement. A new high
  resolution code called RAMSES}.
\newblock {\em \aap}, 385:337--364.

\bibitem[{Thacker} and {Couchman}, 2001]{ThackerCouchman01}
{Thacker}, R.~J. and {Couchman}, H.~M.~P. (2001).
\newblock {Star Formation, Supernova Feedback, and the Angular Momentum Problem
  in Numerical Cold Dark Matter Cosmogony: Halfway There?}
\newblock {\em \apjl}, 555:L17--L20.

\bibitem[{Tissera}, 2000]{Tissera00}
{Tissera}, P.~B. (2000).
\newblock {Analysis of Star Formation in Galaxy-like Objects}.
\newblock {\em \apj}, 534:636--649.

\bibitem[{Tissera} and {Dominguez-Tenreiro}, 1998]{TisseraRosa98}
{Tissera}, P.~B. and {Dominguez-Tenreiro}, R. (1998).
\newblock {Dark matter halo structure in CDM hydrodynamical simulations}.
\newblock {\em \mnras}, 297:177--194.

\bibitem[{Tozzi} and {Norman}, 2001]{TozziNorman01}
{Tozzi}, P. and {Norman}, C. (2001).
\newblock {The Evolution of X-Ray Clusters and the Entropy of the Intracluster
  Medium}.
\newblock {\em \apj}, 546:63--84.

\bibitem[{Tresse} and {Maddox}, 1998]{tm98}
{Tresse}, L. and {Maddox}, S.~J. (1998).
\newblock {The H alpha Luminosity Function and Star Formation Rate at Z
  approximately 0.2}.
\newblock {\em \apj}, 495:691.

\bibitem[{Treyer} et~al., 1998]{tre98}
{Treyer}, M.~A., {Ellis}, R.~S., {Milliard}, B., {Donas}, J., and {Bridges},
  T.~J. (1998).
\newblock {An ultraviolet-selected galaxy redshift survey: new estimates of the
  local star formation rate}.
\newblock {\em \mnras}, 300:303--314.

\bibitem[{Tyson} et~al., 1998]{Tyson98}
{Tyson}, J.~A., {Kochanski}, G.~P., and {dell'Antonio}, I.~P. (1998).
\newblock {Detailed Mass Map of CL 0024+1654 from Strong Lensing}.
\newblock {\em \apjl}, 498:L107.

\bibitem[{Tytler} et~al., 1996]{Tytler96}
{Tytler}, D., {Fan}, X.-M., and {Burles}, S. (1996).
\newblock {Cosmological baryon density derived from the deuterium abundance at
  redshift Z = 3.57.}
\newblock {\em \nat}, 381:207--209.

\bibitem[{Valageas} and {Silk}, 1999]{vs99}
{Valageas}, P. and {Silk}, J. (1999).
\newblock {The entropy history of the universe}.
\newblock {\em \aap}, 350:725--742.

\bibitem[{Valluri} and {Jog}, 1990]{ValluriJog90}
{Valluri}, M. and {Jog}, C.~J. (1990).
\newblock {Collisional removal of H I from the inner disks of Virgo Cluster
  galaxies}.
\newblock {\em \apj}, 357:367--372.

\bibitem[{van den Bosch} et~al., 2002]{bosch}
{van den Bosch}, F.~C., {Abel}, T., {Croft}, R.~A.~C., {Hernquist}, L., and
  {White}, S.~D.~M. (2002).
\newblock {The Angular Momentum of Gas in Proto-Galaxies I. Implications for
  the Formation of Disk Galaxies}.
\newblock {\em \tt(astro-ph/0201095)}.

\bibitem[{van den Bosch} and {Swaters}, 2001]{BoschSwaters01}
{van den Bosch}, F.~C. and {Swaters}, R.~A. (2001).
\newblock {Dwarf galaxy rotation curves and the core problem of dark matter
  haloes}.
\newblock {\em \mnras}, 325:1017--1038.

\bibitem[{van Haarlem} et~al., 1997]{Haarlem97}
{van Haarlem}, M.~P., {Frenk}, C.~S., and {White}, S.~D.~M. (1997).
\newblock {Projection effects in cluster catalogues}.
\newblock {\em \mnras}, 287:817--832.

\bibitem[{van Leer}, 1977]{...vanLeer79}
{van Leer}, B. (1977).
\newblock {Towards the ultimate conservative difference scheme: IV. A new
  approach to numerical convection}.
\newblock {\em Journal of Computational Physics}, 23:276--299.

\bibitem[{Weinberg}, 1972]{Weinberg72}
{Weinberg}, S. (1972).
\newblock {\em {Gravitation and cosmology: Principles and applications of the
  general theory of relativity}}.
\newblock New York: Wiley, |c1972.

\bibitem[{Weisskopf} et~al., 2000]{Weisskopf00}
{Weisskopf}, M.~C., {Tananbaum}, H.~D., {Van Speybroeck}, L.~P., and {O'Dell},
  S.~L. (2000).
\newblock {Chandra X-ray Observatory (CXO): overview}.
\newblock In {\em Proc. SPIE Vol. 4012, p. 2-16, X-Ray Optics, Instruments, and
  Missions III, Joachim E. Truemper; Bernd Aschenbach; Eds.}, volume 4012,
  pages 2--16.

\bibitem[{West} et~al., 1995]{West95}
{West}, M.~J., {Jones}, C., and {Forman}, W. (1995).
\newblock {Substructure: Clues to the Formation of Clusters of Galaxies}.
\newblock {\em \apjl}, 451:L5.

\bibitem[{White}, 2000]{White00}
{White}, D.~A. (2000).
\newblock {Deconvolution of ASCA X-ray data - II. Radial temperature and
  metallicity profiles for 106 galaxy clusters}.
\newblock {\em \mnras}, 312:663--688.

\bibitem[{White}, 1976]{White76}
{White}, S.~D.~M. (1976).
\newblock {The dynamics of rich clusters of galaxies}.
\newblock {\em \mnras}, 177:717--733.

\bibitem[{White}, 1984]{White84}
{White}, S.~D.~M. (1984).
\newblock {Angular momentum growth in protogalaxies}.
\newblock {\em \apj}, 286:38--41.

\bibitem[{White} et~al., 1987]{White87}
{White}, S.~D.~M., {Davis}, M., {Efstathiou}, G., and {Frenk}, C.~S. (1987).
\newblock {Galaxy distribution in a cold dark matter universe}.
\newblock {\em \nat}, 330:451--453.

\bibitem[{White} and {Frenk}, 1991]{WhiteFrenk91}
{White}, S.~D.~M. and {Frenk}, C.~S. (1991).
\newblock {Galaxy formation through hierarchical clustering}.
\newblock {\em \apj}, 379:52--79.

\bibitem[{White} et~al., 1983]{White83}
{White}, S.~D.~M., {Frenk}, C.~S., and {Davis}, M. (1983).
\newblock {Clustering in a neutrino-dominated universe}.
\newblock {\em \apjl}, 274:L1--L5.

\bibitem[{White} et~al., 1993]{White93}
{White}, S.~D.~M., {Navarro}, J.~F., {Evrard}, A.~E., and {Frenk}, C.~S.
  (1993).
\newblock {The Baryon Content of Galaxy Clusters - a Challenge to Cosmological
  Orthodoxy}.
\newblock {\em \nat}, 366:429.

\bibitem[{White} and {Rees}, 1978]{WhiteRees78}
{White}, S.~D.~M. and {Rees}, M.~J. (1978).
\newblock {Core condensation in heavy halos - A two-stage theory for galaxy
  formation and clustering}.
\newblock {\em \mnras}, 183:341--358.

\bibitem[{White} and {Zaritsky}, 1992]{WhiteZaritsky92}
{White}, S.~D.~M. and {Zaritsky}, D. (1992).
\newblock {Models for Galaxy halos in an open universe}.
\newblock {\em \apj}, 394:1--6.

\bibitem[{Xu} et~al., 2001]{Xu01}
{Xu}, H., {Jin}, G., and {Wu}, X. (2001).
\newblock {The Mass-Temperature Relation of 22 Nearby Clusters}.
\newblock {\em \apj}, 553:78--83.

\bibitem[{Xue} and {Wu}, 2000]{XueWu00}
{Xue}, Y. and {Wu}, X. (2000).
\newblock {The $L_{X}-T$, $L_{X}$-{$\sigma$}, and {$\sigma$}-T Relations for
  Groups and Clusters of Galaxies}.
\newblock {\em \apj}, 538:65--71.

\bibitem[{Yan} et~al., 1999]{yan99}
{Yan}, L., {McCarthy}, P.~J., {Freudling}, W., {Teplitz}, H.~I., {Malumuth},
  E.~M., {Weymann}, R.~J., and {Malkan}, M.~A. (1999).
\newblock {The Halpha Luminosity Function and Global Star Formation Rate from
  Redshifts of 1-2}.
\newblock {\em \apjl}, 519:L47--L50.

\bibitem[{Yepes}, 2001]{Yepes01}
{Yepes}, G. (2001).
\newblock {Cosmological numerical simulations: past, present and future}.
\newblock In {\em The Restless Universe}, page 217.

\bibitem[{Yepes} et~al., 1997]{YK3}
{Yepes}, G., {Kates}, R., {Khokhlov}, A., and {Klypin}, A. (1997).
\newblock {Hydrodynamical simulations of galaxy formation: effects of supernova
  feedback}.
\newblock {\em \mnras}, 284:235--256.

\bibitem[{Yoshida} et~al., 2000]{Yoshida00}
{Yoshida}, N., {Springel}, V., {White}, S.~D.~M., and {Tormen}, G. (2000).
\newblock {Weakly Self-interacting Dark Matter and the Structure of Dark
  Halos}.
\newblock {\em \apjl}, 544:L87--L90.

\bibitem[{Yoshikawa} et~al., 2000]{Yoshikawa00}
{Yoshikawa}, K., {Jing}, Y.~P., and {Suto}, Y. (2000).
\newblock {Cosmological Smoothed Particle Hydrodynamic Simulations with Four
  Million Particles: Statistical Properties of X-Ray Clusters in a Low-Density
  Universe}.
\newblock {\em \apj}, 535:593--601.

\bibitem[{Zabludoff} and {Mulchaey}, 1998]{ZabludoffMulchaey98}
{Zabludoff}, A.~I. and {Mulchaey}, J.~S. (1998).
\newblock {The Properties of Poor Groups of Galaxies. I. Spectroscopic Survey
  and Results}.
\newblock {\em \apj}, 496:39.

\bibitem[{Zabludoff} et~al., 1996]{Zabludoff96}
{Zabludoff}, A.~I., {Zaritsky}, D., {Lin}, H., {Tucker}, D., {Hashimoto}, Y.,
  {Shectman}, S.~A., {Oemler}, A., and {Kirshner}, R.~P. (1996).
\newblock {The Environment of ``E+A'' Galaxies}.
\newblock {\em \apj}, 466:104.

\bibitem[{Zaroubi} and {Hoffman}, 1993]{ZH93}
{Zaroubi}, S. and {Hoffman}, Y. (1993).
\newblock {Gravitational Collapse in an Expanding Universe: Asymptotic
  Self-similar Solutions}.
\newblock {\em \apj}, 416:410.

\bibitem[{Zaroubi} et~al., 1996]{Zaroubi96}
{Zaroubi}, S., {Naim}, A., and {Hoffman}, Y. (1996).
\newblock {Secondary Infall: Theory versus Simulations}.
\newblock {\em \apj}, 457:50.

\bibitem[{Zwicky}, 1937]{zwicky37}
{Zwicky}, F. (1937).
\newblock {On the Masses of Nebulae and of Clusters of Nebulae}.
\newblock {\em \apj}, 86:217--246.

\end{thebibliography}

% Incluir en tesis.bbl
%-------------------------
% \addcontentsline{toc}{chapter}{\ ~~ Bibliography}
% 
% \bibitem[{Jethro Tull}, 1995]{.}
% \begin{quote}{\em\ \\
% Words get written. Words get twisted.\\
% Old meanings move in the drift of time.\\
% Lift the flickering torches. See gentle shadows change\\
% the features of the faces cut in unmoving stone.}
%   
% -- Jethro Tull : {\em Roots to Branches} (1995) --\\
% \end{quote}
%-------------------------

\newpage
\pagestyle{empty}
\ 

\vfill
\tiny Y sobre todo, a Pilar y a Petri.
\end{document}